%% file: 3DQLC.tex
\documentclass[aps,prb,superscriptaddress,twocolumn,tightenlines,floatfix,amsmath,amssymb]{revtex4}
\pdfoutput=1

\usepackage{graphicx}
\usepackage[caption=false]{subfig}
\usepackage{color}
\graphicspath{{./img/}{./mathematicaimg/}{./}}

\usepackage{hyperref}

\usepackage{multirow}
\usepackage{bigdelim}

\newcommand{\td}{\mathrm{d}}
\newcommand{\te}{\mathrm{e}}
\newcommand{\ti}{\mathrm{i}}
\newcommand{\ts}{\mathrm{s}}
\newcommand{\ft}{\mathfrak{t}}

\newcommand{\tL}{\mathrm{L}}
\newcommand{\tT}{\mathrm{T}}

\newcommand{\tR}{\mathrm{R}}
\newcommand{\tS}{\mathrm{S}}

\begin{document}

\title {Dual gauge field theory of quantum liquid crystals in three dimensions}

\author{Aron J. Beekman}
\email{aron@phys-h.keio.ac.jp}
\affiliation{Department of Physics, and Research and Education Center for Natural Sciences,
Keio University, Hiyoshi 4-1-1, Yokohama, Kanagawa 223-8521, Japan}
\author{Jaakko Nissinen}
\affiliation{Institute-Lorentz for Theoretical Physics, Leiden University, PO Box 9506, NL-2300 RA Leiden, The Netherlands}
\affiliation{Low Temperature Laboratory, Aalto University, P.O. Box 15100, FI-00076 Aalto, Finland}
\author{Kai Wu}
\affiliation{Stanford Institute for Materials and Energy Sciences, SLAC National Accelerator Laboratory and Stanford University, Menlo Park, CA 94025, USA}
\author{Jan Zaanen}
\affiliation{Institute-Lorentz for Theoretical Physics, Leiden University, PO Box 9506, NL-2300 RA Leiden, The Netherlands}

\date {\today}
\begin{abstract}
The dislocation-mediated quantum melting of solids into quantum liquid crystals is extended from two to three spatial dimensions, using a generalization of boson--vortex or Abelian-Higgs duality. Dislocations are now Burgers-vector-valued strings that trace out worldsheets in spacetime while the phonons of the solid dualize into two-form (Kalb--Ramond) gauge fields. We propose an effective dual Higgs potential that allows for restoring translational symmetry in either one, two or three directions, leading to the quantum analogues of columnar, smectic or nematic liquid crystals. In these phases, transverse phonons turn into gapped, propagating modes while compressional stress remains massless. Rotational Goldstone modes emerge whenever translational symmetry is restored. We also consider electrically charged matter, and find amongst others that as a hard principle only two out of the possible three rotational Goldstone modes are observable using electromagnetic means.
\end{abstract}

\maketitle

\tableofcontents

\section{Introduction}\label{sec:Introduction}
\input{sec_introduction.tex}

\section{Symmetry principles of quantum liquid crystals}\label{sec:Symmetry principles of quantum liquid crystals}
\input{sec_symmetry.tex}

\section{Preliminaries}\label{sec:Preliminaries}
\input{sec_preliminaries.tex}

\section{Dual elasticity in three dimensions}\label{sec:Dual elasticity in three dimensions}
\input{sec_dualelasticity.tex}

\section{Dislocation-mediated quantum melting}\label{sec:Dislocation-mediated quantum melting}
\input{sec_quantummelting.tex}

\section{Nematic phases}\label{sec:nematic}
\input{sec_nematic.tex}

\section{Smectic phases}\label{sec:smectic}
\input{sec_smectic.tex}

\section{Columnar phases}\label{sec:columnar}
\input{sec_columnar.tex}

\section{Charged quantum liquid crystals}\label{sec:Charged quantum liquid crystals}
\input{sec_charged.tex}

\section{Conclusions}\label{sec:Conclusions}
\input{sec_conclusions.tex}

\begin{acknowledgments}
 A.J.B. is supported by the MEXT-Supported Program for the Strategic Research Foundation at Private Universities ``Topological Science'' (Grant No. S1511006). This work was supported by the Netherlands foundation for Fundamental Research of Matter (FOM). 
\end{acknowledgments}

\appendix
\section{Fourier space coordinate systems}\label{sec:Fourier space coordinate systems}
\input{sec_fouriercoordinates.tex}

\bibliography{qlc3d_references}

\end{document}

%% file: sec_introduction.tex
\subsection{Quantum liquid crystals: the context}

Liquid crystals are ``mesophases'' of matter with a ``vestigial'' pattern of spontaneous symmetry breaking arising at intermediate temperatures or coupling: rotational symmetry is broken while translational invariance partially or completely persists. Classical liquid crystals are formed from highly anisotropic molecular constituents which, upon cooling from the liquid phase, can order their respective orientations while maintaining translational freedom. Only at lower temperatures crystallization sets in. These forms of matter have been known for about a century, and their  theoretical description was established by De Gennes and many others~\cite{DeGennesProst95, SinghDunmur02, ChaikinLubensky00}. Starting from the opposite side, it was long realized that dislocations (the topological defects associated with translational order) are responsible for material degradation and even melting of solids~\cite{Friedel64}. Berezinskii, Kosterlitz and Thouless (BKT) in their landmark papers already suggested that unbinding of dislocations and disclinations (rotational topological defects) will lead to the disordering of two-dimensional solids~\cite{KosterlitzThouless72,KosterlitzThouless73}, he theory of which was further developed and refined by Nelson, Halperin and Young~\cite{HalperinNelson78,NelsonHalperin79,Young79}. We will refer to the topological melting driven by dislocation unbinding  as the {\em KTNHY transition}. Here it was also predicted that an intermediate phase exists as a result of the exclusive proliferation of dislocations in a triangular 2D crystal, dubbed  the {\em hexatic liquid crystal}.  Translational symmetry is fully restored but the rotational symmetry remains  broken down to the $C_6$ point group characterizing the triangular crystal. 

Almost two decades later, Kivelson, Fradkin and Emery~\cite{KivelsonFradkinEmery98} proposed that the spatial ordering of electrons in strongly-correlated electron systems, as realized in 
underdoped high-$T_\mathrm{c}$ superconductors, could feature symmetry properties analogous to classical liquid crystals. The  stripe `crystalline' order is now destroyed at zero temperature by quantum 
fluctuations in the form of proliferating dislocations, such that on macroscopic length scales the system forms a nematic quantum fluid (superconductor) which maintains however the orientational preference of the stripe electronic crystal. 
This signaled the birth of the subject of quantum liquid crystals.  Quite some empirical support was found since then supporting the existence of such forms of quantum liquid crystals. This includes direct evidences for the existence 
of quantum nematic order  in underdoped cuprates, likely related 
to the original context of fluctuating stripes~\cite{AndoEtAl02, Vojta09, OganesyanKivelsonFradkin01, BorziEtAl07,HinkovEtAl08, FradkinEtAl10,FradkinKivelson10,Fradkin12}. This theme flourished in the context of the iron superconductors where quite some evidence surfaced for the prominent role of orientational symmetry breaking driven by the 
electron system as being central to their physics~\cite{ChuangEtAl10,ChuEtAl12,FernandesChubukovSchmalian14}. An ambiguity in these condensed matter systems is that the crystal formed by the atoms is already breaking space translations and rotations while the electron and ion systems are coupled. 
The quasi two-dimensional electron systems in the iron and copper superconductors are typically realized in tetragonal square lattices where the rotational symmetry is broken to a point group characterized by a fourfold axis. This fourfold symmetry is broken to an orthorhombic crystal structure characterized by a two-fold rotational symmetry $C_4 \to C_2$, dubbed the ``Ising nematic phase''. Given that symmetry-wise the purely electronic and crystalline tendencies to lower the point group symmetry cannot be distinguished, one does face a degree of ambiguity that cannot be avoided, giving rise to ongoing debates about the origin of the electronic nematicity in these materials~\cite{FernandesChubukovSchmalian14}.
 
Inspired by the initial suggestion by Kivelson {\em et al.} one of the authors (J.Z.) initiated a program to extend the KTNHY topological melting ideas to the quantum realms, initially in two space dimensions.  
The emphasis has been here all along on the fundamental, theoretical side based on the symmetries and associated defects. The main restriction is that it only deals with matter formed from bosons: the constructions rest on the machinery of statistical 
physics being mobilized in the $D+1$ dimensional Euclidean spacetime, turning into the quantum physics of bosons  after Wick rotation.  This matter lives in the Galilean continuum and the point of departure
is the spontaneous breaking of space translations and rotations into a crystal. The KTNHY transition is just one particular example of a Kramers--Wannier  (or weak--strong) duality and it was found out in the 1980s how to extend this to 
three dimensions when dealing with Abelian symmetries. In the context of crystalline elasticity one can rest on strain--stress duality, where phonon degrees of freedom are mapped to {\em dual stress gauge fields}. 
This amounts to a generalization of the famous vortex--boson or Abelian-Higgs duality, as pioneered by Kleinert~\cite{Kleinert89b}. Using the well-known mapping of a $D$-dimensional quantum to a $D+1$-dimensional classical system, the 2+1D quantum liquid crystals were investigated by stress-strain duality starting with Ref.~\onlinecite{ZaanenNussinovMukhin04}. The relation is essentially the same as the KTNHY case in two 
dimensions. One first establishes the structure of the weak--strong duality by focusing on the minimal $U(1)$-case associated with vortex melting, to then extend it to the richer theater of the space groups 
underlying the crystalline symmetry breaking, profiting from the fact that the restoration of translational invariance by dislocations is associated with an Abelian symmetry.   

The essence is that this duality language is geared to describe the physics of a quantum fluid (in fact, a superfluid or superconductor) that is in the limit of {\em maximal} correlation, being as close to the solid as possible. Only the collective excitations are important here. It is assumed at the onset that the particles forming the crystal continue to be bound: the `construction material' of the quantum liquid consists of local crystalline order supporting phonons disrupted by a low density 
of topological defects: the dislocations. At length scales smaller than the distance between the dislocations the liquid behaves still like the solid. However, at larger distances the translational symmetry is restored 
by a condensate formed out of the quantized dislocations. To a certain degree the liquid crystal aspect is a convenience.
The {\em Bose condensate of dislocations} restoring the translational symmetry is straightforwardly described in terms of a ``dual stress superconductor''. The rotational topological defects, disclinations, that restore the rotational symmetry, are just harder to deal with technically and by ``keeping disclinations out of the vacuum'' rotational symmetry continues to be broken, describing the quantum liquid crystal. 
The isotropic quantum fluid is realized when these disclinations proliferate as well~\cite{Kleinert83}.

This program resulted in a series of papers that gradually exposed the quite extraordinary physics of such maximally correlated quantum liquid 
crystals in 2+1 dimensions~\cite{ZaanenNussinovMukhin04,KleinertZaanen04,Cvetkovic06,CvetkovicNussinovZaanen06,CvetkovicZaanen06a,CvetkovicZaanen06b,CvetkovicNussinovMukhinZaanen08,ZaanenBeekman12,BeekmanWuCvetkovicZaanen13,LiuEtAl15}. 

Recently, we have written an extensive review that comprehensively details the dual gauge field theory of these quantum liquid crystals in two dimensions~\cite{QLC2D}, to which we shall hereafter refer as QLC2D. 
The present work is the extension of this theory to three spatial dimensions and we recommend the novice to the subject to have a close look  at QLC2D first. We will often refer back to those results, while we do not hesitate to skip 
derivations and explanations provided there when these are representative for the way things work in 3+1D as well. We also refer the reader to the introduction of QLC2D for more background on the history of and the physical interest in quantum liquid crystals.

\subsection{From two to three dimensions: weak--strong duality and the string condensate}\label{subsec:From two to three dimensions}

Our universe has three spatial dimensions and therefore the most natural quantum states of matter are formed in 3+1 dimensions. The generalization of the theory to 3+1D has been quite an ordeal ---
we are even not completely confident that the solution we present here is really watertight.  Wherein lies the difficulty? 
This is rooted in the fundamentals of Abelian  weak--strong dualities which are very well understood in both 1+1/2D (KT topological melting) and 2+1D/3D (Abelian-Higgs duality~\cite{FisherLee89,Kleinert89a,KiometzisKleinertSchakel95,HerbutTessanovic96,CvetkovicZaanen06a,NguyenSudbo99,HoveSudbo00,HoveMoSudbo00,SmisethSmorgravSudbo04,SmisethEtAl05,SmorgravEtAl05}) while it is much less settled in 3+1D for
quite deep reasons.  At the heart of these dualities is the notion that given a particular form of spontaneous symmetry breaking, the unique agents associated with restoring the symmetry are the topological excitations.

Let us first consider a broken global $U(1)$-symmetry, where the vortex is the topological workhorse. 
In the zero-temperature ordered phase these only occur in the form of bound vortex--antivortex pairs since a single 
free vortex suffices to destroy long-range order. In 1+1D they are point-like entities (instantons) in spacetime having a logarithmic interaction, subjected to the famous BKT vortex-unbinding transition. In 
2+1D vortices are `particle-like' objects characterized by worldlines forming closed loops in spacetime in the ordered phase. At the quantum phase transition these loops `blow out', forming a {\em tangle of worldlines} corresponding 
to a Bose condensate of vortices. In the ordered phases vortices are subjected to long-range interactions which work in exactly the same way in this particular dimension as electromagnetic interactions, namely by coupling to vector gauge fields. In the disordered phase, this gauged vortex condensate is therefore a {\em dual superconductor} (Higgs phase). In the context of quantum elasticity, the dislocations take the role of vortices forming the {\em dual stress superconductor}.  There is however much more additional structure and the outcome is the rich world described in QLC2D. 

The complication coming in at 3+1D is that dislocations (or vortices) are `line-like', forming loops in space that trace out worldsheets, not worldlines, in spacetime. In other words, they are {\em strings}. In 2+1D we are dealing with an ordinary Bose condensate of particles, constructed using the second-quantization procedure. Second quantization is however not applicable to strings in 3+1D and a fool-proof procedure to write down the effective field theory associated with the `foam' formed in spacetime from proliferated dislocation strings is just not available. Here we have to rely on a guess based on symmetry considerations that was first proposed by Rey~\cite{Rey89} in the context of fundamental string field theory. Let us present here a crude sketch of the essence of this affair in the minimal setting of the
Abelian-Higgs/vortex duality associated with the topological melting of the superfluid. 

The point of departure is the relativistic Josephson action $\mathcal{L} \sim (\partial_{\mu} \varphi)^2$ describing the phase mode of the superfluid $\varphi$ in imaginary time. 
The $U(1)$-field is compact and vortices arise as the topological excitations. The elementary dualization in 2+1D maps the phase mode $\varphi$ onto a vector gauge field $a_\mu$ and the vortex onto a particle current $J^\mathrm{V}_\mu$, while the action is recast as $f_{\mu \nu} f_{\mu \nu} + a_{\mu} J^\mathrm{V}_\mu$. This describes the worldlines of isolated vortices in terms of the vortex current $J^\mathrm{V}_\mu$, being subjected to a long-range interaction mediated by an effective $U(1)$-gauge field $a_{\mu}$ with field strength $f_{\mu \nu} = \partial_mu a_\nu - \partial_\nu a_\mu$. This is identical to electrodynamics in this particular dimension; one may interpret the superfluid as the Coulomb phase of an electromagnetic system sourced by conventional currents $J^\mathrm{V}_\mu$. The gauge fields $a_{\mu}$ arise as a way to impose the conservation of the supercurrent (field strength): $j_{\mu} = \epsilon_{\mu\kappa\lambda} f_{\kappa\lambda}$ is conserved $\partial_{\mu} j_{\mu} = 0$ when the original phase field $\psi$ is smooth. This continuity equation can be identically imposed by parameterizing the currents in terms of the gauge fields as $j_{\mu} = \epsilon_{\mu \nu \lambda} \partial_{\nu} a_{\lambda}$, and $a_\mu$ is directly sourced by the vortex currents $J^\mathrm{V}_\mu$.

The duality is easily extended in this ordered, Coulomb phase to 3+1D. The only difference is that one has to invoke {\em two-form} gauge fields $b_{\mu \nu}$. Namely, the supercurrent continuity equation $\partial_\mu j_\mu = 0$ is imposed by expressing it as the `four-curl' of a two-form field:  $j_{\mu} = \epsilon_{\mu \nu \kappa \lambda} \partial_{\nu} b_{\kappa \lambda}$. At the same time, the vortex is a worldsheet in spacetime, parametrized by $J^\mathrm{V}_{\mu\nu}$. The action for an isolated piece of vortex world sheet has the form $\mathcal{L} \sim  h_{\mu \nu \kappa}h_{\mu \nu \kappa} + b_{\mu \nu} J^\mathrm{V}_{\mu \nu}$, where $h_{\mu \nu \kappa} = \epsilon_{\mu\nu\kappa\lambda} j_\lambda$ is the field strength associated 
with the gauge field $b_{\mu \nu}$. This is well known in string theory where such two-form fields arise naturally and are known as Kalb--Ramond fields~\cite{KalbRamond74}.

This dual description of the ordered phase is only the beginning of the story. We have just summarized the dual version of the interaction between isolated vortices deep in the ordered, superfluid phase. Towards the disordering quantum phase transition, in 2+1D vortex worldline loops grow and proliferate (vortices condense). This disordered state is relativistic {\em superconductor} (Higgs phase) formed out of vortex matter. Namely, the dual gauge fields $a_\mu$ couple minimally to a complex scalar field $\Phi = | \Phi | \te^{\ti \phi}$, representing the second-quantized collective vortex condensate degrees of freedom. In the London limit where the amplitude $|\Phi|$ is frozen, this leads to the Ginzburg--Landau form $\mathcal{L} \sim |\Phi|^2 (\partial_\mu \phi - a_{\mu})^2 + f_{\mu \nu} f_{\mu \nu}$.

It is here that the great difficulty of the duality in 3+1D is found. The vortex strings of 3+1D proliferate (condense) into a `foam' of worldsheets in spacetime, and the question arises: what is the universal form of the effective action describing such a `string condensate'? This is a fundamental problem: the construction of string field theory. As a matter of fact, presently it is just not known how to generalize second quantization to stringy degrees of freedom. One can however rely on symmetry.  Deep in the dual superconductor, the minimal coupling principle appears to insist that there is only a single consistent way of writing a Josephson action. As Rey pointed out~\cite{Rey89}, see also Ref.~\onlinecite{Franz07}, the two-form gauge field $b_{\mu \nu}$ has to be Higgsed completely and this is  accomplished  by a Lagrangian of the form $ \mathcal{L} \sim |\Phi|^2( \partial_{\mu} \phi_{\nu} -   \partial_{\nu} \phi_{\mu} - b_{\mu \nu})^2$. One is now led to accept that the `string foam' is characterized by a vector-valued phase field $\phi_{\nu}$, having more degrees of freedom than the simple scalar $\phi$ in 2+1D.
 
 As we discussed elsewhere, problems of principle arise with this construction in the context of this disordered superfluid/dual superconductor in 3+1D~\cite{BeekmanSadriZaanen11}.  The dual superconductor can be interpreted as a boson-Mott insulator and it appears that the vectorial phase field  $\phi_{\nu}$ {\em overcounts} the number of degrees of freedom. The Anderson-Higgs mechanism transfers the condensate degrees of freedom to the longitudinal polarizations of the photon (dual gauge) field. The scalar field $\phi$ has one degree of freedom but the vectorial phase field $\phi_\nu$ contains two degrees of freedom that, together with the single Goldstone mode of the superfluid, end up forming a triplet of degenerate massive modes in the 3+1D disordered superfluid. Conversely, the boson-Mott insulator is known to possess two massive propagating modes, the ``doublon and holon'' excitations. We proposed a resolution to repair this overcounting~\cite{BeekmanSadriZaanen11,BeekmanZaanen12}.
 
 How does this play out in the current context of quantum liquid crystals? As we will see below, translational symmetry can be restored `one direction at a time', and the disorder field theory consists basically of three more-or-less independent $U(1)$-fields. These cause the shear degrees of freedom to be gapped, leading to the `liquid behavior' of liquid crystals. Furthermore, up to three rotational Goldstone modes emerge once translational symmetry is restored. 
 All these degrees of freedom are a priori accommodated in the ordinary, linear stress operators of elasticity---these are not the condensate phase degrees of freedom that are transferred by the Anderson--Higgs mechanism to the longitudinal polarizations of the dual gauge field. However, we benefit from the additional structure of elasticity, which contains not only linear stress, the canonical conjugate to displacements, but also torque stress, which is conjugate to local rotations. Torque stress cannot be unambiguously defined as long as shear rigidity is present, but it becomes a good physical quantity in the quantum liquid crystals. We find below that the condensate phase degrees of freedom do leave their mark on torque stresses. As we shall identify in Sec.~\ref{subsec:Torque stress in the quantum nematic} there are components, corresponding to the longitudinal two-form gauge fields, which are visible in the torque stress linear response. This is not only a clear sign that the problems outlined in Ref.~\onlinecite{BeekmanSadriZaanen11} do not arise, but also a great, and possibly first, way to test the existence of a condensate of the form proposed in Ref.~\onlinecite{Rey89} in condensed matter.

\subsection{Overview and summary of results}

As we just argued, assuming that we can rely on the minimal coupling construction for the `stringy' condensate of the dual stress superconductor, the theory of the quantum liquid crystals in 3+1D becomes a as-straightforward-as-possible
generalization of this physics in 2+1D. We have accordingly organized this paper closely following the 2+1D template~\cite{QLC2D}.  In the next three chapters we set the stage by reviewing general symmetry principles, and
generalities of elasticity theory as of relevance to the remainder. In the remaining sections we will then develop step-by-step the theory of the various forms of quantum liquid-crystalline order.

The main difference in three dimensions is the nature of rotational symmetry; its ramifications for the universal 
features associated with the order parameter theory will be reviewed in Sec.~\ref{sec:Symmetry principles of quantum liquid crystals}.
 For empirical reasons, nearly all nematic liquid crystals of the soft matter tradition are of a very special kind: the uniaxial nematics formed from the `rod-like molecules' that orient their long axis in the same direction. As we will briefly review in the next section, these are only a part of a very large class of {\em generalized nematics} characterized by the $O(3)$ rotational symmetry of isotropic three-dimensional space, broken 
down to some point group. In two dimensions all rotational proper point groups are Abelian while in 3D the point groups are generally non-Abelian. As a consequence the order parameter theory of these 3D generalized nematics is a very rich and complex affair~\cite{LiuEtAl16b,LiuEtAl16,NissinenEtAl16}. The uniaxial nematic has the point-group symmetry $D_{\infty\mathrm{h}}$, which breaks only two out of three rotational symmetries and the proper rotational part of which is Abelian; it is therefore not a good representative of rotational symmetry breaking in three dimensions.

In order to render the duality construction as simple and transparent as possible we depart from
a maximally symmetric  setting: the `isotropic nematic'. In 2+1D this is literally realized by the hexatic liquid crystal, where one departs from a triangular crystal characterized by isotropic elasticity as far as its long-distance properties are concerned, and this isotropic nature is carried over to the `quantum hexatic'.  In 3+1D there is no space group associated with isotropic elasticity. Instead one can consider a cubic crystal and assert that the cubic anisotropies can be approximately ignored: this is our point of departure. The $O_\mathrm{h}$ point group of the cubic crystal is however non-Abelian with far-reaching consequences for disclination defects. Nevertheless, as long as we are not interested in condensation of disclinations into the liquid (superfluid) phase, these complications can be ignored. The `isotropic quantum nematic' breaks three rotational symmetries and should carry three rotational Goldstone modes, which we shall verify explicitly with dual gauge fields. As we already discovered in QLC2D, smectic type phases have a particular elegant description in the duality setting in terms of a partial condensation of dislocations. As we will further elucidate in this section, in 3+1D this implies that both quantum smectic and columnar phases arise naturally. 

In Sec.~\ref{sec:Preliminaries} we review some basic material: the field theory of quantum elasticity, stress--strain duality, rotational elasticity and static topological defect lines in solids. Quantum elasticity is just the classical theory of elasticity with an added quantum kinetic energy in imaginary time, promoted to the path integral formulation of the quantum partition function. This is a linear theory of deformations that simply describes acoustic phonons. Usually elasticity theory is expressed in term of strain fields but by employing stress--strain duality it can be formulated as well in terms of stress tensors, which are in turn the field strengths in the dual-gauge-field-theoretical formulations in the remainder. The theory governing the low-energy excitations of a translationally symmetric but rotationally rigid medium can be called {\em rotational elasticity}, which is shortly reviewed. The topological defects, the agents destroying the crystalline order of the solid state, are dislocations and disclinations with Burgers resp. Frank vectors as topological charge.

\begin{figure*}
 \subfloat[crystal -- $\mathbb{Z}^3$]{\includegraphics[height=3cm]{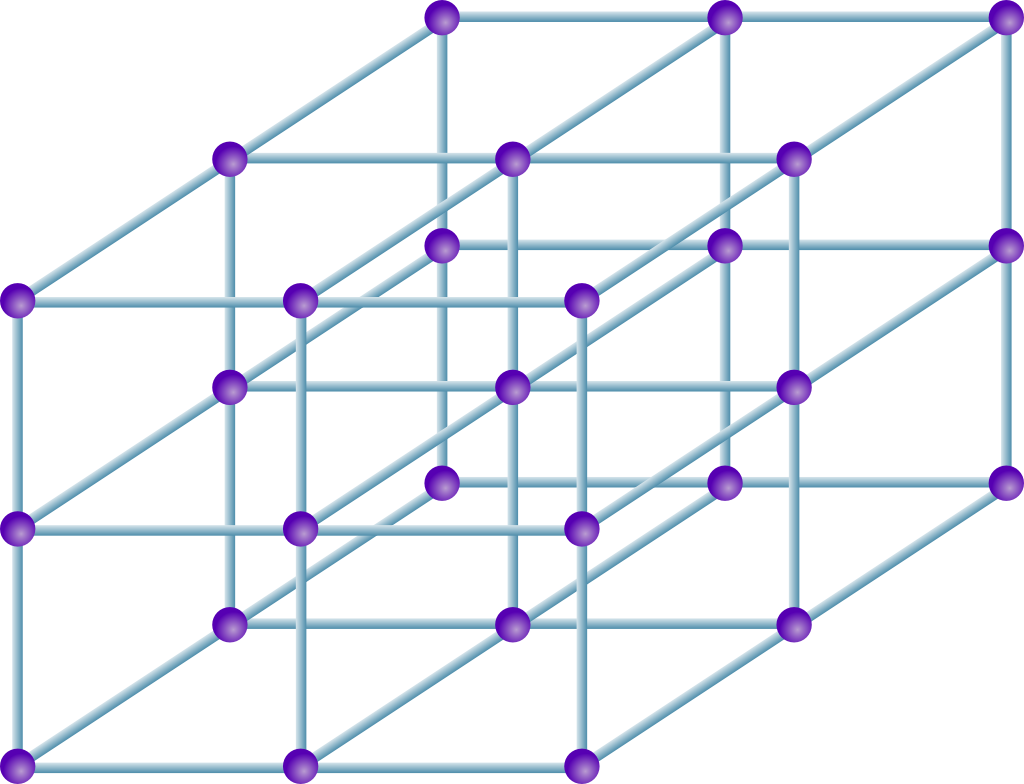}\label{subfig:crystal}}
 \hfill
 \subfloat[columnar -- $\mathbb{Z}^2 \times \mathbb{R}$]{\includegraphics[height=3cm]{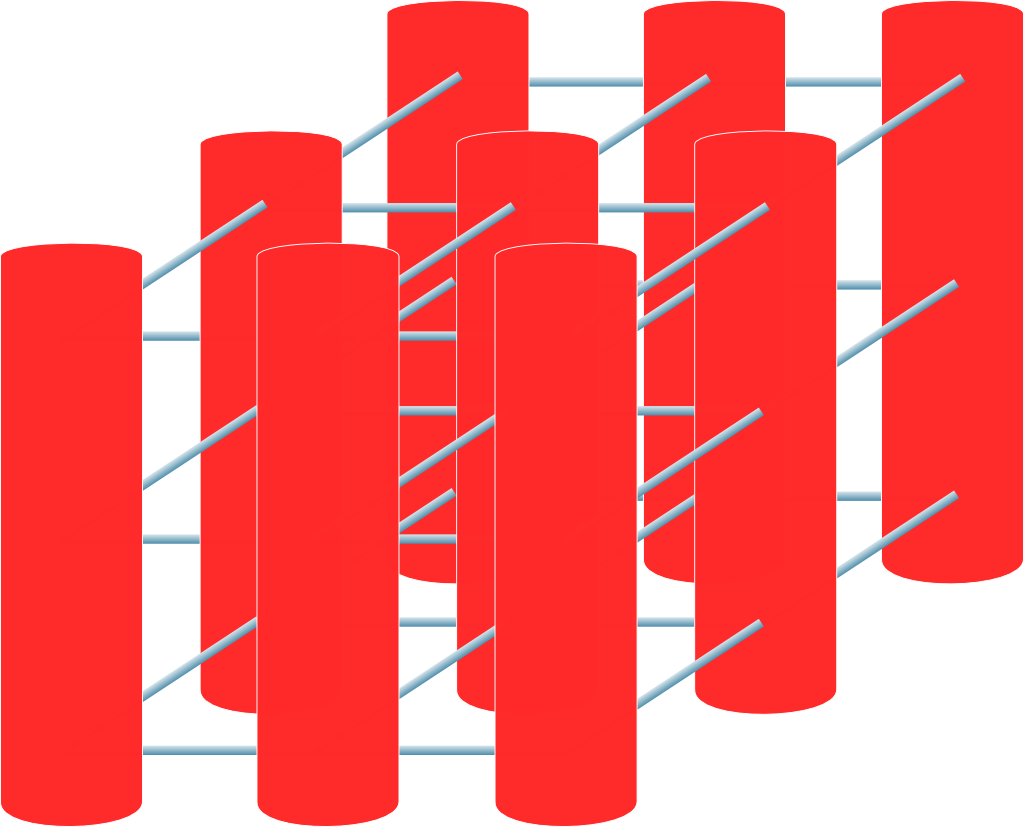}\label{subfig:columnar}}
 \hfill
 \subfloat[smectic -- $\mathbb{Z} \times \mathbb{R}^2$]{\includegraphics[height=3cm]{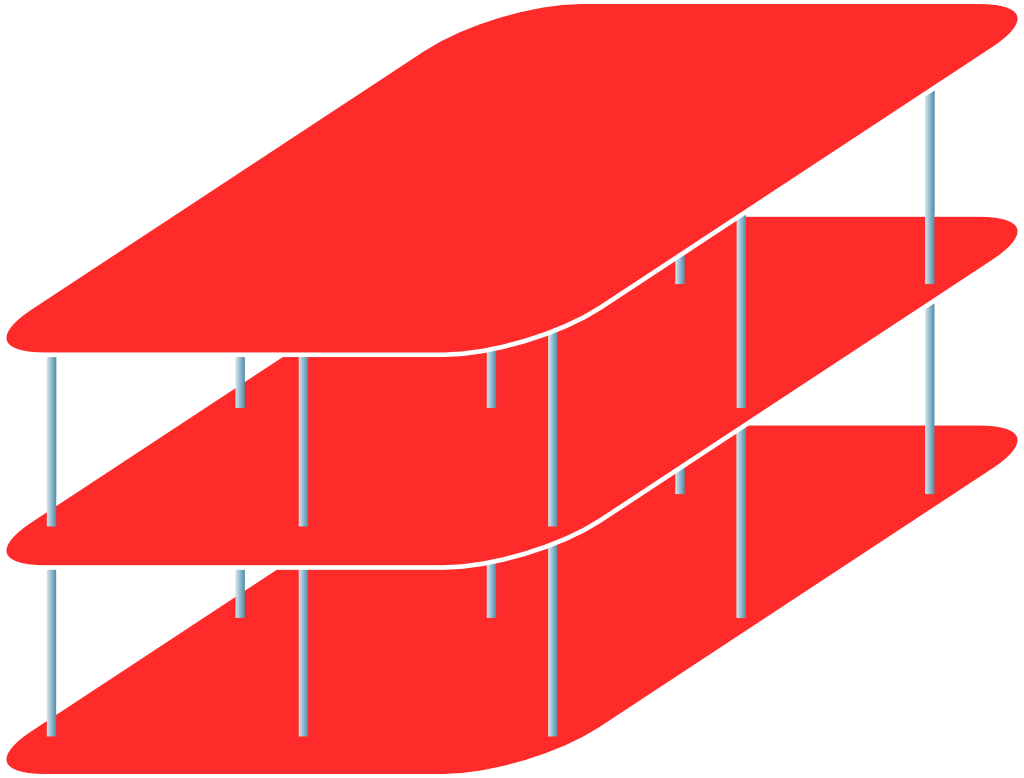}\label{subfig:smectic}}
 \hfill
 \subfloat[nematic -- $\mathbb{R}^3$]{\includegraphics[height=3cm]{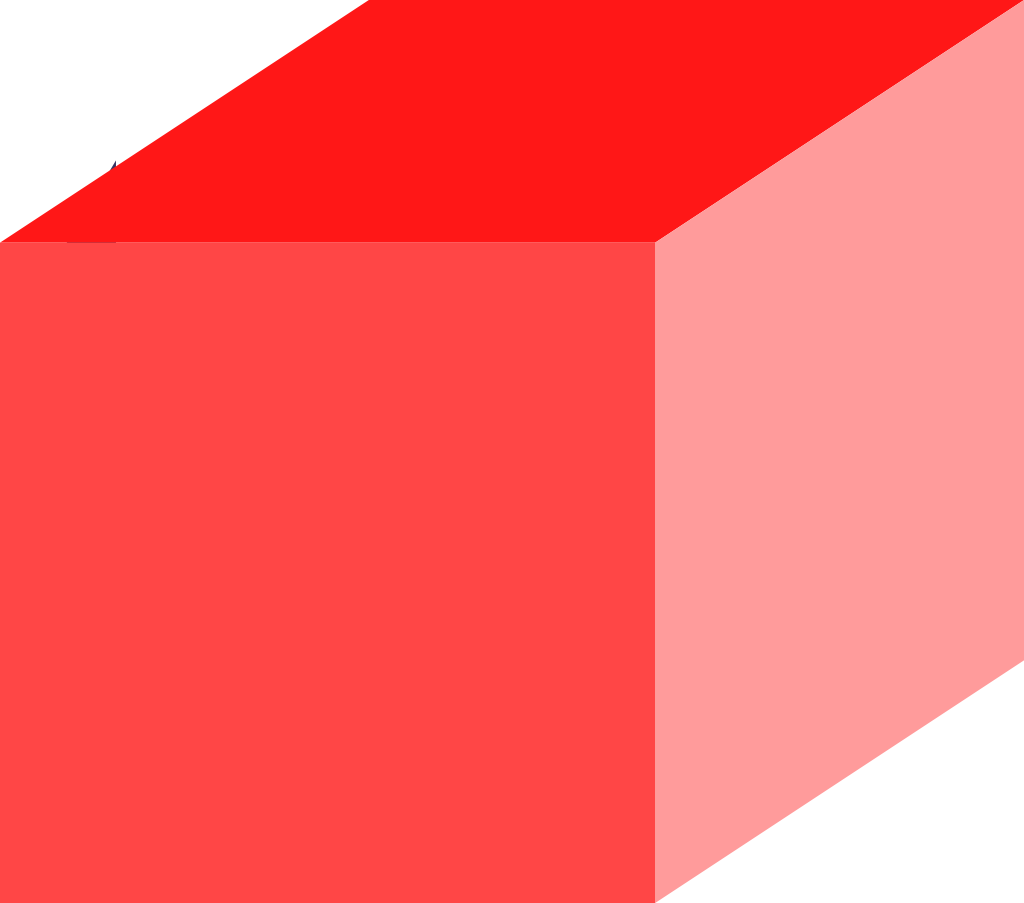}\label{subfig:nematic}}
 \caption{
 Sketch of the symmetry of the solid and liquid crystals. Features in red denote translational symmetry, liquid-like in that direction. The translation group is also indicated in the captions. Rotational symmetry is broken to a discrete point group $\bar{P}$ in all cases. \protect\subref{subfig:crystal} For simplicity we start with a cubic crystal with translational symmetry completely broken down to a discrete subgroup. \protect\subref{subfig:columnar} Restoring translations in one direction leads to a regular 2D array of 1D liquid lines, the columnar phase. In the superconductivity jargon, this is the stripe (``river of charge'') phase. \protect\subref{subfig:smectic} Restoring translations in two dimensions yields the smectic phase; the liquid plane has the features of a 2D nematic. This is the (highly simplified) envisaged scenario in high-$T_\mathrm{c}$ superconductors. \protect\subref{subfig:nematic} Restoring translations all directions leads to a generalized nematic phase; the remaining anisotropy depends on the details of the rotational symmetry breaking. In this work we only consider explicitly the simplified case of the `isotropic' nematic, with only a single rotational modulus, see Sec.~\ref{subsec:Torque stress in the quantum nematic}.  }\label{fig:phases}
\end{figure*}

Resting on the seminal work of Kleinert~\cite{Kleinert89b}, we found that in 2+1D the theory can be rewritten in terms of {\em stress gauge fields} that enumerate the capacity of the solid medium to propagate forces between external stresses as well as the internal stresses sourced by the dislocations~\cite{QLC2D}. This gauge theory corresponds to a `flavored' version of quantum electrodynamics in 2+1D, in terms of the usual one-form $U(1)$-gauge fields identifying phonons with `stress photons'. This is drastically different in 3+1D, which we shall extensively explain in Sec.~\ref{sec:Dual elasticity in three dimensions}. The topological defects are now worldsheets in spacetime. Since these act as stress sources, the gauge fields that propagate the stress are two-form gauge fields of the kind encountered in string theory.  By working through the two-form gauge field formalism we do show that at least on the level of description of the elastic medium, 
the correct phonon propagators are impeccably reproduced: compare Eqs.~\eqref{eq:longitudinal propagator}, \eqref{eq:transverse propagator} with Eqs.~\eqref{eq:longitudinal propagator stress propagator relation}, \eqref{eq:transverse propagator gauge field}. 

Sec.~\ref{sec:Dislocation-mediated quantum melting} is the core of the development in this paper. The quantum liquid crystals are described as solids subjected to a proliferation (condensation) of dislocations. In 2+1D this is in principle 
a straightforward affair because the dislocations are fundamentally like bosonic particles and the tangle of dislocation lines in spacetime is just a Bose condensate that is `charged' under the stress gauge fields: this is a plain Higgs 
condensate and the quantum liquid crystals are therefore called {\em stress superconductors} similar to the dual superconductors in the context of the Abelian-Higgs duality~\cite{NguyenSudbo99,HoveSudbo00,HoveMoSudbo00}. 
As we discussed in Sec.~\ref{subsec:From two to three dimensions}, this path gets slippery in 3+1D because we have now to rely on an effective field theory description of the `string foam' formed in spacetime by the proliferation of the dislocations. This section will be devoted to a careful formulation of the Higgs action, with the bottom line that all gauge field components obtain a Higgs gap as usual. We also highlight the complications encountered in the construction of the 
dislocation condensate that were already on the foreground in the 2+1D case~\cite{QLC2D} which straightforwardly generalize to 3+1D: the {\em glide} and {\em Ehrenfest constraints} as well as the population of distinct Burgers vectors that is behind the difference between the columnar-, smectic- and nematic-type orders, see Fig.~\ref{fig:phases}.
  
The machinery is now in place and can be unleashed on the various kinds of quantum liquid crystals. We start with the {\em quantum nematic} order in Sec.~\ref{sec:nematic}. This is defined as a condensate where all 
  Burgers vector directions contribute equally, completely restoring the translational symmetry while the rotational symmetry is still broken. Resting on the prescription of Sec.~\ref{sec:Dislocation-mediated quantum melting} we find that this 3+1D quantum nematic shares all the traits of the 2+1D version. This acts as a sanity check confirming that the `Higgsing' of Sec.~\ref{sec:Dislocation-mediated quantum melting} does make sense. As in the 2+1D case, we find that the transverse phonons of the solid acquire a mass, indicating that shear stresses can no longer propagate through the liquid at length scales larger than the {\em shear penetration depth}, in close analogy with the way that magnetic forces cannot propagate in an electromagnetic superconductor. In addition, the quantum nematic is also a regular superfluid. 
  It is the same mechanism as in 2+1D: the glide constraint encodes for the fact that dislocations ``do not carry volume'' and therefore the compressional stress is not affected by the dislocation condensate. The result 
  is that the longitudinal phonon of the solid turns into the second sound/phase mode of the superfluid. Last but not least, a new feature in 3+1D is the way that the rotational Goldstone bosons (or `torque photons' in stress
  language) arise in the quantum nematic. The gross mechanism is the same as in 2+1D; using the `dynamical Ehrenfest constraint' formulation~\cite{BeekmanWuCvetkovicZaanen13,QLC2D}  it becomes manifest that these modes are quite literally {\em confined} in  the solid, while deconfining and becoming massless in the quantum liquid crystal with a rigidity that is residing in the dislocation condensate itself. The novelty is that in 3+1D we find according to expectations three such modes, that separate in two degenerate `transverse' modes and a `longitudinal' one, characterized by a parametrically different velocity. 
 
 As we discovered in 2+1D, the topological melting view offers a most elegant way of also dealing with the quantum smectic type of order. This just exploits the freedom to choose preferential directions for the 
 Burgers vectors in the dislocation condensate. In the nematic all Burgers directions contribute equally, while in the 2+1D smectic dislocations proliferate with their Burgers vectors oriented in one particular spatial 
 direction, only restoring translations in that particular dimension. We found that the long-wavelength physics of such quantum smectics is surprisingly rich. Intuitively one expects that a smectic is a system that 
 is one direction behaving like a liquid, remembering its solid nature in the other direction. However, we found that matters are quite a bit more interesting with the solid and liquid features being `intertwined'
 in the literal sense of the word. We show in Sec.~\ref{sec:smectic} that much of the same pattern occurs in 3+1D . This landscape is now enriched by the fact that the dislocations can proliferate with Burgers vectors in 
 one or two directions, defining the {\em columnar} and {\em smectic} quantum phases. There is room for even more richness to occur. Dealing with the quantum smectic (``stacks of liquid planes''), when the momentum 
 of the propagating modes lie precisely in the liquid-like plane we find that the response is indistinguishable from a 2D quantum nematic, except for small, dimension-dependent differences in the velocities of the massless modes. 
When the momentum lies in a solid--liquid plane it instead behaves like 2D quantum smectic.  Precisely along the solid direction a longitudinal phonon is recovered which is at first sight surprising since 
the shear modulus is contributing despite the fact that the transverse directions are liquid-like. Last but not least, we find one rotational Goldstone mode associated to the plane where translational symmetry 
is restored, in accordance with recent predictions~\cite{WatanabeMurayama13}. 
 
In Sec.~\ref{sec:columnar} we deal with  the 3D quantum columnar phase with its two solid directions (``arrays of liquid lines''). We find that the longitudinal phonon and one transverse phonon remain massless, while a second transverse phonon picks up a Higgs mass. 
 There is also a massive mode due to the fluctuations of the dislocation condensate itself, although these two massive modes are coupled for almost all directions of momentum. 
 In the special cases that momentum lies exactly in the plane orthogonal to the liquid-like direction, or in a plane with one solid-like and the liquid-like direction, one obtains response 
 similar to the 2D solid and 2D smectic respectively. Since there is no plane with vanishing shear rigidity, rotational Goldstone modes are absent.
 
As we showed in QLC2D it is straightforward to extend the theory from neutral substances to electrically charged ones, which is the subject of Sec.~\ref{sec:Charged quantum liquid crystals}. We now depart from a charged `Wigner crystal' keeping track of the coupling to electromagnetic fields 
when the duality transformation is carried out.  There is now the technical difference that the stress gauge fields have a two-form and the EM gauge fields a one-form nature; the effect is that not all stress fields couple to the electromagnetic fields. As a novelty we find that the `longitudinal' rotational Goldstone mode is a purely neutral entity. Different from its transverse partners, it stays electromagnetically quiet even in the finite-momentum regime where all collective modes turn into electromagnetic observables in 2+1D. 
Notwithstanding, the highlights of the 2+1D case all carry over to 3+1D. Most importantly, we show that the quantum nematics are characterized by a genuine electromagnetic Meissner effect proving directly that these are literal superconductors, while smectic and columnar phases have strongly-anisotropic superconductivity.

In Sec.~\ref{sec:Conclusions} we shall discuss the relevance of this work for real-world materials, and highlight roads for future research.

Finally a brief explanation of of our conventions regarding units and terminology. We work almost always in Euclidean time $\tau = \ti t$, and the quantum partition function at zero temperature is expressed as an Euclidean path integral $Z = \int
\exp (- \mathcal{S} ) = \int \exp ( - \int \td \tau \td^3 x \, \mathcal{L})$. We employ relativistic notation in which the temporal component $\ft = c \tau$ has units of length, where $c$ is an appropriate velocity, usually the shear velocity $c_\tT$. Greek indices $\mu, \nu,\ldots$ run over space and time while Roman indices $m,n,\ldots$ run over space only. Like in QLC2D, we will almost always work in one of two Fourier--Matsubara coordinate systems, where the axes are parallel or orthogonal to momentum. In the first system $(\ft,\tL,\tR,\tS)$, the temporal coordinate $\ft$ is unchanged, but the three spatial coordinates are divided into one longitudinal $\tL$, and two transverse directions $\tR,\tS$ with respect to the spatial momentum $\mathbf{q}$. The directions $\tR,\tS$ are orthogonal but otherwise arbitrary. The second system $(0,1,\tR,\tS)$, has one direction, $0$, parallel to spacetime momentum $p_\mu = (\frac{1}{c} \omega_n,\mathbf{q})$, where $\omega_n$ is a Matsubara frequency. The second direction, $1$ is orthogonal to $p_\mu$, but within the $(\ft\tL)$-plane, while the transverse direction $\tR,\tS$ are as before. The explicit coordinate transformations are given in Appendix~\ref{sec:Fourier space coordinate systems}, where we make, without loss of generality, one particular choice of axes.
We set $\hbar \equiv 1$ everywhere.

%% file: sec_symmetry.tex
The quantum liquid-crystalline phases which are the focus of this paper are ordered, in principle zero-temperature states of matter that spontaneously break  a symmetry. The symmetry at stake is the rotational invariance (isotropy) of space itself that is broken by the medium itself. Since only spatial and no temporal dimensions are involved, there is no sharp distinction between zero-temperature and thermal states of matter accomplishing the same feat. As we will see, the only difference of principle between classical liquid crystals and the bosonic variety of quantum liquid crystals that we consider here is in the `liquid part'. Classical liquid crystals are at the same time behaving as dissipative classical fluids while our quantum version is a {\em superfluid}, or superconductor in the charged case. Alluding to the universal long-wavelength properties associated with the order, this in turn implies a single novelty in the superfluid case.  A highly peculiar breach of established symmetry breaking wisdom occurs which is not as famous as it should be. Breaking a continuous symmetry usually implies a propagating Goldstone mode, like the phonon of a crystal. Accordingly, one would expect that a nematic crystal that breaks the isotropy of space should be characterized by `rotational phonons'. 
However, it has been shown a long time ago that even in the long-wavelength limit this rotational Goldstone mode has a finite coupling to the circulation of the normal, hydrodynamical fluid with the effect that this mode is overdamped 
even for its momentum tending to zero~\cite{DeGennesProst95, SinghDunmur02, ChaikinLubensky00}. This is different in the zero-temperature superfluid/superconductor: now the circulation of fluid is `massive' (quantized vorticity) and the rotational Goldstone modes are protected, as usual.

\subsection{Generalizing nematic order: `isotropic' versus `cubic' nematics}\label{subsec:Generalizing nematic order}

Another issues is the form of the order parameter theory associated with liquid crystals in general. The reader should be familiar with the textbook cartoon, revolving around the kinetics of ``rod-like molecules''. In the isotropic fluid
these rods are both translationally and rotationally disordered with the rods pointing in all space directions. In the nematic phase these rods line up while they continue to be translationally disordered. Upon further lowering temperature
these rods may form liquid layers, that stack in a periodic array in the direction perpendicular to the layer: the smectic. At the lowest temperatures full crystalline order may set in. This cartoon is quite representative for much of 
the classical liquid crystals; for deep reasons of chemistry, stiff, rod-like molecules are abundant and nearly all existent liquid crystals are of this `uniaxial kind'. However, viewed from a general symmetry breaking perspective these 
uniaxial nematics are highly special and even pathological to a degree. Group theory teaches that the symmetry group describing the isotropy of Euclidean space $O(3)$ encompasses {\em all} three-dimensional point groups as its subgroups.
The uniaxial nematics are associated with the $D_{\infty \mathrm{h}}$ point group that is special in  the regard that it only breaks the rotational isotropy in two of the three rotational planes of the $O(3)$ group. One ramification is
that it is characterized by only {\em two} rotational Goldstone modes. More generic 3D point groups break the isotropy in all three independent rotational planes and the Goldstone modes count in a way similar to the phonons
of the crystal: there are two `transverse' and one `longitudinal' acoustic modes associated with the rotational symmetry breaking, see Sec.~\ref{subsec:Rotational elasticity} and \ref{subsec:Torque stress in the quantum nematic}. 

In the present duality setting we depart from the maximally symmetry breaking state: a crystal breaking both translations and rotations, characterized by one of the 230 space groups. By 
proliferating the topological defects we {\em restore} the symmetry step-by-step.  The principle governing the existing vestigial liquid-crystalline phases is that a priori, the topological defects associated with the restoration 
of translational symmetry (the dislocations) can be sharply distinguished from those that govern the restoration of the isotropy of space -- the disclinations. Given the right microscopic circumstances, the disclinations can `stay massive' (not proliferating in the vacuum), while the dislocations have proliferated and condensed forming our dual `stress superconductor' with restored translational invariance and a liquid nature of the state of matter. Since these liquid crystals are `descendants' of the crystal, they are characterized by the `leftover' point group symmetry of the crystal. Point groups that are not compatible with the crystalline breaking of translations (encapsulated by the space groups) involving e.g. 5-fold rotations are therefore excluded. 

It is now merely a matter of technical convenience to begin with the {\em most} symmetric space groups. In fact, to avoid as much as possible the details coming from crystalline anisotropies that 
just obscure the essence we will look at from the simplest possible solid: the one described by the theory of isotropic elasticity in three space dimensions. This is similar in spirit to the famous 
KTNHY theory of topological melting in 2D, which considers the special case of a {\em triangular} lattice, which is unique in the regard that its long-wavelength theory is precisely isotropic elasticity in two dimensions. 
Upon proliferating the dislocations a nematic-type liquid crystal is formed that was named the ``hexatic'' since it is characterized by the six-fold rotational symmetry ($C_6$ point group) descending from the crystal. For the long-wavelength properties the precise form of the remnant discrete rotational symmetry is insignificant, the only thing that matters is that there is rotational rigidity. This is the reason we group all these states under the umbrella ``nematics'' (see also below).

In 3D there is no 
space group that is described precisely by isotropic elasticity, characterized by merely a bulk (compression) and a shear modulus. This of course has influence on the descendant liquid crystals. The `rotational elasticity' theory of generalized nematics (characterized by any 3D point group) has been systematically enumerated~\cite{StallingaVertogen94} and it follows that even the most symmetric point groups such as the $O_\mathrm{h}$-group describing `cube-like' nematics (instead of the `rod-like' uniaxial ones) are characterized by three independent moduli. As we will see, departing from the isotropic solid there is only room for a single rotational modulus. Accordingly, the reader
should appreciate our `isotropic nematics' as being like a `cubic nematic' where we have switched off the moduli encoding for the cubic anisotropies by hand. 

In fact, inspired by the considerations in the previous paragraph some of the authors felt a need to understand better the order parameter theory of such `generalized' (beyond uniaxial) nematics~\cite{LiuEtAl16b,LiuEtAl16,NissinenEtAl16}. They found out
that a systematic classification is just missing in the soft-matter literature, actually for a good reason. As it turns out, one is dealing with quite complex tensor order parameters involving tensors up to rank 6 for the most 
symmetric point groups! It was subsequently found that discrete, non-Abelian gauge theory can be mobilized to compute both the explicit order parameters as well as the generic statistical physics associated with this 
symmetry breaking in a relatively straightforward way. With regard to the latter, it was found that in  case of the most symmetric point groups one runs into thermal fluctuation effects of an unprecedented magnitude~\cite{LiuEtAl16b}. 
In the present context we just ignore these complications. We are primarily interested in the infinitesimal fluctuations around the ordered states and these are not sensitive to the intricacies of the `big-tensor' order parameters. In fact, all one needs to know is that our isotropic nematic is breaking rotations much like a cubic nematic, with the ramification that it should be characterized by two transverse and one longitudinal rotational Goldstone boson, see Sec.~\ref{subsec:Rotational elasticity} and \ref{subsec:Torque stress in the quantum nematic}.  

\subsection{Quantum smectics: neither crystals nor superfluids}

In the vestigial order hierarchy the next state one meets is the smectic type (translational order in $D-1$ dimensions), sandwiched in between the crystal and the nematic type states. Yet again the textbook version is, from the viewpoint of general symmetry principles, of a very special kind.  It is entirely focused on the `rod-like' $D_{\infty \mathrm{h}}$-molecules that now first arrange in liquid two-dimensional layers, which in turn stack in an array periodic perpendicular to these layers, breaking translations in this direction. Even more so than for the nematics a truly general `effective field theory description' departing  from tight symmetry principles is lacking. This deficit becomes on the foreground especially when dealing with the zero-temperature {\em quantum} smectic states of matter. The `liquid nature' becomes now associated with superfluidity, and there should be a well-defined sector of long-wavelength Goldstone-type excitations. Are these like phonons resp. superfluid phase modes (second sound) depending on whether one looks along the `solid'  resp. `liquid' directions? We shall see that these characteristics do shimmer through, but this is only a small part of the story. We found in the 2+1D case a remarkably complex assortment of collective modes reflecting the truly intertwined nature of superfluid and elastic responses~\cite{CvetkovicZaanen06b,QLC2D}. In part, this is already understood in the soft-matter literature in the form of the {\em undulation mode}: the transverse mode propagating in the liquid direction 
acquires a {\em quadratic dispersion} since the lowest-order interactions between the liquid layers are associated with their curvature~\cite{DeGennesProst95,ChaikinLubensky00}. These are impeccably reproduced in our smectics seen as dual stress superconductors 
of a particular kind. Yet again, in 3+1D there is even more to explore than in 2+1D; much of the sections on {\em quantum smectic} (\ref{sec:smectic}) and {\em columnar} (\ref{sec:columnar}) order are dedicated to charting this rich landscape. 

\begin{figure*}
 \hfill
 \subfloat[hexagonal crystal]{\includegraphics[height=5cm]{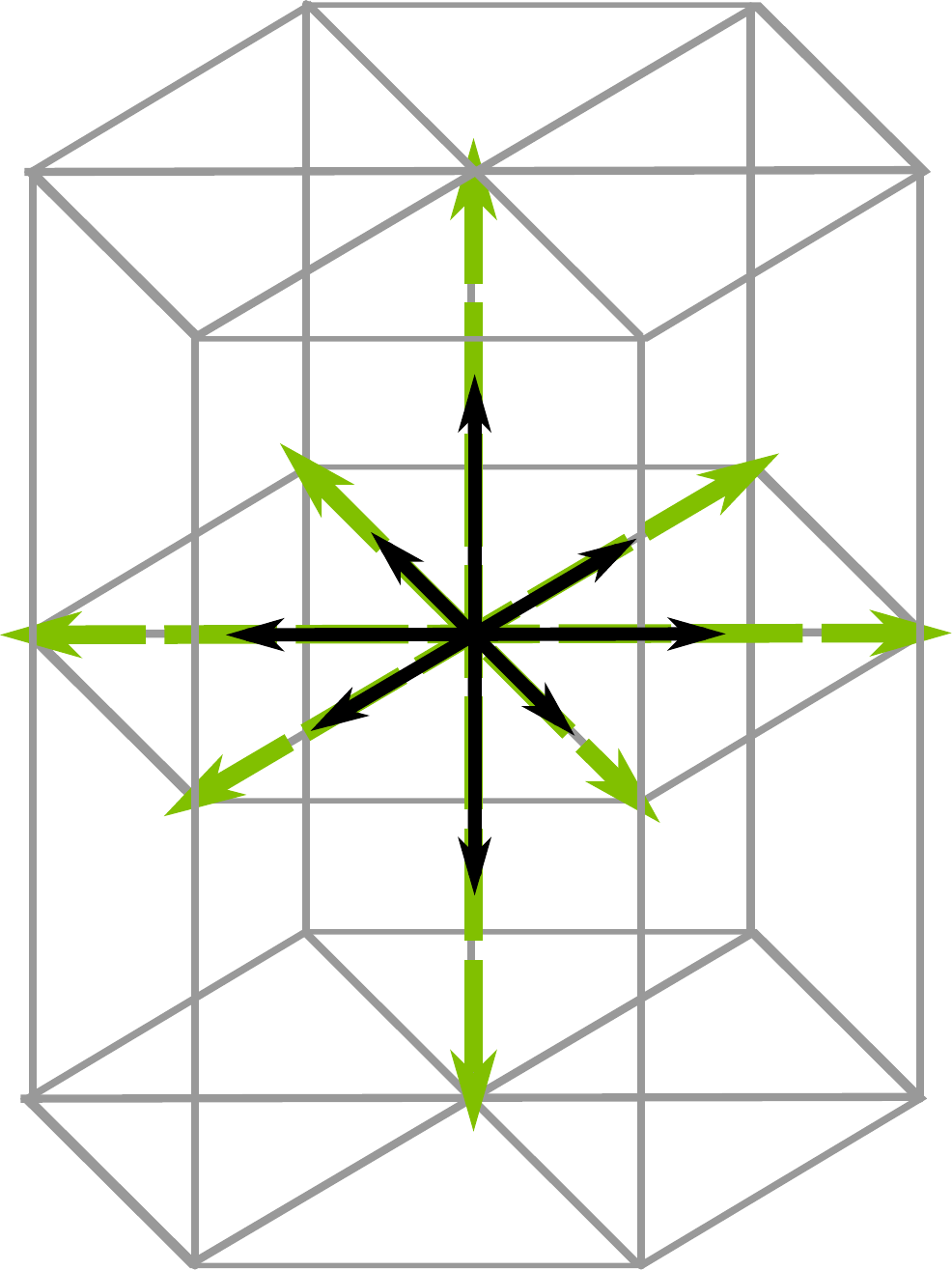}\label{subfig:hexagonal melting crystal}}
 \hfill
 \subfloat[columnar]{\includegraphics[height=5cm]{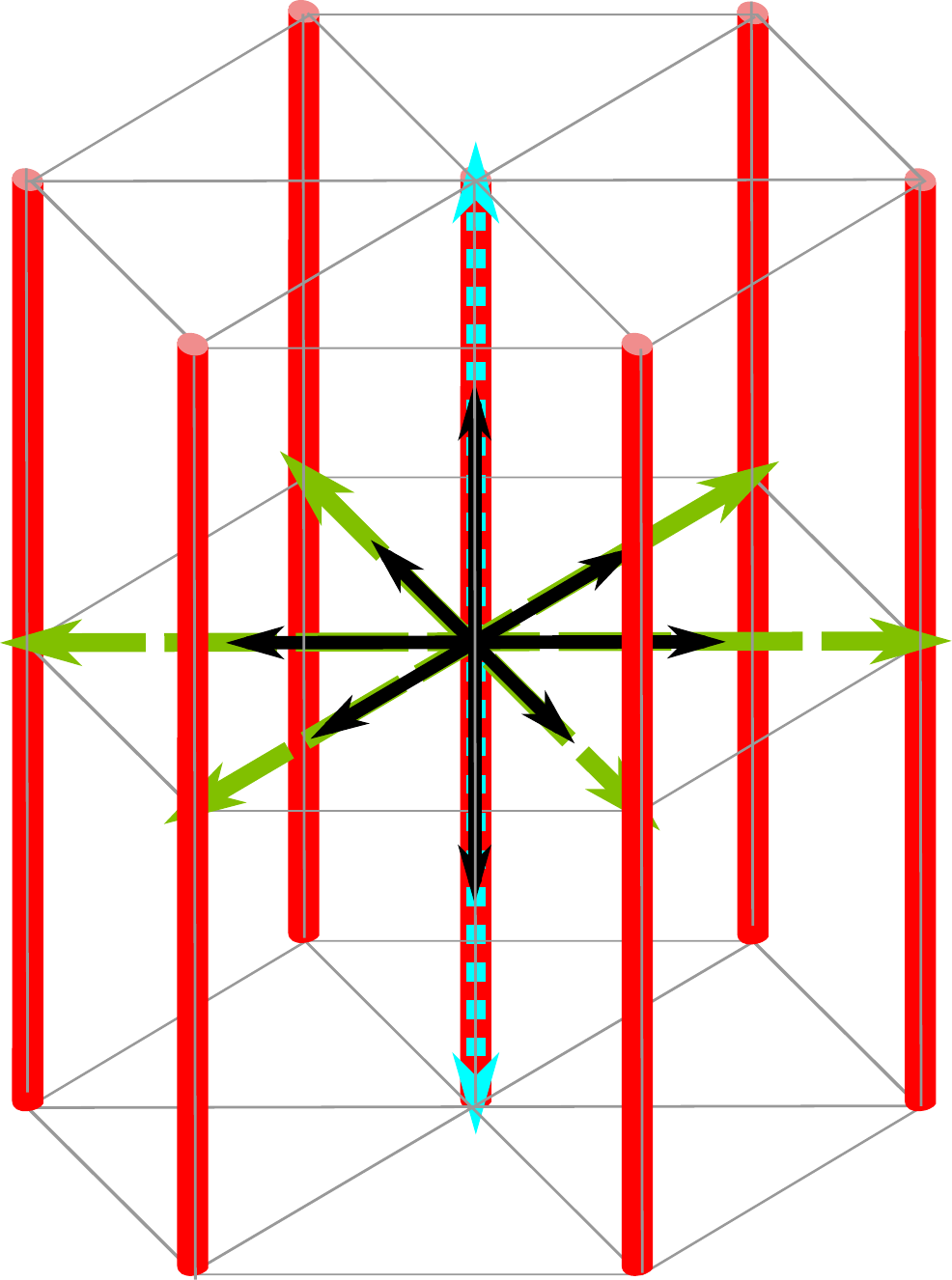}\label{subfig:hexagonal melting columnar vertical}}
 \hfill
  \subfloat[smectic]{\includegraphics[height=5cm]{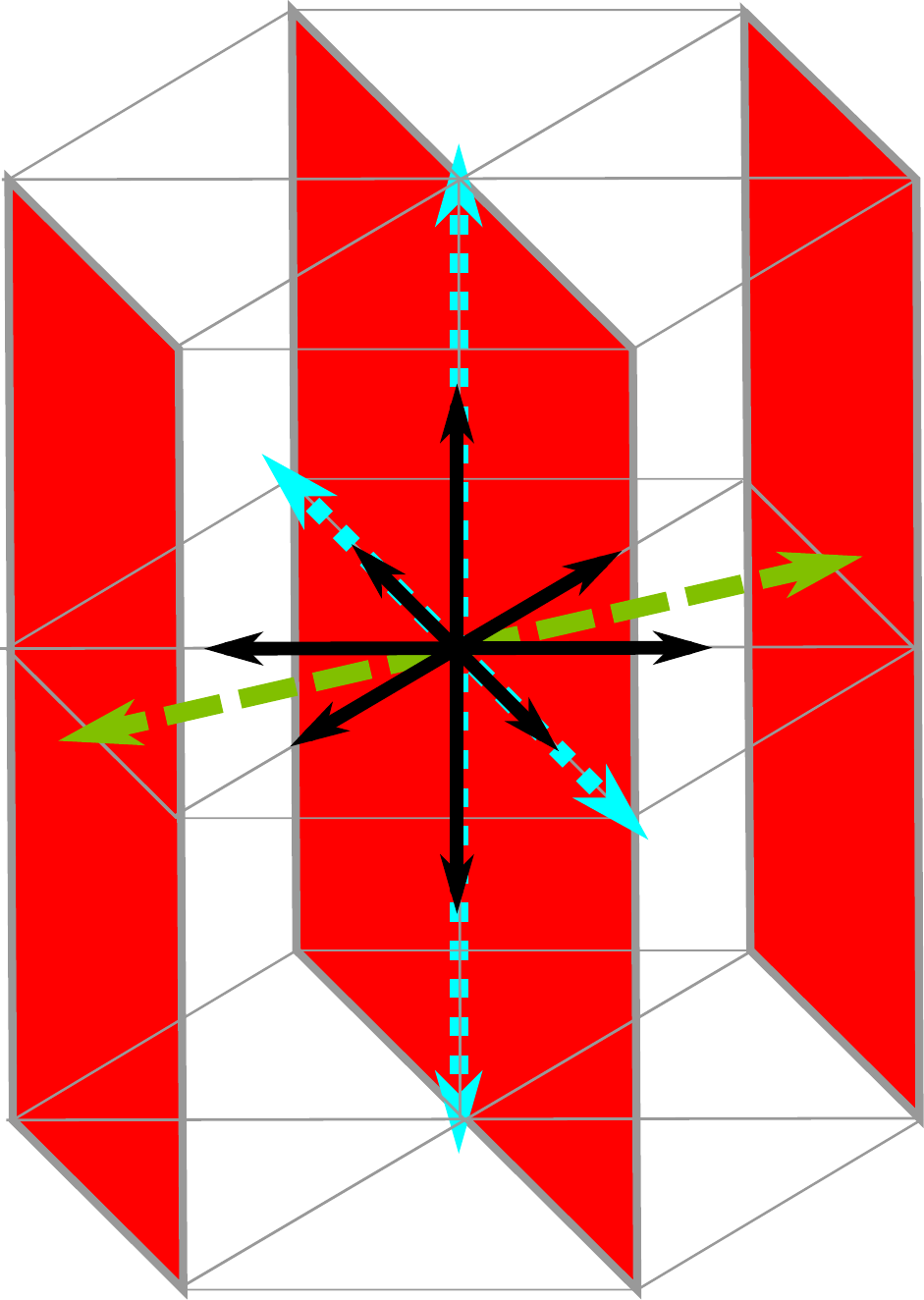}\label{subfig:hexagonal melting smectic vertical}}
 \hfill\null
 \\
 \hfill
 \subfloat[columnar]{\includegraphics[height=5cm]{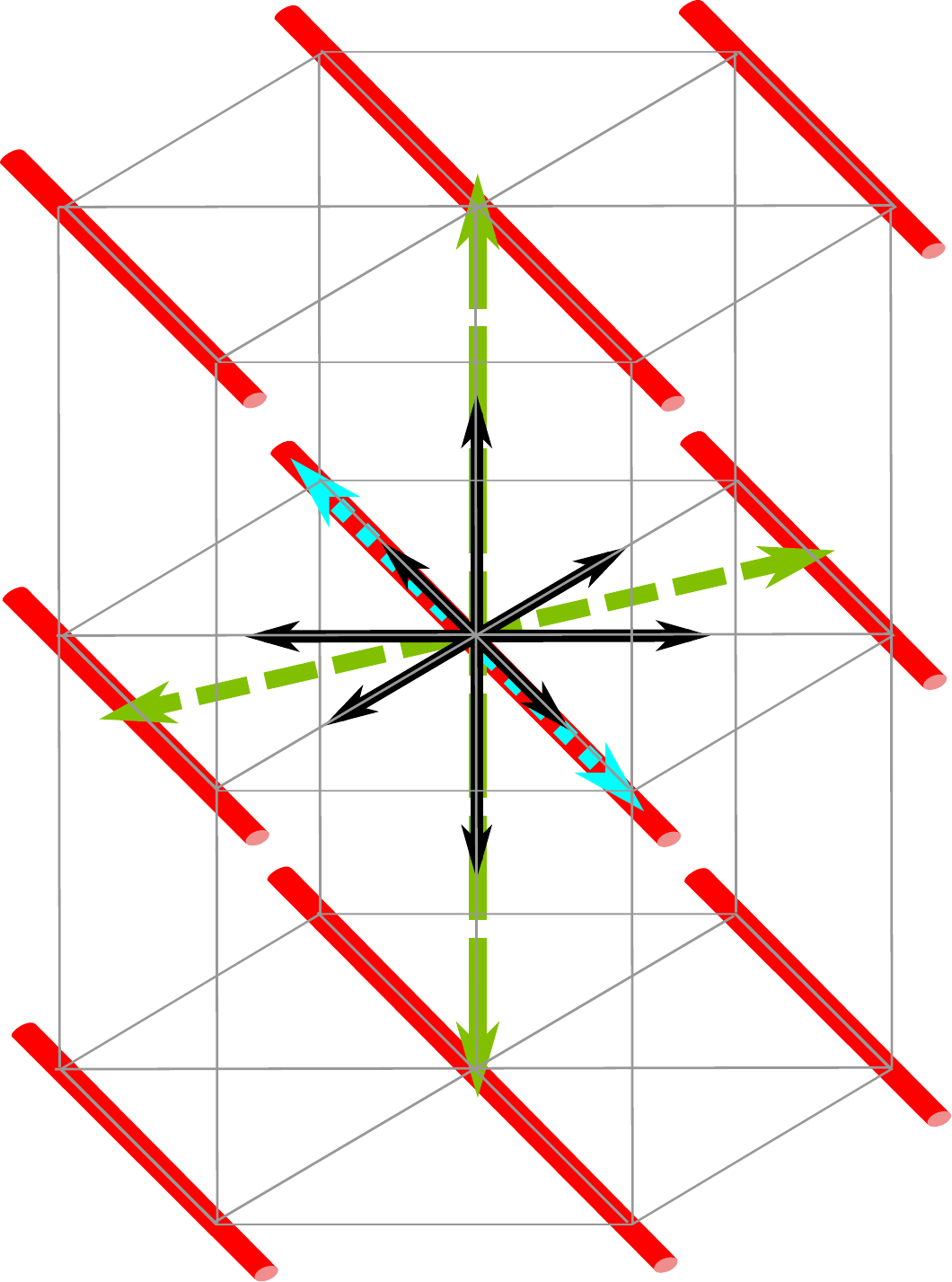}\label{subfig:hexagonal melting columnar horizontal}}
 \hfill
  \subfloat[smectic]{\includegraphics[height=5cm]{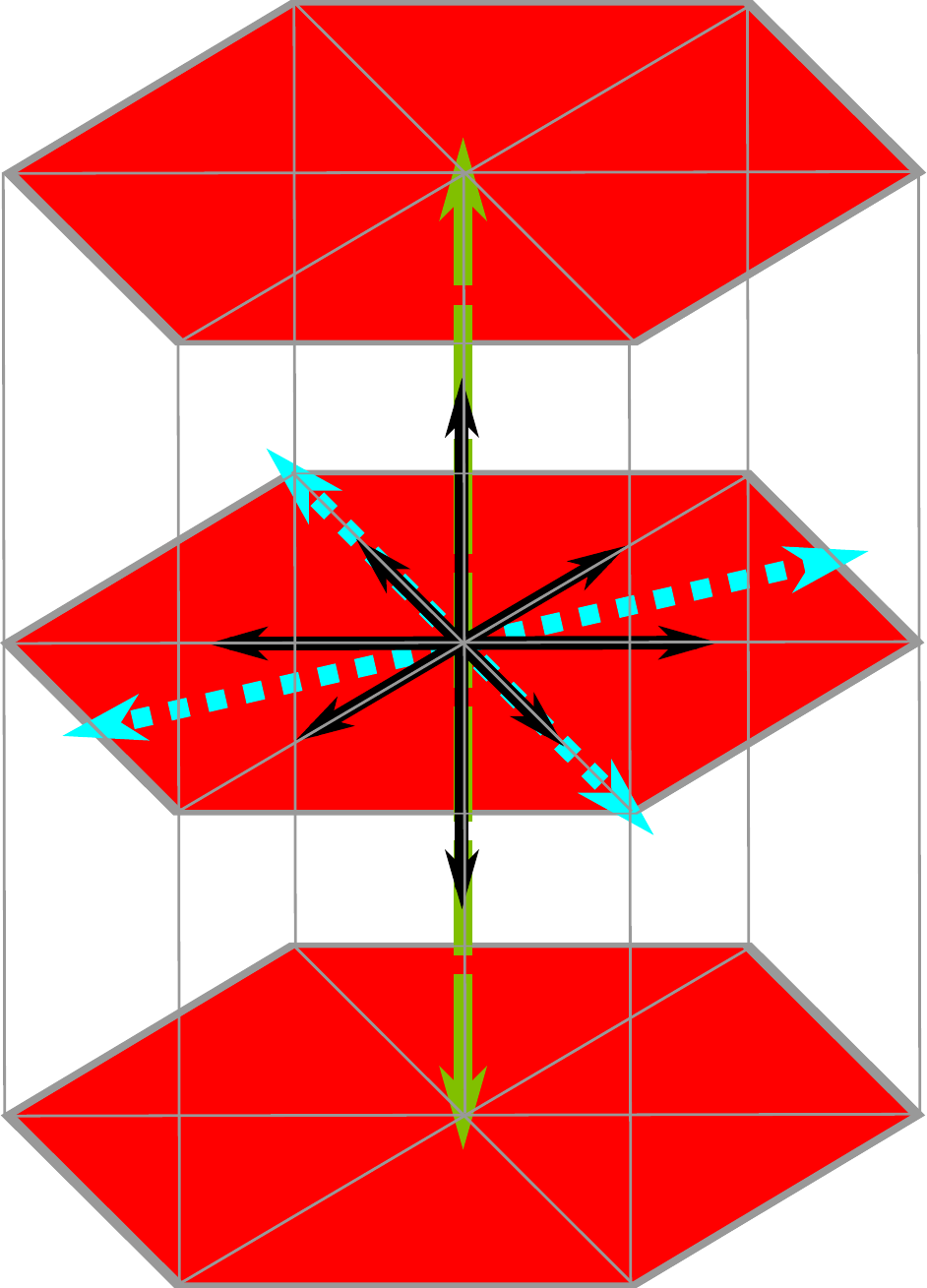}\label{subfig:hexagonal melting smectic horizontal}}
 \hfill
 \subfloat[nematic]{\includegraphics[height=5cm]{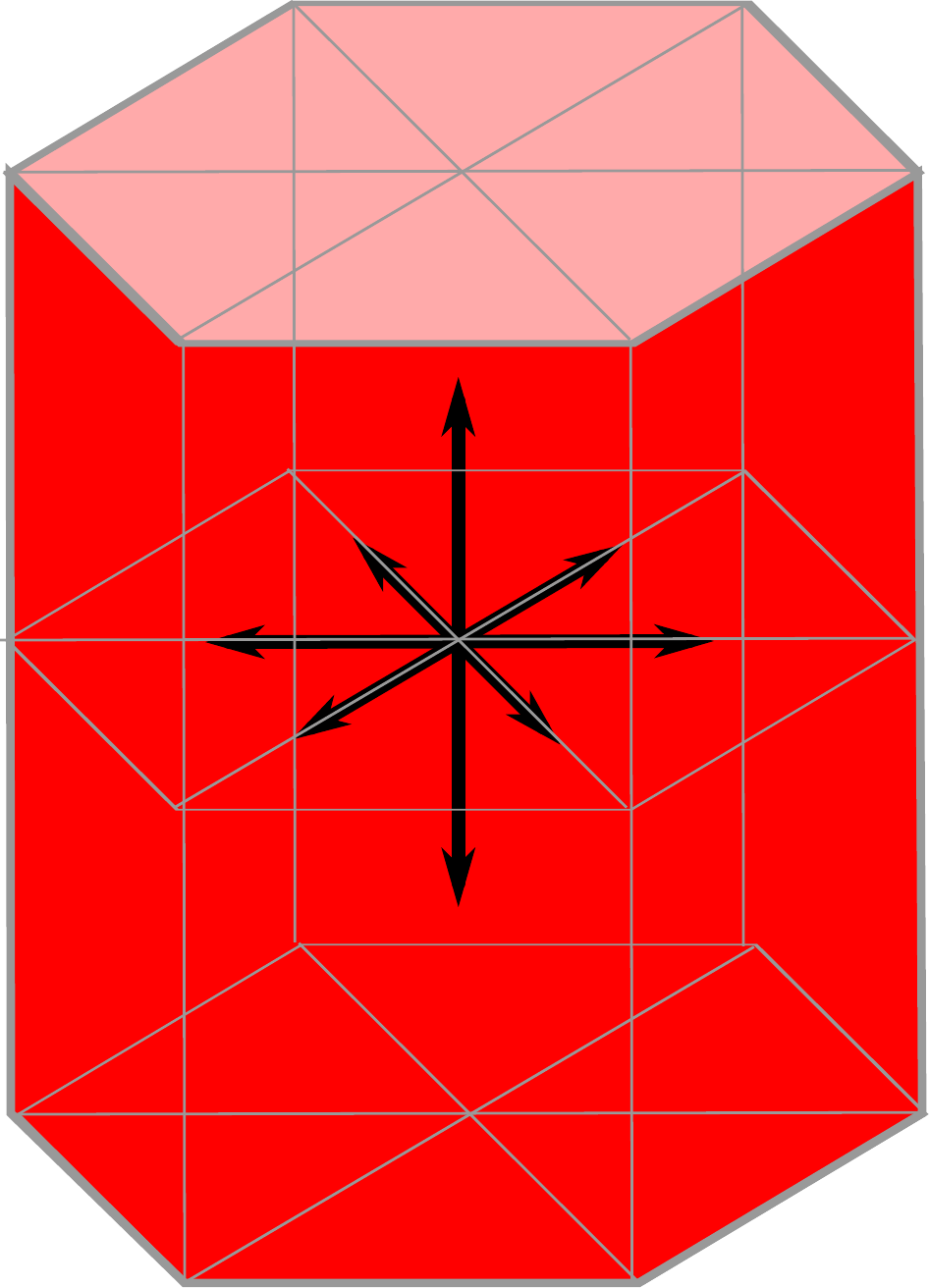}\label{subfig:hexagonal melting nematic}}
 \hfill\null
 \caption{Sequential dislocation-mediated melting of a hexagonal crystal with $D_{6\mathrm{h}}$ point group. Grey lines are bonds in the original hexagonal crystal and are guides to the eye only in the other phases. Green dashed lines indicate the elementary Burgers vectors of dislocations. Red color indicates lines/planes/volumes with translational symmetry due to condensation of dislocations with Burgers vectors in blue dotted lines. The black arrows are the `rotational cross' indicating the broken rotational symmetry that is the same throughout all the phases. \protect\subref{subfig:hexagonal melting crystal} All symmetry is broken, and the elementary Burgers vectors can point in six in-plane and two out-of-plane directions. \protect\subref{subfig:hexagonal melting columnar vertical} Dislocations with Burgers vectors in the vertical direction (blue) condense and restore translational symmetry, resulting in a 2D array of liquid lines. Since the remaining translational order is orthogonal to the liquid directions, the remaining Burgers vectors (green) match the original in-plane Burgers vectors of the crystal. \protect\subref{subfig:hexagonal melting smectic vertical} If we furthermore melt along an in-plane direction, the result is a periodic stack of liquid planes: the quantum smectic. Note that the Burgers vectors (green) in this smectic do no longer point along the original crystal axes. Because points that are separated by vectors along the liquid directions (vectors in blue) are equivalent, the remaining Burgers vectors must be orthogonal to the liquid planes.
 \protect\subref{subfig:hexagonal melting columnar horizontal} Alternatively, translational symmetry restoration can take place in the 6-fold plane. In-plane Burgers vectors in this columnar phase are not parallel to those of the parent crystal.  \protect\subref{subfig:hexagonal melting smectic vertical} Melting all in-plane translational order leads to stacks of liquid planes with $C_6$ in-plane rotational order, i.e. a stack of hexatics.
 \protect\subref{subfig:hexagonal melting nematic} Proliferation of all dislocations restores translational symmetry completely. Rotational symmetry remains broken because disclinations are forbidden. The rotational order is remembered from the original crystal point group (black cross).
 }\label{fig:hexagonal melting}
\end{figure*}

Although a Landau-style `direct' order parameter theory is lacking for ``generalized (quantum) smectics'' (i.e. going beyond $D_{\infty \mathrm{h}}$), the topological principles beyond our weak--strong 
duality are sufficiently powerful to formulate such a theory in the dual language of stress superconductivity. Like for the nematics, the main limitation is that we have formulated this theory departing from isotropic elasticity. The effects of the anisotropies associated with the real 3D space groups are presently unexplored and may be taken up as an open challenge. 
It was realized in the classic literature on thermal topological melting that smectic-type order is actually a natural part of this agenda~\cite{OstlundHalperin81}. It appears that is was first addressed in the quantum context 
independently in the early work by us~\cite{ZaanenNussinovMukhin04}, and by Bais \& Mathy who studied the possible liquid crystal phases with the fanciful Hopf symmetry breaking formalism~\cite{BaisMathy06,MathyBais07}.  This works as follows:
as before, we depart from the crystal with a particular point group embedded in its space group. The dislocations are characterized by their topological charge: the Burgers vector. 
These are associated with the deficient translations in the crystal lattice and accordingly they point only in lattice directions and are equivalent under the point-group transformations. 
In a cubic lattice, for instance, Burgers vectors point in orthogonal spatial $x$- ,$y$- and $z$-directions, while in a hexagonal crystal these point in the $z$ direction or in are six equivalent 
directions in the $xy$-plane associated with the sixfold axis, see Fig.~\ref{subfig:hexagonal melting crystal}. 

The master principle governing both the smectic- and nematic-type vestigial phases is that the dislocations are allowed to proliferate while keeping the disclinations ``out of the vacuum''. The point group symmetry of the crystal is maintained while translational symmetry is restored. But we just learned that  there is quite a variety of Burgers vectors; how should these be arranged
in the dislocation condensate? This is governed by precise topological rules. The first rule is that {\em the Burgers vectors of dislocations have to be `locally antiparallel'.}  A disclination is topologically
identical to a macroscopic number of dislocations with parallel Burgers vectors~\cite{Kleinert89b,ZaanenNussinovMukhin04,QLC2D}. These are not allowed in the vacuum and therefore we have to insist that on the microscopic scale a dislocation 
with Burgers vector pointing in the $\vec{B}$-direction of the lattice is always accompanied one pointing in precisely the opposite $-\vec{B}$-direction. The second rule is that {\em the translational symmetry gets restored precisely in the direction of the Burgers vectors}. In other words, points that differ by a (not necessarily integer) multiple of the Burgers vector become equivalent. 

In the generalized nematic, translation symmetry is restored in {\em all} spatial directions and this implies 
that all Burgers vector directions are populated equally in the dislocation condensate of the dual stress superconductor. One notices that this dislocation condensate remembers the point group of the 
crystal through the requirement that it is formed out of ``Burgers vectors pointing in the allowed directions''. In fact, as we will discuss in more detail in section~\ref{subsec:Torque stress in the quantum nematic} the rotational elasticity of the nematic is carried by the dislocation condensate itself.  

However, this ``equal Burgers vector population'' need not to be the case: it is perfectly compatible with the topological rules to populate only the pair-antiparallel Burgers vectors in e.g. one particular direction. Accordingly, translational symmetry is restored in that one of the three space dimensions and this 
is the topological description of the {\em columnar} state, Fig~\ref{subfig:columnar}. In a next step, the condensate can pick Burgers vectors such that the translational symmetry is restored in two orthogonal 
space directions, leaving the third axis unaffected: this is the {\em smectic} state in three dimensions, Fig.~\ref{subfig:smectic}. One notices a peculiar tension between the point group of the crystal and the way that the `liquid directions' emerge. Translational symmetry can only be independently restored in the three orthogonal ($x,y,z$) spatial directions since points that differ by a vector in a liquid direction are equivalent.
Accordingly, the `liquid' can occur either in one direction (the columnar phase), one plane (the smectic) or in all three directions (the nematic). In a cubic crystal this is straightforward; the three cubic axes are coincident with  the three orthogonal `translational' directions and by proliferating dislocations in either one or two directions one obtains immediately the smectic and columnar phases shown in the cartoons Fig.~\ref{fig:phases}.

However, dealing with e.g. a hexagonal crystal this gets more confusing, see Fig.~\ref{fig:hexagonal melting}. The first melting transition to a columnar phase takes pairs of antiparallel Burgers vectors along one of the crystal axes. For instance, one can choose the direction perpendicular to the 6-fold plane, Fig.~\ref{subfig:hexagonal melting columnar vertical}. The result is a regular triangular array of liquid lines. The dislocations in this columnar phase are still along the original crystal axes. If dislocation condensation takes place with Burgers vectors in a second direction, a smectic is obtained, Fig.~\ref{subfig:hexagonal melting smectic vertical}. This is a periodic stack of liquid planes. Note that the periodicity is no longer along an axis of the parent crystal, but obviously perpendicular to the planes. Accordingly, the dislocations in this smectic have Burgers vectors in this perpendicular direction, not commensurate with the Burgers vectors of the parent crystal. Here we see the two important consequences of the rules mentioned above:
1) dislocation melting always takes place restoring translations symmetry in orthogonal directions, even though the elementary Burgers vectors of the parent crystal may not be orthogonal. 
2) the remnant rotational order is {\em independent} of the translational symmetry restoration and is completely inherited from the parent crystal. 
This can be clearly seen in e.g. Fig.~\ref{subfig:hexagonal melting smectic vertical}.

Conversely, we could first melting along an in-plane direction as in Fig.~\ref{subfig:hexagonal melting columnar horizontal}. Now we have `three kinds of physics': liquid-like in one in-plane direction, solid-like in the orthogonal in-plane direction, and solid-like in the out-of-plane direction, which was already inequivalent due to the original crystal anisotropy. Again the Burgers vectors have to be perpendicular to the liquid lines, not necessarily parallel to the original crystal axes. If the next melting step is again in-plane, we end up with a periodic stack of liquid layers, see Fig.~\ref{subfig:hexagonal melting smectic horizontal}. Each layer is like a 2D hexatic phase with $C_6$-symmetry in the plane. We will verify this explicitly in Sec.~\ref{sec:smectic}. The overall structure is a particular 3D smectic. 

In all cases, once all translational symmetry has been restored due to melting of dislocations with Burgers vectors in all direction, a generalized nematic is obtained, see Fig.~\ref{subfig:hexagonal melting nematic}. The rotational order is the point group of the parent crystal, $D_{6\mathrm{h}}$ in this case.

It takes some special care to precisely formulate the equations describing this `Burgers vector population' affair in the construction of the dual dislocation condensates.  For the 2+1D quantum liquid crystals
this was for the first time put in correct form in  QLC2D --- although the Higgs terms in the effective dual actions were correct in earlier work, the derivation was flawed. As it turns out, this procedure straightforwardly generalizes to the 3+1D case, which is the topic of Sec.~\ref{sec:Dislocation-mediated quantum melting}. In the sections dealing with the smectic (\ref{sec:smectic}) and columnar (\ref{sec:columnar}) phases we will expose the remarkably rich landscape of `intertwined' liquid--solid responses of these systems. Once again, given our specialization to the strictly isotropic case this description is far from complete and we leave it to future work to find out a complete inventory of the long-wavelength physics that follows from this peculiar interplay of partial translational and full rotational symmetry breaking.  

%% file: sec_preliminaries.tex
\subsection{Elasticity as a quantum field theory}\label{subsec:Field-theoretic elasticity}
In QLC2D we provided an exposition of the quantum-field-theoretic formulation of linear elasticity. Let us here summarize the highlights. The principal quantities are displacement fields $u^a(\mathbf{x})$, referring to the deviation in direction $a$ from the equilibrium position of the constituent particle at position $\mathbf{x}$ in the coarse-grained continuum limit. The long-wavelength finite-energy configurations are enumerated in terms  of the gradients of the displacement field $\partial_m u^a$.  Departing from equilibrium the  potential energy density of solids takes the familiar form known from elasticity theory~\cite{Kleinert89b}
\begin{align}
 e^{(1)}_\mathrm{solid}(\mathbf{x}) &= \tfrac{1}{2} \partial_m u^a C_{mnab} \partial_n u^b,\label{eq:first order elastic energy}\\
 e^{(2)}_\mathrm{solid}(\mathbf{x}) &= \tfrac{1}{2} \partial_m \partial_r u^a C^{(2)}_{mrnsab} \partial_n \partial_s u^b.\label{eq:second order elastic energy}
\end{align}
Here $C_{mnab}$ is called the elastic tensor and its independent non-zero components are called elastic constants, while $C^{(2)}_{mrnsab}$ represents the second-order contributions in the gradient expansion. 
The elastic tensor is subjected to a number of symmetries and constraints. Importantly, antisymmetric combinations
\begin{equation}\label{eq:rotation field definition}
 \omega^{ab} = \frac{1}{2} (\partial_a u^b - \partial_b u^a),
\end{equation}
represent local rotations that must vanish to first order since these cannot change the energy of the crystal. Accordingly,  $C_{mnab}$ must be symmetric in $m,a$ and in $n,b$ and Eq.~\eqref{eq:first order elastic energy} 
contains only the symmetric combinations called {\em strains}:
\begin{equation}
 u^{ab} = \frac{1}{2} ( \partial_a u^b + \partial_b u^a).
\end{equation}
The crystalline symmetry in terms of its {\em space group} further reduces the number of independent elastic constants.

We extend this well-known theory of elasticity to the quantum regime by taking into account the quantum kinetic energy~\cite{ZaanenNussinovMukhin04}. We shall employ the Euclidean coherent-state path-integral formalism in an expansion of fluctuations around the maximally-correlated crystalline state, 
defined by the partition function
\begin{align}
 Z_\mathrm{solid}                         &= \int \mathcal{D} u^a\; \te^{- \mathcal{S}_\mathrm{solid} },\label{eq:solid partition function}\\
 \mathcal{S}_\mathrm{solid}                        &= \int \td \tau \td^D x\; \mathcal{L}_\mathrm{solid} ,\label{eq:solid action}\\
 \mathcal{L}_\mathrm{solid}              &= \mathcal{L}_\mathrm{kin} + \mathcal{L}_\mathrm{pot},\\
 \mathcal{L}_\mathrm{kin} &= \frac{1}{2\rho} (\partial_\tau u^a)^2,\label{eq:elasticity kinetic term}\\
 \mathcal{L}_\mathrm{pot} &=  e^{(1)}_\mathrm{solid}(x) + e^{(2)}_\mathrm{solid}(x).\label{eq:elasticity potential term}
\end{align}
Here the argument $x$ of the displacement fields $u^a(x)$ is understood to contain both space and time $x = (\tau, \mathbf{x})$, and the sign of the potential energy is consistent with our convention for imaginary time~\cite{ZaanenNussinovMukhin04,QLC2D}.

Although the formalism is valid for general elastic tensors, we shall treat explicitly only the case of the {\em isotropic solid}. Even though the crystal breaks rotational symmetry, the long-distance physics may still be effectively isotropic, as is the case for for instance the triangular lattice in 2D. In 3D, solids consisting of many crystalline and glasses (``amorphous solids'') are effectively isotropic~\cite{ChaikinLubensky00,Kleinert89b}. Isotropic solids are described by only two elastic constants: the {\em bulk} or {\em compression modulus} $\kappa$ and the {\em shear modulus} $\mu$. In contrast, in liquids or gases there is only a compression modulus, while  for instance crystals with cubic symmetry are characterized  
 by three independent elastic constants.

The potential energy for the isotropic solid in $D$ space dimensions is defined in terms of the elastic moduli
\begin{equation}\label{eq:isotropic solid elastic constants}
 C_{mnab}  = D \kappa P^{(0)}_{mnab} + 2\mu P^{(2)}_{mnab}.
\end{equation}
where the projectors of `angular momentum' $s = 0,1,2$ on the space of (1,1)-tensors under $SO(D)$-rotations~\cite{Kleinert89b}:
\begin{align}
 P^{(0)}_{mnab} &= \frac{1}{D} \delta_{ma} \delta_{nb},\label{eq:P0}\\
 P^{(1)}_{mnab} &= \frac{1}{2} ( \delta_{mn}\delta_{ab} - \delta_{mb} \delta_{na}),\label{eq:P1}\\
 P^{(2)}_{mnab} &= \frac{1}{2} ( \delta_{mn}\delta_{ab} + \delta_{mb} \delta_{na}) - \frac{1}{D} \delta_{ma} \delta_{nb}.\label{eq:P2}
\end{align}
These projectors satisfy $P^{(s)}_{mnab} P^{(s')}_{nkbc} = \delta_{ss'} P^{(s)}_{mkac}$ and
\begin{equation}
 P^{(0)}_{mnab} + P^{(1)}_{mnab} + P^{(2)}_{mnab} = \delta_{mn} \delta_{ab}.
\end{equation}
The absence of a term proportional to $P^{(1)}$ in Eq.~\eqref{eq:isotropic solid elastic constants} signifies that local rotations Eq.~\eqref{eq:rotation field definition} cannot change the energy of the crystal. The strain component that is singled out by $P^{(0)}$ is called {\em compression strain} while the components in the  $P^{(2)}$-subspace are called {\em shear strain}. In $D$ dimensions there are $\tfrac{1}{2} D^2 + \tfrac{1}{2}D - 1$ shear components, in particular 
there are  2 shears in $D=2$ and 5 in $D=3$.

The relation between the compression and shear modulus can be expressed using the Poisson ratio $\nu$ via
\begin{align}\label{eq:Poisson ratio definition}
 \kappa &= \mu \frac{2}{D} \frac{1+\nu}{1 - (D-1)\nu},&
 \nu &= \frac{ D \kappa - 2 \mu}{D(D-1) \kappa + 2\mu}.
\end{align}
The Poisson ratio takes values in $-1 \le \nu \le 1/(D-1)$, and is usually positive. Another quantity used frequently is the Lam\'e constant $\lambda = \kappa - \frac{2}{D} \mu$. Combining the 
kinetic and potential terms Eqs.~\eqref{eq:elasticity kinetic term},\eqref{eq:elasticity potential term} we define
\begin{align}\label{eq:relativistic solid Lagrangian}
\mathcal{L}_\mathrm{solid} 
   &= \frac{1}{2} \partial_\mu u^a C_{\mu\nu ab} \partial_\nu u^b,\nonumber\\
C_{\mu\nu ab} 
   &= \frac{1}{\mu} \delta_{\mu \ft} \delta_{\nu \ft} \delta_{ab} + C_{mnab}.
\end{align}
Throughout this paper  we will use the `relativistic' time $\ft = c_\tT \tau = \sqrt{\mu/\rho}\ \tau$ with the unit of length, while $c_\tT$ is the shear velocity such that  $\partial_\mu = (\frac{1}{c_\tT} \partial_\tau, \partial_m)$. Since there cannot be a displacement in the time direction $u^\tau \equiv 0$, the strains $\partial_\mu u^a$ are characterized by a relativistic `spacetime' index $\mu$ and a purely spatial `lattice' index $a$.

The second-order term Eq.~\eqref{eq:second order elastic energy} reduces greatly due to the symmetry of the isotropic solid~\cite{Kleinert89b}:
\begin{align}
  e_2 (\mathbf{x}) &= \frac{1}{2} 2\mu \big \lbrack \tfrac {1 - (D-2) \nu}{1 - (D-1) \nu}
 \ell'^2 \partial_m \partial_j u^j \partial_m \partial_k u^k \nonumber\\
 &\phantom{mmm} + \ell^2 \partial_m
  \omega^{ab} \partial_m \omega^{ab} \big \rbrack. \label{eq:isotropic solid second gradient energy}
\end{align}
Here $\ell$ {is the {\em length scale of rotational stiffness}: at length scales smaller than $\ell$, contributions} due to local rotations become important. Similarly, $\ell'$ is the length scale below which second-order compressional contributions become important, but these do not change anything qualitatively and will be ignored in the remainder of this work.

The dynamical properties of the solid can be found by applying infinitesimal external stresses and measuring the responses. In other words, we are interested in the Green's function (propagator) $\langle u^a  \;u^b \rangle$. 
For the isotropic solid, these have the simple form
\begin{equation}
 \langle u^a \; u^b \rangle = \frac{1}{\rho} \left[ \frac{P^\tL_{ab}}{\omega_n^2 + c_\tL^2 q^2(1 + \ell^{\prime 2} q^2) } + \frac{P^\tT_{ab}}{\omega_n^2 + c_\tT^2 q^2(1 + \ell^2 q^2)} \right].\label{eq:displacement propagator}
\end{equation}
using the longitudinal and transverse projectors $P^\tL_{ab} = q_a q_b /q^2$, $P^\tT_{ab} = \delta_{ab} - P^\tL_{ab}$. In  addition,  the longitudinal and transverse velocity are, respectively:
\begin{align}
c_\tL &= \sqrt{ \frac{ \kappa + 2\frac{D-1}{D}\mu }{\rho}} = \sqrt{\frac{2\mu}{\rho} \frac{1- (D-2)\nu}{1 - (D-1)\nu}} ,\label{eq:longitudinal velocity definition}\\ 
c_\tT &= \sqrt{\frac{\mu}{\rho}}.
\end{align}
From Eq.~\eqref{eq:displacement propagator} we see that there is one longitudinal acoustic phonon with velocity $c_\tL$ and $D-1$ transverse acoustic  phonons with velocity $c_\tT$. These correspond of course to 
 the Goldstone modes due to spontaneous breaking of $D$ translational symmetries.

After the dislocation-unbinding phase transition, the displacement fields $u^a$ are no longer well defined, and these propagators lose their meaning. We can however still consider the strain propagators $\langle \partial_m u^a \; \partial_n u^b\rangle$, that have a well defined meaning both in the ordered and disordered phases~\cite{ZaanenNussinovMukhin04,CvetkovicZaanen06a}. We are particularly interested in the longitudinal ($\tL$) and transverse ($\tT$) propagators.
In the solid these correspond to,
\begin{align}
 G_\tL &= \langle \partial_a u^a \; \partial_b u^b \rangle =   \frac{1}{\mu} \frac{c_\tT^2 q^2}{\omega_n^2 + c_\tL^2 q^2(1 + \ell^{\prime 2} q^2)}, \label{eq:longitudinal propagator} \\
 G_\tT &= 2 \langle \omega^{ab} \; \omega^{ab} \rangle = \frac{1}{\mu} \frac{ (D-1) c_\tT^2 q^2}{\omega_n^2 + c_\tT^2 q^2(1 + \ell^2 q^2)}.\label{eq:transverse propagator}
\end{align}
Here the factor $D-1$ in the transverse propagator arises from summing the contributions of the $D-1$ transverse phonons.

\subsection{Stress--strain duality}\label{subsec:Stress--strain duality}
Following QLC2D, the first step in the dualization procedure is to define the canonical four-momenta conjugate to the displacement field $u^a$ via
\begin{equation}\label{eq:stress tensor definition}
 \sigma^a_\mu = - \ti \frac{\delta \mathcal{S}}{\delta (\partial_\mu u^a)} = -\ti C_{\mu\nu ab} \partial_\nu u^b.
\end{equation}
where we used Eq.~\eqref{eq:relativistic solid Lagrangian} while  $-\ti$ follows from the standard conventions in the Euclidean formalism~\cite{Kleinert89b,ZaanenNussinovMukhin04,QLC2D}.  The quantity $\sigma^a_\mu$ is called the {\em stress tensor}. In static elasticity the stress tensor has only spatial components $\sigma^a_m$ while it is symmetric under $a \leftrightarrow m$ since only symmetric strains $u^{ab}$ are allowed. Similar to the strain fields, in the 
imaginary-time extension of the quantum theory the upper (Latin) labels are purely spatial while the lower (Greek) indices are referring to spacetime 
since there are no displacements in the time direction, $u^\tau \equiv 0$. The absence of antisymmetric stress components is known as  {\em Ehrenfest constraints}~\cite{QLC2D}:
\begin{equation}\label{eq:Ehrenfest constraints}
 \epsilon_{cma} \sigma_m^a = 0 \quad \forall \; c.
\end{equation}

Let us now focus on the dual Lagrangian,  where the principal variables are the stresses $\mathcal{L}_\mathrm{dual} = \mathcal{L}_\mathrm{dual}[\sigma^a_\mu]$. This can be derived equivalently 
by a Legendre transformation or by a Hubbard--Stratonovich transformation of the original Lagrangian. In both cases we need to `invert' the elastic tensor $C_{\mu\nu ab}$. However, due to the absence of antisymmetric strains, the elastic tensor has zeros
 amongst its eigenvalues and it cannot be inverted directly.  However, the Lagrangian surely contains only physical fields and the dualization operation can be carried out   `component-by-component'. It is most useful to bring the original 
Lagrangian into a block-diagonal form, to then invert the respective non-zero blocks. For the isotropic solid with elastic tensor Eq.~\eqref{eq:isotropic solid elastic constants}, these correspond to  the $P^{(0)}$- and  $P^{(2)}$-parts
as well as the kinetic energy. In QLC2D we already derived the dual stress action in arbitrary spatial dimension $D$, 
\begin{align}
 \mathcal{Z}_\mathrm{solid}  &= \int \mathcal{D} \sigma^a_\mu \mathcal{D} u^b \ \te^{-\mathcal{S}_\mathrm{dual}}, \label{eq:dual partition sum} \\
 \mathcal{S}_\mathrm{dual} &= \int_0^\beta \td \tau \int \td^D {\mathbf{x}} \ \mathcal{L}_\mathrm{dual} + \ti \sigma^a_\mu \partial_\mu u^a,\label{eq:dual action} \\
\mathcal{L}_\mathrm{dual} &=  \frac{1}{2\mu} \lvert\sigma^a_\ft\rvert^2 + \frac{1}{2} \sigma^a_\mu \big( \frac{1}{D \kappa} P^{(0)}_{\mu\nu ab} + \frac{1}{2\mu}P^{(2)}_{\mu\nu ab} \big) \sigma^b_\nu \nonumber\\
&= \frac{1}{2\mu} \lvert\sigma^a_\ft\rvert^2 + \frac{1}{8\mu} \big[ \sigma^a_m \sigma^a_m + \sigma^a_m \sigma^m_a - \frac{2\nu}{1 +\nu} \sigma^a_a \sigma^b_b \big]. \label{eq:dual solid Lagrangian}
\end{align}
with the  Poisson ratio $\nu$ given by Eq.~\eqref{eq:Poisson ratio definition}.

In appendix ~\ref{sec:Fourier space coordinate systems} we introduce a quite convenient Fourier space coordinate system. In this system, the direction $\tL$ is parallel to the momentum $\mathbf{q}$ while $\tR,\tS$ are two transverse directions perpendicular to $\tL$ and to each other. All fields in the action are real-valued, and we demand that $\sigma^\dagger(p) = \sigma(-p)$ in momentum space~\cite{Kleinert89a,ZaanenNussinovMukhin04,QLC2D}. 

The dual Lagrangian is then block diagonal, containing five sectors:
\begin{align}\label{eq:solid stress Lagrangian components}
  \mathcal{L}_\mathrm{dual} &= \mathcal{L}_{\tT 1} + \mathcal{L}_{\tT 2} +\mathcal{L}_{\tT 3} + \mathcal{L}_{\tL 1} + \mathcal{L}_{\tL 2},\\
  \mathcal{L}_{\tT 1} &= \frac{1}{8\mu} 
\begin{pmatrix} \sigma^{\tR \dagger}_\ft \\ \sigma^{\tR \dagger}_\tL \\ \sigma^{\tL \dagger}_\tR  \end{pmatrix}^\tT
 \begin{pmatrix} 
 4 & 0  & 0\\ 
 0 & 1 & 1 \\
 0 & 1 & 1 
 \end{pmatrix}
 \begin{pmatrix} \sigma^\tR_\ft \\  \sigma^{\tR }_\tL \\ \sigma^{\tL}_\tR \end{pmatrix},\nonumber\\
 \mathcal{L}_{\tT 2} &= \frac{1}{8\mu} 
 \begin{pmatrix} \sigma^{\tS \dagger}_\ft \\ \sigma^{\tS \dagger}_\tL \\ \sigma^{\tL\dagger}_\tS  \end{pmatrix}^\tT
 \begin{pmatrix} 
 4 & 0  & 0\\ 
 0 & 1 & 1 \\
 0 & 1 & 1 
 \end{pmatrix}
 \begin{pmatrix} \sigma^\tS_\ft \\  \sigma^{\tS }_\tL \\ \sigma^{\tL}_\tS \end{pmatrix},\nonumber\\
 \mathcal{L}_{\tT 3} &= \frac{1}{8\mu} 
 \begin{pmatrix} \sigma^{\tR \dagger}_\tS \\ \sigma^{\tS \dagger}_\tR  \end{pmatrix}^\tT
 \begin{pmatrix} 
  1 & 1 \\
  1 & 1 
 \end{pmatrix}
 \begin{pmatrix} \sigma^{\tR }_\tS \\ \sigma^{\tS}_\tR \end{pmatrix},\nonumber\\
 \mathcal{L}_{\tL 1} &=   \frac{1}{8\mu} \frac{2}{1+\nu}
 \begin{pmatrix} \sigma^{\tL \dagger}_\ft \\ \sigma^{\tL \dagger}_\tL \\ \sigma_-  \end{pmatrix}^\tT
 \begin{pmatrix} 
 \scriptstyle 2(1+\nu) & 0 & 0 & 0\\ 
 0 & 1 & \sqrt{2}\nu \\
 0 & \sqrt{2}\nu & 1-\nu
 \end{pmatrix}
 \begin{pmatrix} \sigma^\tL_\ft \\ \sigma^{\tL }_\tL \\ \sigma_- \end{pmatrix},\nonumber\\
\mathcal{L}_{\tL 2} &=   \frac{1}{4\mu} \lvert \sigma_+ \rvert^2.
\end{align}
We will soon find out that $\mathcal{L}_{\tL 1}$ contains the longitudinal phonon, while $\mathcal{L}_{\tT 1}$ and $\mathcal{L}_{\tT 2}$ contain the transverse phonons in the $\tR$- resp. $\tS$-transverse directions. Here we have defined 
\begin{align}
 \sigma_+ &= \frac{1}{\sqrt{2}}(\sigma^\tR_\tR + \sigma^\tS_\tS), &
 \sigma_- &= \frac{1}{\sqrt{2}}(\sigma^\tR_\tR - \sigma^\tS_\tS).
 \label{eq:sigma plus minus definition}
\end{align}
In fact, we could have defined similar symmetry and antisymmetric combinations for the transverse sectors, but we refrain from doing so because we need second-order contributions as we shall explain just below. For more context about this division into sectors, see Sec.~\ref{subsec:Interpretation of stress components}.

To express the transverse propagator Eq.~\eqref{eq:transverse propagator} in the dual stress fields, we need the contributions from second-gradient elasticity since  
the first three matrices in Eq.~\eqref{eq:solid stress Lagrangian components} are not invertible (see Sec.~\ref{subsec:Dual propagator relations}). 
From the second-gradient contribution Eq.~\eqref{eq:isotropic solid second gradient energy}, we will only use the rotational part. Expressed in $\omega^c = \tfrac{1}{2} \epsilon^{cab} \omega^{ab}$ these become,
\begin{equation}\label{eq:second-gradient elasticity}
 \mathcal{L}^{(2)} = \frac{1}{2} 4 \mu \ell^2 (\partial_c\omega^c)^2.
\end{equation}
The canonical momentum conjugate to the rotation field $\omega^c = \tfrac{1}{2} \epsilon^{cab} \omega^{ab}$ is the {\em torque stress} $\tau^c_m$:
\begin{equation}\label{eq:torque stress definition}
 \tau^c_m = -\ti \frac{\delta \mathcal{S} }{\delta (\partial_m \omega^c) } = -\ti 4 \mu \ell^2 \partial_m\omega^c.
\end{equation}
There is no separate temporal component, since the rotations $\omega^c$ are descendant from displacements $u^a$ via Eq.~\eqref{eq:rotation field definition} and do not have their own dynamics.
The dual second-gradient Lagrangian is then~\cite{QLC2D}
\begin{equation}\label{eq:dual torque stress Lagrangian}
 \mathcal{L}^{(2)}_\mathrm{dual} =  \frac{1}{8 \mu \ell^2} (\tau^c_m)^2.
\end{equation}
In the presence of torque stress, the Ehrenfest constraints are softened, and read~\cite{QLC2D}
\begin{equation}\label{eq:Ehrenfest constraint torque stress}
 \epsilon_{cma} \sigma_m^a = \partial_m \tau_m^c.
\end{equation}
substituting his equation in Eq.~\eqref{eq:dual torque stress Lagrangian} yields, 
\begin{align}\label{eq:second gradient Lagrangian in stress tensor}
 \mathcal{L}^{(2)}_\mathrm{dual} &= \frac{1}{8 \mu \ell^2} \frac{1}{q^2} \epsilon_{cma} \sigma_m^{a\dagger} \epsilon_{cnb} \sigma^b_n\\
 &=\frac{1}{8\mu \ell^2 q^2} \Big[
 \begin{pmatrix}  \sigma^{\tR \dagger}_\tL \\ \sigma^{\tL \dagger}_\tR  \end{pmatrix}^\tT
 \begin{pmatrix} 
  1 & -1 \\
 -1 & 1 
 \end{pmatrix}
 \begin{pmatrix} \sigma^{\tR }_\tL \\ \sigma^{\tL}_\tR \end{pmatrix}\nonumber\\
 &\phantom{mmmm} + 
 \begin{pmatrix} \sigma^{\tS \dagger}_\tL \\ \sigma^{\tL\dagger}_\tS  \end{pmatrix}^\tT
 \begin{pmatrix} 
  1 & -1 \\
  -1 & 1 
 \end{pmatrix}
 \begin{pmatrix} \sigma^{\tS }_\tL \\ \sigma^{\tL}_\tS \end{pmatrix}\nonumber\\
 &\phantom{mmmm} + 
 \begin{pmatrix} \sigma^{\tR \dagger}_\tS \\ \sigma^{\tS \dagger}_\tR  \end{pmatrix}^\tT
 \begin{pmatrix} 
  1 & -1 \\
  -1 & 1 
 \end{pmatrix}
 \begin{pmatrix} \sigma^{\tR }_\tS \\ \sigma^{\tS}_\tR \end{pmatrix}
 \Big].
\end{align}
To find the propagators of Eqs.~\eqref{eq:longitudinal propagator}, \eqref{eq:transverse propagator} on the dual (stress) side, one needs to introduce stress gauge fields, which we do in Sec.~\ref{subsec:Stress gauge fields} below.

\begin{figure*}
 \null\hfill
 \subfloat[longitudinal normal $\omega^\tL$]{\includegraphics[scale=.3]{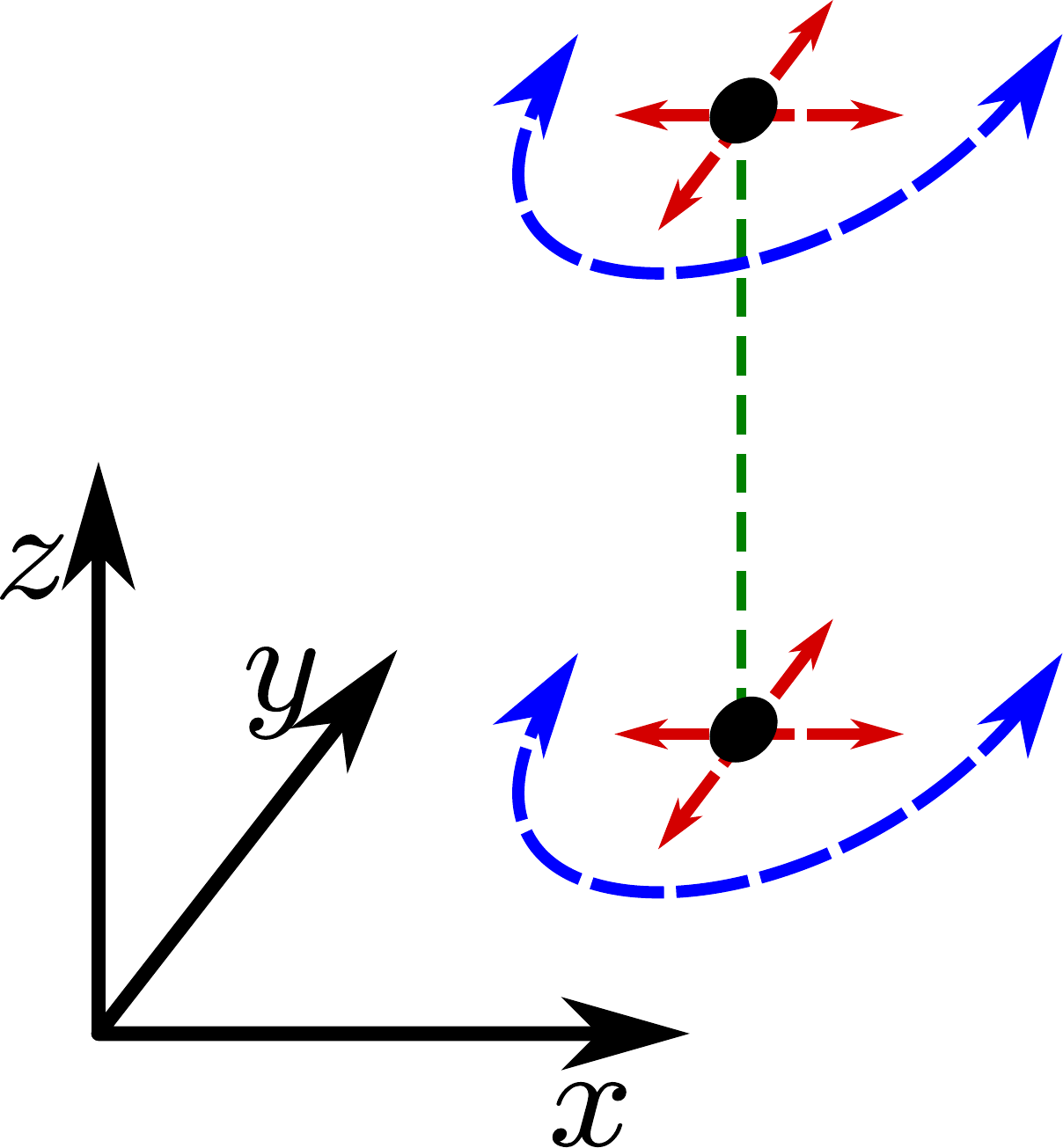}\label{subfig:longitudinal rotational Goldstone}}
 \hfill
 \subfloat[transverse normal $\omega^\tR$, $\omega^\tS$]{\includegraphics[scale=.3]{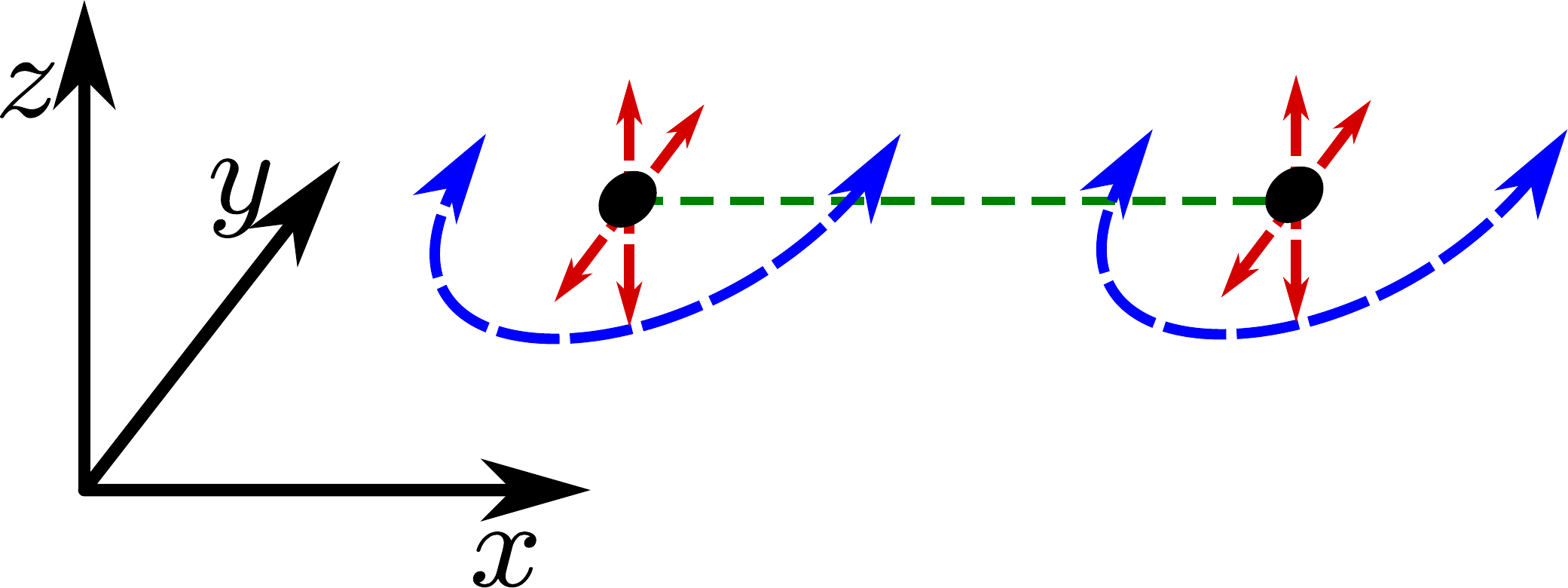}\label{subfig:transverse rotational Goldstone}}
  \hfill\null
 \caption{Rotational modes geometry. Depicted are rotational modes between two torque sources indicated as black dots. The momentum is in the direction of the dotted green line connecting these dots, in the $z$- (left) and $x$-direction (right). The blue dashed semicircle line denotes the rotational plane, normal to the index $c$ of $\omega^c$, in this case the $z$-direction, which is longitudinal (left) and transverse (right). The rotational field $\omega^c$ dualizes to the two-form torque stress gauge field $h^c_{\kappa\lambda}$, see Sec.~\ref{subsec:Torque stress gauge fields}. The red cross denotes the transverse polarization of this gauge field $(\kappa\lambda) = (\tR \tS)$, perpendicular to momentum. These transverse components of the dual torque stress gauge field represent the rotational (Goldstone) mode. 
 }\label{fig:rotational Goldstone modes}
\end{figure*}

\subsection{Rotational elasticity}\label{subsec:Rotational elasticity}
In a medium that is completely translationally symmetric, but which has rotational rigidity, one can also write down the general form of the elastic energy of the long-distance, low-energy excitations. The displacement fields $u^a$ are now ill-defined, but the rotation fields $\omega^c = \tfrac{1}{2} \epsilon^{cab} \omega^{ab}$ are good quantities, and are the fundamental Goldstone fields of this medium. They can be derived using differential geometry as the small fluctuations of the contortion tensor~\cite{Kleinert89b,BohmerDownesVassiliev11,BohmerTamanini15}

We are of course thinking of a nematic liquid crystal where all translational symmetry is restored, but this theory would hold for any translation-invariant medium with spontaneously broken rotational symmetry. For an isotropic medium, like our `isotropic nematic phase', this problem has been studied previously, see for instance Refs.~\onlinecite{BohmerDownesVassiliev11,BohmerTamanini15} and references therein. For systems with broken rotational symmetry, both discrete and continuous, the enumeration of elastic moduli was derived in Ref.~\onlinecite{StallingaVertogen94}. Here we focus on the isotropic case, which as before should be thought of as $O_\mathrm{h}$-symmetry in the limit of vanishing cubic anisotropy.

In close analogy to Sec.~\ref{subsec:Field-theoretic elasticity}, the general form of the rotational-elastic Lagrangian is~\cite{BohmerTamanini15}
\begin{align}\label{eq:general rotational Lagrangian}
 \mathcal{L}_\mathrm{rot} &=  \tfrac{1}{2}  \rho_\mathrm{rot} (\partial_\tau \omega^c)^2 \nonumber\\
 &\phantom{mm} + \tfrac{1}{2}  \partial_m \omega^a ( \kappa_0 P^{(0)}_{mnab} + \kappa_1 P^{(1)}_{mnab} + \kappa_2 P^{(2)}_{mnab} ) \partial_n \omega^b.
\end{align}
Here $\rho_\mathrm{rot}$ is the `density' of the rotationally rigid medium, rather to be thought of as moment of inertia. The constants $\kappa_{0,1,2}$ define the rotationally elastic properties. Note that there is now no reason why the antisymmetric sector $P^{(1)}$ should be absent. Using relations Eqs.~\ref{eq:P0}--\ref{eq:P2} it can be shown that
\begin{equation}\label{eq:rotational sector dependence}
 \int \partial_m \omega^a P^{(2)}_{mnab} \partial_n \omega^b = \int \partial_m \omega^a (2P^{(0)}_{mnab} + P^{(1)}_{mnab}) \partial_n \omega^b.
\end{equation}
Here we performed partial integrations and assumed the surface term vanishes. Then the rotational Lagrangian can be rewritten as
\begin{align}
 \mathcal{L}_\mathrm{rot} 
 &= \tfrac{1}{2} \rho_\mathrm{rot} (\partial_\tau \omega^c)^2 + \tfrac{1}{2} \partial_m \omega^a \Big( (\kappa_0 + 2 \kappa_2) P^{(0)}_{mnab} 
  \nonumber\\
 &\phantom{mmmmmmmmmmmmm} + (\kappa_1 + \kappa_2) P^{(1)}_{mnab} \Big) \partial_n \omega^b \nonumber\\
 &= \tfrac{1}{2} \rho_\mathrm{rot} (\partial_\tau \omega^c)^2 + \tfrac{\kappa_0 + 2 \kappa_2}{6} (\partial_c \omega^c)^2 + \tfrac{\kappa_1 + \kappa_2}{4} (\epsilon_{abc} \partial_b \omega^c)^2\nonumber\\
 &= \tfrac{1}{2} ( \rho_\mathrm{rot} \omega_n^2  + \tfrac{\kappa_0 + 2 \kappa_2}{3} q^2)  (\omega^\tL)^2  \nonumber\\
 &\phantom{mm} + \tfrac{1}{2} ( \rho_\mathrm{rot} \omega_n^2  + \tfrac{\kappa_1 + \kappa_2}{2}  q^2) \big( (\omega^\tR)^2 +  (\omega^\tS)^2\big).
\end{align}
We see that there are again longitudinal and transverse velocities, given by
\begin{align}\label{eq:rotational velocities}
 c_\tL^\mathrm{rot} &= \sqrt{\frac{\kappa_0 + 2 \kappa_2}{3\rho_\mathrm{rot}}}, &
 c_\tT^\mathrm{rot} &= \sqrt{\frac{\kappa_1 + \kappa_2}{2\rho_\mathrm{rot}}}.
\end{align}
Note however that the interpretation of these propagating modes is slightly subtle. The vector $\omega^c$ describes rotational deformations in the plane {\em perpendicular} to $c$. So the field $\omega^\tL$, where $\tL$ is parallel to spatial momentum, describes rotations in the plane perpendicular to the propagating direction. This is counterintuitive when thinking of phonons or photon polarizations, and the reader should take caution when using these terms in the rotational context. For clarity, we have illustrated this in Fig.~\ref{fig:rotational Goldstone modes}.

The longitudinal and transverse velocities are related via the Poisson ratio $\nu$ according to Eq.~\eqref{eq:longitudinal velocity definition}. Therefore one can define a `rotational Poisson ratio' as~\cite{BohmerTamanini15}:
\begin{equation}\label{eq:rotational Poisson ratio}
 \nu_\mathrm{rot} = \frac{\kappa_0 - 3 \kappa_1 - \kappa_2}{2\kappa_0 - 3\kappa_1 + \kappa_2}.
\end{equation}
The interpretation of this quantity is as follows: if one perturbs the system by an external rotational torque such as $\tau^x_x$, there can be rotational strain in both the parallel direction ($\partial_x \omega^x$) as well as the perpendicular directions ($\partial_y \omega^y$ and $\partial_z \omega^z$). The negative ratio between the longitudinal and the transverse response is the rotational Poisson ratio. In an ordinary solid, the Poisson ratio is usually positive, meaning that a longitudinal elongation is accompanied by a transverse contraction. Translating this to the rotational context, a positive rotational Poisson ratio means a positive response parallel to the external torque would be accompanied by negative rotational strain in the orthogonal rotational planes. We will come back to this when discussing the rotational Goldstone modes of the quantum nematic in Sec.~\ref{subsec:Torque stress in the quantum nematic}.

\begin{figure*}
 \subfloat[edge dislocation]{\includegraphics[height=2.7cm]{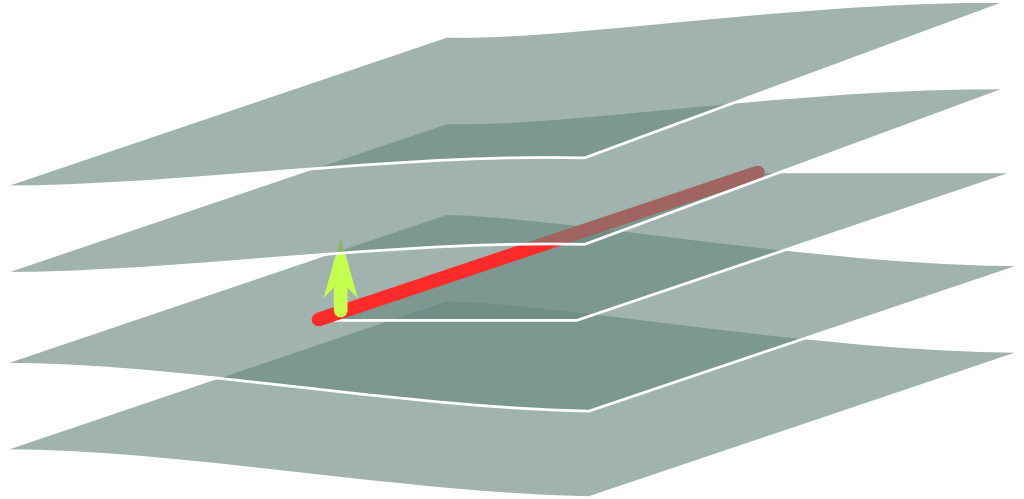}\label{subfig:edge dislocation}}
 \hfill
 \subfloat[screw dislocation]{\includegraphics[height=2.7cm]{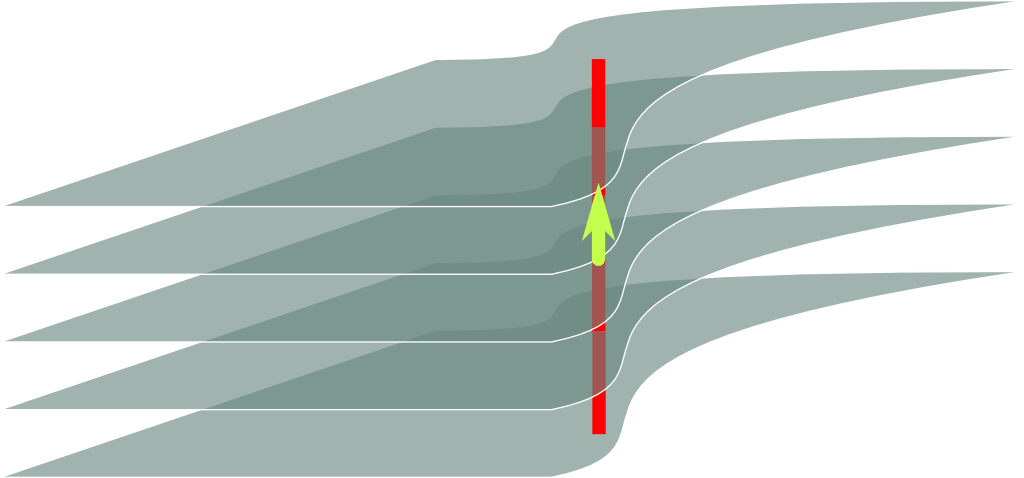}\label{subfig:screw dislocation}}
 \hfill
  \subfloat[wedge disclination]{\includegraphics[height=2.7cm]{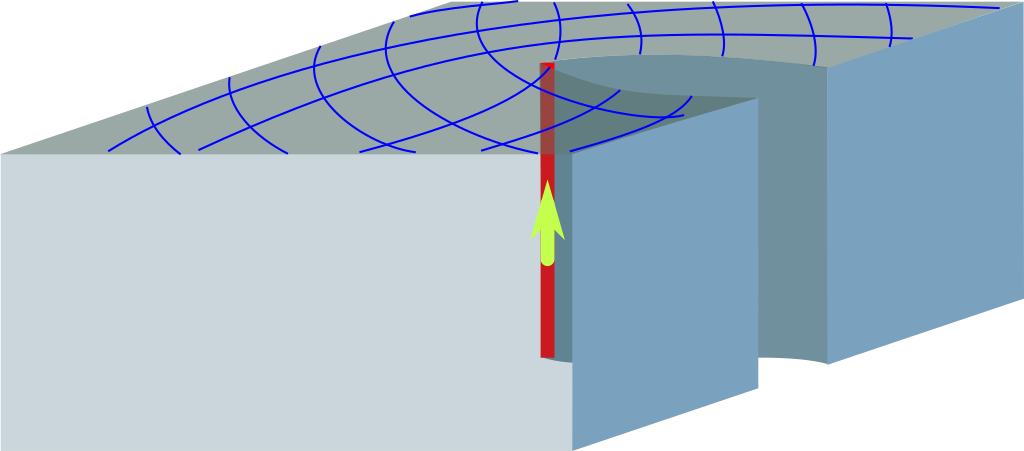}\label{subfig:wedge disclination}}
 \caption{Elementary topological defects in solids. The defects are always linelike, indicated by the red line. The topological charge, Burgers vector for dislocations and Frank vector for disclinations, is indicated by an arrow. \protect\subref{subfig:edge dislocation} Edge dislocation, the Burgers vector (light green arrow) is perpendicular to the dislocation line (red).  \protect\subref{subfig:screw dislocation} Screw dislocation, the Burgers vector is parallel to the dislocation line.  \protect\subref{subfig:wedge disclination} Wedge disclination, with Frank vector parallel to the disclination line. }\label{fig:topological defects}
\end{figure*}

\subsection{Dislocations and disclinations}\label{subsec:Dislocations and disclinations}

The very idea of topological melting/weak-strong duality is that the (quantum) phase transition from the solid to the (quantum) liquid corresponds to a proliferation of free topological defects.  
In general topological defects depend on the topological structure of order parameter space, which is the coset $G/H$ where $G$ is the symmetry group of the Lagrangian, and $H \subset G$ is the subgroup of unbroken symmetries. Topological defects can come in any dimensionality that is lower than the dimension of space. For instance, in three space dimensions, there are two-dimensional defects (like domain walls), one-dimensional line defects and zero-dimensional point defects. These $\bar{D}$-dimensional defects are classified by the $(D-1 -\bar{D})$th homotopy group of the space $G/H$ denoted by $\pi_{D-1-\bar{D}}(G/H)$. 
When dealing with the breaking of the Euclidean group $G = E = \mathbb{R}^D \rtimes O(D)$ to a discrete subgroup (space group) $H = \mathbb{Z}^D \rtimes \bar{P}$, one finds $\pi_2\big((\mathbb{R}^D \rtimes O(D))/(\mathbb{Z}^D \rtimes \bar{P})\big) \simeq \pi_1(\mathbb{Z}^D \rtimes \bar{P}) \simeq \pi_0(1) \simeq 1$~\cite{Mermin79} (since $O(D)$ is not simply connected, the actual homotopy group sequence is slightly different, but in any case as long as $H$ is discrete $\pi_0 \simeq 1$). In other words, topological monopoles do not occur. The $\pi_1$-defects are the dislocations and disclinations~\cite{Mermin79,Kleinert89b,ChaikinLubensky00,SinghDunmur02,QLC2D}.  Dislocations are the defects associated with the translational symmetry breaking. These can be pictured as inserting or removing a half-plane of material (this mental cutting and gluing procedure is called {\em Volterra process}). Traversing a contour around the dislocation core will result in a deficient lattice vector, which is called the {\em Burgers vector} $B^a$. It is the topological charge of the dislocation, since it does not depend on the details of the contour. In three dimensions, the dislocation is a line defect. When the Burgers vector is orthogonal to this line, it is called an {\em edge dislocation}, whereas a defect line parallel to the Burgers vector is called a {\em screw dislocation}, see Fig.~\ref{fig:topological defects}. Later, we consider closed dislocation loops of constant Burgers vector, which must be of edge-type somewhere along the loop.

The defects associated with rotational symmetry breaking are called {\em disclinations} and the Volterra process consists of inserting or removing wedges of material, which lead to deficient rotations, see Fig.~\ref{subfig:wedge disclination}. The magnitude is characterized by a deficit angle $\Omega$ and the topological charge is a tensor normal to the plane of rotation with $D-2$ indices called the {\em Frank tensor}. In 3D this becomes the Frank vector $\Omega^c$. If the Frank vector is parallel to the disclination line, it is called a {\em wedge disclination}, and otherwise a {\em bend} or {\em twist disclination}~\cite{Kleinert89b,ChaikinLubensky00,SinghDunmur02}. In 3D, the definitions of the Burgers and Frank vectors are:
\begin{align}
 B^a 
    &= \oint_{\partial \mathcal{S}} \td x^m \; \partial_m u^a,\label{eq:Burgers vector definition}\\
 \Omega^c
    &= \oint_{\partial \mathcal{S}} \td x^m \; \partial_m \tfrac{1}{2}\epsilon^{cab} \omega^{ab}. \label{eq:Frank vector definition}
\end{align}
Here ${\partial \mathcal{S}}$ is an arbitrary closed contour encircling the core of the topological defect, such that the surface area $\mathcal{S}$ enclosed by $\partial \mathcal{S}$ is pierced by the defect line. An arbitrary defect  can 
have both translational and rotational character, but this can always be decomposed into multiple elementary defects that are pure dislocations and disclinations. Furthermore, dislocations and disclinations are not independent, see Eq.~\eqref{eq:dislocation disclination interdepence} below. One of the consequences is that a disclination--anti-disclination pair is not topologically trivial but equivalent to a dislocation line~\cite{Kleinert89b,QLC2D}. 
In fact, upon pulling apart such a disclination-antidisclination pair in the solid the energy increases with the square of the distance: disclinations are `quadratically confined' in the solid, whereas these dipole pairs attain finite energy in their separation and `deconfine' in the (quantum) nematic.  The precise (de)confinement mechanism is a highlight of the weak--strong duality in the 2+1D case~\cite{BeekmanWuCvetkovicZaanen13,QLC2D}, and in Secs.~\ref{subsec:Torque stress gauge fields}, \ref{subsec:Torque stress in the quantum nematic} we will see that it 
also applies to 3+1D. Notice that although the gross physical meaning of (de)confinement is similar to that found in non-Abelian Yang--Mills theory, it appears to be due to a different mechanism.  
  
Turning to  the 3+1D quantum theory, the spatial dislocation/disclination loops on a time slice turn into worldsheets in spacetime. In our quantum field theoretical setting, 
these topological defects take the form of non-critical (Nielsen--Olesen) bosonic strings because we depart from solids formed from bosonic constituents. These 
in turn interact with each other via the excitations of the ordered background, the phonons, and these long-range interactions can be represented by effective gauge fields as we will discuss in great detail in Sec.~\ref{subsec:Stress gauge fields}.
On the time slice, the density of these strings can be enumerated in terms of defect density fields, defined by
\begin{align}
 J^a_n (x) &= \epsilon_{nkl} \partial_k \partial_l u_a(x) = \delta_n (L,x) B^a,\label{eq:dislocation density definition}\\
 \Theta^c_n (x) &= \epsilon_{nkl} \partial_k \partial_l  \tfrac{1}{2} \epsilon^{cab} \omega^{ab}(x) = \delta_m (L,x) \Omega^c.
\end{align}
for the dislocations ($J$) and disclinations ($\Theta$). The displacement fields  $u^a$ and $\omega^{ab}$ are singular at the core of the defects and become multivalued fields with non-commuting partial derivatives~\cite{Kleinert89b,Kleinert08}. Eqs.~\eqref{eq:Burgers vector definition}, \eqref{eq:Frank vector definition} can be retrieved by integrating these densities over the surface $\mathcal{S}$ and using Stokes' theorem. On the right-hand side, we use the definition of the delta function on the defect line $L$ parametrized by $s$, given by~\cite{Kleinert89b}:
\begin{equation}\label{eq:line delta function}
 \delta_n (L,x) = \int_L \td s\ \partial_{s} x^L_n(s)  \delta^{(D)}\big(x - x_k^L(s) \big).
\end{equation}

There is a dynamical constraint acting on the motion of edge dislocations. In a crystal they only move in the direction of their Burgers vector, and this is called {\em glide motion}~\cite{Friedel64,CvetkovicNussinovZaanen06}. The reason is that this motion is only a rearrangement of constituent particles, whereas motion orthogonal to the Burgers vector ({\em climb motion}) entails addition or removal of interstitial particles. In real crystals these are energetically very costly
and accordingly their density is small at not too high temperatures. In quantum crystals at zero temperature these occur only as virtual fluctuations involving a finite energy scale with the effect that the glide constraint becomes absolute in the deep IR. In fact, our limit of ``maximal crystalline correlations'' can be viewed as being equivalent to the demand that such constituent particles are infinite-energy excitations.  The precise formulation of this {\em glide constraint} will be given in Eq.~\eqref{eq:glide constraint} after we have discussed in more detail the nature of the dislocation worldsheets. 

%% file: sec_dualelasticity.tex
In this section we develop the description of a 3+1D quantum solid in terms of dual variables: the stresses $\sigma^a_\mu$ and the two-form stress gauge fields $b^a_{\mu\nu}$. This follows closely the development in  2+1D as outlined in QLC2D, but the novelty in 3+1D are the two-form gauge fields, see Sec.~\ref{subsec:Two-form gauge fields}. Additionally, the number of elastic degrees of freedom is larger, with the effect that the expressions get more elaborate. We will first obtain the explicit dual action for the isotropic solid, to then proceed to re-derive the phonon propagators using dual variables only. Most importantly, the dual formalism can then be directly applied to the quantum liquid crystals in the later sections.

\subsection{Dislocation worldsheets}\label{subsec:Dislocation worldsheets}

\begin{figure}
 \includegraphics[width=7cm]{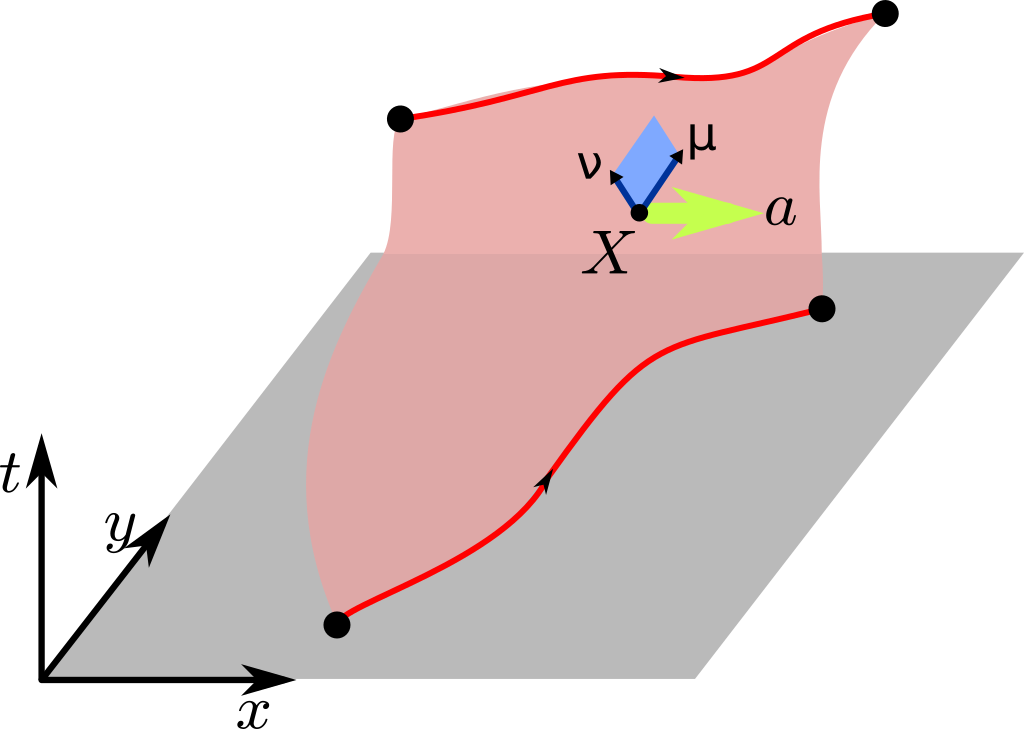}
 \caption{Section of a dislocation worldsheet, depicting a dislocation line (red) moving in time. In blue is represented the surface element $J^a_{\mu\nu}(X)$ at point $X$ with Burgers vector $B^a$.}\label{fig:dislocation worldsheet}
\end{figure}

Before we dualize the action of the solid, let us first discuss dislocation lines in the imaginary time setting of the quantum theory. The dislocation density $J^a_n(x)$ in Eq.~\eqref{eq:dislocation density definition} is a static quantity. In spacetime, the dislocation line along $n$ with Burgers vector $a$ can move in direction $\mu$ where $\mu$ contains both temporal and spatial components. The dislocation line $J^a_n$ traces out a {\em worldsheet} $J^a_{\mu\nu}$ in spacetime, see Fig.~\ref{fig:dislocation worldsheet}. The density of the line is represented by $J^a_{tn} = J^a_n$ and the flow or current in direction $m$ of the line along $n$ is represented by $J^a_{mn}$. The worldsheet element $J^a_{\mu\nu}(x)$ at $x$ is a two-form quantity in the differential geometry sense, cf. Sec.~\ref{subsec:Two-form gauge fields}, and is antisymmetric in its indices $\mu,\nu$. The {\em dislocation worldsheet element} is defined by
\begin{equation}\label{eq:dislocation current definition}
 J^a_{\mu\nu} (x) =  \epsilon_{\mu\nu\kappa\lambda} \partial_\kappa \partial_\lambda\tfrac{1}{2} u^a(x).
\end{equation}
Here $\epsilon_{\mu\nu\kappa\lambda}$ is the completely antisymmetric Levi-Civita symbol and $\epsilon_{txyz} = 1$. This is the 3+1D generalization of Eq.~\eqref{eq:dislocation density definition}. Note that the Burgers vector $a$ is always spatial, since fields are always smooth in the time direction. Therefore Lorentz symmetry is still badly broken; close to the critical point, we are at most dealing with an emergent relativistic theory where the `speed of light' is actually a material speed such as the phonon velocity $c_\tT$. We will  call $J^a_{\mu\nu}$ the {\em dislocation current} that couples to the dual stress gauge field $b^a_{\mu\nu}$ defined below, in analogy with the particle current $j_\mu$ sourcing a vector gauge field $A_\mu$ in Maxwell electrodynamics. By definition $J^a_{t n}$ is an edge dislocation if $a \neq n$ and a screw dislocation if $a = n$. Note that because of the antisymmetry in the lower indices $J^a_{an}$ (no sum) equivalently represents the current in direction $a$ of an edge dislocation or the opposite of the current in direction $n$ of a screw dislocation.

In the bulk of the solid and in absence of disclinations the dislocation current is conserved:
\begin{equation}\label{eq:dislocation conservation}
 \partial_\mu J^a_{\mu\nu} = 0 \quad \forall a,\nu.
\end{equation}
This implies that a dislocation line cannot begin or end in the material: dislocation lines must be closed loops, and dislocation worldsheets must be closed surfaces.
 The worldsheet picture of defect lines can be very useful even in condensed matter physics. For instance, we derived all important dynamical electromagnetic effects of Abrikosov vortices in superconductors by regarding them as worldsheets in Ref.~\onlinecite{BeekmanZaanen11}.

In Sec.~\ref{subsec:Dislocations and disclinations} we mentioned the {\em glide constraint} which states that edge dislocations can only move in the direction of their Burgers vector. The precise statement in terms of the dislocation currents was derived in Ref.~\onlinecite{CvetkovicNussinovZaanen06}:
\begin{equation}\label{eq:glide constraint}
 \epsilon_{tamn} J^a_{mn} = 0.
\end{equation}
Recall that $J^a_{mn}$ is the current in direction $m$ of the dislocation line along $n$ with Burgers vector $a$. Eq.~\eqref{eq:glide constraint} states that edge dislocations ($a\neq n$) cannot move in the direction orthogonal to the Burgers vector. For screw dislocations ($a = n$) there is no such constraint: their motion perpendicular to the dislocation line does not involve addition or removal of constituent particles. The glide constraint is in fact a consequence of conservation of particle number. This can be seen by inserting the definition Eq.~\eqref{eq:dislocation current definition}:
\begin{align}
 0 = \epsilon_{tamn} \epsilon_{mn\kappa\lambda} \partial_\kappa \partial_\lambda u^a(x) = (\partial_\ft \partial_a - \partial_a \partial_\ft) u^a = 0.
\end{align}
To lowest order, fluctuations of the mass density $\rho(x)$ are $\rho_0 \partial_a u^a$ while the mass current is $j_a(x) = \rho_0 \partial_t u^a$. Thus the glide constraint is equivalent to the conservation law $\partial_\ft \rho + \partial_a j_a =0$~\cite{CvetkovicNussinovZaanen06,QLC2D}. The glide constraint is active during dislocation condensation, and amazingly turns out to protect the compression mode, in turn related to the conservation law, from obtaining a dual Higgs mass, see Sec.~\ref{sec:Dislocation-mediated quantum melting}. 

There is a similar generalization for the disclination worldsheet, defined by
\begin{equation}\label{eq:disclination worldsheet}
  \Theta^c_{\mu\nu} (x) = \epsilon_{\mu \nu \kappa\lambda } \partial_\kappa \partial_\lambda  \tfrac{1}{2} \epsilon^{cab} \omega^{ab}(x).
\end{equation}
The interpretation is that $\Theta^c_{tn}(x)$ is the disclination density at $x$ of the line along $n$ with rotational plane orthogonal to $c$, while $\Theta^c_{mn}(x)$ is the flow or current of that line in direction $m$. For $\Theta^c_{tn}(x)$, if $n = c$ it is the density of a segment of a {\em wedge} disclination while it is the density of an {\em twist} disclination if $n \neq c$. A closed disclination line will typically be of wedge or twist nature at different positions. 

In the presence of disclinations, the dislocation current is no longer conserved. Instead of Eq.~\eqref{eq:dislocation conservation} we have~\cite{Kleinert89b,QLC2D}
\begin{equation}\label{eq:dislocation disclination interdepence}
 \partial_\mu J^a_{\mu\nu} = \epsilon_{\ft abc} \Theta^c_{b\nu}.
\end{equation}
This equation implies that a disclination line can source dislocations. If $\nu = \ft$, then the left-hand side is the divergence of dislocation density, which can only be non-zero if the dislocation line ends. In other words, a dislocation line can end on a static twist disclination line. If $\nu = n$, the right-hand side denotes the current or flow of a disclination line. Then this equation implies that a moving disclination leaves dislocations in its wake~\cite{QLC2D,KlemanFriedel08}. This equation is a consequence of the fact that translations and rotations are not independent. In the space group the point group operations (including rotations) are in semidirect relation with the translations; locally, a rotation is equivalent to two {\em finite} translations, which in topological context turns into the statement that disclinations can be formed from a finite density of dislocations with equal Burgers vector.  As we will discuss in more detail later, the liquid crystals can be topologically defined
by insisting that disclinations are massive (i.e. absent) which in turn implies that even locally the Burgers vectors have to be antiparallel in the dislocation condensate, since a finite `Burgers vector magnetization' is the same as a finite disclination density in the vacuum.

Since we are treating temporal and spatial dimensions on the same footing, a velocity is needed to compare quantities with different units. Certainly, we are in the idealized limit devoid of interstitials, disorder and other influences that could 
dissipate the motion of the phonons and the topological defects.  Everything moves ballistically without scattering or drag. In the vortex--boson duality~\cite{Franz07,BeekmanSadriZaanen11}, upon approaching the quantum critical point, scale invariance sets in as well as emergent Lorentz invariance. There can be only one velocity governing both sides of the phase transition: the velocity associated with the vortices  coincides precisely with the phase velocity of the superfluid. In the same vein, the velocity associated with the defect-condensate is also the phase velocity. In the present context of elasticity, the same argument holds in principle, but there is a complication in the form of the glide constraint which restricts the motion of edge dislocations. Since screw dislocations do not suffer from this, Friedel argued that the dislocation speed should equal the material speed, given by the transverse or shear velocity $c_\tT$~\cite{Friedel64}. In real-world solids, there is some evidence that edge dislocations can move `transonically' with speeds up to the longitudinal velocity $c_\tL$, see e.g. Ref.~\onlinecite{RuestesEtAl15}. A difficulty is that the arguments for emergent Lorentz invariance become precise near the 
continuous quantum phase transition, while deep in the solid `irrelevant' operators such as the nature of the chemical bond may become  important; e.g. dislocations in covalent solids are immobile while simple metals are malleable 
because of the rather isotropic nature of their electronic binding forces. Weak--strong dualities acquire their universal meaning in any case only close to the continuous quantum phase transition where one should become insensitive 
to microscopic details. In  principle, the scale of the characteristic velocity of the dislocations and the dislocation condensate is therefore assumed to be the transverse phonon velocity and the only complication arises from the glide
constraint.  This also suggests that the velocity of edge dislocations $c_\mathrm{e}$ and of screw dislocations $c_\mathrm{s}$ can a priori be different. 

In QLC2D we found that it is actually very helpful to treat the velocity of dislocations as different from the shear velocity, since it enables one to track the degrees of freedom originating in the dislocation condensate. On the other hand, below in Sec.~\ref{subsubsec:Dislocation worldsheet condensation} we will find that differing velocities for edge and screw dislocations considerably complicate the computations. Although our formalism is able in principle to handle the general case, with the exception of the end of Sec.~\ref{subsubsec:Dislocation worldsheet condensation} we will set $c_\mathrm{e} = c_\mathrm{s} \equiv c_\mathrm{d}$ as the uniform dislocation velocity, which in turn should be of the order of the shear velocity $c_\tT$.

\subsection{Two-form gauge fields}\label{subsec:Two-form gauge fields}

Given  a conserved current in four dimensions $j_\mu$,  the associated conservation law (continuity equation) $\partial_\mu j_\mu = 0$ can be imposed by expressing the current as the four-curl of a two-form gauge field $b_{\kappa\lambda}$ (sometimes called Kalb--Ramond field~\cite{KalbRamond74})
\begin{equation}\label{eq:two-form gauge field definition}
 j_\mu(x)  = \epsilon_{\mu\nu\kappa\lambda} \partial_\nu \tfrac{1}{2} b_{\kappa\lambda} (x).
\end{equation}
 The only relevant components of $b_{\kappa\lambda}$ are those antisymmetric in $\kappa,\lambda$, and this defines an antisymmetric two-form field. By expressing physical quantities in terms of $b_{\kappa\lambda}$ the constraint $\partial_\mu j_\mu=0$ is thereby identically satisfied. In addition, the field $b_{\kappa\lambda}$ is a gauge field since the addition of the gradient of an arbitrary smooth vector field $\varepsilon_\lambda(x)$ leaves the current $j_\mu$ in Eq.~\eqref{eq:two-form gauge field definition} invariant:
\begin{equation}\label{eq:two-form gauge transformations}
 b_{\kappa\lambda} (x) \to b_{\kappa\lambda}(x) + \partial_\kappa  \varepsilon_\lambda(x) - \partial_\lambda  \varepsilon_\kappa(x).
\end{equation}

Let us now focus on the counting of the physical, propagating degrees of freedom of a two-form gauge field. First, recall that the number of propagating degrees of freedom of a one-form gauge field $A_\mu(x)$ (e.g. the photon field in electromagnetism) equals the number of spatially transverse components. In $D$ space dimensions, there is one longitudinal and $D-1$ transverse directions. Accordingly,  such a one-form field represents one physical `photon'
in 2+1 dimensions and two such degrees of freedom in 3+1 dimensions. A two-form gauge field has instead two indices, and the components representing physical, propagating degrees of freedom (`photons') are the ones for which both indices are transverse. In 3+1 dimensions, there are two transverse directions, and since the field is antisymmetric in its two indices, there is a only one independent component with purely transverse indices. In $D$ space dimensions, 
the number of propagating  physical degrees of freedom is 
\begin{equation}
 \begin{pmatrix} D-1 \\ 2 \end{pmatrix} = \frac{(D-1)!}{2! (D-3)!} = \frac{(D-1)(D-2)}{2}.
 \end{equation}
This result is only valid for $D \ge 3$. In particular, in 3+1 dimensions, the field $b_{\kappa\lambda}$ represents only a single physical `photon'. As we will see later, each phonon independently dualizes into a two-form gauge field, 
corresponding to one such a physical propagating `stress photon'. 

 Another way to establish the correct number of degrees of freedom is to consider the gauge volume due to the gauge transformations Eq.~\eqref{eq:two-form gauge transformations}. The field $b_{\mu\nu}$ in 3+1 dimensions has six independent components due to its antisymmetry. However, the components that transform under gauge transformations do not correspond to physical degrees of freedom, and can be removed by a suitable gauge fixing. Since adding the gradient of an arbitrary scalar field $\varepsilon_\mu(x) \to \varepsilon_\mu(x) + \partial_\mu \xi(x)$ leads to the exact same gauge transformation Eq.~\eqref{eq:two-form gauge transformations}, there is a redundancy within the specification of $\varepsilon_\mu$. This is sometimes called ``gauge-in-the-gauge''. According to the theory of constraints in dynamical systems~\cite{HenneauxTeitelboim92}, the gauge transformation field $\varepsilon_\mu$ consists of three gauge components $\varepsilon_m$ and their time derivatives to function as 6 gauge parameters to be used for gauge fixing. However, due to the redundancy just mentioned we can fix only $6-1=5$ components, leaving one component as a propagating degree of freedom. 

\subsection{Stress gauge fields}\label{subsec:Stress gauge fields}

The original displacement field $u^a$ still features in Eq.~\eqref{eq:dual action} as a result of the Legendre transformation~\cite{QLC2D}. In order to eliminate it in favor of the stresses, we need to perform the path integral over this field by properly taking into account the defects. Topological defects are singularities in the displacement field, see Sec.~\ref{subsec:Dislocations and disclinations}. Therefore we split $u^a$ in a smooth and a singular part:
\begin{equation}\label{eq:displacement field smooth singular}
 u^a(x) = u^a_\mathrm{smooth} + u^a_\mathrm{sing}.
\end{equation}
On the smooth part we are allowed to perform integration by parts, to subsequently integrate out $u^a_\mathrm{smooth}$  in the path integral as a Lagrange multiplier for the constraint 
\begin{equation}\label{eq:stress conservation}
 \partial_\mu \sigma^a_\mu=0.
\end{equation}
This constraint corresponds with the conservation of stress --- external stresses will appear on the RHS. This can be implemented 
by expressing the stress tensor as the four-curl of a two-form gauge field, see Sec~\ref{subsec:Two-form gauge fields}:
\begin{equation}\label{eq:dual stress gauge field definition}
 \sigma^a_\mu  = \epsilon_{\mu\nu\kappa\lambda} \partial_\nu \tfrac{1}{2} b^a_{\kappa\lambda}.
\end{equation}
At this point we specialized to 3+1 dimensions; in 2+1D one would be dealing instead with one-form stress gauge fields, see QLC2D. We call $b^a_{\mu\nu}$ the {\em dual stress gauge field}. As before,  $b^a_{\kappa\lambda}$ is antisymmetric in $\kappa \leftrightarrow \lambda$, 
while the factor of $\tfrac{1}{2}$ is inserted for later convenience. 
 
The stress tensor $\sigma^a_\mu$ (the `field strength') is invariant under the set of gauge transformations, cf. Eq.~\eqref{eq:two-form gauge transformations},
\begin{equation}\label{eq:stress gauge transformation}
  b^a_{\kappa\lambda} (x) \to b^a_{\kappa\lambda}(x) + \partial_\kappa  \varepsilon^a_\lambda(x) - \partial_\lambda  \varepsilon^a_\kappa(x).
\end{equation}
Here $\varepsilon^a_\lambda$ are three independent arbitrary vector fields, one for each $a = x,y,z$. Transforming to the Fourier-Matsubara coordinates $(0,1,\tR,\tS)$ for the spacetime indices (see appendix~\ref{sec:Fourier space coordinate systems}), the components $\sigma^a_0$ are removed by stress conservation. Furthermore, the remaining components are in one-to-one relation with the stress gauge field components: 
\begin{align}
\sigma^a_\ft &= q b^a_{\tR\tS}, &
\sigma^a_\tL &= \ti \frac{\omega_n}{c_\tT} b^a_{\tR \tS}, &
\sigma^a_\tR &= p\; b^a_{1 \tS}, &
\sigma^a_\tS &= p\; b^a_{1\tR}. \label{eq:stress gauge field components}
\end{align}
The reason is that the components $(\tR\tS)$, $(1\tR)$, $(1\tS)$ are gauge-invariant, while all other components $(01)$, $(0\tR)$, $(0\tS)$ are pure gauge and do not contribute the stress tensor in Eq.~\eqref{eq:dual stress gauge field definition}. This is equivalent to imposing the Lorenz gauge fix $\partial_\mu b^a_{\mu\nu} = 0\ \forall\, \nu,a$.

We still have to deal with the singular displacement field $u^a_\mathrm{sing}$. Using the definition Eq.~\eqref{eq:dual stress gauge field definition} we derive, 
\begin{align}\label{eq:stress dislocation minimal coupling}
 \ti \sigma^a_\mu \partial_\mu u^a_\mathrm{sing} &=   \ti (\epsilon_{\mu\nu\kappa\lambda} \partial_\nu \tfrac{1}{2} b^a_{\kappa\lambda})( \partial_\mu u^a_\mathrm{sing})\nonumber\\
 &=  \ti  b^a_{\kappa\lambda}\tfrac{1}{2}\epsilon_{\mu\nu\kappa\lambda} \partial_\mu \partial_\nu u^a_\mathrm{sing} =  \ti b^a_{\kappa\lambda} J^a_{\kappa\lambda}.
\end{align}
Here we performed integration by parts on $b^a_{\kappa\lambda}$ which is smooth everywhere, and used the definition of the dislocation worldsheet Eq.~\eqref{eq:dislocation current definition}. 
It follows that the topological defects source the dual gauge fields, though these sources are now worldsheets the gauge fields themselves are of the two-form kind. In 2+1D 
dislocations are instead `particles' (worldlines in 2+1D) and just source the stress gauge fields in the same way as electrically charged particles source electromagnetic gauge fields.

As discussed at length in QLC2D, the essence of stress--strain duality is that one translates the way that the elastic medium vibrates (phonons) into its capacity to propagate forces between the dislocations which are the natural 
internal sources of stress. In the language of forces, the theory just takes the shape of a gauge theory: in this sense, phonons turn into literal (stress) photons. This stress gauge theory formulation is highly convenient in the
further development of the duality since it revolves around matter formed from the dislocations, and the gauge theory is the easy way to describe the interactions between the dislocations. In 2+1D quantum elasticity turns 
into a `flavored' form of  electromagnetism, but that is a coincidence characteristic for this dimension. A scalar field theory in 3+1D is not dual to a one-form gauge field (electromagnetism); instead we find the dual two-form gauge theory. Furthermore we just learned that it counts the physical degrees of freedom correctly (``1 phonon $=$ 1 stress photon'') at the same time doing justice to the fact that the sources of internal stress are worldsheets, Eq.~\eqref{eq:stress dislocation minimal coupling}.  
For systems with one broken symmetry generator (like the broken $U(1)$-symmetry of superfluid), there is a single two-form field, as explained in Sec.~\ref{subsec:Two-form gauge fields}. In solids, there are $D$ broken generators when breaking $\mathbb{R}^D \to \mathbb{Z}^D$, and accordingly there are $D$ phonons turning into $D$ `flavors' of two-form gauge fields. This goes hand-in-hand with the topological charge of dislocations, the Burgers vector $B^a$, which is clearly not simply an integer winding number. The symmetry between the flavor index $a$ and the spacetime indices $\kappa$, $\lambda$ in $b^a_{\kappa\lambda}$ is far from perfect, and in fact it is often useful to regard them as completely separated. However, as can be inferred from Eq.~\eqref{eq:dual solid Lagrangian}, there is some `mixing' between the Burgers and spacetime `sectors'. 

Finally, let us rewrite the partition sum Eq.~\eqref{eq:dual partition sum} in terms of the dual gauge fields:
\begin{align}
 \mathcal{Z}_\mathrm{solid} = \int \mathcal{D}b^a_{\kappa\lambda} \mathcal{D} J^a_{\kappa\lambda} \mathcal{F}(b^a_{\kappa\lambda})\; \te^{-\int \td \tau \td^3x\; \mathcal{L}_\mathrm{dual}}.
\end{align}
Here $\mathcal{F}(b^a_{\kappa\lambda})$ is a gauge-fixing factor enforcing for instance a Lorenz gauge fix $\mathcal{F}(b^a_{\mu\nu}) = \delta(\partial_{\mu} b^a_{\mu\nu})$. At this stage of the development,
the action $\mathcal{L}_{\rm dual}$ describes describes the quantum mechanics of a isolated dislocations and 
 $\mathcal{D} J^a_{\kappa\lambda}$ denotes the sum over worldsheet elements of closed worldsheets associated with dilute dislocation--antidislocation loops on the time slice. The summation over the dislocation worldsheets in the path-integral cannot be performed and the action $\mathcal{L}_{\rm dual}$ should not be confused with the description of a dense system (`foam' in spacetime) of dislocations which is the subject of string field theory

 More specifically, $\mathcal{L}_\mathrm{dual}$ is the sum of Eqs.~\eqref{eq:solid stress Lagrangian components} and \eqref{eq:second gradient Lagrangian in stress tensor} in which we have substituted Eq.~\eqref{eq:stress gauge field components} to obtain the contributions:
\begin{widetext}
\begin{align}
\mathcal{L}_\mathrm{dual} &= \mathcal{L}_{\tT 1} + \mathcal{L}_{\tT 2} + \mathcal{L}_{\tT 3} + \mathcal{L}_{\tL 1} + \mathcal{L}_{\tL 2} + \ti b^{a\dagger}_{\kappa\lambda} J^a_{\kappa\lambda},\label{eq:solid stress gauge field Lagrangian}\\
 \mathcal{L}_{\tT 1} &= \frac{1}{8\mu}
\begin{pmatrix}b^{\tL\dagger}_{1 \tS} &  b^{\tR\dagger}_{\tR\tS} \end{pmatrix}
\begin{pmatrix} 
 p^2(1 + \frac{1}{\ell^2 q^2}) & \ti \frac{1}{c_\tT}\omega_n p(1 - \frac{1}{\ell^2 q^2}) \\
 -\ti \frac{1}{c_\tT}\omega_n p(1 - \frac{1}{\ell^2 q^2}) & \frac{1}{c_\tT^2}\omega_n^2(1 + \frac{1}{\ell^2 q^2}) + 4 q^2 
 \end{pmatrix}
\begin{pmatrix}b^{\tL}_{1 \tS} \\  b^{\tR}_{\tR\tS} \end{pmatrix},\label{eq:L1} \\
\mathcal{L}_{\tT 2} &= \frac{1}{8\mu} 
\begin{pmatrix}b^{\tL\dagger}_{1 \tR} &  b^{\tS\dagger}_{\tR\tS} \end{pmatrix}
\begin{pmatrix} 
 p^2(1 + \frac{1}{\ell^2 q^2}) & \ti \frac{1}{c_\tT}\omega_n p(1 - \frac{1}{\ell^2 q^2}) \\
 -\ti \frac{1}{c_\tT}\omega_n p(1 - \frac{1}{\ell^2 q^2}) & \frac{1}{c_\tT^2}\omega_n^2(1 + \frac{1}{\ell^2 q^2}) + 4 q^2 
 \end{pmatrix}
\begin{pmatrix}b^{\tL}_{1 \tR} \\  b^{\tS}_{\tR\tS} \end{pmatrix}, \\
\mathcal{L}_{\tT 3}&= \frac{1}{8\mu}
\begin{pmatrix}b^{\tR\dagger}_{1 \tR} & b^{\tS\dagger}_{1\tS} \end{pmatrix}
\begin{pmatrix} 
 p^2(1 + \frac{1}{\ell^2 q^2}) & p^2(1 - \frac{1}{\ell^2 q^2}) \\
 p^2(1 - \frac{1}{\ell^2 q^2}) & p^2(1 + \frac{1}{\ell^2 q^2})
 \end{pmatrix}
\begin{pmatrix}b^{\tR}_{1 \tR} \\  b^{\tS}_{1 \tS} \end{pmatrix},\label{eq:L3}\\
\mathcal{L}_{\tL 1} &= \frac{1}{8\mu}  \frac{2}{1+\nu}
\begin{pmatrix}b^\dagger_{1 -} &  b^{\tL \dagger}_{\tR\tS} \end{pmatrix}
\begin{pmatrix} 
 (1-\nu) p^2 &  \ti\sqrt{2} \nu \frac{1}{c_\tT}\omega_n p \\
  -\ti \sqrt{2}\nu \frac{1}{c_\tT}\omega_n p &  \frac{1}{c_\tT^2}\omega_n^2 + 2 (1 + \nu) q^2 
 \end{pmatrix}
\begin{pmatrix}b_{1 -}\\ b^{\tL }_{\tR\tS} \end{pmatrix}. \label{eq:solid Lagrangian gauge field L1}\\
\mathcal{L}_{\tL 2} &=  \frac{1}{4\mu}  p^2 \lvert b_{1+} \rvert^2\label{eq:solid Lagrangian gauge field L2}
\end{align}
\end{widetext}
Here $b^{a\dagger}_{\kappa\lambda} J^a_{\kappa\lambda}$ is a shorthand for $\frac{1}{2} b^{a\dagger}_{\kappa\lambda} J^a_{\kappa\lambda} + \frac{1}{2} J^{a\dagger}_{\kappa\lambda} b^a_{\kappa\lambda}$, and in analogy to Eq.~\eqref{eq:sigma plus minus definition} we have defined
\begin{align}
 b_{1+} &= \frac{1}{\sqrt{2}} (b^\tR_{1\tS} + b^\tS_{1\tR}), &
 b_{1-} &= \frac{1}{\sqrt{2}} (b^\tR_{1\tS} - b^\tS_{1\tR}).
\end{align}

\subsection{The interpretation of stress components}\label{subsec:Interpretation of stress components}

We will derive the propagators of Sec.~\ref{subsec:Field-theoretic elasticity} on the dual side. Let us however first find out  what can be learned regarding the spectrum of excitations from the Lagrangians Eqs.~\eqref{eq:solid stress Lagrangian components}, \eqref{eq:solid stress gauge field Lagrangian}. At first, there are 12 components $\sigma^a_\mu$ as $\mu = \ft,x,y,z$ and $a = x,y,z$. Let us first consider the static, 3+0D limit where the three temporal components are absent: $\sigma^a_\ft = 0$. Stress conservation $\partial_m \sigma^a_m = -q \sigma^a_\tL =0$ removes another three components. Furthermore we have the three Ehrenfest constraints $\sigma^a_m = \sigma^m_a$, so that the only physical components are $\sigma^\tR_\tS + \sigma^\tS_\tR, \sigma_+$ and $\sigma_-$, and these must be interpreted as static elastic `Coulomb' forces. From Eq.~\eqref{eq:solid stress Lagrangian components} we can see that the first two have a correlation function proportional to the shear modulus $\mu$ and correspond to shear forces, while in this limit the Lagrangian $\mathcal{L}_{\tL 1}$ reduces to
\begin{align}\label{eq:electric shear stress 3+0D}
 \mathcal{L}_{\tL 1} \text{(3+0D)} &= \frac{1}{4\mu} \frac{1 -\nu}{1 +\nu} \lvert \sigma_- \rvert^2.
\end{align}
This involves the compression modulus $\kappa$ through the Poisson ratio $\nu$, and this contribution represents the compressional force. In Ref.~\cite{ZaanenNussinovMukhin04}, in analogy to electromagnetism, this force was called `electric', related to longitudinal stress, while the other two are purely transverse and can be called `magnetic'. These equations reproduce the results of classical elasticity, see for instance Ref.~\onlinecite{Kleinert89b}. Note that the sectors $\mathcal{L}_{\tT 1}$ and $\mathcal{L}_{\tT 2}$ are completely absent in this limit.

What happens when we include quantum dynamics by adding $\sigma^a_\tau$? We still have three conservation laws and three Ehrenfest constraints: there are $12 -3 -3 = 6$ physical components. Three of these are the same elastic Coulomb forces as above although they get modified at finite energies, while the three new degrees of freedom are the propagating phonons, two transverse and one longitudinal. The conservation laws are now $\partial_\mu \sigma^a_\mu = 0$ as in Eq.~\eqref{eq:stress conservation}, so the components $\sigma^a_\ft$ and $\sigma^a_\tL$ together represent one phonon per $a$. Once the Ehrenfest constraints are lifted through adding second-order terms Eq.~\eqref{eq:second gradient Lagrangian in stress tensor}, three more forces are added, which fall off exponentially with length scale $\ell$. These are short-ranged, rotational forces, and can be classified as either longitudinal or transverse as explained in Sec.~\ref{subsec:Rotational elasticity}. Together, there are 9 physical degrees of freedom, three of which are the propagating phonons. The breakdown of this classification for the sectors of Eq.~\eqref{eq:solid stress Lagrangian components} is given in Table~\ref{table:stress degrees of freedom}.

\newlength{\oldcolsep}
\setlength{\oldcolsep}{\tabcolsep}
\setlength{\tabcolsep}{.5em}
\begin{table}
 \begin{tabular}{ccccc}
 \toprule
 sector & & phonon & shear force & rotational force \\
 \cline{1-1}  \cline{3-5} 
  $\tT 1$  & & transverse & -- & transverse \\
  $\tT 2$  & & transverse & -- & transverse \\
  $\tT 3$  & & -- & magnetic & longitudinal \\
  $\tL 1$  & & longitudinal & electric & -- \\
  $\tL 2$  & & -- & magnetic & -- \\
  \botrule
 \end{tabular}
\caption{Classification of stress degrees of freedom of the 3+1D isotropic solid into the five sectors identified in Eq.~\eqref{eq:solid stress Lagrangian components}. We distinguish three types: phonons, which are propagating degrees of freedom; shear forces which have a Coulomb-like nature; and rotational forces which are only present once the Ehrenfest constraints are softened and which fall off exponentially. }\label{table:stress degrees of freedom}
\end{table}
\setlength{\tabcolsep}{\oldcolsep}

\begin{table}
 \begin{tabular}{ccl}
 \toprule
  component & spin & interpretation \\
  \hline
  $-\sigma^\tL_\tL + \sigma^\tR_\tR - \sigma^\tS_\tS$ & (0,0) & longitudinal phonon \\
  \rule{0pt}{3ex}
  $\sigma^\tR_\tS -\sigma^\tS_\tR$                    & (1,0) & longitudinal rotational force \\
  $\sigma^\tR_\tL -\sigma^\tL_\tR$                    & (1,1) & transverse rotational force \\
  $\sigma^\tS_\tL -\sigma^\tL_\tS$                    & (1,-1) & transverse rotational force \\
  \rule{0pt}{3ex}
  $2\sigma^\tL_\tL + \sigma^\tR_\tR - \sigma^\tS_\tS$ & (2,0) & electric shear\\
  $\sigma^\tL_\tR + \sigma^\tR_\tL$                   & (2,1) & transverse phonon \\
  $\sigma^\tL_\tS + \sigma^\tS_\tL$                   & (2,-1) & transverse phonon \\
  $\sigma^\tR_\tR + \sigma^\tS_\tS$                   & (2,2) & magnetic shear\\
  $\sigma^\tR_\tS + \sigma^\tS_\tR$                   & (2,-2) & magnetic shear\\
  \botrule
 \end{tabular}
\caption{Helicity decomposition of the 3+0D stress tensor with respect to the momentum $\mathbf{q}$ which sets the longitudinal direction $\tL$, where it is understood that the phonons will emerge once dynamics is added. The components are not yet normalized as to not clutter the notation.}\label{table:stress helicity decomposition}
\end{table}

Even more insight can be gained by the so-called helicity decomposition of the stress tensor~\cite{Kleinert89b}. Again we restrict ourselves to 3+0D where the temporal components $\sigma^a_\ft$ are absent. However, we will keep around the longitudinal components $\sigma^a_\tL$ which are subject to stress conservation, because we know these will contain the phonons once we switch on the time axis. The stress tensor $\sigma^a_m$ can now be decomposed by regarding its behavior under spatial 3-rotations. Since the rotation group generates the angular momentum operators, this classification can be assigned `spin quantum numbers' $(s,m)$. Because $\sigma^a_m$ is a 2-tensor, the `total spin' number $s$ takes values in $0,1,2$, while the `magnetic' number takes values in $(-s, \ldots, s)$. The $s$-sectors where also employed in Eqs.~\eqref{eq:P0}--\eqref{eq:P2}. Performing the helicity decomposition, one ends up with Table~\ref{table:stress helicity decomposition}.

The single spin-0 component is invariant under rotations; this is clearly the compressional, longitudinal phonon. The three spin-1 components are absent when the Ehrenfest constraint is imposed, so these correspond to the rotational forces. In the spin-2 sector, the $m=\pm 1$-components are the two transverse phonons, while the other three are the shear Coulomb forces. The $s=2,m=0$-component is the `electric shear' related to longitudinal stress, while the $m= \pm 2$-components are the `magnetic' shear forces. In a forthcoming article, we will show that this spin-2 character becomes very apparent when an elastic medium is coupled to (linearized) gravity.

Up to know we have considered classification of the stress tensor components. As we already emphasized, a key insight of vortex--boson duality is that the dual gauge fields themselves have a direct physical meaning as the mediators of interactions between the topological defects. This can be seen by looking at the Lagrangian Eq.~\eqref{eq:solid stress gauge field Lagrangian}, which is of the form
\begin{equation}\label{eq:general dual Langrangian Coulomb phase}
 \mathcal{L}_\mathrm{dual} = (\partial b )^2 + \ti b \cdot J.
\end{equation} 
reducing to the Maxwell action of electromagnetism in 2+1D~\cite{CvetkovicZaanen06a,QLC2D} while in 3+1D one runs into the two-form gauge theory which we showed elsewhere to be 
very convenient dealing with interactions between superfluid and Abrikosov vortices~\cite{BeekmanSadriZaanen11,BeekmanZaanen11}.  It is now natural 
to focus on the stress gauge field propagators, which can be determined in the usual way from the generating functional with external (dislocation) sources~\cite{QLC2D}
\begin{equation}
 \langle b^{a \dagger}_{\mu\nu} \; b^b_{\kappa\lambda} \rangle = \frac{1}{\mathcal{Z}[0]} \frac{\delta}{\delta J^{b\dagger}_{\kappa\lambda} } \frac{\delta}{\delta J^{a\phantom{\dagger}}_{\mu\nu}} \mathcal{Z}[J] \Big\rvert_{J=0}.
\end{equation}
This amounts to integrating out the $b$-fields in Eq.~\eqref{eq:solid stress gauge field Lagrangian}, which boils down to inverting the matrices, since the Lagrangian is quadratic.
It is now insightful to use the Coulomb gauge fix $\partial_m b^a_{m \nu} = 0$ instead of the Lorenz gauge fix, since it removes all occurrences of the $\tL$-components in $\mu$, $\nu$. 
The Lagrangian in this gauge fix is easily obtained from Eq.~\eqref{eq:solid stress gauge field Lagrangian} by substituting $p b^a_{1 \nu}  = -q b^a_{\ft \nu}$ according to Eq.~\eqref{eq:tLRS to 01RS transformation}. This also implies $p b^{1 -}  = -q b^{\ft -}$.  After this substitution and inverting the matrices, we read off the diagonal components to find for the longitudinal sector $\mathcal{L}_{\tL 1}$:
\begin{align}
\langle b^\dagger_{\ft -} \; b_{\ft -} \rangle &= \frac{\mu}{q^2} \frac{2}{1-2\nu} \frac{\omega^2 - (1+\nu)c_\tT^2 q^2}{\omega^2 - c_\tL^2 q^2},\label{eq:stress gauge field propagator electric shear}\\
\langle b^{\tL \dagger}_{\tR\tS} \; b^\tL_{\tR\tS} \rangle &= -\mu \frac{ c_\tL^2}{\omega^2 - c_\tL^2 q^2 }.\label{eq:stress gauge field propagator longitudinal phonon}
\end{align}
where we have performed the Wick rotation to real time $\omega_n \to \ti \omega - \delta$ (ignoring the infinitesimal $\delta$).
The second equation obviously describes  the propagating longitudinal phonon with velocity $c_\tL$. In the limit $\omega \to \infty$ the first equation is proportional to $1/q^2$, indicating that this is an instantaneous force like the Coulomb force. In the static limit $\omega \to 0$, the propagator is $2\mu \frac{1- \nu}{1+\nu} \frac{1}{q^2}$, consistent with Eq.~\eqref{eq:electric shear stress 3+0D}.
This was already identified in QLC2D: the static force between edge dislocation sources is carried by the temporal components of the dual stress gauge field in the longitudinal sector (carrying a transverse Burgers  index). The novelty in 3+1D is that there are two more, `magnetic' shear forces, carried by $b_{\ft +}$ resp. $b^\tR_{\ft \tR} + b^\tS_{\ft \tS}$ in the $\tL 2$- resp. $\tT 3$-sectors. Their propagator is simply $\frac{2\mu}{q^2}$.

Let us now turn to the transverse sectors $\mathcal{L}_{\tT 1, \tT 2, \tT 3}$. Integrating out  the dual stress gauge fields in the Coulomb gauge fix yields the propagators,
\begin{align}
 \langle b^{\tL\dagger}_{\ft \tR} \; b^\tL_{\ft \tR} \rangle = \langle b^{\tL\dagger}_{\ft \tS} \; b^\tL_{\ft \tS} \rangle 
 &= \mu \frac{1}{q^2} \frac{ \omega^2 (1 + \ell^2 q^2) - 4 c_\tT^2 \ell^2 q^4}{\omega^2 - c_\tT^2 q^2(1+ \ell^2 q^2)},\label{eq:stress gauge field propagator rotational force}\\
 \langle b^{\tR\dagger}_{\tR\tS} \; b^\tR_{\tR \tS} \rangle = \langle b^{\tS\dagger}_{\tR\tS} \; b^\tS_{\tR \tS} \rangle
 &= -\mu \frac{c_\tT^2(1 + \ell^2 q^2)}{\omega^2 - c_\tT^2 q^2(1+\ell^2q^2) },\label{eq:stress gauge field propagator transverse phonons}\\
 \langle b^{\tR\dagger}_{\ft \tR} \; b^\tR_{\ft \tR} \rangle = \langle b^{\tS\dagger}_{\ft \tS} \; b^\tS_{\ft \tS} \rangle
 &= \mu \frac{1 + \ell^2 q^2}{q^2}.\label{eq:stress gauge field propagator magnetic shear}
\end{align}
The second equation describes the two transverse phonons propagating with velocity $c_\tT$. The third equation, stemming from the $\tT 3$-sector shows a static shear force, that persists in the limit $\ell \to 0$; it is the partner of the magnetic shear force in the $\tL 2$-sector. However, for $q \gg 1/\ell$, this propagator does not vanish but becomes proportional to $\ell^2$, showing that at short length scales it probes the second-order, rotational elasticity. Therefore the $T3$-sector contains magnetic shear and a rotational force. The first equation, in the limit $\omega \to \infty$, reduces to the third equation, and one would be led to think it also contains a shear force next to a rotational force. That this is not the case can be seen in the limit $\omega \to 0$, where we have 
\begin{equation}
 \langle b^{\tL\dagger}_{\ft \tR} \; b^\tL_{\ft \tS} \rangle = \langle b^{\tL\dagger}_{\ft \tS} \; b^\tL_{\ft \tR} \rangle  (\omega \to 0) = 4\mu \frac{1}{q^2 + \frac{1}{\ell^2}}.\label{eq:rotational force suppression}
\end{equation}
These Coulomb forces decay exponentially on the {\em rotational length scale} $\ell$ of Eq.~\eqref{eq:isotropic solid second gradient energy}. In the limit $\ell \to 0$ of a crystal, these forces disappear altogether. This result is also found in 2+1D (see QLC2D) where one such short-ranged force is present which we called the {\em rotational force} since it is only present when local rotations are permitted in the elastic medium. All in all, we end up with the classification in Table~\ref{table:stress degrees of freedom}.

\subsection{Dual propagator relations}\label{subsec:Dual propagator relations}

Given the dual action, let us now derive the strain (phonon) propagators $G_\tL$ and $G_\tT$ from Eqs.~\eqref{eq:longitudinal propagator}, \eqref{eq:transverse propagator} but this time using the stress photon formalism. 
One first has to  translate these strain propagators to the stress gauge field language.  This is more subtle than one may naively expect, and it was for the first time accomplished in 2+1D in Ref.~\onlinecite{ZaanenNussinovMukhin04}. 
The main issue is that the external sources in the path integral used to derive the two-point functions must themselves be carried through the dualization procedure. Here we will introduce a method for the transverse
 propagator which is slightly modified compared to the 2+1D case~\cite{QLC2D}.

Let us first consider the longitudinal propagator Eq.~\eqref{eq:longitudinal propagator}, defined as,
\begin{align}
 G_\tL &= \langle \partial_a u^{a} \; \partial_b u^b \rangle 
 = \frac{1}{Z[0]} \frac{\delta}{\delta \mathcal{K}} \frac{\delta}{\delta \mathcal{K}} Z[\mathcal{K}] \big\rvert_{\mathcal{K} =0},\label{eq:longitudinal propagator from path integral}\\
 \mathcal{Z}[\mathcal{K}] &= \int \mathcal{D} u^a \exp(-\mathcal{S}_\mathrm{solid}-\int \td \tau \td^D x\; \mathcal{K} \partial_a u^a\big).
\end{align}
According to Eq.~\eqref{eq:stress tensor definition}, the stress tensor changes under the introduction of $\mathcal{K}$. It only affects the compression component, 
\begin{equation}
   P^{(0)}_{mnab} \sigma_n^b =  -\ti D \kappa P^{(0)}_{mnab} \partial_n u^b -\ti \mathcal{K} \delta_{ma}, \label {eq:P0 stress with source}
\end{equation}
In the dualization procedure, the dual Lagrangian is modified from Eq.~\eqref{eq:dual solid Lagrangian} to~\cite{QLC2D},
\begin{align}
  \mathcal{L}_\mathrm{dual} = \frac{1}{2} \sigma^a_\mu C^{-1}_{\mu\nu ab} \sigma^b_\nu - \frac{1}{2\kappa}  \mathcal{K}^2 +  \ti \frac{1}{D\kappa}\mathcal{K} \sigma^a_a + \ti \sigma^a_\mu \partial_\mu u^a. 
\end{align}
From this expression and Eq.~\eqref{eq:longitudinal propagator from path integral} we find
\begin{equation}\label{eq:longitudinal propagator stress propagator relation}
 G_\tL = \frac{1}{\kappa} - \frac{1}{(D\kappa)^2} \langle \sigma^{a}_a\; \sigma^b_b \rangle.
\end{equation}
Using Eq.~\eqref{eq:stress gauge field components} the compressional stress can be written as,
\begin{align}
 \sigma^a_a = -\sigma^\tL_\tL + \sqrt{2} \sigma_-
 = -\ti\frac{\omega_n}{c_\tT} b^\tL_{\tR\tS} + \sqrt{2} p\; b_{1-}.
\end{align}
Note that $G_\tL$ only depends on $\mathcal{L}_{\tL 1}$ Eq.~\eqref{eq:solid Lagrangian gauge field L1} and not $\mathcal{L}_{\tL 2}$ Eq.~\eqref{eq:solid Lagrangian gauge field L2}. In terms of the stress gauge fields, the explicit expression for the stress term in  the longitudinal propagator becomes
\begin{align}
 \langle \sigma^{a}_a\;  \sigma^b_b \rangle &= 
 \frac{\omega_n^2}{c_\tT^2} \langle b^{\tL\dagger}_{\tR \tS}\;  b^\tL_{\tR \tS} \rangle
 + 2p^2 \langle b^\dagger_{ 1 -}\;  b_{1 -} \rangle \nonumber \\
& \phantom{m} + \ti \sqrt{2} \frac{\omega_n}{c_\tT} p \big( \langle b^{\tL\dagger}_{\tR \tS}\;  b_{1 -} \rangle 
 - \langle  b^\dagger_{1 -}\; b^{\tL}_{\tR \tS} \rangle \big) 
 .\label{eq:compression stress propagator}
\end{align}
The great advantage of this expression is  that it is valid both in the ordered (solid) but also the  disordered (quantum liquid crystal) phases.
The propagators of the stress gauge fields can be obtained as usual from the generating functional with external sources $\mathcal{Z}[J^a_{\kappa\lambda}]$, i.e. integrating out the dual stress fields and taking functional derivatives with respect to $J^a_{\kappa\lambda}$. Since the Lagrangian is Gaussian in $b^a_{\kappa\lambda}$, this amounts to inverting the matrix in Eq.~\eqref{eq:solid Lagrangian gauge field L1}. It can be verified that after this inversion and inserting the correct contributions in Eq.~\eqref{eq:compression stress propagator} and then Eq.~\eqref{eq:longitudinal propagator stress propagator relation}, we obtain the correct expression Eq.~\eqref{eq:longitudinal propagator} (ignoring the contributions from second-gradient elasticity)
We emphasize again that although this dual route is much more laborious in case of the solid, it can be taken in the quantum liquid crystals as well.

The derivation of the transverse propagator is along the same lines. For 2+1 dimensions, there is only one transverse phonon and the longitudinal and transverse sectors there are rather similar~\cite{QLC2D}. In 3+1 dimensions however, there are two transverse phonons instead of one. More importantly, there are three rotational planes instead of one. In fact, inspecting Eq.~\eqref{eq:transverse propagator}, the transverse propagator is a sum over three terms
\begin{equation}
 G_\tT = 2 \sum_{ab} \langle \omega^{ab} \; \omega^{ab} \rangle =  4 \sum_c \langle \omega^c \; \omega^c \rangle.
\end{equation}
It will turn out to be quite convenient to consider these three terms separately in terms of the Fourier space components for the index $c$. Let us therefore consider
\begin{align}
 G_\tT &= G_{\tT 1} + G_{\tT 2} + G_{\tT 3} \nonumber\\
 &= 4 \langle \omega^{\tS\dagger} \; \omega^\tS \rangle + 4 \langle \omega^{\tR\dagger} \; \omega^\tR \rangle +4 \langle \omega^{\tL\dagger} \; \omega^\tL \rangle.\label{eq:transverse propagator from torque correlators}
\end{align}
We will see that the labels $\tT 1$, $\tT 2$, $\tT 3$ match those of the solid Lagrangian Eqs.~\eqref{eq:L1}--\eqref{eq:L3}. These terms can in fact be straightforwardly computed calculated in terms of the  strains, Eq.~\eqref{eq:displacement propagator}. The result is,
\begin{align}
 G_{\tT 1} &=  G_{\tT 2}  =  \frac{1}{\mu} \frac{ c_\tT^2 q^2}{\omega_n^2 + c_\tT^2 q^2(1 + \ell^2 q^2)},\label{eq:GT1 GT2 solid}\\
 G_{\tT 3} &= 0.\label{eq:GT3 solid}
\end{align}
$G_{\tT 1}$, $G_{\tT 2}$ describe the transverse phonons in the $\tR$- resp. $\tS$-direction while $G_{\tT 3}$ is just unphysical. In Sec.~\ref{sec:nematic} we will show that in the isotropic nematic each of these three propagators will feature a massless rotational Goldstone mode. To compare the result of Eqs.~\eqref{eq:GT1 GT2 solid}, \eqref{eq:GT3 solid} with the definitions Eq.~\eqref{eq:transverse propagator from torque correlators}: for the component $\omega^\tL$  the momentum is perpendicular to the rotational plane. Apparently the propagator between a source and sink of the rotational field which are separated from each other perpendicular to the rotational plane vanishes identically, while forces can be 
exchanged when they are separated within the rotational plane. These latter forces  are obviously just shear forces. In the presence of second-order elasticity Eq.~\eqref{eq:second gradient Lagrangian in stress tensor}, all three propagators are finite, although the interactions associated with these second-order terms are short ranged.

To derive the dual relations for the transverse propagators, let us add sources $\mathcal{J}^c$ coupling to each of the rotation fields $\omega^c$. Similar to Eq.~\eqref{eq:longitudinal propagator from path integral}, we then have
\begin{align}
 4\langle \omega^c  \; \omega^{c} \rangle 
 &= \frac{1}{\mathcal{Z}[0]} \frac{\delta}{\delta \mathcal{J}^{c}}  \frac{\delta}{\delta \mathcal{J}^c} \mathcal{Z}[\mathcal{J}] \big\rvert_{\mathcal{J} =0}, \ {\text{no sum } c} \label{eq:transverse propagator from path integral}\\
 \mathcal{Z}[\mathcal{J}] &= \int \mathrm{D} \omega^c \exp\big(-\mathcal{S}_\mathrm{solid} - \int \td \tau \td^D x\; 2 \mathcal{J}^c \omega^c \big).
\end{align}

Since the antisymmetric components of $\sigma^a_m$ are absent from first-order elasticity, we need second-gradient terms to derive the dual propagators. Starting from Eq.~\eqref{eq:second-gradient elasticity} with the source term in Eq.~\eqref{eq:transverse propagator from path integral} added, the torque stress is modified from Eq.~\eqref{eq:torque stress definition} to 
\begin{equation}
 \tau^c_m = -\ti 4 \mu \ell^2 \partial_m\omega^c + 2\ti \frac{\partial_m}{(\partial_n)^2} \mathcal{J}^c.
\end{equation}
Here we used the formal equality $1 = (\partial_m/\partial_n^2) \partial_m$ and integration by parts. The terms in the dual Lagrangian involving torque stress become,
\begin{align}
 \mathcal{L}^{(2)}_\mathrm{dual}[J] &= \frac{1}{4\mu \ell^2} \Big( \frac{1}{2} \lvert \tau^c_m \rvert^2  
 - 2\ti \frac{\partial_m \tau^c_m}{q^2} \mathcal{J}^c - 2 \mathcal{J}^{c\dagger}  \frac{1}{q^2} \mathcal{J}^c
 \Big) \nonumber\\
 &\phantom{mm} +\ti \tau^c_m \partial_m \omega^c.
\end{align}
We find subsequently from Eq.~\eqref{eq:transverse propagator from path integral} the identities for each $c$  (no sum over $c$):
\begin{equation}
 4\langle \omega^c \; \omega^{c} \rangle
 = \frac{1}{ \mu  \ell^2 q^2} - \frac{1}{(2 \mu \ell^2 q^2)^2} \langle \partial_m \tau^c_m\; \partial_n \tau^{c }_n \rangle.
\end{equation}
Finally we can substitute the Ehrenfest constraints in the presence of torque stress Eq.~\eqref{eq:Ehrenfest constraint torque stress} to find for each component $c$ separately,
\begin{equation}
 4\langle \omega^c \; \omega^{c } \rangle
 = \frac{1}{ \mu  \ell^2 q^2} - \frac{1}{(2 \mu \ell^2 q^2)^2} \langle \epsilon_{cma} \sigma^a_m\; \epsilon_{cnb} \sigma^{b}_n \rangle.\label{eq:transverse propagator stress propagator relation}
\end{equation}
This expression is valid even in the limit $\ell \to 0$: the factors of $\ell$ will cancel out to leave a non-divergent expression. Substituting the dual stress gauge field using Eq.~\eqref{eq:stress gauge field components},
\begin{widetext}
 \begin{align}
 G_{\tT 1} &= \frac{1}{\mu \ell^2 q^2} - \frac{1}{(2 \mu \ell^2 q^2)^2} \Big[
 p^2 \langle b^{\tL\dagger}_{1 \tS} \; b^{\tL}_{1 \tS} \rangle + \frac{\omega_n^2}{c_\tT^2} \langle b^{\tR\dagger}_{\tR\tS}\;  b^\tR_{\tR \tS}\rangle + \ti \frac{\omega_n}{c_\tT} p \Big( \langle b^{\tR\dagger}_{\tR\tS} \; b^\tL_{1 \tS}\rangle - \langle b^{\tL\dagger}_{1\tS}\;  b^\tR_{\tR \tS}\rangle \Big) 
 \Big],\nonumber\\
 G_{\tT 2} &= \frac{1}{\mu \ell^2 q^2} - \frac{1}{(2 \mu \ell^2 q^2)^2} \Big[
 p^2 \langle b^{\tL\dagger}_{1 \tR}\;  b^{\tL}_{1 \tR} \rangle + \frac{\omega_n^2}{c_\tT^2} \langle b^{\tS\dagger}_{\tR\tS}\;  b^\tS_{\tR\tS}\rangle + \ti \frac{\omega_n}{c_\tT} p \Big( \langle b^{\tS\dagger}_{\tR\tS} \; b^\tL_{1 \tR}\rangle - \langle b^{\tL\dagger}_{1\tR}\;  b^\tS_{\tR\tS}\rangle \Big)
 \Big],\nonumber\\
 G_{\tT 3} &= \frac{1}{\mu \ell^2 q^2} - \frac{1}{(2 \mu \ell^2 q^2)^2}  \Big[ p^2 \langle b^{\tR\dagger}_{1 \tR} \; b^{\tR}_{1 \tR} \rangle + p^2 \langle b^{\tS\dagger}_{1 \tS} \; b^\tS_{1\tS }\rangle -  p^2 \langle b^{\tS\dagger}_{1\tS} \; b^\tR_{1 \tR}\rangle -  p^2\langle b^{\tR\dagger}_{1\tR}\;  b^\tS_{1\tS }\rangle \Big].\label{eq:transverse propagator gauge field}
\end{align}
\end{widetext}
This makes clear why we choose these specific labels. In the isotropic solid, the propagator $G_{\tT 1}$ is a result of the Lagrangian contribution $\mathcal{L}_{\tT 1}$ etc. while the longitudinal propagator is related to $\mathcal{L}_{\tL 1}$. Again, it can be verified directly by inverting the matrices in Eqs.~\eqref{eq:L1}--\eqref{eq:L3} that these expressions reproduce Eqs.~\eqref{eq:GT1 GT2 solid}, \eqref{eq:GT3 solid}.

\begin{figure*}
 \begin{center}
   \includegraphics[width=7.7cm]{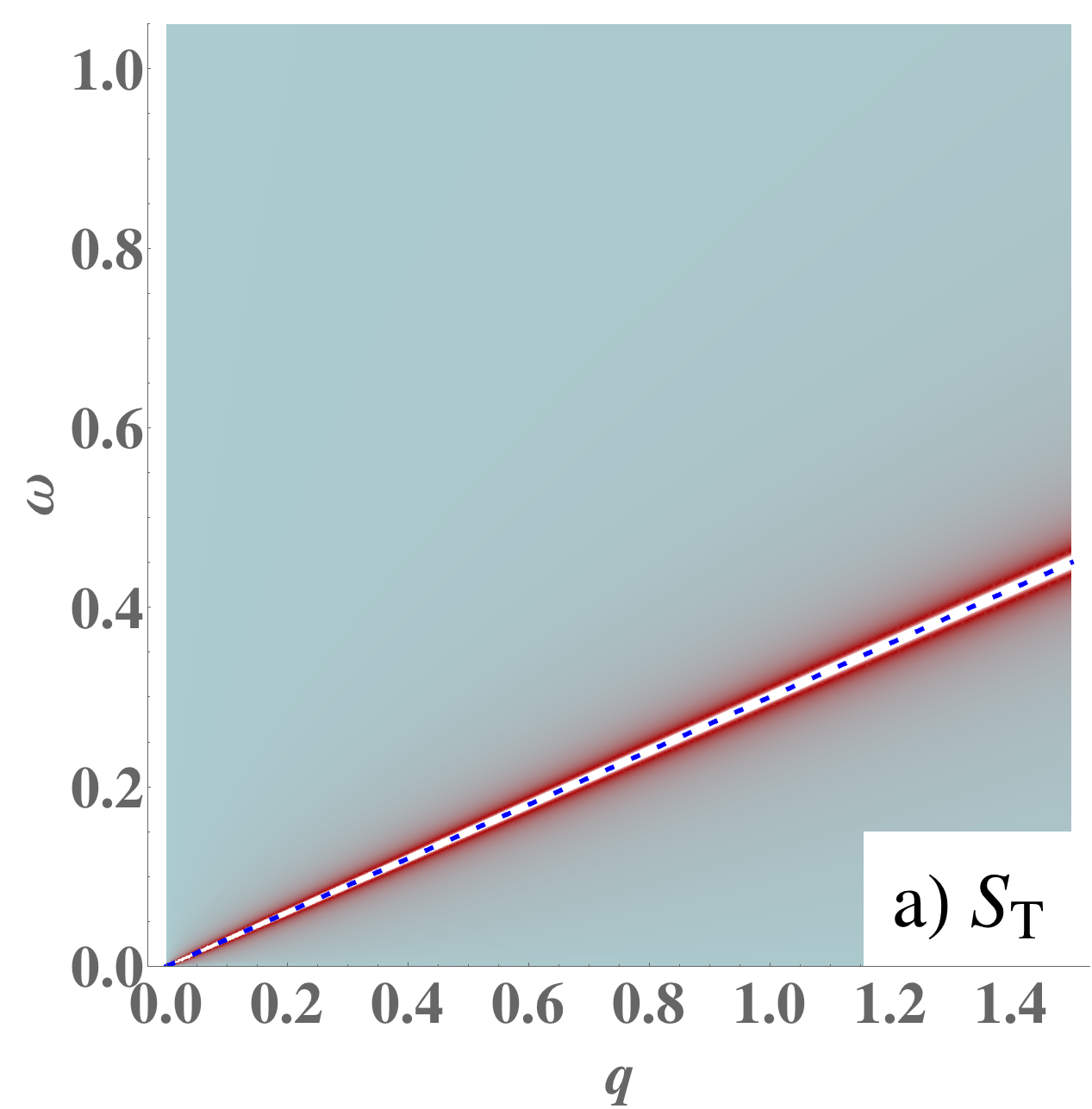}
 \hfill
\includegraphics[width=7.7cm]{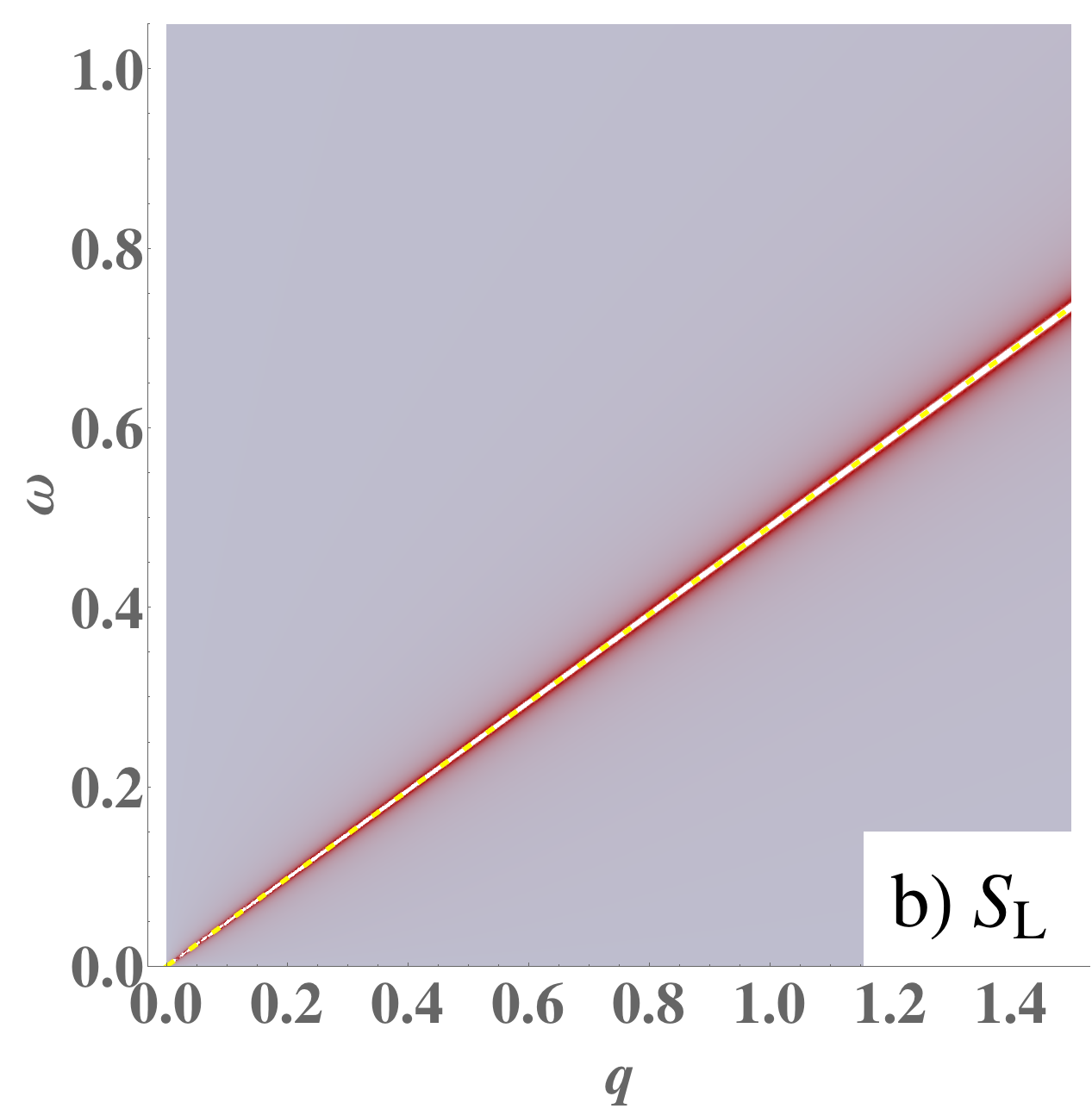}\\
{\centering
\includegraphics[scale=0.6]{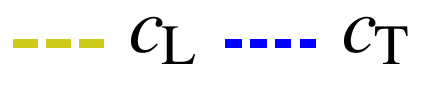}
}\\
 \caption{Spectral functions Eq.~\eqref{eq:spectral function definition} (left: transverse; right: longitudinal) of the isotropic solid in units of the inverse shear modulus $1/\mu \equiv 1$, with Poisson ratio $\nu = 0.2$. The width of the poles is artificial and denotes the relative pole strengths: these ideal poles are actually infinitely sharp. The propagating modes, phonons, are massless (zero energy as $q \to 0$) and have linear dispersion $\omega = cq$. The velocities are the transverse $c_\tT$ resp. longitudinal velocity $c_\tL$. The pole in the transverse sector is doubly degenerate as there are two transverse phonons.}\label{fig:solid spectral functions}
 \end{center}
\end{figure*}

Upon Wick rotating to real time $ \omega_n \to \ti \omega - \delta$ one can compute the spectral functions
\cite{QLC2D}, 
\begin{equation}\label{eq:spectral function definition}
 S (\omega,q) = \mathrm{Im}\; G (\ti \omega -\delta, q).
\end{equation}
where $G = G(\omega_n ,q)$ is the particular propagator under investigation. The propagating modes show up as poles in the spectral function and for future reference we plot the spectral functions $G_\tL$ and $G_{\tT 1} = G_{\tT 2}$ in Fig.~\ref{fig:solid spectral functions}. One infers the presence of the longitudinal and transverse phonon poles,  linearly dispersing with velocities $c_\tL$ and $c_\tT$, respectively.

\subsection{Torque stress gauge fields}\label{subsec:Torque stress gauge fields}

We have seen that the stress gauge fields $b^a_{\kappa\lambda}$ are dual to the displacement fields $u^a$, coupling minimally to the dislocation sources $J^a_{\kappa\lambda}$. One can wonder whether such a dualization 
exists as well for the rotation fields $\omega^c$ of Eq.~\eqref{eq:rotation field definition}. The answer is affirmative. In Eq.~\eqref{eq:torque stress definition} we have already seen that in the presence of higher-order elasticity terms the momentum canonical to the rotation field is the torque stress $\tau^c_m$. It is as well possible to consider the fate of torque stress in linear elasticity when $\ell = 0$. Define~\cite{BeekmanWuCvetkovicZaanen13,QLC2D}:
\begin{equation}\label{eq:linear torque stress definition}
 \tau^c_\mu = \epsilon_{cba} \epsilon_{b\mu \kappa\lambda}\tfrac{1}{2} b^a_{\kappa\lambda}.
\end{equation}
The reader should be careful that this definition of torque stress  and  Eq.~\eqref{eq:torque stress definition} are of a different origin and they should not be interchanged. Notice that $ \tau^c_\mu $ is not gauge invariant under the gauge transformations Eq.~\eqref{eq:stress gauge transformation}. This is not a problem, since in the present formulation it is at first simply a collection of stress gauge fields. On a deeper level, torque stress is not unambiguously defined as long as there is a rigidity against linear shear stress. In the nematic phases, this latter rigidity is lost and torque stress becomes a genuine physical force, see Sec.~\ref{subsec:Torque stress in the quantum nematic}. Once more, this stems from the fact that translations and rotations are intertwined by their semidirect relationship. With the definition Eq.~\eqref{eq:linear torque stress definition}, the Ehrenfest constraints $\epsilon_{cma} \sigma^a_m = 0$ turn into dynamical constraints
\begin{equation}\label{eq:torque stress conservation}
 \partial_\mu \tau^c_\mu = 0,\qquad \forall \ c.
\end{equation}
This can be seen by taking the divergence of Eq.~\eqref{eq:linear torque stress definition} and using the definition of the stress gauge fields Eq.~\eqref{eq:dual stress gauge field definition}. In other words, for a conserved torque stress 
the linear stress tensor is symmetric in its spatial indices. A non-zero torque stress will induce antisymmetric linear stress. As long as external torque stress is absent, we can enforce it explicitly by defining torque stress as the curl of the {\em torque stress gauge field} $h^c_{\kappa\lambda}$:
\begin{equation}\label{eq:torque stress gauge field definition}
 \tau^c_\mu = \epsilon_{\mu\nu\kappa\lambda} \partial_\nu \tfrac{1}{2} h^c_{\kappa\lambda}.
\end{equation}
Writing this out in components we find
\begin{align}\label{eq:torque stress gauge field components}
  \tau^c_\ft &= q h^c_{\tR\tS}, &
 \tau^c_\tL &= \ti \frac{\omega_n}{c_\tT} h^c_{\tR\tS}, &
 \tau^c_\tR &= p h^c_{1\tS}, &
 \tau^c_\tS &=  p h^c_{1\tR}.
\end{align}

The torque stress tensor is invariant under the gauge transformations
\begin{equation}
 h^c_{\kappa\lambda}(x) \to h^c_{\kappa\lambda}(x) + \partial_\kappa  \chi^c_\lambda(x) - \partial_\lambda  \chi^c_\kappa(x),
\end{equation}
where $\chi^c_\kappa$ is any arbitrary Burgers-flavored vector field. However, there is still the initial ambiguity of defining the torque stress through Eq.~\eqref{eq:linear torque stress definition}

We can now mimic the derivation Eqs.~\eqref{eq:displacement field smooth singular} and \eqref{eq:stress dislocation minimal coupling} to find that the torque stress gauge field couples minimally to the disclination current Eq.~\eqref{eq:disclination worldsheet}:
\begin{align}
  \ti \tau^c_\mu \partial_\mu \omega^c_\mathrm{sing} &=   \ti (\epsilon_{\mu\nu\kappa\lambda} \partial_\nu \tfrac{1}{2} h^c_{\kappa\lambda})( \partial_\mu \omega^c_\mathrm{sing})\nonumber\\
 &=  \ti  h^c_{\kappa\lambda}\tfrac{1}{2}\epsilon_{\mu\nu\kappa\lambda} \partial_\mu \partial_\nu \omega^c_\mathrm{sing} =  \ti h^c_{\kappa\lambda} \Theta^c_{\kappa\lambda}.\label{eq:torque stress disclination minimal coupling}
\end{align}
The interpretation is accordingly: the torque stress gauge fields mediate interactions between disclination sources. It is therefore possible to examine the nature of interactions between disclinations even in the solid, and this is
quite insightful.  One can use Eq.~\eqref{eq:linear torque stress definition} and then Eq.~\eqref{eq:torque stress gauge field definition} to express the relevant part of  the elasticity Lagrangian in terms of the torque stress gauge field, to integrate out these gauge fields and take functional derivatives with respect to the disclination sources. 
Write out Eq.~\eqref{eq:linear torque stress definition} component-by-component:
\begin{align}
\tau^\tL_\ft &= b^\tR_{\tL \tR} - b^\tS_{\tL\tS}, &
\tau^\tR_\ft &= b^\tL_{\tL \tR} - b^\tS_{\tR\tS}, &
\tau^\tS_\ft &= b^\tL_{\tL \tS} - b^\tR_{\tR\tS},\nonumber\\
\tau^\tL_\tL &= b^\tR_{\ft \tR} - b^\tS_{\ft\tS}, &
\tau^\tR_\tL &= b^\tL_{\ft \tR}, &
\tau^\tS_\tL &= b^\tL_{\ft \tS},\nonumber\\
\tau^\tL_\tR &= b^\tR_{\ft \tL} , &
\tau^\tR_\tR &= b^\tL_{\ft \tL} + b^\tS_{\ft\tS}, &
\tau^\tS_\tR &= b^\tR_{\ft \tS},\nonumber\\
\tau^\tL_\tS &= b^\tS_{\ft \tL} , &
\tau^\tR_\tS &= b^\tS_{\ft \tR}, &
\tau^\tS_\tS &= -b^\tL_{\ft \tL} + b^\tR_{\ft\tR}.
\label{eq:torque stress in stress gauge field components}
\end{align}
Not all components of $b^a_{\kappa\lambda}$ have inverse relations with respect to $\tau^a_{\mu}$. Note that there now appear some components $b^E_{\ft \tL}$ which are completely absent in linear elasticity. These will become important in the dislocation condensates.

To simplify the derivation, let us impose the Coulomb gauge  for the stress gauge fields: $\partial_k b^a_{k\lambda} = - q b^a_{\tL \lambda} =0  \ \forall a,\lambda$. The Lagrangian Eq.~\eqref{eq:solid stress gauge field Lagrangian} is easily converted using the relation $b^a_{1 \lambda} = - \frac{q}{p} b^a_{\ft \lambda}$ which holds in the Coulomb gauge. Furthermore, using Eq.~\eqref{eq:torque stress in stress gauge field components} in the Coulomb gauge and Eq.~\eqref{eq:torque stress gauge field components}  we find the explicit relations:
\begin{align}
 b^\tR_{\tR\tS}                    &= - \tau^\tS_\ft   &&= - q h^\tS_{\tR\tS},&\nonumber\\
 b^\tL_{\ft \tS}                   &= \tau^\tS_\tL     &&= \ti \frac{\omega}{c_\tT} h^\tS_{\tR\tS},\nonumber\\
 b^\tS_{\tR\tS}                    &= - \tau^\tR_\ft   &&= - q h^\tR_{\tR\tS},\nonumber\\
 b^\tL_{\ft \tR}                   &= \tau^\tR_\tL     &&= \ti \frac{\omega}{c_\tT} h^\tR_{\tR\tS},\nonumber\\
 b^\tR_{\ft \tR} + b^\tS_{\ft \tS} &= \tau^\tR_\tR + \tau^\tS_\tS &&= p h^\tR_{1\tS} + p h^\tS_{1\tR},\nonumber\\
 b^\tR_{\ft \tR} - b^\tS_{\ft \tS} &= \tau^\tL_\tL     &&= \ti \frac{\omega}{c_\tT} h^\tL_{\tR\tS}.
\end{align}
We can now express the transverse sectors of the elasticity Lagrangian directly in torque stress gauge fields, finding:
\begin{align}
 \mathcal{L}_{\tT 1} &= \frac{1}{2\mu}q^2 p^2 \left\lvert h^\tS_{\tR\tS} \right\rvert^2 ,\nonumber\\
 \mathcal{L}_{\tT 2} &= \frac{1}{2\mu}q^2 p^2 \left\lvert h^\tR_{\tR\tS} \right\rvert^2,\nonumber\\
 \mathcal{L}_{\tT 3} &= \frac{1}{2\mu} q^2 p^2 \left \lvert h^\tR_{1\tS} + h^\tS_{1 \tR} \right\rvert^2.
 \label{eq:solid Lagrangian torque stress gauge field}
\end{align}
Notice that the terms from second-order elasticity $\propto \ell^2$ drop out. This is because by definition Eq.~\eqref{eq:second gradient Lagrangian in stress tensor} these are proportional to $\partial_\mu \tau^c_\mu$, and as such should vanish if we use the torque stress gauge fields Eq.~\eqref{eq:torque stress gauge field definition}. 

These expressions have the following physical meaning: the torque stress gauge fields mediate interactions between disclination sources. As usual, the purely transverse components $h^c_{\tR\tS}$ represent the propagating components, while the temporal components $h^c_{\ft \tR}$, $h^c_{\ft \tS}$ (which are equal to $-\frac{q}{p}h^c_{1 \tR}$, $-\frac{q}{p}h^c_{1 \tS}$ in the Coulomb gauge) represent static `Coulomb' forces. Due to the additional factor of $q^2$ in Eq.~\eqref{eq:solid Lagrangian torque stress gauge field}, the propagators behave as $\langle h^\tR_{\tR \tS} h^{\tR\dagger}_{\tR \tS}\rangle \propto 1/q^2 p^2$ etc., falling off much quicker than $1/p^2$. In 3D, this implies that the energy 
of a disclination--antidisclination pair increases quadratically with their separation. This is physically the same phenomenon as the confinement of quarks,
i.e. that the gluons of QCD cannot occur in isolation below the confinement scale $\Lambda_{\rm QCD}$, with the only difference that in QCD the energy between the quark sources increases linearly. In the same vein one can claim that the `torque photons' are confined in the solid at scales less than the rotational confinement scale, which in the solid is associated with the rotational stiffness length scale $\ell$~\cite{BeekmanWuCvetkovicZaanen13,QLC2D}. In Sec.~\ref{sec:nematic} we will see that these rotational forces and associated rotational Goldstone modes are deconfined in the dislocation condensates,
turning into mediators of physical long-range interactions between the (deconfined) disclinations.  Notice that the component related to the third rotational Goldstone mode $h^\tL_{\tR\tS}$ is completely absent in the solid. It should originate from the components $\tau^\tL_\ft$ and $\tau^\tL_\tL$, which are present only in the second-gradient term in $\mathcal{L}_{\tT 3}$. In Sec.~\ref{sec:nematic} we will see that a Goldstone mode does emerge in this sector in the nematic phase.

Finally, it is also possible to substitute $b_{\ft\pm} = p (h^\tR_{1\tR} \pm h^\tS_{1 \tS})$ in the longitudinal sectors $\mathcal{L}_{\tL 1,2}$. This shows that also the Coulomb forces mediated by these components are confined. The component $b^\tL_{\tR\tS}$ representing the longitudinal phonon has no counterpart in the torque sector.

%% file: sec_quantummelting.tex
With the dualization of strains into stresses and the formulation of the stress-gauge theory  of the previous section we have identified an efficient formalism for the description of 
quantum elastic matter. However, we did not learn new physics in the process. The true power of the formalism lies in its ability to allow for a description of the quantum liquid-crystalline phases as well, in the special
limit of maximal crystalline correlations. In short, when dislocations proliferate they form a condensate, minimally coupled  to the dual stress gauge fields: the {\em dual stress superconductor}. 
In 2+1D this was conceptually a straightforward affair since the dislocations behave like bosonic particles that form a conventional Bose condensate. However, in 3+1D we are facing the problem of 
constructing a `string condensate', a fundamental problem where progress has been slow, as we already discussed in the introduction. As we will elaborate in this section, we know just enough to construct an effective action that appears to be sufficient to describe the most salient properties of the quantum liquid crystals.  
 
\subsection{Condensation of particle-like topological defects}

\subsubsection{Vortex condensation in 2+1D}\label{subsubsec:Vortex condensation}

Before turning  to the dislocation worldsheets in 3+1D, let us shortly review the results of Abelian-Higgs duality in 2+1D. The Berezinskii--Kosterlitz--Thouless phase transition~\cite{Berezinskii70,KosterlitzThouless72,KosterlitzThouless73} takes place in 2+0D at finite temperature where true long-range order is absent due to the Mermin--Wagner--Hohenberg theorem~\cite{MerminWagner66,Hohenberg67}. Thus starting from a state with algebraic long-range order at low temperature (e.g. a two-dimensional superfluid), increasing the temperature introduces vortex--antivortex pairs as thermal excitations. At the critical temperature, such pairs {\em unbind} and vortices and antivortices can appear freely, with the effect that the superfluid order has been destroyed. This is the disordered phase.

In the zero-temperature quantum setting, we must introduce the imaginary time axis and consider the problem in 2+1D spacetime. The quantum phase transition is now tuned by the quantum fluctuations, induced in the superfluid by the charging energy of the constituent bosons.
Departing from  the superfluid phase, with its zero temperate broken global $U(1)$-symmetry the disordering  quantum fluctuations are completely enumerated  in terms of closed loops formed from vortex--antivortex worldlines. 
Assuming that the distance between the vortices is large compared to the lattice cut-off (the low vortex-fugacity limit) the physics is completely captured in terms of the vortices as a system of relativistic bosonic particles, interacting via 
long-range Coulomb-like forces, with a mass that is vanishing at the quantum phase transition.  In the superfluid, defects only appear as small, closed spacetime loops of vortex worldlines, representing the creation and subsequent annihilation of vortex--antivortex pairs. Increasing quantum fluctuations increases the occurrence and the size of such loops, and at the quantum critical point the loops `blow out', becoming as large as the system size and forming a `tangle of free (anti)vortices'. The disordered phase therefore corresponds to a Bose condensate, minimally coupled to the gauge fields mediated by the locally superfluid medium found at distances smaller than the separation between the vortex
particles~\cite{ZaanenBeekman12}. This is known as the {\em vortex--boson} or {\em Abelian-Higgs duality}, in the condensed matter context introduced by Fisher and Lee~\cite{FisherLee89}, where it flourished in studies of the boson-Hubbard model~\cite{FisherEtAl89}. It has been further elaborated on in for instance Refs.~\onlinecite{Kleinert89a,KiometzisKleinertSchakel95,HerbutTessanovic96,CvetkovicZaanen06a} and is supported by very strong numerical evidence~\cite{NguyenSudbo99,HoveSudbo00,HoveMoSudbo00,SmisethSmorgravSudbo04,SmisethEtAl05,SmorgravEtAl05}.

In the ordered (superfluid) phase, the low-energy excitation is simply its Goldstone mode (superfluid second sound), which has the action of a free real scalar field--in the strong-correlation limit amplitude fluctuations are suppressed. This is dualized into a 2+1D vector field $b_\mu$ with vortex sources $J_\mu$ and action~\cite{FisherEtAl89,Kleinert89a,ZaanenNussinovMukhin04,CvetkovicZaanen06a,QLC2D},
\begin{equation}
 \mathcal{L}_\mathrm{superfluid} = \frac{g}{2} (\epsilon_{\mu\nu\lambda} \partial_\nu b_\lambda)^2 + \ti b_\lambda J_\lambda.
\end{equation}
Here $g$ is the coupling constant. This is of the form Eq.~\eqref{eq:general dual Langrangian Coulomb phase}. In the vortex condensate  individual vortex sources $J_\mu$ have lost their identity, being replaced by 
the collective condensate field $\Phi(x)$, the amplitude of which obtains a vacuum expectation value in the disordered phase (the vortex `superfluid density'). In 2+1D this `second-quantization' procedure can be derived rigorously 
employing a  lattice formulation~\cite{Kleinert89a,KiometzisKleinertSchakel95,Cvetkovic06}. Here one obtains the partition function of a grand canonical ensemble of meandering vortex (world)lines as bosonic particles, which are charged under the dual gauge field. In the disordered phase the Lagrangian is given by~\cite{FisherLee89,LeeFisher91,KiometzisKleinertSchakel95,Cvetkovic06}
\begin{equation}\label{eq:vortex-boson Higgs action}
 \mathcal{L}_\mathrm{diso.} = \frac{g}{2} (\epsilon_{\mu\nu\lambda} \partial_\nu b_\lambda)^2 + \frac{1}{2} \lvert (\partial_\mu - \ti b_\mu) \Phi \rvert^2 + \frac{\alpha}{2} \lvert \Phi\rvert^2 + \frac{\beta}{4} \lvert \Phi \rvert^4.
\end{equation}
Here $\alpha$ and $\beta$ are Ginzburg--Landau parameters. While $\beta >0$ always, when $\alpha$ becomes negative the quantum phase transition takes place. This is precisely the Ginzburg--Landau--Wilson action of a 
relativistic superconductor, and the dual gauge fields $b_\mu$ become massive due to the Anderson--Higgs mechanism, implying that the interactions they mediate fall off exponentially with distance: the dual Meissner effect. 
We refer to the second term in Eq.~\eqref{eq:vortex-boson Higgs action} as the {\em Higgs term} which is added to the {\em Coulomb term}.

Summarizing, there are three important requirements for condensation of topological defects to take place:
\begin{enumerate}
 \item This construction only applies to {\em bosonic} forms of matter: when the constituents are bosons, motion of the topological defects does not lead to any fermion interchanges. If the defects are furthermore Abelian (i.e. the belong to an Abelian homotopy group), they behave just as bosonic particles themselves, forming eventually the dual Bose condensate. 
 This is self evident for superfluids but it is unknown how to formulate such a dualization for e.g. a crystal formed from fermions. 
 \item  The strong-correlation or low defect-fugacity limit is hard wired in the construction. Only in this limit is the physics of the disordered state captured entirely in terms of the dual vortex fields.
 \item Equivalently, non-topological defects such as interstitials in crystals are completely ignored which is another way of imposing the strong correlation limit. The effect will be that a variety of 
 propagating massive excitations are found that will get `(over)damped by the interstitials' upon moving away from the limit of maximal correlation.
\end{enumerate}

\subsubsection{Dislocation condensation in 2+1D}\label{subsubsec:Dislocation condensation}

Let us now turn to elasticity. The idea that melting of a solid takes place due to proliferation of dislocations is quite old. Starting with the Kosterlitz--Thouless papers~\cite{KosterlitzThouless72,KosterlitzThouless73}, the classical theory of dislocation-mediated melting in 2D was established in the late 1970s by Nelson, Halperin and Young~\cite{HalperinNelson78,NelsonHalperin79,Young79} so that we now speak of the KTNHY-transition of a 2D crystal to a 2D liquid crystal. This famously includes the prediction of the hexatic phase, which we refer to as $C_6$-nematic~\cite{LiuEtAl15,LiuEtAl16b,NissinenEtAl16,LiuEtAl16,QLC2D}. It was also realized that melting dislocations but not disclinations leads to the liquid crystal, while proliferation of dislocations and disclinations at the same time is in fact the ordinary first-order solid--liquid transition~\cite{Kleinert83}. Furthermore, dislocations can proliferate with Burgers vectors preferentially in one direction which `turns liquid' while the other direction `remains solid'~\cite{OstlundHalperin81}. This is the smectic liquid crystal with partial translational symmetry restoration.

It was the insight of Ref.~\onlinecite{ZaanenNussinovMukhin04} that dislocation-mediated quantum melting in 2+1D at zero temperature can be achieved by a generalization of Abelian-Higgs duality. Here the topological charge of the defects is not an integer in $\mathbb{Z}$ but a Burgers vector in $\mathbb{Z}^D$. Departing from a bosonic crystal,  the dislocations braid as bosons as well. Accordingly,  the dislocations form a conventional Bose condensate and in essence
the melting of the solid in a quantum liquid crystal is governed as well by the Abelian-Higgs duality we just discussed.

So far, a rigorous derivation of the Higgs term for dislocation-mediated quantum melting has not been achieved and no numerical simulations have been performed. However, we can rest on general principles to obtain these terms~\cite{ZaanenNussinovMukhin04,Cvetkovic06,QLC2D}. The condensation of bosons can only be the standard Ginzburg--Landau complex scalar field affair with the $\lvert\Phi\rvert^4$-potential as in Eq.~\eqref{eq:vortex-boson Higgs action}. For dislocations with Burgers vector $B^a$, $a = x,y$, there are two copies of $U(1)$-fields $\Phi^a$. There is {\em a priori} no $SU(2)$ symmetry between these fields, so the symmetry group is simply $U(1) \times U(1)$. The reason is that Burgers vectors can only point in the lattice directions, so we cannot make an arbitrary rotation in Burgers space. Once a condensate with Burgers vector in, say, the $x$-direction is established, all lattice points in the $x$-direction become equivalent. This also implies that any remaining dislocations must point orthogonal to the already-condensed direction, regardless of the original space group of the crystal. We already mentioned this in Sec.~\ref{sec:Symmetry principles of quantum liquid crystals} and in Fig.~\ref{fig:hexagonal melting}.

The reader should 
be aware that the construction of the condensate was for the first time fully understood only in QLC2D --- the earlier papers were not quite correct on this issue.
The bottom line is that the Higgs term of the dislocation condensate is identical to that of a two-component Bose--Einstein condensate~\cite{KasamatsuEtAl05}, having the general form~\cite{QLC2D}
\begin{align}
 \mathcal{L}_\mathrm{Higgs} &=  \frac{1}{2} \sum_{a = x,y} \lvert (\partial_\mu - \ti b^a_\mu) \Phi^a\rvert^2 +
 \tfrac{\alpha_x}{2} | \Phi^x |^2 + \tfrac{\alpha_y}{2} | \Phi^y |^2 
  \nonumber\\ &\phantom{mmm}
  + \tfrac{\beta_x}{4}| \Phi^x |^4  + \tfrac{\beta_y}{4} | \Phi^y |^4 
 + \tfrac{\gamma}{2}  | \Phi^x |^2 | \Phi^y |^2. \label{eq:two-condensate potential}
\end{align}
The first term is the minimal coupling of the dual stress gauge fields to the dislocation condensates, replacing the coupling term $\ti b^a_\mu J^a_\mu$ and is a direct generalization of Eq.~\eqref{eq:vortex-boson Higgs action}. The other terms are the Higgs potential. The parameters $\alpha_{x}$, $\alpha_y$, $\beta_x$, $\beta_y$, $\gamma$ depend on the parent crystal structure and other material details. The last term is the coupling between the two condensates: for large positive $\gamma$ only one of $\Phi^x$, $\Phi^y$ will obtain a vacuum expectation value and only dislocations with Burgers vector in that direction condense; this is the smectic phase characterized by a restored translational symmetry in only one direction. On the other hand, for small $\gamma$ the condensates are `locked together' and condense at the same time: this is the nematic phase restoring translational symmetry in both spatial directions~\cite{QLC2D}. The remnant rotational order in the nematic phase depends on the details of the parameters in Eq.~\eqref{eq:two-condensate potential}, and ultimately on the parent crystal symmetry~\cite{LiuEtAl15}.

The Higgs action Eq.~\eqref{eq:two-condensate potential} together with the original stress action of the solid phase is sufficient to completely capture the low-energy spectrum of the quantum smectic and nematic phases, which was worked out in detail for the 2+1D isotropic solid in QLC2D. We can already anticipate the contents of the next subsection: in 3D there will be three dislocation condensate fields $\Phi^a$ that could condense one-by-one to restore translational symmetry in one, two or three directions respectively. The order and nature of this condensation depend on the Ginzburg--Landau phenomenological parameters, ultimately descendant from the crystal structure and other details of the microscopic physics. 

\subsection{Condensation of line-like topological defects}

We gave now arrived at the key section of this work: the construction of the Higgs terms for dislocation fields in 3+1 dimensions. Unfortunately, we will not derive this term in a systematic way. Ideally, one would procure the partition function of a grand canonical ensemble of dislocation worldsheets. This is however complicated by two factors. First, even in 2+1D the Higgs term for dislocation worldlines with vectorial topological charges has not been derived rigorously. But more importantly, the worldsheet nature of linelike defects is a serious obstacle. The theory describing the quantum mechanics of a single line-like object is string theory. However, the quantum field theory of an ensemble of such strings is {\em string field theory}~\cite{Witten86}, which is far from completely settled. But again, resting on general principles it is possible to argue what the form of the Higgs term associated with the spacetime `string foam' formed out of the proliferated dislocations must be.

\subsubsection{Vortex worldsheet condensation}\label{subsubsec:Vortex worldsheet condensation}

The condensation of extended objects has naturally received substantial attention in the literature. Especially on the lattice, it has been long ago established as a physical phenomenon~\cite{Kogut79,Savit80}. In field theory condensation of two-form fields was considered in Ref.~\onlinecite{Rey89} and in the context of vortex--boson duality in Refs.~\onlinecite{MotrunichSenthil05,Franz07}, and very recently in Ref.~\onlinecite{YeGu15}. The main issue here is how to couple the two-form field $b_{\mu\nu}$ to the phase degree of freedom of the condensate field $\Phi$. These works solve the problem by assigning a `vectorial phase' to the condensate field. The condensate field is assumed to be of the form
\begin{equation}\label{eq:vortex condensate vectorial phase}
 \Phi(x) =  \lvert \Phi(x) \rvert \te^{\ti \int \td X^\mu(\sigma) \phi_\mu[X(\sigma)]}.
\end{equation}
Here $\sigma = \sigma_{1,2}$ are coordinates on the worldsheet, $X(\sigma)$ is a map from the worldsheet to real space (called {\em target space} in this context), and the integral is over the entire worldsheet. We also see immediately that shifting $\phi_{\mu}\to \phi_{\mu}+\partial_{\mu} \zeta$ with a total derivative is a redundancy (a constant phase shift) This is to be contrasted with the usual complex scalar field $\Phi = \lvert \Phi \rvert \te^{\ti \phi}$. One obtains the seemingly straightforward generalization of the minimal coupling  term in the London limit where amplitude fluctuations are suppressed:
\begin{equation}\label{eq:string minimal coupling}
 \mathcal{L}_\mathrm{min.coup.} = \frac{1}{4} \lvert \Phi \rvert^2 (\partial_\mu \phi_\nu - \partial_\nu \phi_\mu - b_{\mu\nu} )^2.
\end{equation}
The extra factor of $\frac{1}{2}$ is to compensate over the sum over antisymmetric indices.

In Ref.~\onlinecite{BeekmanSadriZaanen11} we argued that this form cannot be correct at least for the case of vortex--boson duality in the superfluid--Mott insulator transition due to a too large number of propagating modes. The counting as follows. It is known that a massive two-form field $b_{\mu\nu}$ in 3+1 dimensions has three propagating degrees of freedom. This can be seen as follows: from Sec.~\ref{subsec:Two-form gauge fields} we know that a free real scalar field is dual to a free two-form gauge field in 3+1D, and hence the latter carries a single propagating degree of freedom. The dynamics of vortex condensate of the form Eq.~\eqref{eq:vortex condensate vectorial phase} is carried by $\phi_\mu$, and a free vector gauge field in 3+1D carries two propagating degrees of freedom. When the vortex condensate is coupled to the two-form gauge field, its two degrees of freedom are transferred to (``eaten by'') the gauge field through the Anderson--Higgs mechanism, making a total of three propagating modes. 

However the (boson) Mott insulator has two propagating modes (the `doublon' and `holon' excitations) in any spatial dimension, and for this reason it cannot be described by such a `stringy' condensate. We put forward the possibility that the condensation of vortices with a finite size and a core energy could be different from that of coreless, critical string. We argued therefore that the vortex condensate cannot carry more than one degree of freedom, and must be described by an ordinary scalar phase field $\phi$. We proposed two alternative ways to coupling this phase field to the two-form gauge field in Refs.~\onlinecite{BeekmanSadriZaanen11,BeekmanZaanen12}.

In the current context of dislocation condensation, it turns out that as long as one is interested in correlations of linear stress $\sigma^a_\mu$ only, for instance the longitudinal and transverse propagators in Eqs.~\eqref{eq:longitudinal propagator stress propagator relation}, \eqref{eq:transverse propagator stress propagator relation} any Higgs term that gives a mass to the dual gauge fields $b^a_{\kappa\lambda}$ leads to correct results. The reason is that these quantities are by definition independent of the longitudinal components of $b^a_{\kappa\lambda}$ and as such cannot be influenced by the dislocation condensate phase degrees of freedom at all. However, we are also particularly interested in the fate of torque stress $\tau^c_\mu$, which we announced to become deconfined in the liquid crystals. To properly take this into account, it seems that the dislocation worldsheet condensate must follow the `stringy' form Eq.~\eqref{eq:string minimal coupling}. Therefore, below we shall consider the flavored generalization of Eqs.~\eqref{eq:vortex condensate vectorial phase}, \eqref{eq:string minimal coupling}.

\subsubsection{Dislocation worldsheet condensation}\label{subsubsec:Dislocation worldsheet condensation}

As we just argued, we shall not attempt to derive the full Higgs action of a condensate of dislocation worldsheets, but instead assume that this is governed by the minimal coupling form Eq. (\ref{eq:string minimal coupling}). In all other 
regards, there is no difference of principle compared to the construction in the 2+1D case~\cite{QLC2D}. Let us nevertheless spell out once again the assumptions.

\paragraph{Assumptions.}
As in 2+1D, the quantum liquid crystals in 3+1D can be characterized topologically as quantum liquids with the property that disclinations are massive topological excitations.  There is surely the practical question whether forms of matter exist
where disclinations are sufficiently costly compared to dislocations so that the simultaneous proliferation of dislocations and disclinations can be avoided; the latter case results in the standard first-order quantum  phase transition from the crystal to the isotropic superfluid as realized for instance in  $^4$He. This is however a matter of microscopic details which is beyond the field-theoretical scope: here we just assert that in principle such quantum liquid crystals can be formed in 3+1D, and study their low-energy properties. 

A next concern is the precise symmetry of the resulting quantum liquid-crystalline states. The quantum nematics are in this regard most straightforward. These break the $O(3)$ to some (usually discrete) point group. As we discussed in Sec.~\ref{sec:Symmetry principles of quantum liquid crystals}, the order parameter theory already in the classical case becomes a quite complicated affair when dealing with anything else than the simple uniaxial nematics found in LCD screens and so forth~\cite{LiuEtAl15,LiuEtAl16b,NissinenEtAl16,LiuEtAl16}. 
Here we will just avoid these difficulties by asserting that our quantum nematics are `isotropic', which means ``indistinguishable from the full spontaneous symmetry breaking of $O(3)$ itself'' when it comes to observable properties like the velocities of the rotational Goldstone bosons, or more generally the elasticity theory of the nematic~\cite{StallingaVertogen94}. This is the same `emergent' isotropy as which one encounters in the isotropic solid, see Sec.~\ref{subsec:Field-theoretic elasticity}. In principle this can be improved on by departing from the elasticity theory associated with the real 3D space groups, but this is just much more menial labor while it will 
not affect the basic construction. We leave this as an open problem for follow-up work. 

With these assumptions, we can afford to be rather ignorant regarding the details of the point group symmetries of the crystal and the quantum liquid crystals, given that the main rules for the construction of the dislocation condensate are 
insensitive to these details. As we learned in 2+1D~\cite{QLC2D}, the first two rules are as follows: (i) For every space dimension a separate dislocation condensate is needed, to restore the translation symmetry in that particular direction. In three space dimensions we are dealing with three separate condensates. Depending on the sign and the strength of the interactions between these condensates one obtains a columnar, a smectic 
or a nematic phase corresponding to dislocation condensates switching on in one, two or three spatial directions, see Fig.~\ref{fig:hexagonal melting}. 
(ii) A finite  `Burgers vector magnetization' is coincident with a finite disclination density and this should be forbidden in the quantum liquid crystal phases. Accordingly, the dislocation condensate in every spatial direction is characterized by the requirement that the Burgers vectors are constrained to be locally anti-parallel. 

In fact, by far the most important assumption with regard to the limitations on the physics we can address is the low-fugacity assumption. We have emphasized it already a number of times: it is assumed that all the degrees of 
freedom are collective, either corresponding to the phonons or the dislocations. In order to reach this limit, the distance between the dislocations in the condensate has to be very large compared to the lattice constant. The 
quantum liquid is  described as a system that is locally like a solid carrying propagating phonons. All the `disordering action' is due to the dislocations: the system turns into a liquid at distances larger than the average distance 
between the dislocations. This has the physical consequence that all the collective degrees of freedom are {\em propagating}, including the finite energy `massive shear photons'. This should be considered as just an extreme limit that will never precisely be reached in any physical system. The degrees of freedom that are ignored are the interstitials, in essence loose constituent particles that would be the fundamental degrees of freedom were one departing 
from the weakly-interacting `kinetic gas' limit which is the traditional starting point constructing theories of liquid crystals. These interstitials can be viewed in turn as bound dislocation--antidislocation pairs that could be included perturbatively departing from our weak--strong duality limit. This has not been studied systematically yet, but their qualitative effects are obvious: these would damp our massive propagating modes, to the extent that these modes can be entirely overdamped reaching the gaseous limit. In fact, these interstitials are much more of an issue in 3+1D as compared to the 2+1D case. In 2+1D both dislocations and interstitials are particle-like excitations and at least
in the Euclidean continuum the core energies of interstitials are just much higher than those of dislocations for ubiquitous microscopic reasons. However, in 3+1D interstitials continue to be particle-like, while the dislocations turn into strings. Accordingly, interstitials fluctuate much more easily and their gain in quantum kinetic energy may well overwhelm their high core-energies.  The state of matter formed by a proliferation of interstitials while dislocations stay massive is the {\em supersolid} and it is well understood that for these reasons supersolids are in principle ubiquitous in 3+1D. It depends yet again on the microscopic numbers but it appears for the same reasons to be much more problematic to reach the limit of maximal crystalline correlations in 3+1D as compared to the situation in lower dimensions.   

As a corollary to this ``no interstitial'' rule, the (anti-)dislocations can proliferate preserving the total number of particles: ``dislocations do not carry volume'', with the ramification that (iii) the glide constraint becomes absolute
(Sec.~\ref{subsec:Dislocations and disclinations}), and the ``no climb'' condition has to be imposed on the condensate. Just as in 2+1D~\cite{QLC2D}, this will have the consequence that the dislocations do not couple to compressional stress. Accordingly, sound stays massless in the (nematic) quantum liquid and it can be proven that this system is also a conventional superfluid.  

In summary, the construction of the dual stress superconductor in 3+1D follows in every regard the template we studied throughly in  2+1D, except for the fundamental description of the string condensate. We will assume that the condensate in each Burgers direction is governed by the minimal coupling form  Eq.~\eqref{eq:string minimal coupling}. 

\paragraph{The Higgs potential.}
Given these assumptions, we have to construct a Higgs potential  of $D=3$ complex scalar fields $\Phi_i$ ($i=x,y,z)$ associated with the separate dislocation condensates with Burgers directions in three spatial directions. As argued above and in Fig.~\ref{fig:hexagonal melting}, these directions must be orthogonal. Up to fourth order in the fields, the most general potential involving only density--density couplings is
\begin{align}
 \mathcal{L}_\mathrm{\lvert \Phi \rvert} &=  \tfrac{1}{2} \alpha_x | \Phi^x |^2 + \tfrac{1}{2} \alpha_y | \Phi^y |^2 + \tfrac{1}{2} \alpha_z | \Phi^z |^2 
  \nonumber\\ &\phantom{mm}
  + \tfrac{1}{4} \beta_x | \Phi^x |^4  + \tfrac{1}{4} \beta_y | \Phi^y |^4  + \tfrac{1}{4} \beta_z | \Phi^z |^4 \nonumber\\&\phantom{mm}  
 + \tfrac{1}{2}\gamma_{xy}  | \Phi^x |^2 | \Phi^y |^2
 + \tfrac{1}{4}\gamma_{xz}  | \Phi^x |^2 | \Phi^z |^2 \nonumber\\
 &\phantom{mm}
 + \tfrac{1}{4}\gamma_{yz}  | \Phi^y |^2 | \Phi^z |^2. 
 \label{eq:three-condensate potential}
\end{align}
This potential is identical to the one describing a three-component Bose--Einstein condensate, which was studied in Ref.~\onlinecite{RobertsUeda06}. There it was established that the condensate consists of either one,
 two or three species of bosons, pending the values of the couplings $\gamma_{xy}$, $\gamma_{xz}$, $\gamma_{yz}$. In the present context, it is possible to have dislocation condensates which restore translational symmetry
  in one, two or three directions.
In analogy to the  classical liquid crystals we will call these the {\em columnar}, {\em smectic} and {\em nematic} phases respectively, regardless the nature of the remnant rotational symmetry breaking, see Fig.~\ref{fig:phases} and Fig.~\ref{fig:hexagonal melting}. 

\paragraph{Minimal coupling.}
Having established the form of the Higgs potential, we now need to couple the condensate fields $\Phi^a$ to the dual stress gauge fields $b^a_{\mu\nu}$. Following Sec.~\ref{subsubsec:Vortex worldsheet condensation}, the coupling is
\begin{align}
 \mathcal{L}_\mathrm{min.coup.}  &= \frac{1}{4} \sum_a \lvert \Phi^a \rvert^2 ( \partial_\mu \phi^a_\nu - \partial_\nu \phi^a_\mu - b^a_{\mu\nu} )^2,\label{eq:dislocation condensate minimal coupling step 1}
\end{align}
where $\phi^a_\mu$ are the vectorial phase fields of condensate $\Phi^a$ as in Eq.~\eqref{eq:string minimal coupling}. 

The next issue to address is the velocity of the dislocation condensate. In the vortex--boson duality governing superfluids, the circulation around vortices is obviously ordinary superflow, so the vortex itself also moves with the superfluid phase velocity. From a different viewpoint, at the quantum critical point with emergent relativity there can be only one velocity scale. Nevertheless, it is useful to keep the vortex velocity scale formally distinct from the superfluid phase velocity in the calculations, to determine the nature of the massive modes in the Higgs phase~\cite{CvetkovicZaanen06a,QLC2D}. 

For dislocations, the situation is slightly different. First of all there is the glide constraint restriction certain dislocation motion. Furthermore, it is known that screw and edge dislocations may move at different velocities, see Sec.~\ref{subsec:Dislocation worldsheets}.
Below, we will first treat an idealized case where all dislocations move with a single velocity $c_\td$, as it contains the important physics such as the emergent rotational Goldstone modes while the equations remain tractable. Afterwards, we will discuss the more realistic case with two velocity scales $c_\ts$ and $c_\te$. In the first case, the minimal coupling Eq.~\eqref{eq:dislocation condensate minimal coupling step 1} becomes,
\begin{align}
  \mathcal{L}_\mathrm{min.coup.}  &= \frac{1}{4} \sum_a \lvert \Phi^a \rvert \Big(  \frac{1}{c_\td^2} 2(\partial_\tau \phi^a_n - \partial_n \phi^a_\tau - b^a_{\tau n} )^2 \nonumber\\
  &\phantom{mmmmm} + (\partial_m \phi^a_n - \partial_n \phi^a_m - b^a_{m n} )^2\Big).
  \label{eq:dislocation condensate minimal coupling step 2}
\end{align}
Here summation over the Latin indices $m,n$ is implied and the factor of 2 in the first term accounts for the antisymmetry in $\tau, n$. It is easiest to perform the calculations in the dislocation-Lorenz gauge fix 
\begin{equation}
\frac{1}{c_\td^2} \partial_\tau b^a_{\tau \nu} + \partial_m b^a_{m \nu} = 0,\label{eq:dislocation Lorenz gauge fix} 
\end{equation}
since we can be sure that  the condensate phase degrees of freedom are decoupled. It is then convenient to rewrite the whole expression with temporal components rescaled with $c_\td$ instead of $c_\tT$. 
For this purpose we define fields and derivatives indicated with a tilde, as follows~\cite{QLC2D}:
\begin{align}
 \tilde{b}^a_{\kappa\lambda} &= ( \frac{1}{c_\td} b^a_{\tau l} , b^a_{kl} ) ,& \tilde{\phi}^a_\lambda &= (\frac{1}{c_\td} \phi_\tau,\phi_l), \nonumber\\
 \tilde{\partial}_\mu &= ( \frac{1}{c_\td} \partial_\tau , \partial_m ), &
 \tilde{p}_\mu &= (\frac{1}{c_\td} \omega_n , q_m).\label{eq:tilde fields definition}
\end{align}
The relations between the original and the tilde fields are
\begin{align}
 \tilde{b}^a_{\ft n} &= \frac{c_\tT}{c_\td} b^a_{\ft n} = \frac{1}{c_\td} b^a_{\tau n}, &
 \tilde{p}_\ft = \frac{c_\tT}{c_\td} p_\ft = \frac{1}{c_\td} \omega_n.
\end{align}
Clearly this affects the $0,1,\ft$-components, but the $\tL,\tR,\tS$-components are unchanged: $A_{\tL,\tR,\tS} = \tilde{A}_{\tL,\tR,\tS}$ for any field $A_\mu$. With these redefinitions, the minimal coupling term can be written as
\begin{align}
 \mathcal{L}_\mathrm{min.coup.}  &= \frac{1}{4}\sum_a \lvert \Phi^a \rvert ( \tilde{\partial}_\mu \tilde{\phi}^a_\nu - \tilde{\partial}_\nu \tilde{\phi}^a_\mu - \tilde{b}^a_{\mu\nu} )^2,
  \label{eq:dislocation condensate minimal coupling step 3}
\end{align}

\paragraph{Glide constraint.}
Finally, we need to implement the glide constraint Eq.~\eqref{eq:glide constraint} which dictates that edge dislocations can only move in the direction parallel to their Burgers vector, and which is rooted in particle number conservation. In three dimensions, it is possible that the dislocation line moves perpendicular to its Burgers vector at some part along the line, while it moves in the opposite direction somewhere else along its line, effectively `borrowing' an interstitial particle from itself~\cite{CvetkovicNussinovZaanen06}. This is sometimes called {\em restricted climb motion}. The glide constraint Eq.~\eqref{eq:glide constraint} then holds for the line as a whole, so that we have the integral identity $\int \td^D x\; \epsilon_{\ft mna} J^a_{mn}(x) = 0$. The glide constraint can be enforced in the path integral using a Lagrange multiplier field $\lambda(x)$:
\begin{equation}
 Z_\mathrm{glide} = \int \mathcal{D} \lambda \te^{\int \lambda  \epsilon_{\ft mna} J^a_{mn}}.
\end{equation}
Since the dual stress gauge field is minimally coupled to dislocation sources, this amounts to the shift~\cite{QLC2D},
\begin{equation}\label{eq:glide constraint substitution}
 \tilde{b}^a_{\mu\nu} \to \tilde{b}^a_{\mu\nu} + \lambda  \epsilon_{\ft \mu\nu a}.
\end{equation}
in the minimal coupling term. 
Notice that since the Levi-Civita symbol already contains an entry $\ft$, we can replace $m \to \mu$, $n \to \nu$ with impunity. Also, since the glide constraint only concerns components with spatial indices, we can replace tilde fields as $\tilde{b}^a_{mn} = b^a_{mn}$. With regard to the dislocation condensate, the factors $b^a_{\mu\nu}$ can be simply substituted  by Eq.~\eqref{eq:glide constraint substitution}. This is exactly the same procedure as followed in the 2+1D case~\cite{QLC2D,Cvetkovic06}. Later we will see that this constraint has the effect that the compression mode always stays massless although the shear modes acquire a Higgs mass. 
This of course makes sense: the quantum liquid crystal is at the same time a superfluid, characterized  by a massless sound mode.

\paragraph{Higgs term for a single velocity.}
We have now collected all the pieces and  we can write down the Higgs term of the dislocation condensate coupled to two-form dual stress gauge fields:
\begin{widetext}
 \begin{align}
 \mathcal{L}_\mathrm{Higgs} &= \frac{1}{4}\sum_{a} \lvert \Phi^a \rvert^2
  (\tilde{\partial}_\mu \tilde{\phi}^a_\nu - \tilde{\partial}_\nu \tilde{\phi}^a_\mu - \tilde{b}^a_{\mu\nu} - \lambda \epsilon_{\ft \mu \nu a} ) ^2 \\
  &=
  \frac{1}{4} \sum_a \frac{(\Omega^{a})^2}{c_\tT^2 \mu} \Big[ 
  \lvert \tilde{\partial}_\mu\tilde{\phi}^a_\nu - \tilde{\partial}_\nu\tilde{\phi}^a_\mu \rvert^2 +
    (\tilde{b}^{a\dagger}_{\mu\nu} + \lambda^\dagger \epsilon_{\tau \mu \nu a})  (\delta_{\mu\kappa} - \frac{\tilde{p}_\mu \tilde{p}_\kappa}{\tilde{p}^2} )(\delta_{\nu\lambda} - \frac{\tilde{p}_\nu \tilde{p}_\lambda}{\tilde{p}^2} ) (\tilde{b}^a_{\kappa\lambda} + \lambda \epsilon_{\tau \kappa\lambda a})\Big].
 \label{eq:dislocation Higgs term} 
\end{align}
\end{widetext}
Here we have defined the Higgs mass $\Omega^a = c_\tT \sqrt{\mu} \lvert \Phi^a \rvert$ with units of energy, and imposed the Lorenz gauge fix in the second line, as indicated by the projectors. The dislocation condensate phase modes $\tilde{\phi}^a_\mu$ are clearly decoupled in this gauge fix, and only the gauge-invariant parts of the stress gauge fields appear. The sum over $a$ is the sum over the Burgers directions of the dislocation condensate, which could be in one, two or three spatial directions depending on each of the $\Omega^a$. As argued in QLC2D and in Secs.~\ref{sec:Symmetry principles of quantum liquid crystals}, \ref{subsubsec:Dislocation condensation} these directions are strictly orthogonal to each other, see Fig.~\ref{fig:hexagonal melting}, while at least one of them lies along a crystal axis.

\paragraph{Lagrangian and propagators of the solid in the dislocation gauge fix.}
Eq.~\eqref{eq:dislocation Higgs term} is the Higgs term we shall use in all the calculations below. For future use, let us rewrite the Lagrangian of the elastic medium Eq.~\eqref{eq:solid stress gauge field Lagrangian} in terms of the tilde fields:
\begin{widetext}
\begin{align}
\mathcal{L}_\mathrm{dual} &= \mathcal{L}_{\tT 1} + \mathcal{L}_{\tT 2} + \mathcal{L}_{\tT 3} + \mathcal{L}_\tL + \ti \tilde{b}^{a\dagger}_{\mu\nu} \tilde{J}^a_{\mu\nu},\label{eq:solid stress gauge field Lagrangian tilde}\\
 \mathcal{L}_{\tT 1} &= \frac{1}{8\mu}
\begin{pmatrix}\tilde{b}^{\tL\dagger}_{1 \tS} &  \tilde{b}^{\tR\dagger}_{\tR\tS} \end{pmatrix}
\begin{pmatrix} 
 \frac{c_\td^2}{c_\tT^2}\tilde{p}^2(1 + \frac{1}{\ell^2 q^2}) & \ti \frac{1}{c_\tT}\omega_n \frac{c_\td}{c_\tT}\tilde{p}(1 - \frac{1}{\ell^2 q^2}) \\
 -\ti \frac{1}{c_\tT}\omega_n \frac{c_\td}{c_\tT}\tilde{p}(1 - \frac{1}{\ell^2 q^2}) & \frac{1}{c_\tT^2}\omega_n^2(1 + \frac{1}{\ell^2 q^2}) + 4 q^2 
 \end{pmatrix}
\begin{pmatrix}\tilde{b}^{\tL}_{1 \tS} \\  \tilde{b}^{\tR}_{\tR\tS} \end{pmatrix}, \\
\mathcal{L}_{\tT 2} &= \frac{1}{8\mu} 
\begin{pmatrix}\tilde{b}^{\tL\dagger}_{1 \tR} &  \tilde{b}^{\tS\dagger}_{\tR\tS} \end{pmatrix}
\begin{pmatrix} 
 \frac{c_\td^2}{c_\tT^2}\tilde{p}^2(1 + \frac{1}{\ell^2 q^2}) & \ti \frac{1}{c_\tT}\omega_n \frac{c_\td}{c_\tT}\tilde{p}(1 - \frac{1}{\ell^2 q^2}) \\
 -\ti \frac{1}{c_\tT}\omega_n \frac{c_\td}{c_\tT}\tilde{p}(1 - \frac{1}{\ell^2 q^2}) & \frac{1}{c_\tT^2}\omega_n^2(1 + \frac{1}{\ell^2 q^2}) + 4 q^2 
 \end{pmatrix}
\begin{pmatrix}\tilde{b}^{\tL}_{1 \tR} \\  \tilde{b}^{\tS}_{\tR\tS} \end{pmatrix}, \\
\mathcal{L}_{\tT 3} &= \frac{1}{8\mu}
\begin{pmatrix}\tilde{b}^{\tR\dagger}_{1 \tR} & \tilde{b}^{\tS\dagger}_{1\tS} \end{pmatrix}
\begin{pmatrix} 
 \frac{c_\td^2}{c_\tT^2}\tilde{p}^2(1 + \frac{1}{\ell^2 q^2}) & \frac{c_\td^2}{c_\tT^2}\tilde{p}^2(1 - \frac{1}{\ell^2 q^2}) \\
 \frac{c_\td^2}{c_\tT^2}\tilde{p}^2(1 - \frac{1}{\ell^2 q^2}) & \frac{c_\td^2}{c_\tT^2}\tilde{p}^2(1 + \frac{1}{\ell^2 q^2})
 \end{pmatrix}
\begin{pmatrix}\tilde{b}^{\tR}_{1 \tR} \\  \tilde{b}^{\tS}_{1 \tS} \end{pmatrix}, \label{eq:L3 tilde}\\
\mathcal{L}_{\tL 1} &= \frac{1}{8\mu}  \frac{2}{1+\nu}
\begin{pmatrix}\tilde{b}^\dagger_{1 -} &  \tilde{b}^{\tL \dagger}_{\tR\tS} \end{pmatrix}
\begin{pmatrix} 
 (1-\nu) \frac{c_\td^2}{c_\tT^2}\tilde{p}^2 &  \ti\sqrt{2} \nu \frac{1}{c_\tT}\omega_n \frac{c_\td}{c_\tT}\tilde{p} \\
  -\ti \sqrt{2}\nu \frac{1}{c_\tT}\omega_n \frac{c_\td}{c_\tT}\tilde{p} &  \frac{1}{c_\tT^2}\omega_n^2 + 2 (1 + \nu) q^2 
 \end{pmatrix}
\begin{pmatrix}\tilde{b}_{1 -}\\ \tilde{b}^{\tL }_{\tR\tS} \end{pmatrix}. \label{eq:solid Lagrangian gauge field L1 tilde}\\
\mathcal{L}_{\tL 2} &=  \frac{1}{4\mu} \frac{c_\td^2}{c_\tT^2}\tilde{p}^2 \lvert \tilde{b}_{1+} \rvert^2\label{eq:solid Lagrangian gauge field L2 tilde}
\end{align}

Using the relation $A_1 = \frac{c_\td}{c_\tT}\frac{\tilde{p}}{p} \tilde{A}_1$ for any field $A_\mu$, the stress propagators Eqs.~\eqref{eq:longitudinal propagator stress propagator relation}, \eqref{eq:transverse propagator gauge field} 
become in this gauge fix, 
\begin{align}
 G_\tL &=\frac{1}{\kappa} - \frac{1}{(3\kappa)^2} \Big[ 
  \frac{\omega_n^2}{c_\tT^2} \langle \tilde{b}^{\tL\dagger}_{\tR \tS}\;  \tilde{b}^\tL_{\tR \tS} \rangle
 + 2 \frac{c_\td^2}{c_\tT^2} \tilde{p}^2 \langle \tilde{b}^\dagger_{ 1 -}\;  \tilde{b}_{1 -} \rangle 
 + \ti \sqrt{2} \frac{\omega_n}{c_\tT}  \frac{c_\td}{c_\tT} \tilde{p} 
    \Big( \langle \tilde{b}^{\tL\dagger}_{\tR \tS}\;  \tilde{b}_{1 -} \rangle 
      - \langle \tilde{b}^\dagger_{1 -}\; \tilde{b}^{\tL}_{\tR \tS} \rangle 
    \Big) 
 \Big],
 \label{eq:longitudinal propagator gauge field tilde}\\
  G_{\tT 1} &= \frac{1}{\mu \ell^2 q^2} - \frac{1}{(2 \mu \ell^2 q^2)^2} \Big[
\frac{c_\td^2}{c_\tT^2} \tilde{p}^2 \langle \tilde{b}^{\tL\dagger}_{1 \tS} \; \tilde{b}^{\tL}_{1 \tS} \rangle + \frac{\omega_n^2}{c_\tT^2} \langle \tilde{b}^{\tR\dagger}_{\tR\tS}\;  \tilde{b}^\tR_{\tR \tS}\rangle + \ti \frac{\omega_n}{c_\tT} \frac{c_\td}{c_\tT} \tilde{p} \Big( \langle \tilde{b}^{\tR\dagger}_{\tR\tS} \; \tilde{b}^\tL_{1 \tS}\rangle - \langle \tilde{b}^{\tL\dagger}_{1\tS}\;  \tilde{b}^\tR_{\tR \tS}\rangle \Big) 
 \Big],\nonumber\\
 G_{\tT 2} &= \frac{1}{\mu \ell^2 q^2} - \frac{1}{(2 \mu \ell^2 q^2)^2} \Big[
 \frac{c_\td^2}{c_\tT^2}\tilde{p}^2 \langle \tilde{b}^{\tL\dagger}_{1 \tR}\;  \tilde{b}^{\tL}_{1 \tR} \rangle + \frac{\omega_n^2}{c_\tT^2} \langle \tilde{b}^{\tS\dagger}_{\tR\tS}\;  \tilde{b}^\tS_{\tR\tS}\rangle + \ti \frac{\omega_n}{c_\tT} \frac{c_\td}{c_\tT} \tilde{p} \Big( \langle \tilde{b}^{\tS\dagger}_{\tR\tS} \; \tilde{b}^\tL_{1 \tR}\rangle - \langle \tilde{b}^{\tL\dagger}_{1\tR}\;  \tilde{b}^\tS_{\tR\tS}\rangle \Big)
 \Big],\nonumber\\
 G_{\tT 3} &= \frac{1}{\mu \ell^2 q^2} - \frac{1}{(2 \mu \ell^2 q^2)^2}  \Big[\frac{c_\td^2}{c_\tT^2} \tilde{p}^2  \langle \tilde{b}^{\tR\dagger}_{1 \tR} \; \tilde{b}^{\tR}_{1 \tR} \rangle + \frac{c_\td^2}{c_\tT^2} \tilde{p}^2  \langle \tilde{b}^{\tS\dagger}_{1 \tS} \; \tilde{b}^\tS_{1\tS }\rangle +  \frac{c_\td^2}{c_\tT^2} \tilde{p}^2  \langle \tilde{b}^{\tS\dagger}_{1\tS} \; \tilde{b}^\tR_{1 \tR}\rangle +  \frac{c_\td^2}{c_\tT^2} \tilde{p}^2 \langle \tilde{b}^{\tR\dagger}_{1\tR}\;  \tilde{b}^\tS_{1\tS }\rangle \Big].\label{eq:transverse propagator gauge field tilde}
\end{align}
\end{widetext}

\paragraph{Dislocation condensates with two velocities.}
Eqs.~\eqref{eq:dislocation Higgs term}  and \eqref{eq:solid stress gauge field Lagrangian tilde}  suffice to calculate the long wavelength  properties of the columnar, smectic and nematic phases in the idealized case where there is only one dislocation velocity scale $c_\td$. For completeness we now consider the more general case where screw dislocations move with velocity $c_\ts$ while edge dislocations move with velocity $c_\te$, see Sec.~\ref{subsec:Dislocation worldsheets}. As we mentioned there, the dislocation line density $J^a_{\tau n}$ 
represents screw dislocations when $a = n$ and edge dislocations when $a \neq n$. Since these dislocations couple minimally to dual stress gauge field components $b^a_{\tau n}$, it is clear that in the generalization of Eq.~\eqref{eq:dislocation condensate minimal coupling step 2}, the terms  $\lvert b^a_{\tau n} \rvert^2$ with $a = n$ should come with the screw dislocation velocity $c_\ts$ while those with $a \neq n$ should come with the edge dislocation velocity $c_\te$. Explicitly, the Higgs term is (with implicit summation over $m,n$)
\begin{widetext}
 \begin{align}
  \mathcal{L}_\mathrm{min.coup.}  &= \frac{1}{4} \sum_a \lvert \Phi^a \rvert^2 \Big(  \frac{1}{c_\ts^2} 2(\partial_\tau \phi^a_a - \partial_a \phi^a_\tau - b^a_{\tau a} )^2 
   +  \sum_{b \neq a}\frac{1}{c_\te^2} 2(\partial_\tau \phi^a_b - \partial_b \phi^a_\tau - b^a_{\tau b} )^2
   + (\partial_m \phi^a_n - \partial_n \phi^a_m - b^a_{m n} )^2\Big).
  \label{eq:dislocation condensate minimal coupling step 4}
\end{align} 
\end{widetext}

\paragraph{The Lorenz gauge fix with two velocities.}
It is possible to choose the Lorenz gauge fix $\frac{1}{c^2} \partial_\tau b^a_{\tau \nu} + \partial_m b^a_{m\nu} = 0$ with a separate velocity for each combination of $a,\nu$. That is, the gauge conditions are
\begin{align}
 \frac{1}{c_\ts^2} \partial_\tau b^a_{\tau a} + \partial_m b^a_{m a} &= 0\qquad {\text{(no sum } a \text{)}},\label{eq:screw gauge fix}\\
 \frac{1}{c_\te^2} \partial_\tau b^a_{\tau b} + \partial_m b^a_{m b} &= 0\qquad a \neq b.\label{eq:edge gauge fix}
\end{align}
This can be seen as follows. The gauge transformation is Eq.~\eqref{eq:stress gauge transformation}, which in the presence of the condensate fields is
\begin{align}\label{eq:stress gauge transformation dislocation condensate}
   b^a_{\mu\nu} (x) &\to b^a_{\mu\nu}(x) + \partial_\mu  \varepsilon^a_\nu(x) - \partial_\nu  \varepsilon^a_\mu(x).\nonumber\\
  \phi^a_\nu(x) &\to \phi^a_\nu(x) +  \varepsilon^a_\nu(x) + \partial_\nu \zeta^a(x).
\end{align}
where $\zeta^a$ is any flavored scalar field independent of $\varepsilon^a_\nu$, while the 
Lagrangian Eq.~\eqref{eq:dislocation condensate minimal coupling step 4} is invariant under the addition of the gradient. However, two sets of transformations of $\varepsilon^a_\nu$, $\zeta^a$ that differ as
 follows with respect to a flavored scalar field $\eta^a$
\begin{align}
 \varepsilon^a_\nu &\to \varepsilon^a_\nu + \partial_\nu \eta^a, &
 \zeta^a &\to \zeta^a - \eta^a
\end{align}
will lead to the exact same gauge transformations Eq.~\eqref{eq:stress gauge transformation dislocation condensate}. In others words, there is a redundancy in the gauge transformations themselves, a ``gauge-in-the-gauge''. We can use this freedom to choose $\eta^a$ in such a way that
\begin{equation}
 \frac{1}{c_\ts^2} \partial_\tau \varepsilon^a_{\tau} + \partial_m \epsilon^a_m = 0. \label{eq:gauge-in-the-gauge fix}
\end{equation}
Consider the transformation of the following quantity (no sum over $a$),
\begin{align}
&\frac{1}{c_\ts^2} \partial_\tau b^a_{\tau a} + \partial_m b^a_{m a} \to \nonumber\\
& \frac{1}{c_\ts^2} \partial_\tau b^a_{\tau a} + \partial_m b^a_{m a} + (\frac{1}{c_\ts^2} \partial_\tau^2 + \partial_m^2)\varepsilon^a_a + \partial_a (\frac{1}{c_\ts^2} \partial_\tau \varepsilon^a_\tau + \partial_m \varepsilon^a_m).
\end{align}
The last term vanished by the gauge-in-the-gauge fix, and we see that we can choose $\varepsilon^a_a$ in such a way that Eq.~\eqref{eq:screw gauge fix} holds. Afterwards, 
taking the divergence with respect to the velocity $c_\te$ and performing a gauge transformation leads  for the edge components $b^a_{\mu b}$ ($a \neq b$) to,
\begin{align}
&\frac{1}{c_\te^2} \partial_\tau b^a_{\tau b} + \partial_m b^a_{m b} \to \nonumber\\
& \frac{1}{c_\te^2} \partial_\tau b^a_{\tau b} + \partial_m b^a_{m b} + (\frac{1}{c_\te^2} \partial_\tau^2 + \partial_m^2) \varepsilon^a_b + (1 - \frac{c_\ts^2}{c_\te^2}) \partial_b \partial_m \varepsilon^a_m.
\end{align}
Here we used Eq.~\eqref{eq:gauge-in-the-gauge fix}. Now we can choose $\varepsilon^a_b$ ($a \neq b$) in such a way that this whole expression vanishes. 

After imposing these gauge fixes Eqs.~\eqref{eq:screw gauge fix}, \eqref{eq:edge gauge fix}, the condensate phase degrees of freedom have been decoupled and  do not explicitly contribute to the stress propagators. The recipe to perform the calculations is to add the solid Lagrangian Eq.~\eqref{eq:solid stress gauge field Lagrangian} in the same gauge fix to the Higgs term Eq.~\eqref{eq:dislocation condensate minimal coupling step 4} and use this to calculate the various stress propagators.

%% file: sec_nematic.tex
The hard work of establishing the form of the dislocation condensate Higgs term Eq.~\eqref{eq:dislocation Higgs term} (or Eq.~\eqref{eq:dislocation condensate minimal coupling step 4}) has been accomplished.
We are now ready to study the  long-wavelength physics of the quantum liquid crystal phases by calculating  the stress propagators Eqs.~\eqref{eq:longitudinal propagator gauge field tilde}, 
\eqref{eq:transverse propagator gauge field tilde}. This is now a remarkably straightforward affair, even though the equations may seem unwieldy. We shall look at the nematic phases first. 
Not only are the nematic phases of primary interest, but they are also easier to compute and understand, 
because of their higher symmetry as compared to the smectic and columnar phases. For this reason, we shall treat the liquid crystals in order of decreasing spatial symmetry.

As argued before, we shall focus exclusively on the `isotropic nematic': the phase which originates from dislocation-mediated melting of an isotropic solid, see Sec.~\ref{subsec:Generalizing nematic order}. This is in fact an almost literal `spherical 
cow' simplification. One should depart from the space groups of the crystals, and upon dislocation melting a nematic will form with the associated point group symmetry. Different from the 2+1D triangular lattice and hexatic liquid crystal, 
in 3+1D there are no space and point groups that allow for isotropic elasticity. In Refs.~\onlinecite{LiuEtAl15,LiuEtAl16b,NissinenEtAl16,LiuEtAl16} the systematic  order parameter theory of {\em generalized nematics}
associated with arbitrary 3D point groups is developed, while in Ref.~\onlinecite{StallingaVertogen94} the classical elasticity theories of such nematics are derived. A crucial observation  is that our idealized isotropic nematic is {\em very different} 
from the ubiquitous uniaxial nematics of soft condensed matter. These are characterized by a particular low and in a way pathological symmetry: the  $D_{\infty \mathrm{h}}$ space group which is highly special since it has a continuous $O(2)$ subgroup leading to a simple Abelian $\mathbb{Z}_2$ action on the director order parameter. Different from 2D,  point groups in 3D are generically non-Abelian. One consequence of crucial importance in the present context is that the uniaxial nematic only carries two rotational Goldstone bosons
while generic 3D point groups support  {\em three} Goldstone modes  which are `polarized' like phonons (one longitudinal and two transverse modes). The crystal~\cite{LandauLifshitz86,Kleinert89b} and 
descendant nematic liquid crystal~\cite{StallingaVertogen94}
with the highest degree of rotational symmetry is the cubic one, with three elastic constants~\cite{LandauLifshitz86,Kleinert89b}.  In the remainder one may view our isotropic nematic as a `cubic' $O_\mathrm{h}$ nematic where the elastic constant associated with cubic anisotropy is  set to zero.  
   
We shall only treat the idealized case with a single dislocation velocity $c_\td$, Eq.~\eqref{eq:dislocation Higgs term}, which is representative for the main features of the physics while being mathematically transparent. The nematic phase corresponds to the stress superconductor where the dislocations with Burgers vectors in all (three) directions proliferate. This means that the sum in Eq.~\eqref{eq:dislocation Higgs term} involves all spatial directions, so $\Omega^a \neq 0$, $a = x,y,z$. We shall assume that the Higgs mass is the same in all three directions $\Omega^a = \Omega$. Such a phase is expected from the theory Eq.~\eqref{eq:three-condensate potential} at least 
for certain values of the coupling parameters $\gamma_{xy}$, $\gamma_{xz}$, $\gamma_{yz}$~\cite{RobertsUeda06}, and it is natural when viewing it as a $O_\mathrm{h}$ nematic.  
Because the condensate phase degree of freedom has been decoupled in the dislocation-Lorenz gauge fix Eq.~\eqref{eq:dislocation Lorenz gauge fix}, we disregard those terms from now on. 

Collecting these ingredients, the Higgs term we will consider is (with summation over repeated Greek indices) , 
\begin{align}
 \mathcal{L}_\mathrm{Higgs} &=
  \frac{1}{4} \sum_{a = x,y,z} \frac{\Omega^2}{ c_\tT^2 \mu} \Big[ 
   (\tilde{b}^{a\dagger}_{\mu\nu} + \lambda^\dagger \epsilon_{\tau \mu \nu a})  (\delta_{\mu\kappa} - \frac{\tilde{p}_\mu \tilde{p}_\kappa}{\tilde{p}^2} ) \nonumber\\
   &\phantom{mmmmm} \times (\delta_{\nu\lambda} - \frac{\tilde{p}_\nu \tilde{p}_\lambda}{\tilde{p}^2} ) (\tilde{b}^a_{\kappa\lambda} + \lambda \epsilon_{\tau \kappa\lambda a})\Big].
 \label{eq:dislocation Higgs term nematic} 
\end{align}

\subsection{Incorporating the glide constraint}\label{subsec:nematic glide constraint}

The first thing we need to do is to implement the glide constraint by integrating out the Lagrange multiplier field $\lambda$. This is done in the usual way: complete the square, shift the integrand and perform the path integral to leave an overall constant factor~\cite{QLC2D}.  Eq.~\eqref{eq:dislocation Higgs term nematic} is rewritten as, 
\begin{align}
 \mathcal{L}_\mathrm{Higgs} &=
  \frac{1}{4}\sum_{a=x,y,z} \frac{\Omega^2}{ c_\tT^2 \mu} \Big[ 
 2 \lvert \tilde{b}^a_{1\tR} \rvert^2 + 2\lvert \tilde{b}^a_{1\tS} \rvert^2 + 2\lvert \tilde{b}^a_{\tR\tS} \rvert^2 \nonumber\\
  &\phantom{mm}
  + \lambda^\dagger \lambda \epsilon_{\tau \mu \nu a} \epsilon_{\tau \kappa\lambda a} (\delta_{\mu\kappa} - \frac{\tilde{p}_\mu \tilde{p}_\kappa}{\tilde{p}^2} ) (\delta_{\nu\lambda} - \frac{\tilde{p}_\nu \tilde{p}_\lambda}{\tilde{p}^2} )  \nonumber\\
  &\phantom{mm}
  + \lambda^\dagger \epsilon_{\tau \mu \nu a} (\delta_{\mu\kappa} - \frac{\tilde{p}_\mu \tilde{p}_\kappa}{\tilde{p}^2} ) (\delta_{\nu\lambda} - \frac{\tilde{p}_\nu \tilde{p}_\lambda}{\tilde{p}^2} ) \tilde{b}^a_{\kappa\lambda} \nonumber\\
  &\phantom{mm}
  + \tilde{b}^{a\dagger}_{\mu\nu} (\delta_{\mu\kappa} - \frac{\tilde{p}_\mu \tilde{p}_\kappa}{\tilde{p}^2} ) (\delta_{\nu\lambda} - \frac{\tilde{p}_\nu \tilde{p}_\lambda}{\tilde{p}^2} ) \epsilon_{\tau \kappa\lambda a} \lambda\Big].
 \label{eq:dislocation Higgs term nematic 2} 
\end{align}
As usual, the factors of 2 in the first line arise from summing over antisymmetric components. For the last two lines, we recognize that the projectors $(\delta_{\mu\kappa} - \tilde{p}_\mu\tilde{p}_\kappa/\tilde{p}^2)$ etc.
enforce the Lorenz gauge fix on the dual stress gauge fields. The only remaining terms remaining are a quadratic term $\lambda^\dagger\lambda$ and $\lambda^\dagger \epsilon_{\tau mn a} b^a_{mn}$ and its Hermitian conjugate. For the second line, we calculate
\begin{align}
 &\epsilon_{\tau \mu \nu a} \epsilon_{\tau \kappa\lambda a} (\delta_{\mu\kappa} - \frac{\tilde{p}_\mu \tilde{p}_\kappa}{\tilde{p}^2} ) (\delta_{\nu\lambda} - \frac{\tilde{p}_\nu \tilde{p}_\lambda}{\tilde{p}^2} ) \nonumber\\
 &\phantom{mm}= \epsilon_{\tau \mu \nu a} \epsilon_{\tau \mu\nu a} - 2\epsilon_{\tau \mu \nu a}\epsilon_{\tau \kappa \nu a}  \frac{\tilde{p}_\mu \tilde{p}_\kappa}{\tilde{p}^2} \nonumber\\
 &\phantom{mm}= 2 - 2 \frac{q^2 - q_a^2}{\tilde{p}^2} = 2\frac{\frac{1}{c_\td^2}\omega_n^2 + q_a^2}{\tilde{p}^2}, \qquad \text{no sum } a.\label{eq:glide constraint multiplier field prefactor}
\end{align}
Here we used $\tilde{p}^2 = \frac{1}{c_\td^2}\omega_n^2 + q^2$. For the nematic, we sum over $a= x,y,z$ to find
\begin{equation}
 \sum_{a = x,y,z} 2\frac{\frac{1}{c_\td^2}\omega_n^2 + q_a^2}{\tilde{p}^2} = 2\frac{3 \frac{1}{c_\td^2} \omega_n^2 + q^2}{\tilde{p}^2}.
\end{equation}
All together Eq.~\eqref{eq:dislocation Higgs term nematic 2} becomes
\begin{align}
 \mathcal{L}_\mathrm{Higgs} &=
 \frac{\Omega^2}{4 c_\tT^2 \mu}   \Big[   \sum_{a=x,y,z}\Big(  
  2\lvert \tilde{b}^a_{1\tR} \rvert^2 + 2\lvert \tilde{b}^a_{1\tS} \rvert^2 + 2\lvert \tilde{b}^a_{\tR\tS} \rvert^2 \Big) \nonumber\\
  &\phantom{mmmmm}
  + \lambda^\dagger \lambda  \ 2\frac{3 \frac{1}{c_\td^2} \omega_n^2 + q^2}{\tilde{p}^2}
  \nonumber\\
  &\phantom{mmmmm}
  + 2\lambda^\dagger (\tilde{b}^x_{yz} +\tilde{b}^y_{zx} +\tilde{b}^z_{xy}) + \mathrm{h.c.}\Big] \nonumber\\
  &= \frac{\Omega^2}{2 c_\tT^2 \mu}   \Big[   \sum_{a=x,y,z}\Big(  
  \lvert \tilde{b}^a_{1\tR} \rvert^2 + \lvert \tilde{b}^a_{1\tS} \rvert^2 + \lvert \tilde{b}^a_{\tR\tS} \rvert^2 \Big) \nonumber\\
   &\phantom{mmnmm}-  \frac{\tilde{p}^2}{3 \frac{1}{c_\td^2} \omega_n^2 + q^2} \lvert \tilde{b}^x_{yz} +\tilde{b}^y_{zx} +\tilde{b}^z_{xy} \rvert^2 \Big].\label{eq:dislocation Higgs term nematic 3} 
\end{align}
The last equality  arises after integrating out the Lagrange multiplier field $\lambda$. We will soon verify that the compression mode remains massless in the dislocation condensate 
precisely because of the extra term arising from the glide constraint on the last line of Eq. \eqref{eq:dislocation Higgs term nematic 3}. In the case of the gauge-fixed nematic we can use the relation
\begin{align}
  \tilde{b}^x_{yz} +\tilde{b}^y_{zx} +\tilde{b}^z_{xy} 
  &= \tilde{b}^\tL_{\tR\tS}   + \ti \frac{\omega_n}{c_\td \tilde{p}} \tilde{b}^\tR_{1 \tS}- \ti \frac{\omega_n}{c_\td \tilde{p}} \tilde{b}^\tS_{1 \tR} \nonumber\\
  &= \tilde{b}^\tL_{\tR\tS}   + \ti \sqrt{2} \frac{\omega_n}{c_\td \tilde{p}} \tilde{b}_{1 -}
\end{align}
Furthermore, for any field $A^a$ we have $\sum_{a = x,y,z} \lvert A^a \rvert^2 = \sum_{E = \tL,\tR,\tS} \lvert A^E \rvert^2$. For the nematic, the Higgs term in the Lagrangian splits up 
into the same five sectors as the elastic solid Lagrangian Eq.~\eqref{eq:solid stress gauge field Lagrangian tilde}, where only the $\tL 1$-sector is modified by the glide constraint:
\begin{widetext}
\begin{align}\label{eq:nematic Higgs term gauge fields}
 \mathcal{L}_\mathrm{Higgs} &=
 \frac{\Omega^2}{2 c_\tT^2 \mu}   \bigg[  ( \lvert \tilde{b}^\tL_{1\tS} \rvert^2 + \lvert \tilde{b}^\tR_{\tR\tS}\rvert^2 ) + ( \lvert \tilde{b}^\tL_{1\tR} \rvert^2 + \lvert \tilde{b}^\tS_{\tR\tS}\rvert^2 ) + ( \lvert \tilde{b}^\tR_{1\tR} \rvert^2 + \lvert \tilde{b}^\tS_{1\tS}\rvert^2 ) 
 +\lvert b_{1+} \rvert^2 \nonumber\\
 & \phantom{mmmmm}
   + \begin{pmatrix} \tilde{b}^{\dagger}_{1 -} \\ \tilde{b}^{\tL\dagger}_{\tR \tS } \end{pmatrix}^\tT
   \Big[ \begin{pmatrix} 1 & 0  \\ 0 & 1  \end{pmatrix} -
    \frac{1}{3 \frac{1}{c_\td^2} \omega_n^2 + q^2}
\begin{pmatrix}
2\frac{1}{c_\td^2}\omega_n^2 &   -\ti \sqrt{2} \tfrac{1}{c_\td} \omega_n \tilde{p} \\
\ti \sqrt{2}\tfrac{1}{c_\td} \omega_n \tilde{p} & \tilde{p}^2
+ \lvert b_{1+} \rvert^2                                                                                                                                                            \end{pmatrix} \Big]
\begin{pmatrix} \tilde{b}^{\dagger}_{1 -}  \\ \tilde{b}^{\tL\dagger}_{\tR \tS} \end{pmatrix}\bigg].
 \end{align}
\end{widetext}

\subsection{Collective modes of the quantum nematic}\label{subsec:Collective modes in the quantum nematic}

To obtain the spectrum of modes in the nematic phase, we should add Eq.~\eqref{eq:dislocation Higgs term nematic 3} to Eq.~\eqref{eq:solid stress gauge field Lagrangian tilde} and calculate the propagators Eqs.~\eqref{eq:longitudinal propagator gauge field tilde}, \eqref{eq:transverse propagator gauge field tilde}.  After a Wick rotation to real frequency $\omega$, we obtain the first main result of this paper, the stress propagators of the isotropic 3D quantum nematic:
\begin{widetext}
\begin{align}
 G_\tL &= \frac{1}{\mu} \frac{ - c_\tT^2 q^2 (\omega^2 -  \tfrac{1}{3}c_\td^2 q^2 - \Omega^2)}{(\omega^2 - c_\tL^2 q^2)(\omega^2 -  \tfrac{1}{3}c_\td^2 q^2) - \Omega^2 (\omega^2 - c_\kappa^2 q^2)},\label{eq:nematic longitudinal propagator}\\
 G_{\tT 1} = G_{\tT 2} &= \frac{1}{\mu} \frac{ -c_\tT^2 q^2 (\omega^2 -c_\td^2 q^2) - \Omega^2 ( \omega^2 - 2 c_\tT^2 q^2 - \tfrac{1}{2} c_\td^2 q^2 - \Omega^2 )}{(\omega^2 - c_\tT^2 q^2)(\omega^2 - c_\td^2 q^2) - \Omega^2 (\omega^2 - \tfrac{1}{2} c_\td^2 q^2)},\label{eq:nematic GT1}\\
  G_{\tT 3} &= \frac{1}{\mu} \frac{-\Omega^2}{\omega^2 - c_\td^2 q^2}.\label{eq:nematic GT3}
\end{align}
Recall that $c_\tL$ and  $c_\tT$ are the velocities of the longitudinal and  transverse phonons, respectively, while $c_\td$ is the velocity we assigned to the dislocation condensate.  We defined here $c_\kappa^2 = \frac{\kappa}{\rho} = \frac{2}{D} \frac{ 1+ \nu}{1 - (D-1)\nu} c_\tT^2$ as the {\em compression velocity} depending only on the compression modulus $\kappa$ and not the shear modulus $\mu$. This sets the (second) sound velocity in the quantum liquid. 

It is useful to compare these to the corresponding expressions in two dimensions from QLC2D, where we use the appropriate  definitions of $c_{\tL,\mathrm{2D}}$ and $c_{\kappa,\mathrm{2D}} = \sqrt{\kappa / \rho}$ via Eqs.~\eqref{eq:longitudinal velocity definition}, \eqref{eq:Poisson ratio definition},
\begin{align}
G^\mathrm{2D}_\tL &= \frac{1}{\mu} \frac{-c_\tT^2 q^2 ( \omega^2 - \tfrac{1}{2} c_\td^2 q^2 - \Omega^2)}{(\omega^2 - c_{\tL,\mathrm{2D}}^2 q^2)(\omega^2 - \tfrac{1}{2} c_\td^2 q^2) - \Omega^2 (\omega^2 - c_{\kappa,\mathrm{2D}}^2 q^2)}\label{eq:2D nematic GL},\\
G^\mathrm{2D}_\tT &= \frac{1}{\mu}\frac{-c_\tT^2 q^2 (\omega_n^2 - c_\td^2 q^2) - \Omega^2 ( \omega^2 - 2 c_\tT^2 q^2 - \tfrac{1}{2} c_\td^2 q^2 - \Omega^2 )}{(\omega^2 - c_\tT^2 q^2)(\omega^2 - c_\td^2 q^2) - \Omega^2( \omega^2  - \tfrac{1}{2} c_\td^2 q^2)}.\label{eq:2D nematic GT}
 \end{align}
\end{widetext}
We see that the longitudinal propagator is the same apart from the definitions of $c_\tL$ and $c_\kappa$ and a dimensional factor $1/D$ in front of $c_\td^2$.  Furthermore, the transverse propagators associated with the `remnant' transverse phonons $G_{\tT 1}, G_{\tT 2}$ are seen to be  completely independent of dimensionality.  The only novelty is the $G_{\tT 3}$ propagator which vanishes in the solid 
but now describes a massless mode propagating with the condensate velocity in the quantum nematic.  

 \begin{figure*}
  \begin{center}
    \includegraphics[width=7.7cm]{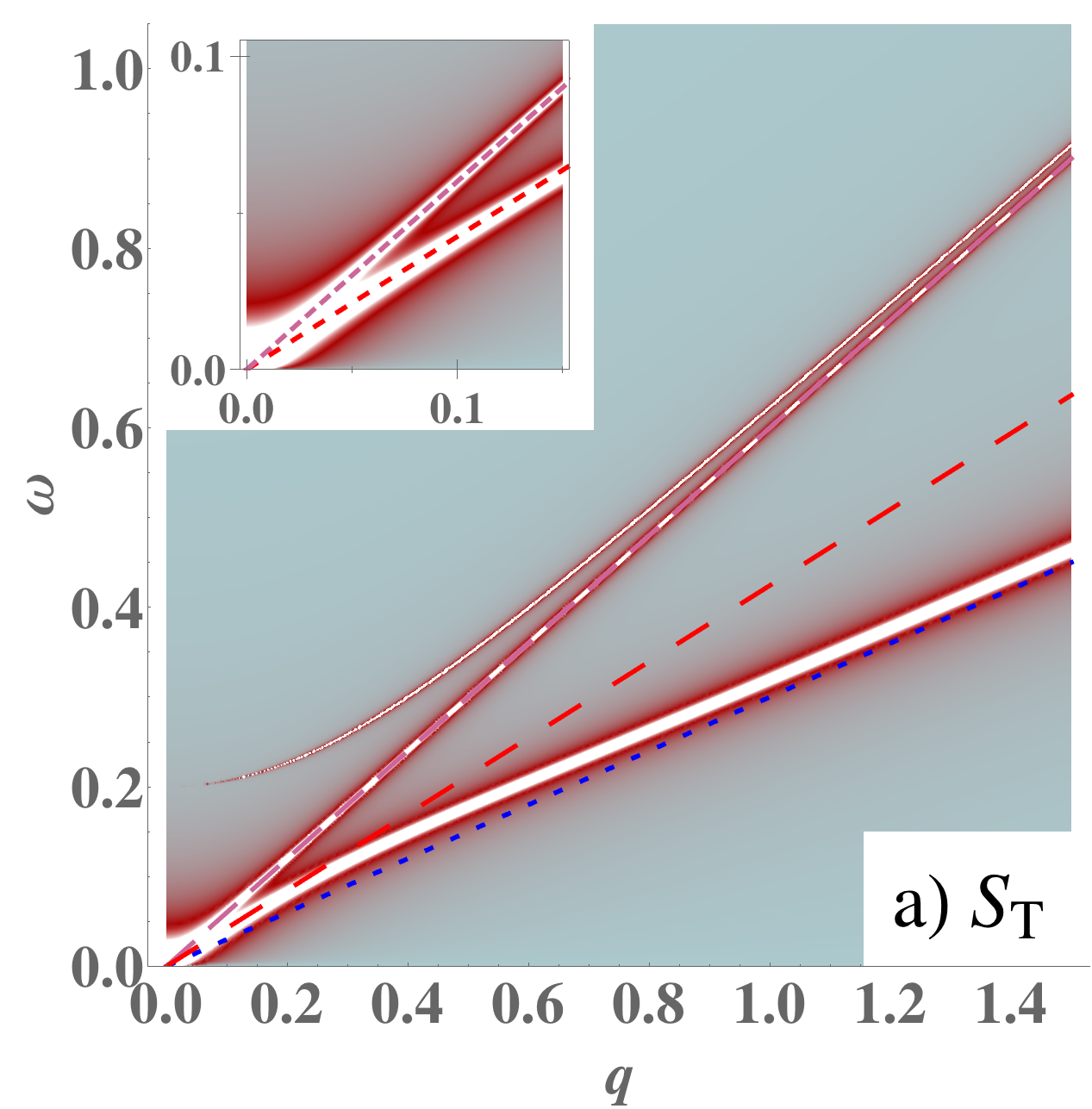}
  \hfill
 \includegraphics[width=7.7cm]{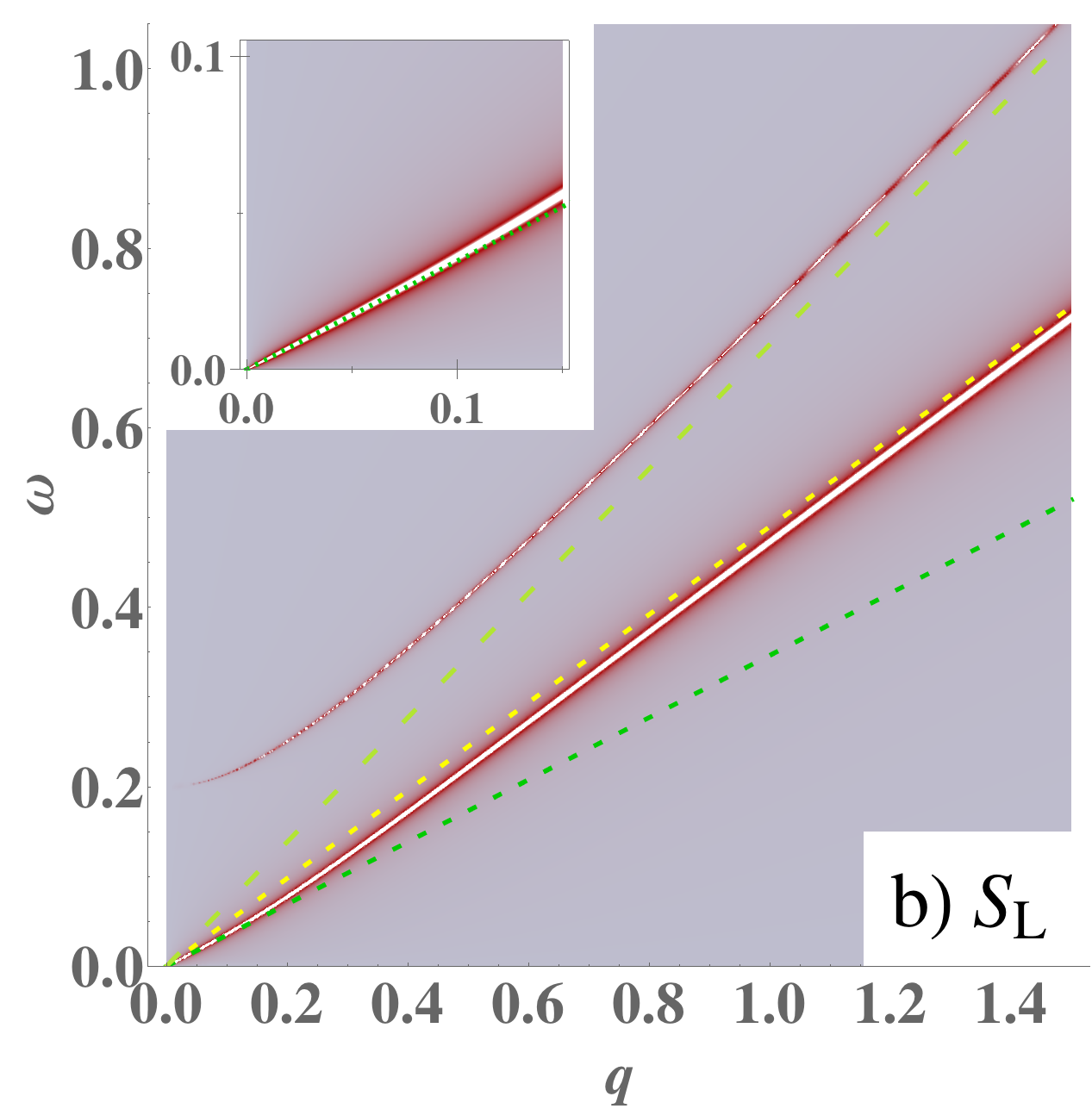}\\
 {\centering
 \includegraphics[scale=.6]{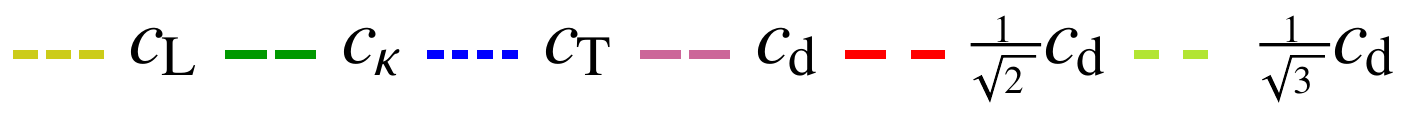}
 }\\
  \caption{Spectral functions of the quantum nematic in units of the inverse shear modulus $1/\mu$, with Poisson ratio $\nu = 0.2$. For a clear picture we have arbitrarily set $c_\td = 2 c_\tT$. The inset is a zoom up near the origin. The width of the poles is artificial and denotes the relative pole strengths: these ideal poles are actually infinitely sharp. a) Transverse spectral function of $G_\tT = G_{\tT 1} + G_{\tT 2} + G_{\tT 3}$ from Eqs.~\eqref{eq:nematic GT1}, \eqref{eq:nematic GT3}. At low energies we find the massless rotational Goldstone modes with velocities $\frac{1}{\sqrt{2}} c_\td$ and $c_\td$ respectively. There is also a massive pole due the gapped shear phonons with Higgs gap $\Omega$, which have vanishing pole strength as $q\to 0$. At high energies we retrieve the  transverse phonons with velocity $c_\tT$ while the condensate modes with velocity $c_\td$ have vanishing pole strength as $q \to \infty$. b) Longitudinal spectral function of $G_\tL$ from Eq.~\eqref{eq:nematic longitudinal propagator}. The massless pole has the pure compression velocity $c_\kappa$ at low energy while it extrapolates to the longitudinal phonon at high momenta. There is also a gapped condensate mode with gap $\Omega$ that extrapolates to a linearly dispersing mode with velocity $\frac{1}{\sqrt{3}} c_\td$ at high momentum, with vanishing pole strength both as $q \to 0$ and as $q \to \infty$.}\label{fig:nematic spectral functions}
  \end{center}
 \end{figure*}

The longitudinal ($\tL$) and transverse ($\tT 1, \tT 2$) propagators were analyzed already in detail in the 2+1D case\cite{QLC2D} but let us repeat this exercise here for completeness.  
The spectral functions are plotted in Fig.~\ref{fig:nematic spectral functions}. Let us first focus on the longitudinal response Eq.~\eqref{eq:nematic longitudinal propagator}. This propagator reveals
one massless and one massive pole with gap $\Omega$. At small energies and to lowest order in momentum their dispersion relations are
\begin{align}
 \omega^{\tL}_1 &=  c_\kappa q + \ldots ,\nonumber \\
 \omega^{\tL}_2 &= \Omega + \frac{c_\td^2 + 4 c_\tT^2}{6 \Omega} q^2 + \ldots . \label{eq:nematic longitudinal dispersions}
\end{align}
The massless  $\omega^{\tL}_1$ pole reveals the second sound mode, characterized by the purely compressional velocity $c_\kappa = \sqrt{\kappa/\rho}$ instead of the longitudinal phonon velocity $c_\tL = \sqrt{(\kappa + 2 \mu)/\rho}$. 
This was a highlight of QCL2D: the dislocation condensate has destroyed the shear rigidity of the solid at long wavelength, while it does not affect the compressional property of the solid since the dislocations ``do not carry volume'' as expressed by the glide constraint. In more detail, the structure of the longitudinal propagator reveals that the longitudinal phonon acquires a mode coupling with a condensate mode that effectively removes its shear rigidity, turning it into a pure sound mode.  The second  mode $\omega^{\tL}_2$ is the massive counterpart; at $q=0$ the mode coupling disappears (the shear component of the longitudinal phonon arises only for finite spatial gradients) and then $\omega^\tL_2$ is characterized by the Higgs mass of the stress superconductor $\Omega$, the only scale in the problem. It propagates with a combination of the transverse and condensate velocities, revealing its mixed origin. 

At very high energies, the two poles disperse linearly with velocities $c_\tL$ and $\sqrt{1/3}c_\td$ respectively. High momentum means small length scales, and we expect to retrieve here the signature of the underlying crystal lattice. This is indeed
the case: at large momenta the mode coupling with the condensate mode switches off and one recovers the longitudinal phonon, while the condensate mode turns `pure', propagating with just $c_\td$. The reader 
should notice that the sound mode is completely universal at long wavelength: this can be continued adiabatically all the way to the gaseous limit described by the Bogoliubov theory of  the weakly interacting Bose gas. However, 
the ``massive shear photon'' is special for the maximal-correlation limit.  As a finite-energy mode, it requires the ``locally solid'' correlations in the liquid to propagate; in approaching the gaseous limit it will get increasingly damped 
to disappear completely in the weakly-interacting limit. One may want to view the roton found in $^4$He as a remnant of this mode in the regime where the crystalline correlation length becomes of the order of the interatomic distance.  

\begin{table}
 \begin{tabular}{ccc}
  \toprule
  sector & massless modes & massive modes \\
  \hline
  $\tL$ & $c_\kappa q $ & $\Omega + \frac{\frac{1}{2}c_\td^2 + 2 c_\tT^2}{3 \Omega} q^2 +\ldots$  \\
  $\tT 1$ & $\frac{1}{\sqrt{2}} c_\td q$ & $\Omega + \frac{\frac{1}{2}c_\td^2 +  c_\tT^2}{2 \Omega} q^2 +\ldots$ \\
  $\tT 2$ & $\frac{1}{\sqrt{2}} c_\td q$ & $\Omega + \frac{\frac{1}{2}c_\td^2 +  c_\tT^2}{2 \Omega} q^2 +\ldots$ \\
  $\tT 3$ & $c_\td q$ & -  \\
  \hline
  total & 4 & 3 \\
  \botrule
 \end{tabular}
\caption{Collective modes in the isotropic quantum nematic. Indicated are the dispersion relations to lowest orders in momentum. The longitudinal phonon is protected by the glide constraint from obtaining a Higgs gap, and turns into a purely compressional mode. The gapped mode could be seen as the shear parts of the longitudinal phonon `being eaten' by the dislocation condensate. In the transverse sectors we see the two transverse phonons picking up a Higgs gap. Furthermore there are three rotational Goldstone modes emerging (deconfining), two with velocity $\frac{1}{\sqrt{2}}c_\td$ and one with velocity $c_\td$.}\label{table:nematic spectrum}
\end{table}

Let us now turn to the transverse sector,
and first consider the $\tT 1, \tT 2$ sectors, which were already identified as identical to the transverse sector in 2+1D. As in the longitudinal sector, these propagators describe two modes with the following dispersions at small momenta:
\begin{align}
 \omega^{\tT 1,2}_1 &= \frac{1}{\sqrt{2}} c_\td q + \ldots ,\label{eq:transverse rotational Goldstone dispersions}\\
 \omega^{\tT 1,2}_2 &= \Omega + \frac{c_\td^2 + 2 c_\tT^2}{4 \Omega} q^2 + \ldots .
\end{align}
The massive modes $\omega^{\tT 1,2}_2$  are  the transverse phonons which have acquired a mass through the Anderson--Higgs mechanism. This comprises one of the main predictions of the dual gauge field theory of these maximally-correlated quantum liquid crystals: transverse phonons do not disappear but should be detectable as massive, propagating modes in the spectrum~\cite{ZaanenNussinovMukhin04,CvetkovicNussinovMukhinZaanen08,QLC2D}. Just as for the longitudinal sector, these modes should get increasingly damped when the solid correlations weaken, to completely disappear in the weakly-coupled, gaseous limit. 

The massless poles $\omega^{\tT 1,2}_1$ are universal: exactly the same mode is found in 2+1D where we identified it as the {\em rotational Goldstone mode} or {\em torque mode} in the stress formalism, since it propagates torque stresses within the quantum nematic~\cite{BeekmanWuCvetkovicZaanen13,QLC2D}. It is an exclusive feature of the zero-temperature superconducting/superfluid nematic. It does not exist in the `high-temperature' classic nematic fluid, the reason being a the `anomaly' of the Goldstone physics in liquid crystals~\cite{DeGennesProst95,ChaikinLubensky00} mentioned in Sec.~\ref{sec:Symmetry principles of quantum liquid crystals}. The trouble is rooted in the fact that the classical nematic fluid is also a regular hydrodynamical fluid. This supports circulation and it turns out that the rotational Goldstone bosons have a finite coupling to this hydrodynamical circulation even in the long-wavelength limit. The effect is that the rotational boson just gets overdamped by the coupling to this circulation. This situation changes drastically in the superfluid/superconductor. Circulation is now massive because of the quantization of the vorticity and at low energies the rotational Goldstone modes cannot be damped, but constitute the propagating modes. 

As will be highlighted in the next subsection \ref{subsec:Torque stress in the quantum nematic}, the recovery of these rotational Goldstone bosons is a highlight of the duality construction.  They are `confined' in the solid as we already discussed in Sec.~\ref{subsec:Torque stress gauge fields},  while they become deconfined (becoming massless and propagating) in the nematic. As elucidated in QLC2D the associated {\em torque rigidity} originates in the dislocation condensate itself: 
is is observed that the velocity of the modes $\omega^{\tT 1,2}_1$ is set by the condensate velocity $c_\td$~\cite{BeekmanWuCvetkovicZaanen13,QLC2D}. Furthermore, the pole strength is proportional to $\Omega^2$, and therefore vanishes when the dislocation condensate is absent.
The only difference between 2+1D and 3+1D is in the number of rotational Goldstone  modes. In two space dimensions there is only one rotational plane where the rotational symmetry is broken ($O(2)$), and accordingly there is one Goldstone boson. In three dimensions there are three rotational planes ($O(3)$) and considering an `isotropic nematic' (or, say, 
a cubic $O_\mathrm{h}$ nematic) the rotational symmetry is broken in all three planes, causing the existence of three rotational Goldstone bosons. The reader should notice that the most common uniaxial nematics are in this regard quite pathological. Their point group is $D_{\infty \mathrm{h}}$, characterized by breaking symmetry in two rotational planes, and therefore only two rotational Goldstones are present. 

Where is the third rotational Goldstone boson? It is found in the $G_{\tT 3}$ propagator. As we stressed in Eq.~\eqref{eq:GT3 solid}, this propagator is vanished identically in the solid but in the nematic it turns into the propagator of a single massless mode, 
\begin{equation}
 \omega^{\tT 3} = c_\td q. \label{eq:longitudinal rotational Goldstone dispersions}
\end{equation}
Compared to the $\omega^{\tT 1,2}$-modes its velocity  larger by a factor $\sqrt{2}$, indicating that this mode is symmetry-wise not equivalent to the other two.  As we will see in a moment, this is due to the fact that the 
`polarizations' of these rotational Goldstone modes count in the same way as for phonons: the ${\tT 1,2}_1$ modes are `transverse' while the ${\tT 3}$ modes turns out to be `longitudinally polarized'. The formalism discussed in the next subsection will yield further insights in these torque stresses.  

For completeness, we ask what happens at large momenta? The two modes described by each of the $G_{\tT 1}$, $G_{\tT 2}$ disperse linearly with velocities $c_\tT$ resp. $c_\td$. On short length scales, we retrieve the transverse phonons of the solid, and the decoupled modes of the dislocation condensate with vanishing spectral weight as $q \to \infty$, just as in the longitudinal sector. The single mode in $G_{\tT 3}$ has the velocity $c_\td$ at all momenta, with spectral weight $\propto 1/q$.

\subsection{Torque stress in the quantum nematic}\label{subsec:Torque stress in the quantum nematic}

We can obtain a better understanding of the rotational Goldstone modes by considering the torque stress and torque stress gauge fields of Sec.~\ref{subsec:Torque stress gauge fields}. Here we are interested in the physics at frequencies 
small compared to the Higgs mass, and we send  $\Omega \to \infty$,  keeping only the Higgs term and ignoring the `phonon part'.

We work with the torque stress fields
\begin{align}
 \tilde{\tau}^c_\mu &= \epsilon_{cba} \epsilon_{b\mu \kappa\lambda}\tfrac{1}{2} \tilde{b}^a_{\kappa\lambda},\label{eq:torque stress definition condensate unitary gauge}\\
 &= \epsilon_{\mu\nu\kappa\lambda} \tilde{\partial}_\nu \tfrac{1}{2} \tilde{h}^c_{\kappa\lambda}.
\end{align}
where all fields are rescaled with respect to the dislocation velocity $c_\td$, as we have done throughout this section.
This definition of the torque stress is not gauge invariant, which in the duality is attributed to the presence of shear rigidity. However, in the dislocation condensate shear rigidity is lost and torque stress becomes a 
physical quantity. In the presence of the dislocation condensate phase degrees of freedom $\tilde{\phi}^a_\kappa$, the torque stress is rather defined as
 \begin{equation}\label{eq:torque stress definition condensate}
 \tilde{\tau}^c_\mu = \epsilon_{cba} \epsilon_{b\mu \kappa\lambda} (\tilde{b}^a_{\kappa\lambda} + \tilde{\partial}_\kappa \tilde{\phi}^a_\lambda -\tilde{\partial}_\lambda \tilde{\phi}^a_\kappa).
\end{equation}
being gauge invariant under the transformations Eq.~\eqref{eq:stress gauge transformation dislocation condensate}. Let us take the unitary gauge fix $\tilde{\phi}^a_\lambda = 0$, bringing us back to Eq.~\eqref{eq:torque stress definition condensate unitary gauge}.
The Higgs term becomes Eq.~\eqref{eq:dislocation condensate minimal coupling step 1},  supplemented by the glide constraint, while the Higgs mass is rescaled by a factor of 2 like in Eq.~\eqref{eq:dislocation Higgs term nematic}:
\begin{align}\label{eq:nematic Higgs term unitary gauge fix}
 \mathcal{L}_\mathrm{Higgs}  &= \frac{\Omega^2}{4 c_\tT^2 \mu } \Big[  \lvert  \tilde{b}^a_{\mu\nu} \rvert^2 - \frac{\tilde{p}^2}{3 \frac{\omega_n^2}{c_\td^2} + q^2} \lvert \tilde{b}^x_{yz} + \tilde{b}^y_{zx} + \tilde{b}^z_{xy} \Big]\nonumber\\
 &= \frac{\Omega^2}{4 c_\tT^2 \mu } \Big[  (2 \lvert \tilde{b}^E_{\ft G} \rvert^2 +  \lvert  \tilde{b}^E_{F G} \rvert^2 \nonumber\\
 &\phantom{mmmmm} - \frac{\tilde{p}^2}{3 \frac{\omega_n^2}{c_\td^2} + q^2} \lvert \tilde{b}^\tL_{\tR\tS} - \tilde{b}^\tR_{\tS\tL} + \tilde{b}^\tS_{\tL\tR} \rvert^2 \Big].
\end{align}
Here the capital indices sum over Fourier components $E,F,G = \{ \tL,\tR,\tS\}$ and we have used the identity $\tilde{b}^x_{yz} + \tilde{b}^y_{zx} + \tilde{b}^z_{xy} = \tilde{b}^\tL_{\tR\tS} - \tilde{b}^\tR_{\tS\tL} + \tilde{b}^\tS_{\tL\tR}$. Some components of $\tilde{b}^a_{\kappa\lambda}$, which were completely absent in the solid Lagrangian, now appear. These could be said to originate in the dislocation condensate itself, transferred to the dual gauge fields via the Anderson--Higgs mechanism.
In the unitary gauge fix, the explicit correspondence between $\tilde{\tau}^c_\mu$ and $\tilde{b}^a_{\kappa\lambda}$ is just Eq.~\eqref{eq:torque stress in stress gauge field components} with all fields replaced by their tilde-equivalents. We can now express the Higgs term explicitly in the torque stress:
\begin{widetext}
 \begin{align}
  \mathcal{L}_\mathrm{Higgs} &= \mathcal{L}_{\mathrm{T1, Higgs}} + \mathcal{L}_{\mathrm{T2, Higgs}} + \mathcal{L}_{\mathrm{T3, Higgs}} + \mathcal{L}_{\mathrm{L1, Higgs}} + \mathcal{L}_{\mathrm{L2, Higgs}} + \mathcal{L}_{\mathrm{X, Higgs}}, \\
  \mathcal{L}_{\mathrm{T1, Higgs}}&= \frac{\Omega^2}{2 c_\tT^2 \mu } \Big[ \frac{1}{2} \lvert \tilde{\tau}^\tS_\ft\rvert^2 +  \lvert \tilde{\tau}^\tS_\tL\rvert^2 + \frac{1}{2} \lvert \tilde{b}^\tL_{\tL\tS} + \tilde{b}^\tR_{\tR\tS}\rvert^2\Big],\\
  \mathcal{L}_{\mathrm{T2, Higgs}}&= \frac{\Omega^2}{2 c_\tT^2 \mu } \Big[ \frac{1}{2} \lvert \tilde{\tau}^\tR_\ft\rvert^2 +  \lvert \tilde{\tau}^\tR_\tL\rvert^2 + \frac{1}{2} \lvert \tilde{b}^\tL_{\tL\tR} + \tilde{b}^\tS_{\tR\tS}\rvert^2\Big],\\
  \mathcal{L}_{\mathrm{T3, Higgs}} &= \frac{\Omega^2}{2 c_\tT^2 \mu } \Big[ \frac{1}{2} \lvert \tilde{\tau}^\tL_\ft\rvert^2 
  + \frac{1}{2} \lvert \tilde{b}^\tR_{\tL\tR} + \tilde{b}^\tS_{\tL\tS}\rvert^2
  + \frac{3}{4} \lvert \tilde{\tau}^\tL_\tL \rvert^2 
  + \frac{3}{4} \lvert \tilde{\tau}^\tR_\tR \rvert^2 
  + \frac{3}{4} \lvert \tilde{\tau}^\tS_\tS \rvert^2 \nonumber\\
 &\phantom{mmmmmmm}
  + \frac{1}{4} (\tilde{\tau}^{\tL\dagger}_\tL \tilde{\tau}^\tR_\tR +\tilde{\tau}^{\tR\dagger}_\tR \tilde{\tau}^\tL_\tL)
  + \frac{1}{4} (\tilde{\tau}^{\tR\dagger}_\tR \tilde{\tau}^\tS_\tS +\tilde{\tau}^{\tS\dagger}_\tS \tilde{\tau}^\tR_\tR)
  - \frac{1}{4} (\tilde{\tau}^{\tL\dagger}_\tL \tilde{\tau}^\tS_\tS +\tilde{\tau}^{\tS\dagger}_\tS \tilde{\tau}^\tL_\tL)
  \Big],\\
  \mathcal{L}_{\mathrm{L1, Higgs}}&= \frac{\Omega^2}{2 c_\tT^2 \mu } \Big[ 
    \frac{1}{2} \lvert \tilde{\tau}^\tS_\tR - \tilde{\tau}^\tR_\tS\rvert^2 
    + \lvert \tilde{b}^\tL_{\tR\tS}\rvert^2  
    + \frac{1}{2}\lvert\tilde{b}^\tR_{\tL\tS} - \tilde{b}^\tS_{\tL\tR} \rvert^2  
    - \frac{\tilde{p}^2}{3 \frac{\omega_n^2}{c_\td^2} + q^2} \lvert \tilde{b}^\tL_{\tR\tS}  - \tilde{b}^\tR_{\tL\tS} + \tilde{b}^\tS_{\tL\tR} \rvert^2 
  \Big],\\
 \mathcal{L}_{\mathrm{L2, Higgs}}&= \frac{\Omega^2}{2 c_\tT^2 \mu } \Big[ 
 \frac{1}{2} \lvert \tilde{\tau}^\tS_\tR + \tilde{\tau}^\tR_\tS\rvert^2 +
 \frac{1}{2}\lvert\tilde{b}^\tR_{\tL\tS} + \tilde{b}^\tS_{\tL\tR} \rvert^2
  \Big],\\
  \mathcal{L}_{\mathrm{X, Higgs}}&= \frac{\Omega^2}{2 c_\tT^2 \mu } \Big[\lvert \tilde{\tau}^\tL_\tR\rvert^2 +  \lvert \tilde{\tau}^\tL_\tS\rvert^2 \Big].  
 \end{align}
\end{widetext}
Here the naming of different sectors follows that of Eq.~\eqref{eq:solid stress gauge field Lagrangian}, although this classification is slightly ambiguous. In particular the components $\tilde{b}^E_{\ft\tL}$ do not have a counterpart in linear elasticity. This also leads to the introduction of a new sector $\mathcal{L}_{\mathrm{X}}$ which is decoupled from all others. We will comment on  the interpretation of these degrees of freedom below. The next step is to substitute the torque stress gauge field Eq.~\eqref{eq:torque stress gauge field definition}, in the Lorenz gauge fix Eq.~\eqref{eq:torque stress gauge field components}. This leads to
\begin{widetext}
 \begin{align}
  \mathcal{L}_{\mathrm{T1, Higgs}}&= \frac{\Omega^2}{2 c_\tT^2 \mu } \Big[ (\frac{\omega_n^2}{c_\td^2}  + \frac{1}{2} q^2) \lvert \tilde{h}^\tS_{\tR\tS} \rvert^2 + \frac{1}{2} \lvert \tilde{b}^\tL_{\tL\tS} + \tilde{b}^\tR_{\tR\tS}\rvert^2\Big],\\
  \mathcal{L}_{\mathrm{T2, Higgs}}&= \frac{\Omega^2}{2 c_\tT^2 \mu } \Big[ (\frac{\omega_n^2}{c_\td^2}  + \frac{1}{2} q^2) \lvert \tilde{h}^\tS_{\tR\tS} \rvert^2 + \frac{1}{2} \lvert \tilde{b}^\tL_{\tL\tR} + \tilde{b}^\tS_{\tR\tS}\rvert^2\Big],\\
  \mathcal{L}_{\mathrm{T3, Higgs}} &= \frac{\Omega^2}{2 c_\tT^2 \mu } \Big[  \frac{1}{4}
  \begin{pmatrix}
   \tilde{h}^{\tL\dagger}_{\tR\tS} & \tilde{h}^{\tR\dagger}_{1 \tS} & \tilde{h}^{\tS\dagger}_{1 \tR} 
  \end{pmatrix}
\begin{pmatrix}
 3 \frac{\omega_n^2}{c_\td^2} + 2 q^2 & - \ti \frac{\omega_n}{c_\td} \tilde{p} &  \ti \frac{\omega_n}{c_\td} \tilde{p} \\
  \ti \frac{\omega_n}{c_\td} \tilde{p} & 3 \tilde{p}^2 & \tilde{p}^2 \\
  - \ti \frac{\omega_n}{c_\td} \tilde{p} & \tilde{p}^2 & 3 \tilde{p}^2 
\end{pmatrix}
 \begin{pmatrix}
   \tilde{h}^{\tL}_{\tR\tS} \\ \tilde{h}^{\tR}_{1 \tS} \\ \tilde{h}^{\tS}_{ 1 \tR} 
  \end{pmatrix}
  + \frac{1}{2} \lvert \tilde{b}^\tR_{\tL\tR} + \tilde{b}^\tS_{\tL\tS}\rvert^2
  \Big],\\
   \mathcal{L}_{\mathrm{L1, Higgs}}&= \frac{\Omega^2}{2 c_\tT^2 \mu } \Big[ 
   \frac{1}{2} \tilde{p}^2 \lvert \tilde{h}^\tS_{1\tS} - \tilde{h}^\tR_{1\tR} \rvert^2  
   +  \lvert \tilde{b}^\tL_{\tR\tS}\rvert^2  
  + \frac{1}{2}\lvert\tilde{b}^\tR_{\tL\tS} -  \tilde{b}^\tS_{\tL\tR} \rvert^2
  - \frac{\tilde{p}^2}{3 \frac{\omega_n^2}{c_\td^2} + q^2} \lvert \tilde{b}^\tL_{\tR\tS} - \tilde{b}^\tR_{\tL\tS} + \tilde{b}^\tS_{\tL\tR} \rvert^2 
  \Big],\\
  \mathcal{L}_{\mathrm{L2, Higgs}}&= \frac{\Omega^2}{2 c_\tT^2 \mu } \Big[ 
   \frac{1}{2} \tilde{p}^2 \lvert \tilde{h}^\tS_{1\tS} + \tilde{h}^\tR_{1\tR} \rvert^2  
  + \frac{1}{2}\lvert\tilde{b}^\tR_{\tL\tS} +  \tilde{b}^\tS_{\tL\tR} \rvert^2
  \Big],\\
  \mathcal{L}_{\mathrm{X, Higgs}}&= \frac{\Omega^2}{2 c_\tT^2 \mu } \Big[ \tilde{p}^2 \lvert \tilde{h}^\tL_{1\tS}\rvert^2 +  \tilde{p}^2 \lvert \tilde{h}^\tL_{1\tR} \rvert^2 \Big].  
 \end{align}
\end{widetext}
We have arrived at a point where the interpretation of the nature of the torque stress carried by the quantum nematic becomes clear. From Sec.~\ref{subsec:Torque stress gauge fields} we know that $h^c_{\tR\tS}$ represent the propagating rotational (Goldstone) modes while $h^c_{1 \tR}$, $h^c_{1 \tS}$ represent static forces. In $\mathcal{L}_{\mathrm{T1, Higgs}}$, $\mathcal{L}_{\mathrm{T2, Higgs}}$ we find two rotational Goldstone modes $h^\tS_{\tR\tS}$,  $h^\tR_{\tR\tS}$ with velocity $\frac{1}{\sqrt{2}} c_\td$. Inverting the matrix in $\mathcal{L}_{\mathrm{T3, Higgs}}$, we find a third rotational Goldstone mode $h^\tL_{\tR\tS}$ with velocity $c_\td$. This confirms our findings 
summarized in  Eqs.~\eqref{eq:transverse rotational Goldstone dispersions}, \eqref{eq:longitudinal rotational Goldstone dispersions}. Similarly, we can see that the static forces are also deconfined, and mediate long-range interactions between disclination sources (as usual, one should 
mobilize the Coulomb gauge  to find out that for instance $\langle \tilde{h}^{\tR\dagger}_{\ft\tR} \tilde{h}^{\tR}_{\ft\tR}\rangle \propto 1/q^2$).

Let us reconsider the discussion regarding the number of degrees of freedom that we started in Sec.~\ref{subsec:Interpretation of stress components}. In the 3+1D solid, we start with 12 stress components $\sigma^a_\mu$. There are three conservation laws $\partial_\mu \sigma^a_\mu =0$ and three Ehrenfest constraints $\sigma^a_m = \sigma^m_a$, such that we are left with six physical stress components: three phonons and three Coulomb forces. Going to 
the dual stress gauge fields $b^a_{\kappa\lambda}$, there are at first 18 independent components due to antisymmetry under $\kappa \leftrightarrow \lambda$. In the solid, by accounting for the gauge freedom and Ehrenfest constraints, these still encode for the same six degrees of freedom in Eqs.~\eqref{eq:stress gauge field propagator electric shear}--\eqref{eq:stress gauge field propagator magnetic shear} in the limit $\ell \to 0$ Eq.~\eqref{eq:rotational force suppression}. In the Higgs phase however, {\em all} components of $b^a_{\kappa\lambda}$ obtain a physical meaning. The Anderson-Higgs mechanism transfers the dislocation phase degrees of freedom to the dual stress gauge field in Eq.~\eqref{eq:dislocation condensate minimal coupling step 1} when taking the unitary gauge fix as in Eq.~\eqref{eq:nematic Higgs term unitary gauge fix}. As long as one is solely interested in correlations in linear stress $\sigma^a_\mu$, only the nine gauge-invariant components of $b^a_{\kappa\lambda}$, which are explicitly employed in Eqs.~\eqref{eq:solid stress gauge field Lagrangian},\eqref{eq:nematic Higgs term gauge fields}, are accessible. These contain the spectrum as enumerated in Table~\ref{table:nematic spectrum}. However, as we have seen just now, by looking at correlations in torque stress $\tau^c_\mu$, other degrees of freedom become activated, which originate in the dislocation condensate. This is the best evidence we have for the applicability of the `stringy' form of the minimal couping Eq.~\eqref{eq:dislocation condensate minimal coupling step 1}.

We can also elucidate the difference between the rotational Goldstone modes in $\mathcal{L}_{\tT 1}$, $\mathcal{L}_{\tT 2}$ on the one hand and $\mathcal{L}_{\tT 3}$ on the other hand. The Goldstone mode $h^\tL_{\tR\tS}$ couples to disclinations $\Theta^\tL_{\tR\tS}$ such that the Frank vector is parallel to the momentum. Apparently the longitudinal Goldstone mode, propagating in the direction of the Frank vector, has a different, typically higher, velocity than those propagating perpendicular to the Frank vector, i.e. within the rotational plane. This Goldstone mode is excited by probing the medium with a torque in the plane of the surface and measuring that torque on the opposite side of the medium, see Fig.~\ref{fig:rotational Goldstone modes}.

Let us clarify the origin of the $\sqrt{2}$ difference between the longitudinal and transverse velocity. The torque stress gauge fields $h^c_{\kappa\lambda}$ are dual to rotational fields $\omega^c$. We have summarized the theory of rotational elasticity in Sec.~\ref{subsec:Rotational elasticity}, where we noted there are longitudinal and transverse velocities in Eq.~\ref{eq:rotational velocities}. We see that for the relation $c_\tL^\mathrm{rot} = \sqrt{2} c_\tT^\mathrm{rot}$ is satisfied when the rotational Poisson ratio in Eq.~\eqref{eq:rotational Poisson ratio} is $\nu^\mathrm{rot} = 0$. This makes perfect sense: a non-zero Poisson ratio would mean that external `longitudinal' torques would also excite transverse rotational modes. In the nematic liquid crystal we expect no such couplings: all the modes should be strictly independent to lowest order.

%% file: sec_smectic.tex
\begin{figure}[t]
\hfill
 \includegraphics[width=5cm]{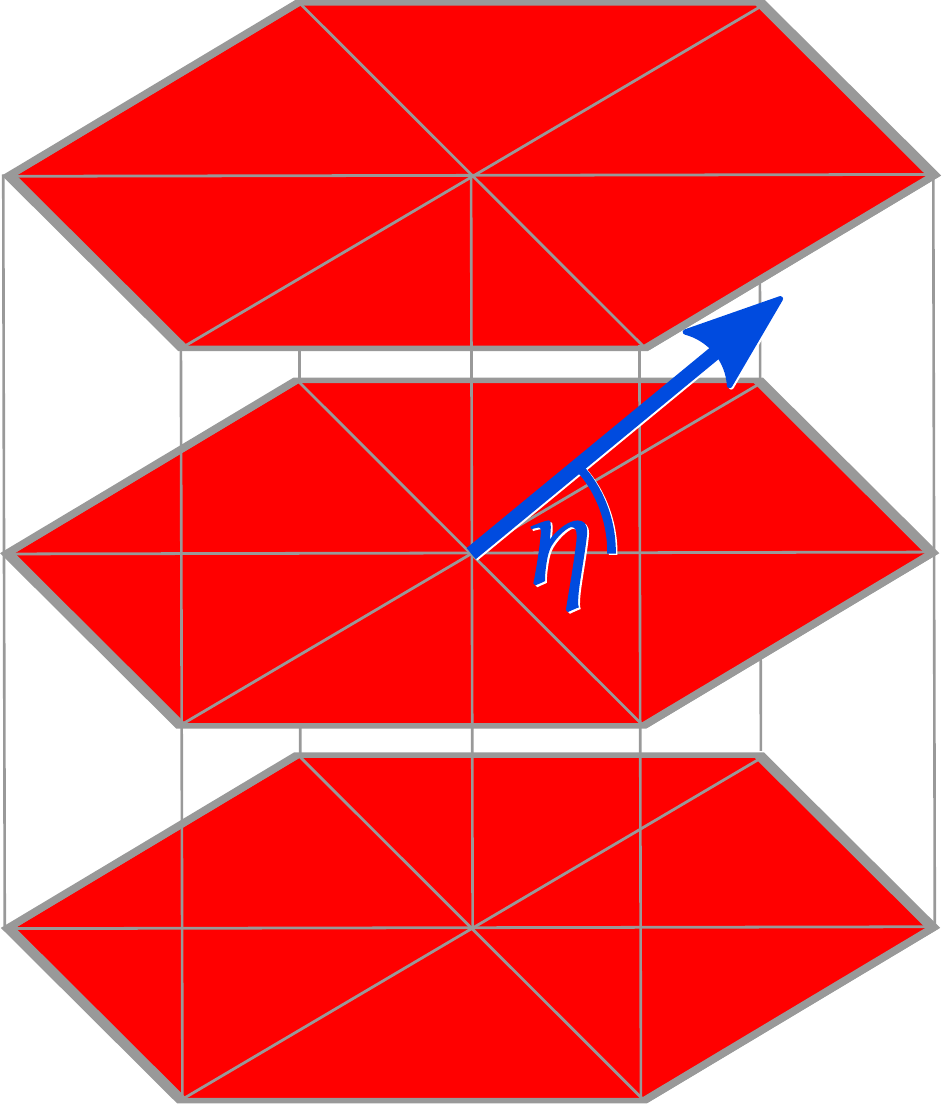}
 \hfill\null
\caption{Definition of the angle $\eta$ between the momentum vector (blue) and the liquid planes (red). $\eta = 0$ is parallel to the planes, $\eta = \pi/2$ is perpendicular to the planes.}\label{fig:smectic eta angle}
\end{figure}

While the nematics are arguably the most relevant liquid crystals, the smectic phases also turn out to be rather interesting. Here translations are restored in two directions, leaving ``periodically stacked liquid layers'', see Fig.~\ref{subfig:smectic}. In QLC2D we have already seen that the low-energy physics is not directly that of a ``solid $\times$ liquid'', but the mode spectrum is rather a coupled mixture of modes from both phases and highly dependent on the interrogation angle: the angle $\eta$ between momentum $\mathbf{q}$ and the liquid plane. Let us shortly review these results of 2+1D.

Naively one would expect shear rigidity to be lost in the liquid direction while it should persist in the solid direction. Then one is led to think that when probing the system with momentum along the liquid direction ($\eta = 0$), the longitudinal response should be like a superfluid while the transverse response should be like the solid. Similarly, with momentum in the solid direction ($\eta = \pi/2$) one would expect in the longitudinal direction a phonon with reduced velocity since the effective shear modulus is vanishing, while the transverse propagator should not excite any phonon. Nature turns out to be more intricate. In both of these special directions, the longitudinal propagator is identical to that of the solid with velocity $c_\tL$: the fact that one direction does not have shear rigidity anymore does not matter at all! Meanwhile,  the transverse propagator for $\eta=\pi/2$ is completely gapped, which is in line with the loss of shear rigidity. For $\eta=0$, there is a transverse phonon with quadratic dispersion $\omega \propto q^2 +\ldots$, which mimics the undulation mode of classical smectics~\cite{DeGennesProst95, ChaikinLubensky00}. However, the ``solid $\times$ liquid'' nature of the smectic shows up at the special angle $\eta = \pi/4$. Here the transverse sector has a phonon with velocity $c_\tT$ like in a solid, while the longitudinal mode has velocity $c_\kappa = \sqrt{\kappa/\rho}$ which is the sound velocity of a superfluid having no contribution from the shear modulus. For general angles $\eta$, the longitudinal velocity varies smoothly from $c_{\tL}$ at $\eta = 0$ down to $c_{\kappa}$ at $\eta = \pi/4$ back to $c_{\tL}$ at $\eta = \pi/2$; the transverse velocity is vanishing at $\eta=0$, $\eta=\pi/2$ while smoothly rising up to $c_\tT$ at $\eta = \pi/4$.

It is interesting to see whether and how these features are reproduced in three spatial dimensions. There are a few important differences. First of all, the smectic is now a stack of two-dimensional liquid planes. While we should formally specify the momentum by two angles, a polar angle out of the liquid plane and an azimuthal one within the plane, the smectic phase obtained from an isotropic solid is invariant under 2D rotation around the solid axis. Therefore we can suffice with only the polar angle $\eta$ which interpolates between momentum completely within the liquid plane $\eta = 0$ and completely in the solid direction $\eta = \pi/2$, see Fig.~\ref{fig:smectic eta angle}. Second, there is one more transverse direction and two more rotational planes than in 2+1D. In particular, translational symmetry exists within the liquid plane and we expect a rotational Goldstone mode to deconfine in that plane (see Sec.~\ref{subsec:Rotational Goldstone mode smectic}). We also expect that the two transverse phonons will have differing behavior depending on which one of them picks up the remnant translational order. Lastly, in 2+1D we had the accidental identity $c_{\kappa,\mathrm{2D}} = \sqrt{ c_{\tL,\mathrm{2D}}^2 - c_\tT^2}$, which does not hold for the corresponding 3D velocities. We will keep our eyes out to see whether this difference will pop up in the smectic phenomenology.

\subsection{Higgs term and glide constraint}
Again, our starting point is Eq.~\eqref{eq:dislocation Higgs term} where the sum over $a$ is now over two orthogonal directions which span the plane in which translational symmetry is restored. Since we depart from isotropic elasticity, all planes of dislocation condensation are equivalent. But even in the general case, as long as the condensate amplitudes $\Omega^a$ are the same for all Burgers directions, the form of the Higgs term is independent of the original crystal space group, due to the fact that the directions of translational symmetry restoration must be orthogonal as explained in Sec.~\ref{sec:Symmetry principles of quantum liquid crystals}. Without loss of generality, we can choose the Burgers vectors of condensed dislocations to lie in the $xz$-plane. Below we always assume $\Omega^x = \Omega^z \equiv \Omega$, which can be achieved for judicious choices of Ginzburg--Landau parameters in Eq.~\eqref{eq:three-condensate potential}.
As before, we choose the dislocation Lorenz gauge fix Eq.~\eqref{eq:dislocation Lorenz gauge fix} and disregard the  condensate phase degrees of freedom as they do not couple to stress. 

First we need to take care of the glide constraint in  Eq.~\eqref{eq:dislocation Higgs term}. This follows a derivation similar to that of the nematic in Sec.~\eqref{subsec:nematic glide constraint}. For the prefactor of $\lambda^\dagger \lambda$, we have the result of Eq.~\eqref{eq:glide constraint multiplier field prefactor}, which now must be summed over the two directions. Integrating out the Lagrange multiplier field $
\lambda$ will lead to
\begin{align}
 \mathcal{L}^{(xz)}_\mathrm{glide} =  - 2 \frac{\Omega^2}{4 c_\tT^2 \mu}  \frac{ \tilde{p}^2}{2\frac{1}{c_\td^2} \omega_n^2 + q_x^2 + q_z^2} \left\lvert \tilde{b}^x_{yz} + \tilde{b}^z_{xy} \right\rvert^2.
\end{align}
If one instead chooses the condensation in the $xy$- or $yz$-plane, similar terms are obtained. The factor of 2 in front comes from summing over antisymmetric indices.

The Higgs term Eq.~\eqref{eq:dislocation Higgs term} for the smectic is then given by:
\begin{align}
 \mathcal{L}^{(xz)}_\mathrm{Higgs}
  &= \frac{\Omega^2}{2 c_\tT^2 \mu}   \Big[   \sum_{a=x,z}\Big(  
  \lvert \tilde{b}^a_{1\tR} \rvert^2 + \lvert \tilde{b}^a_{1\tS} \rvert^2 + \lvert \tilde{b}^a_{\tR\tS} \rvert^2 \Big) \nonumber\\
   &\phantom{mmnmm}-  \frac{\tilde{p}^2}{2 \frac{1}{c_\td^2} \omega_n^2 + q_x^2+q_z^2} \lvert \tilde{b}^x_{yz} +\tilde{b}^z_{xy} \rvert^2 \Big].
   \label{eq:dislocation Higgs term smectic xz} 
\end{align}
with analogous expressions for the $xy$- and $yz$-plane condensates with the coordinates $x,y,z$ permuted. In contrast to the nematic, the glide constraint term does not have a nice, short expression in Fourier space coordinates. We will leave the term as it is, while in the calculations we convert it to Fourier space coordinates, leading in general to a $9 \times 9$ matrix in the basis $\tilde{b}^E_{1 \tR}$, $\tilde{b}^E_{1 \tS}$, $\tilde{b}^E_{\tR\tS}$, $E=\tL,\tR,\tS$. Without loss of generality, we can focus on just one of these three choices, the other two can be obtained through a simple coordinate transformation, in our isotropic (or cubic) case.

\begin{figure}[t]
 \includegraphics[width=8cm]{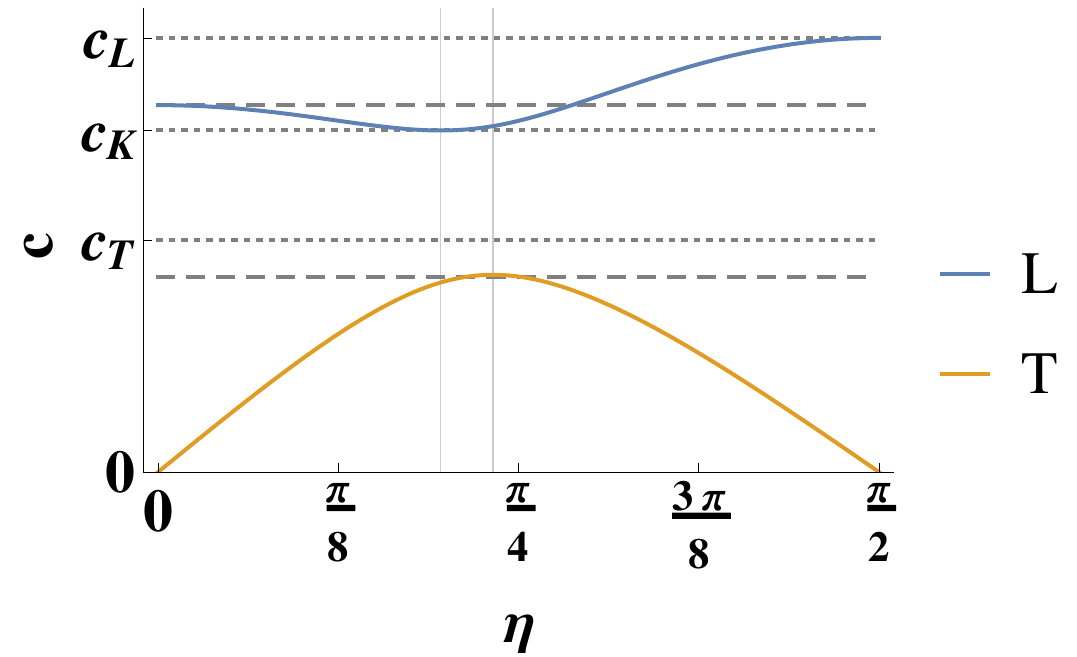}
 \caption{Velocities of the massless modes in the smectic as a function of interrogation angle $\eta$ from Eq.~\eqref{eq:smectic L T1 massless dispersion}. Here we have chosen a representative value of the Poisson ration $\nu = 0.3$. The dotted lines represent $c_\tL$, $c_\kappa$ and $c_\tT$, while the dashed lines represent $\frac{1}{\sqrt{1-2\nu}}c_\tT$ and $\sqrt{\frac{2+2\nu}{3+2\nu}} c_\tT$ as explained in the text. The longitudinal mode (blue) reaches the maximum value $c_\tL$ at $\eta = \pi/2$, but not at $\eta =0$, unlike the 2D case. Its minimum value is $c_\kappa$. The transverse mode (yellow) never reaches the full transverse photon velocity $c_\tT$. Its pole strength vanishes both as $\eta \to 0$ and as $\eta \to \pi/2$.}\label{fig:smectic velocity}
\end{figure}

\begin{figure*}
     \includegraphics[width=7.7cm]{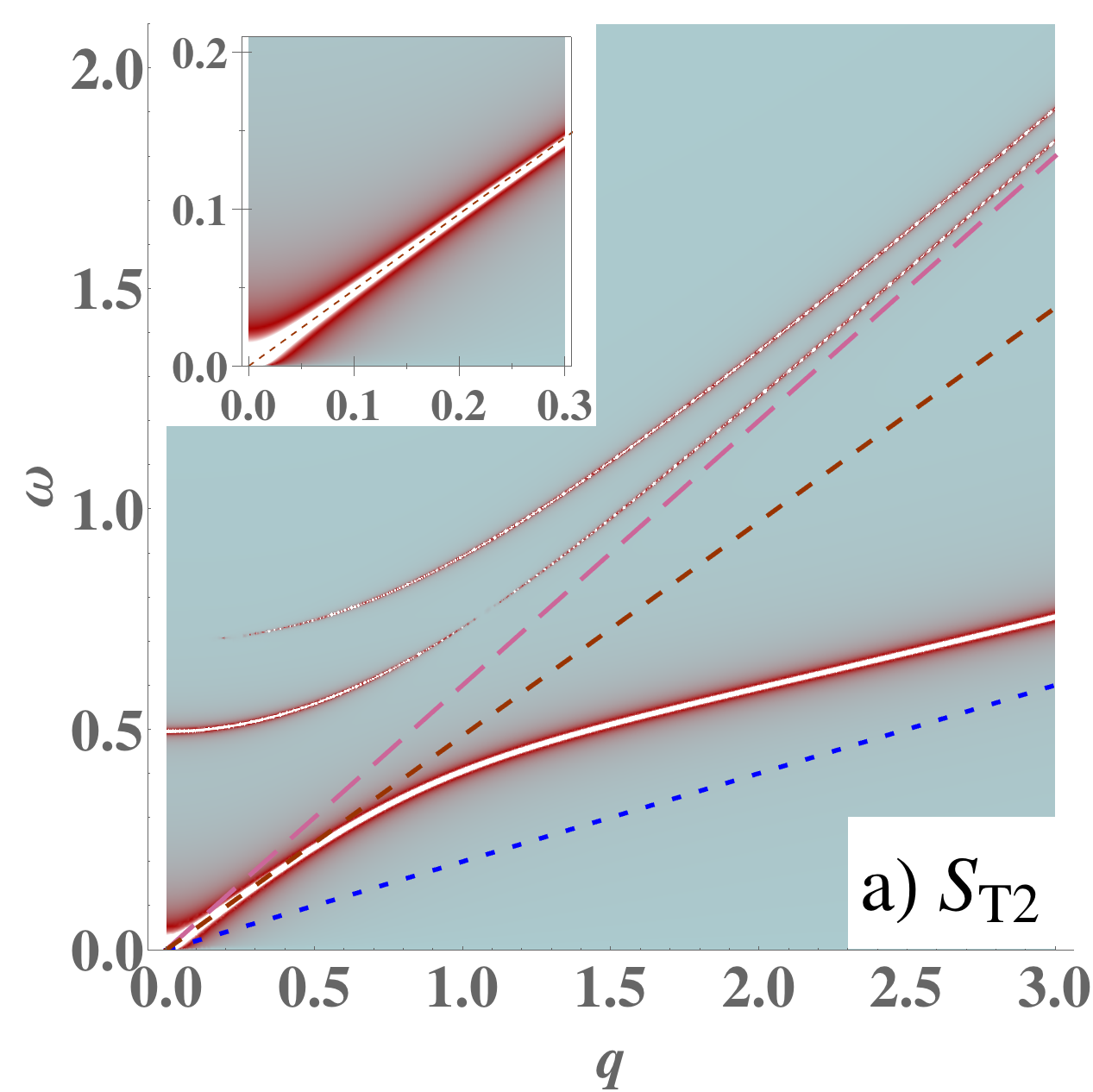}
  \hfill
 \includegraphics[width=7.7cm]{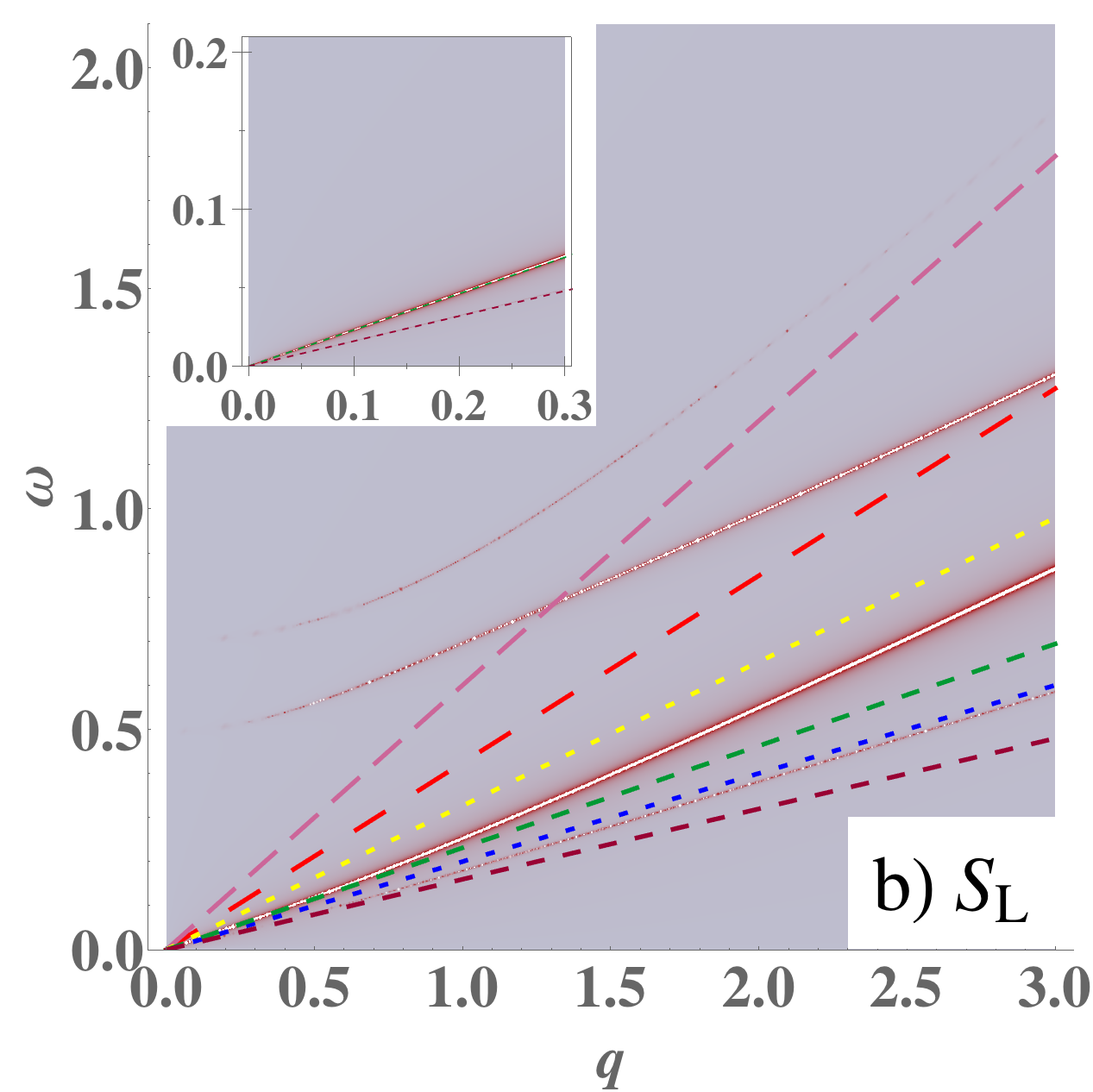}\\
 {\centering
 \includegraphics[scale=.6]{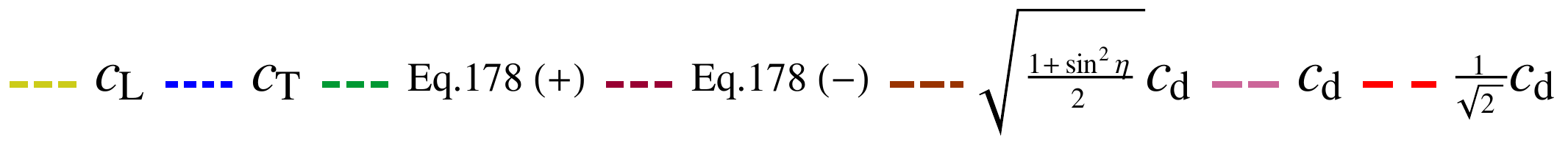}
 }\\
 \caption{Spectral function of the quantum smectic in units of the inverse shear modulus $1/\mu$, with Poisson ratio $\nu = 0.2$, at a representative angle $\eta = 3 \pi/16$. For a clear picture we have arbitrarily set $c_\td = 3 c_\tT$. The inset is a zoom up near the origin. The width of the poles is artificial and denotes the relative pole strengths: these ideal poles are actually infinitely sharp. Now the poles of the sectors $\tL, {\tT 1}$ are mixed, and of $\tT 2, \tT 3$ as well. We plot only $S_{\tT 2}$ and $S_{\tL}$, as $S_{\tT 3}$ resp. $S_{\tT 1}$ have the same poles although with different pole strengths. (a) In the purely transverse response we find the rotational Goldstone mode although with modified velocity. There are two gapped modes with gaps $\Omega$ resp. $\frac{1}{\sqrt{2}} \Omega$. (b) In the $\tL$--$\tT 1$ sector there are two gapless modes with velocities Eq.~\eqref{eq:smectic L T1 massless dispersion}, which extrapolate to the longitudinal and transverse phonon with velocities $c_\tL$ and $c_\tT$ at high energies. Furthermore there are two gapped modes with gaps $\Omega$ and $\frac{1}{\sqrt{2}}\Omega$ that extrapolate to the `transverse' condensate mode with velocity $c_\td$ and the `longitudinal' condensate mode with velocity $\frac{1}{\sqrt{2}}c_\td$ respectively at high energies. }\label{fig:smectic spectral functions}
   \includegraphics[width=7.7cm]{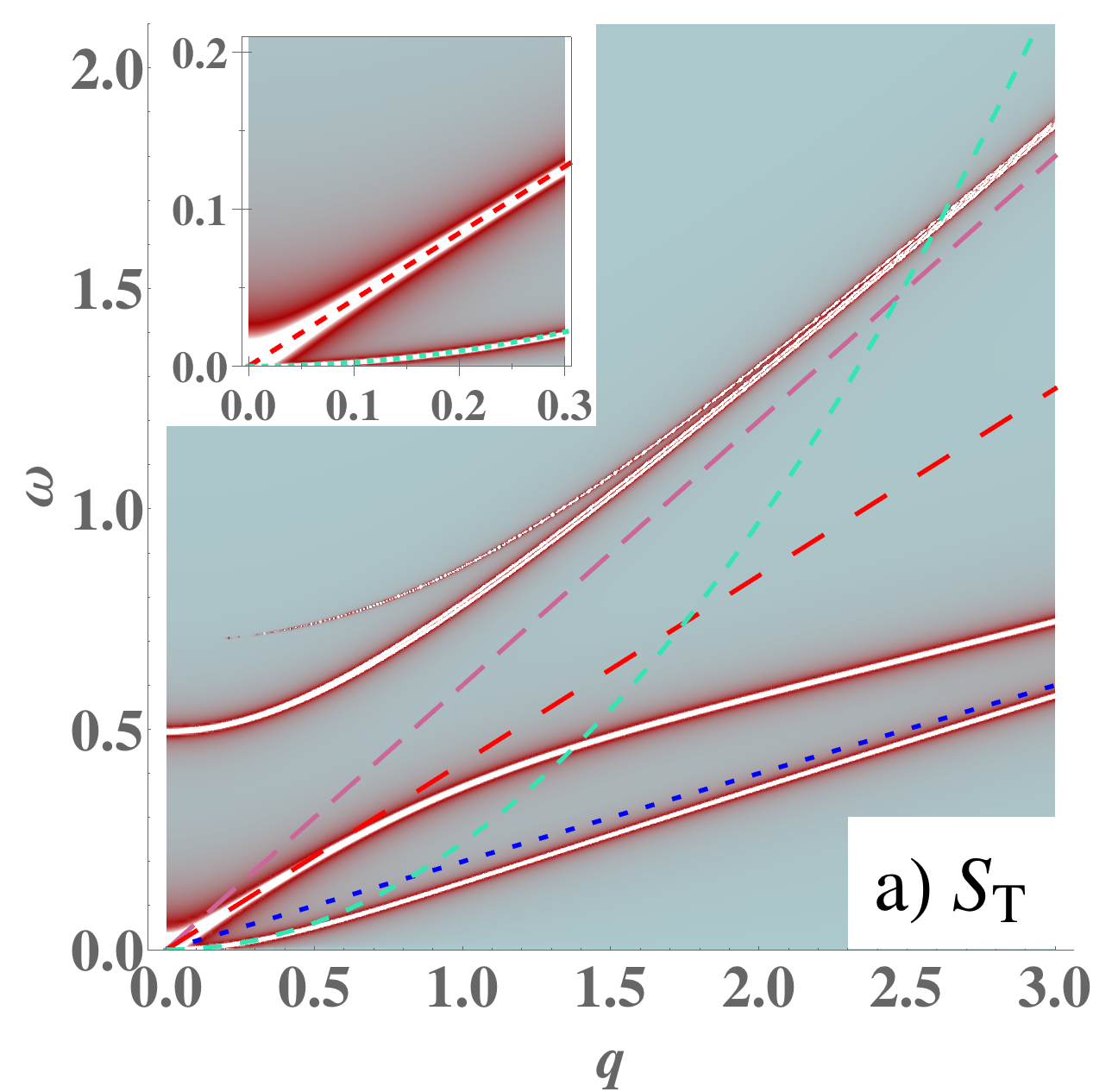}
  \hfill
 \includegraphics[width=7.7cm]{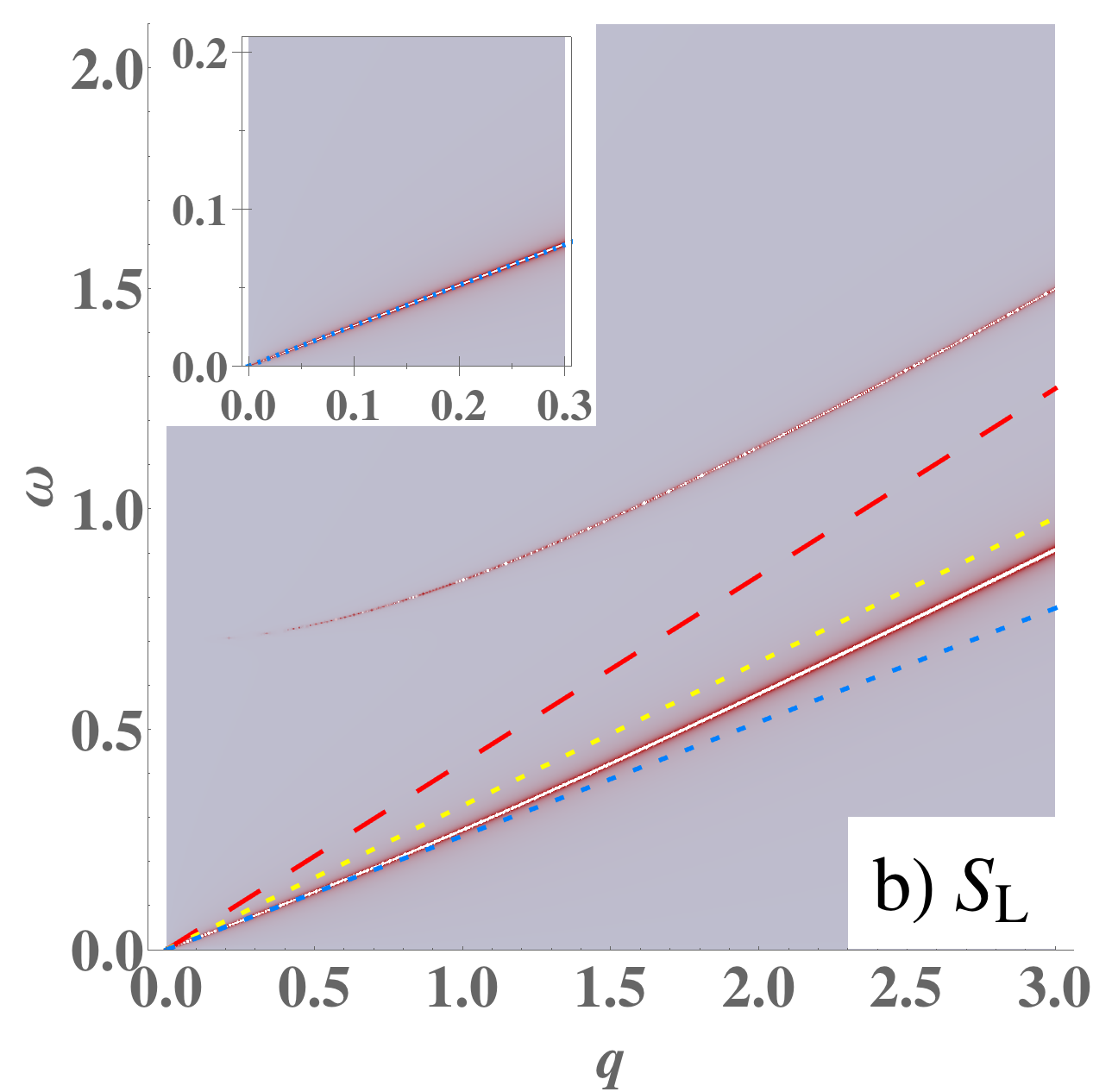}\\
 {\centering
 \includegraphics[scale=.6]{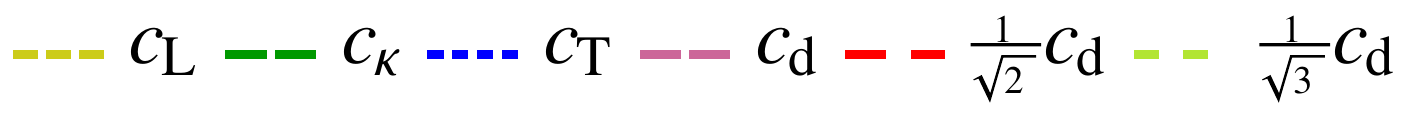}
 }\\
 \caption{Spectral function of the quantum smectic with momentum in the liquid plane ($\eta = 0$), in units of the inverse shear modulus $1/\mu$, with Poisson ratio $\nu = 0.2$.  For a clear picture we have arbitrarily set $c_\td = 3 c_\tT$. The inset is a zoom up near the origin. The width of the poles is artificial and denotes the relative pole strengths: these ideal poles are actually infinitely sharp. The response is independent of the in-plane interrogation angle $\eta$. (a) The transverse response shows the rotational Goldstone mode with velocity $\frac{1}{\sqrt{2}}c_\td$ as well as the quadratically dispersing undulation mode. There are three gapped poles, one with gap $\Omega$ and two with gap $\frac{1}{\sqrt{2}}\Omega$ (the splitting between these latter two modes at intermediate velocities is just barely visible). (b) In the longitudinal response we find a massless mode with velocity $\frac{1}{\sqrt{1-2\nu}} c_\tT = \sqrt{c_\tL^2 -c_\tT^2}$ at low energies extrapolating to the longitudinal phonon at high energies, and a gapped mode with gap $\Omega$.}\label{fig:eta0 smectic spectral functions}
\end{figure*}

\subsection{Collective modes in the quantum smectics}

\begin{table*}
\hfill
 \begin{tabular}{clclc}
 \multicolumn{3}{c}{general $\eta$}\\
  \toprule
  sector & & massless  & &massive  \\
  \hline
  $\tL$ & \rdelim\}{2}{1mm}[]  & \multirow{2}{*}{Eq.~\eqref{eq:smectic L T1 massless dispersion}}  & \rdelim\}{2}{1mm}[] & $\Omega$  \\
  $\tT 1$ &  &  & & $\frac{1}{\sqrt{2}} \Omega$ \\
  $\tT 2$ &\rdelim\}{2}{1mm}[]  &   \multirow{2}{*}{$\frac{1}{\sqrt{2}} \sqrt{1  + \sin^2 \eta} \;c_\td q$}  &\rdelim\}{2}{1mm}[] & $\Omega$ \\
  $\tT 3$ & &  & &  $\frac{1}{\sqrt{2}} \Omega$ \\
  \hline
  total & & 3 & &  4 \\
  \botrule
 \end{tabular}
 \hfill
 \begin{tabular}{cccc}
 \multicolumn{3}{c}{$\eta = 0$}\\
  \toprule
  sector & massless  & & massive  \\
  \hline
  $\tL$ & $\frac{1}{\sqrt{1-2\nu}} c_\tT q $ & & $\Omega + \frac{\frac{1}{2}c_\td^2 +  c_\tT^2}{2 \Omega} q^2$  \\
  $\tT 1$ & $\frac{1}{\sqrt{2}\Omega} c_\td c_\tT q^2$ & &  $\frac{1}{\sqrt{2}}\Omega + \frac{c_\td^2 +  c_\tT^2}{\sqrt{2} \Omega} q^2 $ \\
  $\tT 2$ & $\frac{1}{\sqrt{2}} c_\td q$ & &$\Omega + \frac{\frac{1}{2} c_\td^2 + c_\tT^2}{2 \Omega} q^2 $ \\
  $\tT 3$ & - & & $\frac{1}{\sqrt{2}}\Omega + \frac{c_\td^2}{\sqrt{2} \Omega} q^2$  \\
  \hline
  total & 3 & & 4 \\
  \botrule
 \end{tabular}
  \hfill
 \begin{tabular}{cccc}
 \multicolumn{3}{c}{$\eta = \pi/2$}\\
  \toprule
  sector & massless  & & massive  \\
  \hline
  $\tL$ & $c_\tL q $ & & - \\
  $\tT 1$ & - & & $\sqrt{\frac{1}{2} \Omega^2 + c_\tT^2 q^2} $ \\
  $\tT 2$ & - & & $\sqrt{\frac{1}{2} \Omega^2 + c_\tT^2 q^2} $ \\
  $\tT 3$ & $c_\td q$ & & -  \\
  \hline
  total & 2 & & 2 \\
  \botrule
 \end{tabular}
 \hfill
 \null\\ 
\caption{Collective modes in the smectic phases.  Indicated are the dispersion relations to lowest orders in momentum. 
The leftmost table shows the case for general angle $0 < \eta < \pi/2$. The propagators $G_\tL$ and $G_{\tT 1}$ share their poles: two highly $\eta$-dependent massless poles with velocities depicted in Fig.~\ref{fig:smectic velocity}, and two massive ones. The transverse propagator $G_{\tT 2}$ and $G_{\tT 3}$ contain the rotational Goldstone mode the velocity interpolates between $\frac{1}{\sqrt{2}} c_\td$ and $c_\td$. The middle table shows the results for $\eta = 0$, momentum in the liquid plane. As in the 2D smectic, the $T1$-transverse pole is the undulation mode with quadratic dispersion. The rotational mode is now exclusive found in $G_{\tT 2}$. The rightmost table show the case for momentum in the solid direction, $\eta = \pi/2$. The number of modes is greatly reduced. We find results similar to the 2D smectic: the compression mode obtains the full longitudinal-phonon velocity, while the shear phonons are gapped. The third propagator $G_{\tT 3}$ probes torque correlations in the liquid plane and picks up the rotational Goldstone mode.}\label{table:smectic spectrum}
\end{table*}

To obtain the spectrum of modes in the smectic phase, we add any coordinate of permutation of Eq.~\eqref{eq:dislocation Higgs term smectic xz} to Eq.~\eqref{eq:solid stress gauge field Lagrangian tilde} and calculate the propagators Eqs.~\eqref{eq:longitudinal propagator gauge field tilde}, \eqref{eq:transverse propagator gauge field tilde}. This is most easily performed on a computer. The result is highly dependent on the direction of momentum just as it was in 2+1D. Indeed, one would assume that the parent crystal anisotropy remains influential when not all translational symmetry is restored. However, in the present case of an isotropic crystal, matters do simplify quite a bit, as we can suffice with the polar angle $\eta$ while setting the azimuthal angle $\zeta = 0$ , see Eq.~\eqref{eq:eta zeta angles} and Fig.~\ref{fig:smectic eta angle}. Then the momentum $\mathbf{q} = ( q \cos \eta, q \sin \eta ,0)$ is in the $xy$-plane, and the polar angle $\eta$ tunes between momentum completely in the liquid plane $\eta = 0$, and parallel to the solid direction $\eta = \pi/2$. With these choices, the transverse $\tS$-direction is parallel to the $z$-axis, while $\tL$ and $\tR$ lie in the $xy$-plane. As before, we perform analytic continuation to real time $\omega_n \to \ti \omega - \delta$, where we drop the infinitesimal factors $\ti \delta$ at the last step of the calculation for ease of notation.

The general form of the propagators is too complicated to write down explicitly. However, we can inspect the poles of the propagators to lowest order in momentum. As we have seen before in QLC2D, the longitudinal and transverse sectors mix at general angle $\eta$, and the characteristics of the responses are hybrid. In the 3D case, for the $xz$-condensate, the pair $G_\tL$--$G_{\tT 1}$ share the same poles, as do the pair $G_{\tT 2}$--$G_{\tT 3}$. The pole strengths all differ, however. In Fig.~\ref{fig:smectic spectral functions} we have plotted the spectral functions of $G_\tL$ and $G_{\tT 2}$ respectively. In the $\tL$--$\tT 1$-sector we find four poles. Two of the poles are massive with gaps $\Omega$ and $\Omega/\sqrt{2}$ respectively, to be interpreted as coming from the dislocation condensate in two Burgers directions (recall there were three massive modes in the nematic, see Table~\ref{table:nematic spectrum}). The two other poles are massless with dispersions
\begin{widetext}
 \begin{align}
  (\omega^{\tL, \tT 1}_{1,2})^2 = \frac{(3-2\nu) -  (1-2\nu)\cos 2 \eta \pm \sqrt{ 1 - 4\nu + 20\nu^2 - 2 (3 - 8\nu + 4\nu^2) \cos 2 \eta  + 3(3 - 4\nu - 4\nu^2) \cos^2 2 \eta} }{4(1-2\nu)}c_\tT^2 q^2 + \ldots
  \label{eq:smectic L T1 massless dispersion}
 \end{align}
\end{widetext}

Overall this result is very comparable to 2+1D: there are two massless modes, one extrapolating to the longitudinal phonon and one to the transverse phonon, showing up in both propagators. There is obviously a complicated angle-dependence here. We have plotted the velocity as a function of $\eta$ for a representative value $\nu = 0.3$ in Fig.~\ref{fig:smectic velocity}. We can identify the pole with the plus sign as having longitudinal character while the minus sign has transverse character. Recall that in 2D, the longitudinal velocity varies from $c_\tL$ at $\eta = 0$ to $c_\kappa$ at $\eta = \pi/4$ back to $c_\tL$ at $\eta =\pi/2$. Conversely, the 2D transverse pole has vanishing pole strength and vanishing velocity at $\eta = 0,\pi/2$, while obtaining the full transverse velocity $c_\tT$ at $\eta = \pi/4$~\cite{QLC2D}. Here in 3D the behavior is somewhat different. For the longitudinal mode, the maximum velocity is $c_\tL$ at $\eta =\pi/2$ and the minimum velocity is $c_\kappa$ at $\eta = \tfrac{1}{2} \arccos \tfrac{1}{3}$. At $\eta =0$ the velocity is smaller than $c_\tL$, namely $\frac{1}{\sqrt{1-2\nu}}c_\tT = \sqrt{c_\tL^2 - c_\tT^2}$. The transverse pole has again vanishing pole strength and vanishing velocity at $\eta = 0,\pi/2$. However, the maximum velocity is smaller than $c_\tT$, namely $\sqrt{\frac{2+2\nu}{3+2\nu}} c_\tT$ attained at $\eta = \tfrac{1}{2} \arccos \frac{1-2\nu}{3+2\nu}$. Note that $c_\tL,c_\kappa \to \infty$ as $\nu \to 0.5$.

Compared to the 2D smectics we have two more propagators, namely $G_{\tT 2}$ and $G_{\tT 3}$. 
Again we find is that these two propagators share their poles although the poles strengths are different for each case. For $G_{\tT 2}$ and $G_{\tT 3}$ we find three modes with low-energy dispersions:
\begin{align}
 \omega^{\tT 2, \tT 3}_1 &=  \frac{1}{\sqrt{2}} \sqrt{1 + \sin^2 \eta} \;c_\td q + \ldots ,\nonumber \\
 (\omega^{\tT 2, \tT 3}_2)^2 &= \Omega^2  + \big(  \frac{1}{2} c_\td^2( 1 + \sin^2 \eta) + c_\tT^2 \cos^2 \eta \big) q^2 + \ldots,\nonumber\\
 (\omega^{\tT 2, \tT 3}_3)^2 &= \frac{1}{2} \Omega^2  + ( c_\td^2 \cos^2\eta + c_\tT^2 \sin^2 \eta   ) q^2 + \ldots.\label{eq:smectic T2 T3 poles}
\end{align}
The first mode is a rotational Goldstone mode! We will say much more about this below in Sec.~\ref{subsec:Rotational Goldstone mode smectic}. Its velocity depends on the interrogation angle, interpolating between the `longitudinal torque' and `transverse torque' characteristics explained in Sec.~\ref{subsec:Torque stress in the quantum nematic}. This mode is somehow `divided' between $G_{\tT 2}$ and $G_{\tT 3}$. Next to that there are two gapped modes, mixing the condensate phase mode with the gapped shear phonons.

Eqs.~\eqref{eq:smectic L T1 massless dispersion}, \eqref{eq:smectic T2 T3 poles} are not strictly valid for the special angles $\eta = 0$ (momentum parallel to the `liquid' $x$-direction) and $\eta=\pi/2$ (momentum parallel to the `solid' $y$-direction), although the limiting behavior is correct. In fact, the form of the propagators change and the number of poles is reduced, to such an extent that we can write down the results explicitly. 

For $\eta = 0$, momentum in the liquid plane, the propagators read
\begin{widetext}
\begin{align}
G^{(\eta = 0)}_\tL &= \frac{1}{\mu} \frac{-c_\tT^2 q^2 ( \omega^2 - \tfrac{1}{2} c_\td^2 q^2 - \Omega^2)}{(\omega^2 - c_\tL^2 q^2)(\omega^2 - \tfrac{1}{2} c_\td^2 q^2) - \Omega^2 (\omega^2 -\frac{1}{1-2\nu} c_\tT^2 q^2)},
\label{eq:GL eta0 smectic}\\
G^{(\eta = 0)}_{\tT 1} &= \frac{1}{\mu} \frac{ \frac{1}{2}\omega^2 \Omega^2  - c_\tT^2 q^2 ( \omega^2 - c_\td^2 q^2 -2\Omega^2)}{(\omega^2 - c_\tT^2 q^2)(\omega^2 - c_\td^2 q^2) - \omega^2 \frac{1}{2}\Omega^2},\label{eq:GT1 eta0 smectic}\\
G^{(\eta = 0)}_{\tT 2} &= \frac{1}{\mu} \frac{ - c_\tT^2 q^2 ( \omega^2 - c_\td^2 q^2) - \Omega^2 (  \omega^2- 2c_\tT^2 q^2 - \frac{1}{2}c_\td^2 q^2  -  \Omega^2)}{(\omega^2 - c_\tT^2 q^2)(\omega^2 - c_\td^2 q^2) - \Omega^2( \omega^2 - \frac{1}{2}c_\td^2 q^2)},\label{eq:GT2 eta0 smectic}\\
G^{(\eta = 0)}_{\tT 3} &= \frac{1}{\mu} \frac{ -\frac{1}{2}\Omega^2}{\omega^2 - c_\td^2 q^2 - \frac{1}{2}\Omega^2}.\label{eq:GT3 eta0 smectic}
\end{align}
\end{widetext}
It is instructive to compare this to the propagators of the 2D nematic in  Eqs.~\eqref{eq:2D nematic GL}, \eqref{eq:2D nematic GT}. Indeed the longitudinal propagator is identical except for the form of the longitudinal velocity, due to the different dimensionality in Eq.~\eqref{eq:longitudinal velocity definition}. Also, the velocity $\frac{1}{\sqrt{1-2\nu}} c_\tT = \sqrt{c_\tL^2 -c_\tT^2}$ is different from $c_\kappa$ in 3D. (However, in 2D there is the relation $c_\kappa^\mathrm{2D} = \sqrt{(c_\tL^\mathrm{2D})^2 - c_\tT^2}$.) Notice as well that Eq.~\eqref{eq:GL eta0 smectic} is different from the 3D nematic Eq.~\eqref{eq:nematic longitudinal propagator} in particular considering the velocity of the massive condensate mode: it contains a prefactor of $\frac{1}{2}$ instead of $\frac{1}{3}$. The longitudinal propagator contains 
two modes, with dispersion relations
\begin{align}
 \omega^{\tL}_1(\eta = 0) &=  \frac{1}{\sqrt{1 - 2\nu}} c_\tT  q + \ldots ,\nonumber \\
 \omega^{\tL}_2 (\eta = 0)&= \Omega + \frac{\frac{1}{2}c_\td^2 +  c_\tT^2}{2 \Omega} q^2 + \ldots . \label{eq:eta0 smectic longitudinal dispersions}
\end{align}
The first, massless mode is a mixture of the 2D compression mode and the 3D longitudinal phonon, while the second, massive mode is identical to the 2D nematic condensate mode.

The transverse propagator $G_{\tT 2}$ Eq.~\eqref{eq:GT2 eta0 smectic} is identical to that of the 2D nematic Eq.~\eqref{eq:2D nematic GT}. This makes sense: from Eq.~\eqref{eq:transverse propagator from torque correlators} we know that $G_{\tT 2}$ measures the shear correlations in the $\tL\tS$-plane which for $\eta,\zeta = 0$ is exactly the liquid $xz$-plane. This propagator does not know anything about the third dimension, and it only sees a 2D dislocation condensate: a 2D nematic. This propagator also contains one massless and one massive mode, with dispersion relations,
\begin{align}
 \omega^{\tT 2}_1(\eta = 0) &=  \frac{1}{\sqrt{2}} c_\td  q + \ldots ,\nonumber \\
 \omega^{\tT 2}_2(\eta = 0) &= \Omega + \frac{\tfrac{1}{2} c_\td^2 +  c_\tT^2}{2 \Omega} q^2 + \ldots . \label{eq:eta0 smectic T2 dispersions}
\end{align}
These are the rotational Goldstone mode and a massive condensate--shear mode. 

Next, look at $G_{\tT 1}$ in Eq.~\eqref{eq:GT1 eta0 smectic}, and compare this with the transverse propagator of the 2D smectic at interrogation angle $\eta_{\rm 2D}=0$ in the limit $\ell \to 0$, which we take from QLC2D:
\begin{equation}
 G^\mathrm{2D smec}_{\tT} (\eta_{\rm 2D} = 0) = \frac{1}{\mu}\frac{ -\tfrac{1}{2}\omega^2 \Omega^2 - c_\tT^2 q^2 (\omega^2 - c_\td^2 q^2 + 2 \Omega^2)}{(\omega^2 - c_\tT^2 q^2) (\omega^2 - c_\td^2 q^2 ) - \tfrac{1}{2} \omega^2 \Omega^2  }.
\end{equation}
Here we have rescaled the Higgs mass by a factor of $\sqrt{2}$ due to differing definitions in QLC2D. This equation is indeed identical to Eq.~\eqref{eq:GT1 eta0  smectic}. Again, this makes perfect sense. The 2D smectic propagator at interrogation angle $\eta_{\rm 2D}=0$ has momentum along the liquid direction while the transverse direction is in the solid direction. Now $G_{\tT 2}$ is the propagator of shear correlations orthogonal to the $\tR$ direction, that is in the $\tL\tS$-plane. For $\zeta = 0$, $\tS$ is along the solid $z$-axis, while the longitudinal direction is within the liquid plane. So this is a perfect match for the 2D smectic propagator. We recall the dispersion relations from QLC2D:
\begin{align}
 \omega^{\tT 1}_1 (\eta = 0) &=  \frac{c_\td c_\tT }{\frac{1}{\sqrt{2}}\Omega}  q^2 + \ldots ,\nonumber \\
 \omega^{\tT 1}_2(\eta = 0) &= \frac{1}{\sqrt{2}}\Omega + \frac{c_\td^2 +  c_\tT^2}{\sqrt{2}\Omega} q^2 + \ldots . \label{eq:smectic xy T2 dispersions}
\end{align}
The first mode is massless with quadratic dispersion. This is the {\em undulation mode}~\cite{CvetkovicZaanen06b,QLC2D} known from classical smectic liquid crystals~\cite{DeGennesProst95,SinghDunmur02,ChaikinLubensky00}: the smectic can still support a reactive response to shear stress, but (at low energies) it is less strong than that of a true solid. This is reproduced here in the 3D smectic.

Finally we have the propagator $G_{\tT 3}$ of Eq.~\eqref{eq:GT3 eta0 smectic}, representing shear correlations in the purely transverse plane $\tR\tS$ (the $yz$-plane for $\eta=0$), which has no counterpart in 2D liquid crystals. It has one massive mode with the exact dispersion
\begin{align}
 (\omega^{\tT 3})^2(\eta = 0) &= \tfrac{1}{2}  \Omega^2  + c_\td^2 q^2.\label{eq:eta0 smectic T3 dispersion}
 \end{align}
Is it clearly a massive condensate mode, not influenced by any phonon degrees of freedom, i.e. it does not depend on $c_\tT$. Resoundingly, even though the dislocation condensate seems not to couple to stress in our gauge choice, this mode does show up in the physical transverse stress propagator.

In short at $\eta = 0$, for the in-plane response we recover the 2D nematic in the $\tL$--$\tT 1$ sector apart from some slightly different velocities, while smectic behavior including the undulation mode shows up for momenta transverse to the liquid plane.

The final special case is $\eta = \pi/2$, for which the momentum lies in the solid $y$-direction. At this angle, all four propagators contain only a single mode, with the following dispersion relations:
\begin{align}
\omega^{\tL}(\eta = \pi/2) &= c_\tL q,\label{eq:smectic longitudinal phonon}\\
 (\omega^{\tT 1})^2 (\eta = \pi/2)&= \frac{1}{2} \Omega^2 + c_\tT^2 q^2,\label{eq:smectic pi/2 gapped shear}\\
 (\omega^{\tT 2})^2(\eta = \pi/2) &= \frac{1}{2} \Omega^2 + c_\tT^2 q^2,\\
 \omega^{\tT 3} (\eta = \pi/2)&= c_\td q. 
\end{align}
For the longitudinal direction coinciding with the solid direction, we retrieve the longitudinal phonon in Eq. \eqref{eq:smectic longitudinal phonon}! Even though the two transverse directions do not support shear rigidity, the velocity of the compressional mode at this angle is identical to that of a crystal. just as it was in 2+1D. Meanwhile, the two transverse phonons in the liquid plane have picked up the Higgs gap $\Omega/\sqrt{2}$, while not being mixed with the condensate phase degree of freedom as indicated by the absence of $c_\td$. The final mode is the rotational Goldstone mode which exists in the transverse, liquid plane. We had already seen in Eq.~\eqref{eq:longitudinal rotational Goldstone dispersions} of the nematic that this Goldstone mode has a different velocity than those appearing in $G_{\tT 1}$ and $G_{\tT 2}$.

In conclusion, we have derived the spectra of the 3D `isotropic' smectic liquid crystal and uncovered the collective modes that are expected on grounds of the symmetries.
We find for the 3D smectic a similar pattern as in the 2D smectic: depending on the interrogation angle, i.e. the angle of the momentum relative to the smectic plane, we retrieve some solid-like (massless phonons) and some liquid-like (gapped shear, compression mode) features. However, the modes do not in general exhibit a decomposition of the form `solid $\times$ liquid' , but instead show a mixture of the two characteristics through mode coupling. Assuredly, at the special angle $\eta =0$, the undulation mode of classical smectics is reproduced. From QLC2D, we know that $\eta_{\rm 2D} = \pi/4$ (or $[11]$-direction) for the 2D-condensate is a special angle displaying a perfect `solid $\times$ liquid' response, but we have not been able to detect similar `magic angle' in 3D, cf. Fig.~\ref{fig:smectic velocity}, implying that it could be accidental for 2D as there is only one transverse direction. 
Naturally, a full analysis incorporating the space group symmetries of the 3D crystal should further affect the mode spectrum in terms of $\eta$ and even $\zeta$. We leave this for future considerations.

\subsection{Rotational Goldstone mode}\label{subsec:Rotational Goldstone mode smectic}
We have seen the emergence of a rotational Goldstone mode of rotations within the liquid plane. We can also see this from the torque stress in the deep Higgs limit $\Omega \to \infty$, similar to Sec.~\ref{subsec:Torque stress in the quantum nematic}.

There is a complementary insight into this phenomenon from symmetry principles. It is of course well established that spontaneous breaking of continuous global symmetries gives rise to massless Goldstone modes, which can be viewed as the low-energy fluctuations along the flat directions of order parameter space. However, it is equally well known that in solids, which break both translational and rotational symmetry, only the translational Goldstone modes, the phonons, are ever seen. This apparent contradiction to the Goldstone theorem has been featured even in modern textbooks~\cite{Sethna06}. In recent years, a good understanding of possibilities for having a reduced number of Goldstone modes has developed. First of all, it is possible that two broken symmetry generators excite the same Goldstone mode. This occurs when the vacuum expectation value of the commutator of (the densities of) these symmetry generators is non-vanishing~\cite{WatanabeBrauner11,WatanabeMurayama12,Hidaka13}. This is for instance the case in Heisenberg ferromagnets, where there is only one spin wave although there are two broken generators.

The current issue is however a different one. Recall that the Goldstone modes are infinitesimally small deviations from the preferred order parameter value (the Goldstone theorem is a statement about the limit of energy going to zero). Now, an infinitesimal rotation $\omega^{ab} = \partial_a u^b - \partial_b u^a$ can not locally be distinguished from an infinitesimal translation $u^a$~\cite{LowManohar02}. This simple observation can be extended into a quite general statement according to Ref.~\onlinecite{WatanabeMurayama13}. Namely, when the densities of broken symmetry generators are linearly dependent on each other, they do not excite independent Goldstone modes. In our case, the density of the generator of rotations around the $c$-axis $R^c(x)$ depends upon the generators of translations in the $a$-direction $T^a(x)$ in the Lie algebra of the Poincar\'e group via~\cite{WatanabeMurayama13,Kleinert08}:
\begin{equation}\label{eq:rotations depend on translations}
 R^c (x) = \epsilon^{cba} x^b T^a(x).
\end{equation}
Roughly speaking, if one tries to excite a rotational Goldstone mode in a solid, one will instead excite (transverse) phonons.

In Ref.~\onlinecite{BeekmanWuCvetkovicZaanen13} we showed that in two dimensions indeed the rotational Goldstone mode emerges when translational symmetry is restored (the 2D nematic phase). This is completely consistent with the discussion above: as long as translational symmetry is broken, the relation Eq.~\eqref{eq:rotations depend on translations} prevents the rotational Goldstone mode to emerge as an independent excitation. But now in 3D an interesting scenario crops up. Clearly, if all translational symmetry is restored, three rotational Goldstone modes emerge as we have seen in Sec.~\ref{sec:nematic}. However, if translational symmetry is restored in two directions, according to Eq.~\eqref{eq:rotations depend on translations} one rotational degree of freedom is independent of any other broken symmetries. In this section we have confirmed that indeed one rotational Goldstone mode emerges, precisely in the plane where translational symmetry is restored.

%% file: sec_columnar.tex
\begin{figure}[t]
\hfill
 \includegraphics[width=5cm]{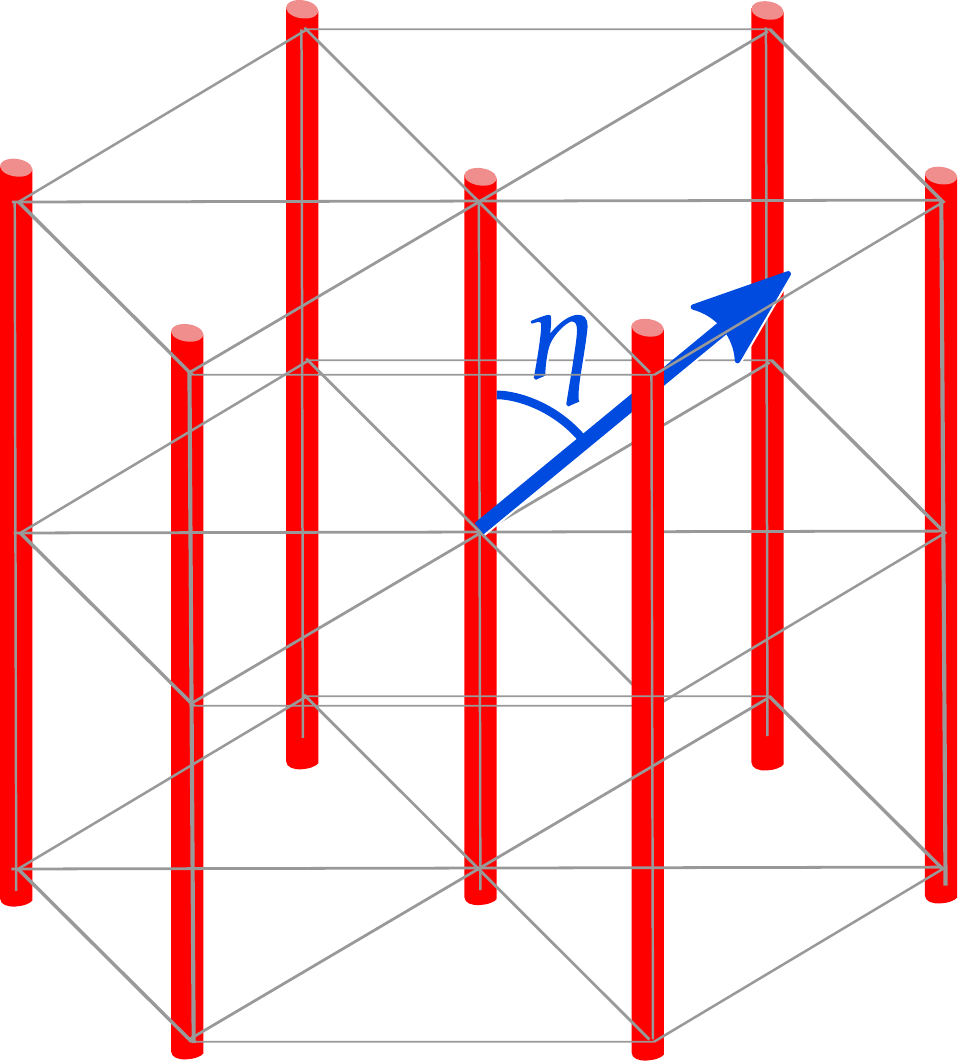}
 \hfill\null
\caption{Definition of the angle $\eta$ between the momentum vector (blue) and the liquid lines (red). $\eta = 0$ is parallel to the lines, $\eta = \pi/2$ is perpendicular to the lines.}\label{fig:columnar eta angle}
\end{figure}

After considering the condensation of dislocations with Burgers vectors in three (nematics) and two (smectic) directions, we now discuss single-Burgers direction condensates, which we call {\em columnar phases} in analogy with classical liquid crystals. One can picture a regular 2D array (corresponding to a planar projection of the original crystal lattice) of 1D liquid lines, see Figs.~\ref{subfig:columnar}, \ref{subfig:hexagonal melting columnar vertical} and \ref{subfig:hexagonal melting columnar horizontal}. 
The recipe is exactly the same as the previous two sections: consider the appropriate Higgs term Eq.~\eqref{eq:dislocation Higgs term}, resolve the glide constraint, add the Higgs term to the Coulomb term and calculate the propagators.

A columnar phase does not exist strictly in 2D but we expect some results to be similar to the 2D smectic, when the momentum is at least partially in the liquid direction.
In general the result will again be highly dependent on the direction of momentum with respect to the crystal axes. However, starting from an isotropic solid, after restoring translational symmetry in one dimension, there is axial symmetry around this liquid direction. Therefore we can capture the general case by considering only the polar angle between the liquid direction and the solid plane, see Fig.~\ref{fig:columnar eta angle}. This angle $\eta$ has therefore the same meaning as for the smectic: $\eta = 0$ is parallel to the liquid direction while $\eta = \pi/2$ is parallel to a solid direction.

When the momentum is in a solid--liquid plane, we expect results similar to the 2D smectic, although there are of course two more transverse propagators. There is the interesting special case of the momentum completely in the solid plane, where we expect some sort of 2D solid response. Furthermore, according to the principles mentioned in Sec.~\ref{subsec:Rotational Goldstone mode smectic}, we do not expect any rotational Goldstone modes in the columnar phase.

\subsection{Higgs term and glide constraint}
The starting point is Eq.~\eqref{eq:dislocation Higgs term} where the sum over $a$ is removed and we select only a single Burgers direction $x$, which can be done without loss of generality when starting from an isotropic solid. 
As before, we choose the dislocation Lorenz gauge fix Eq.~\eqref{eq:dislocation Lorenz gauge fix} and disregard the decoupled condensate phase degrees of freedom. As for the glide constraint, following the patterns of the earlier sections, we find the contributions
\begin{align}
 \mathcal{L}^{(x)}_\mathrm{glide} =  - 2 \frac{\Omega^2}{4 c_\tT^2 \mu}  \frac{ \tilde{p}^2}{\frac{1}{c_\td^2} \omega_n^2 + q_x^2 } \left\lvert \tilde{b}^x_{yz} \right\rvert^2,
\end{align}
and similar for the $y$- and $z$-condensates.

The Higgs term for the columnar $x$-condensates is:
\begin{align}
 \mathcal{L}^{(x)}_\mathrm{Higgs}
  &= \frac{\Omega^2}{2 c_\tT^2 \mu}   \Big[   
  \lvert \tilde{b}^x_{1\tR} \rvert^2 + \lvert \tilde{b}^x_{1\tS} \rvert^2 + \lvert \tilde{b}^x_{\tR\tS} \rvert^2  \nonumber\\
   &\phantom{mmnmm}
   -  \frac{\tilde{p}^2}{ \frac{1}{c_\td^2} \omega_n^2 + q_x^2} \lvert \tilde{b}^x_{yz}\rvert^2 \Big],
   \label{eq:dislocation Higgs term columnar x} 
\end{align}
and similar permutations of $x,y,z$ for the other choices $y$ and $z$.

\begin{figure}
 \includegraphics[width=8cm]{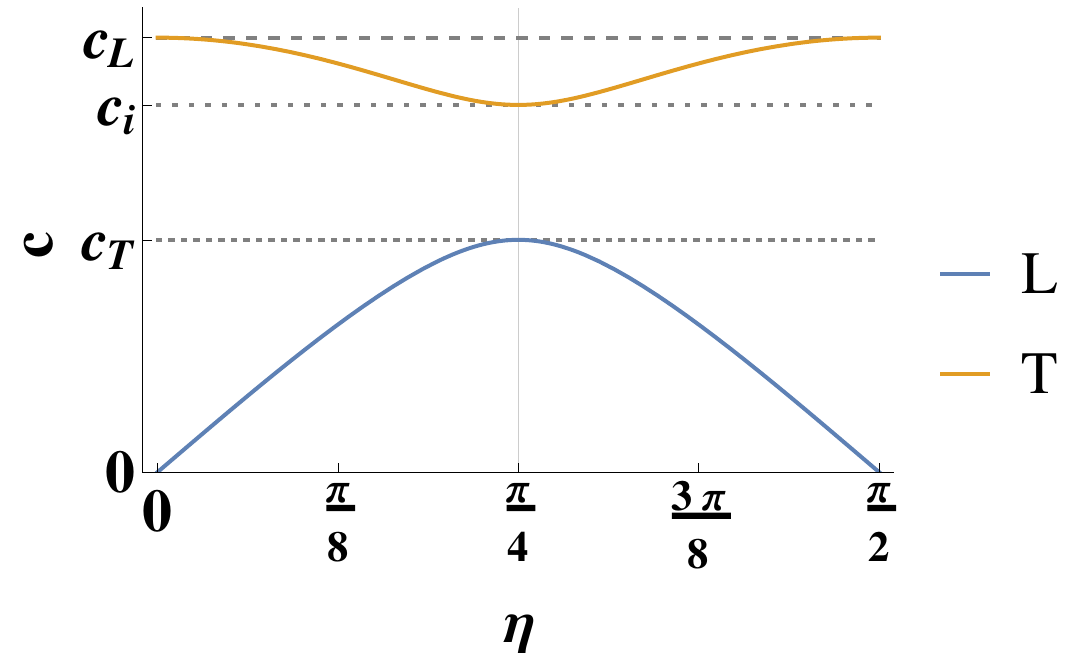}
 \caption{Velocities of the massless modes in the columnar phase as a function of interrogation angle $\eta$ from Eq.~\eqref{eq:columnar L T1 massless dispersion}. Here we have chosen a representative value of the Poisson ration $\nu = 0.3$. The velocity of the longitudinal mode varies from that of the longitudinal phonon $c_\tL$ at $\eta =0 ,\pi/2$ while it is reduced to $c_\mathrm{i} = \sqrt{c_\tL^2 - c_\tT^2}$ at $\eta =\pi/4$. The velocity of the transverse mode is maximal at $\eta =\pi/4$ where it reaches that of the transverse phonon $c_\tT$, while it is reduced to zero at $\eta =0,\pi/2$, where it also has vanishing pole strength. This pattern is identical to that of the 2D smectic~\cite{QLC2D}. }\label{fig:columnar velocity}
\end{figure}

\subsection{Collective modes in the quantum columnar phases}

As mentioned before, departing from an isotropic solid there is axial symmetry around the liquid direction, and therefore we can capture the response of the columnar phase by choosing translational symmetry restoration in the $x$-direction, and setting the azimuthal angle $\zeta= 0$ without loss of generality. The angle $\eta$ again interpolates between momentum completely in the liquid ($\eta =0$) and completely in the solid direction ($\eta = \pi/2$). 
Like for the smectics, the general form of the propagators is complicated and it is not insightful to write them down explicitly. We shall study them through the dispersion relation of their poles and the visualization of their spectral functions.

We find that for general angle $\eta$, the sectors $G_{\tL}$ and $G_{\tT 1}$, and the sectors $G_{\tT 2}$ and $G_{\tT 3}$ are mixed and share the same poles (although the pole strengths are different). For $G_{\tL}$ and $G_{\tT 1}$ there is one massive pole and two massless poles, with dispersion relations,
\begin{align}
 (\omega^{\tL, \tT 1}_{1,2})^2 &=  \frac{1 - \nu \pm \sqrt{\nu^2  - (1-\nu) \sin^2 2 \eta}}{-1 + 2\nu} c_\tT  q + \ldots ,\label{eq:columnar L T1 massless dispersion}\\
 (\omega^{\tL, \tT 1}_3)^2 &= \frac{1}{2}\Omega^2 + \big(c_\tT^2 +  c_\td^2 \cos^2 \eta) \big) q^2 + \ldots .
\end{align}
The last mode is related to gapped shear from the liquid direction while the massless modes interpolate between the longitudinal and transverse phonons in the solid-like directions. The velocity of these massless poles as a function of $\eta$ at a representative value of $\nu = 0.3$ is plotted in Fig.~\ref{fig:columnar velocity}. It mimics the 2D smectic~\cite{QLC2D}: the longitudinal pole has maximum velocity at $\eta = 0,\pi/2$ and minimum velocity $c_{\rm min} = \sqrt{c_\tL^2 - c_\tT^2}$ at $\eta = \pi/4$. The transverse pole has vanishing pole strength and vanishing velocity at $\eta =0,\pi/2$ and a maximum velocity of $c_\tT$ at $\eta =\pi/4$. As we mentioned in QLC2D, remarkably at $\eta = \pi/4$ the ultimate `solid $\times$ liquid' behavior is attained: the shear mode has the full transverse velocity like a transverse phonon, while the longitudinal mode has the pure compression velocity $c^\mathrm{2D}_\kappa = \sqrt{ (c^\mathrm{2D}_\tL)^2 - c_\tT^2}$.

The propagators $G_{\tT 2}$ and $G_{\tT 3}$ share their poles, although the pole strengths differ. They each have one massive and one massless pole, with dispersion relations:
\begin{align}
 \omega^{\tT 2,3}_1  &= \sin \eta\;  c_\tT  q + \ldots ,\\
 \omega^{\tT 2,3}_{2}  &=  \frac{1}{\sqrt{2}} \Omega + \frac{c_\td^2 + \cos^2 \eta\; c_\tT^2}{\sqrt{2} \Omega} q^2 + \ldots .
\end{align}
The first mode is the second transverse phonon, the velocity of which vanishes at $\eta = 0$ and $\eta =\pi/2$, see below. The gapped mode is a condensate mode `partially' coupled with the transverse phonon depending on the angle $\eta$.

\begin{table*}
 \begin{tabular}{clclc}
 \multicolumn{3}{c}{general $\eta$}\\
  \toprule
  sector & & massless  & &massive  \\
  \hline
  $\tL$ & \rdelim\}{2}{1mm}[]  & \multirow{2}{*}{Eq.~\eqref{eq:columnar L T1 massless dispersion}}  & \rdelim\}{2}{1mm}[] & \multirow{2}{*}{$\frac{1}{\sqrt{2}}\Omega + \frac{\cos^2 \eta\; c_\td^2 + c_\tT^2 }{\sqrt{2}\Omega} q^2$} \\
  $\tT 1$ &  &  & & \\
  $\tT 2$ &\rdelim\}{2}{1mm}[]  &   \multirow{2}{*}{$\sin \eta \;c_\tT q$}  & \rdelim\}{2}{1mm}[]& \multirow{2}{*}{$\frac{1}{\sqrt{2}}\Omega + \frac{c_\td^2 + \cos^2 \eta\; c_\tT^2}{\sqrt{2}\Omega} q^2$} \\
  $\tT 3$ & &  & &  \\
  \hline
  total & & 3 & &  2 \\
  \botrule
 \end{tabular}
\hfill
  \begin{tabular}{cccc}
 \multicolumn{3}{c}{$\eta=0$}\\
  \toprule
  sector &  massless  & &massive  \\
  \hline
  $\tL$ &  $c_\tL q$ & & - \\
  $\tT 1$ &$\frac{1}{\sqrt{2}\Omega} c_\td c_\tT q^2$   & & $\frac{1}{\sqrt{2}}\Omega + \frac{c_\td^2 +  c_\tT^2}{\sqrt{2} \Omega} q^2 $\\
  $\tT 2$ &$\frac{1}{\sqrt{2}\Omega} c_\td c_\tT q^2$   & & $\frac{1}{\sqrt{2}}\Omega + \frac{c_\td^2 +  c_\tT^2}{\sqrt{2} \Omega} q^2 $\\
  $\tT 3$ & - &  & - \\
  \hline
  total &  3 & &  2 \\
  \botrule
 \end{tabular}
 \hfill
 \begin{tabular}{cccc}
 \multicolumn{3}{c}{$\eta=\pi/2$}\\
  \toprule
  sector &  massless  & &massive  \\
  \hline
  $\tL$ & $c_\tL q$ & & - \\
  $\tT 1$ & - &  & $\frac{1}{\sqrt{2}} \Omega + \frac{c_\tT}{\sqrt{2} \Omega} q $ \\
  $\tT 2$ & $c_\tT q$ &  & - \\
  $\tT 3$ & - &  & $\frac{1}{\sqrt{2}} \Omega + \frac{c_\td}{\sqrt{2} \Omega} q$  \\
  \hline
  total &  2 & &  2 \\
  \botrule
 \end{tabular}
\caption{Collective modes in the columnar phases.  Indicated are the dispersion relations to lowest orders in momentum. For translational symmetry restoration in the $x$-directions, $G_\tL$ and $G_{\tT 1}$ share their poles as do  $G_{\tT 2}$ and $G_{\tT 3}$. The first pair contain two massless modes which extrapolate to the a longitudinal and transverse phonon, and a gapped shear mode, which is identical to the 2+1D smectic. The second pair contain a second transverse phonon whose velocity increases smoothly from zero to $c_\tT$ as $\eta$ goes from $0$ to $\pi/2$. There is also a gapped mode which has contribution from the dislocation condensate as well as from gapped shear. At $\eta =0$ we find two undulation modes in the $G_{\tT 1,2}$-sectors, while $G_{\tT 3}$ vanishes. At $\eta = \pi/2$, there is a  2D solid in the ${\tL}$--${\tT 2}$-sectors, while the responses in the other two sectors are gapped.}\label{table:columnar spectrum}
\end{table*}

For the special angles $\eta = 0$ and $\eta = \pi/2$ the propagators obtain a simple form. For momentum in the liquid direction $\eta =0$, we find that $G_{\tL}$ is like a pure solid, Eq.~\eqref{eq:longitudinal propagator}, while $G_{\tT 1}$ and $G_{\tT 2}$ both take exactly the smectic form Eq.~\eqref{eq:GT1 eta0 smectic}. The third transverse propagator $G_{\tT 3}$ vanishes identically. The columnar phase looks just like a 2D smectic since the two transverse phonons are gapped in same way due to the axial symmetry. For momentum in the solid direction $\eta = \pi/2$, $G_\tL$ and $G_{\tT 2}$ are the pure solid longitudinal resp. transverse propagators Eqs.~\eqref{eq:longitudinal propagator},\eqref{eq:transverse propagator}, while $G_{\tT 1}$ is again identical to the smectic and is a pure gapped shear mode as in Eq.~\eqref{eq:smectic pi/2 gapped shear}. The third sector $G_{\tT 3}$ contains a pure gapped condensate mode. Indeed, we do not find a rotational Goldstone mode here, as expected.

%% file: sec_charged.tex
So far we have only discussed the quantum liquid crystal phases of electrically neutral bosonic matter. As we will now demonstrate, it  it is rather straightforward to couple the bosonic medium to the electromagnetic (photon) field thereby describing charged (liquid-)crystalline matter. In this scenario, we depart from a bosonic `Wigner crystal' formed out of electrically charged bosons, and proceed with the duality construction keeping track of the electromagnetic fields. After integrating out the stress variables, the end result is the effective electromagnetic response of the medium, including the quantum liquid-crystalline phases.
This machinery was already established in the first paper dealing with the 2+1D case~\cite{ZaanenNussinovMukhin04}, and an expanded up-to-date discussion can be found in QLC2D.  In 2+1D the `stress photons' are described in terms of one-form 
gauge fields and these combine naturally with the electromagnetic one-form gauge fields. In 3+1D the only difference in the formalism is that the stress photons become two-form fields and one faces the question 
of how these consistently couple to the one-form EM fields. We explored this problem already in the context of how vortices in 3+1D superconductors interact with EM fields, discovering a surprisingly efficient and economic 
formalism to describe the electrodynamics of Abrikosov vortices~\cite{BeekmanZaanen11}. We will see in this section that the same machinery applies equally well in the present context.   

We will first find out of how to incorporate electromagnetism in the theory of quantum elasticity in 3+1D. This contains the hard work, after which the EM fields can be rather effortlessly ``pulled through the duality''. In 2+1D we found a series of interesting surprises, but it turns out that with regard to the gross physics, 3+1D follows the pattern we found in 2+1D~\cite{QLC2D}. For this reason we have kept this section  rather concise. We discuss
in some detail the formalism, and when we arrive at the physics we just highlight the novelties tied to three space dimensions, referring the reader to QLC2D for further details.
 
\subsection{Charged dual elasticity}\label{subsec:Charged dual elasticity}

In QLC2D we have shown how to incorporate the electromagnetic interactions in quantum elasticity, leading to the coarse-grained long-wavelength description of the Wigner crystal. This qualifies to be textbook material. One profits here optimally from the stress representation of elasticity, yielding a highly transparent description. Photons are the carriers of the EM force and one would better dualize the phonons into stress photons so that apples are compared to apples. The outcome is a simple linear mode-coupling affair. 

The starting point is the coupling between displacements $u^a$ and the electromagnetic field, which is derived in QCL2D to be of form, 
\begin{equation}
 S_\mathrm{EM} = - \int \td \tau \td^D x \; j^\imath_\mu A^\imath_\mu \equiv - \int \td \tau \td^D x \; \mathcal{A}^a_\mu \partial_\mu u^a,\label{eq:EM coupling}
\end{equation}
where
\begin{align}
 j^\imath_\ft &= - \ti n e^* c_\tT \partial_m u^m, &
 j^\imath_m   &=   \ti n e^* c_\tT \partial_\ft u^m,\\
 A^\imath_\ft &= \ti \frac{1}{c_\tT} V, &
 A^\imath_m   &= A_m,
 \end{align}
 and defining the vector potential that couples to displacements,
 \begin{align}
 \mathcal{A}^a_\mu &= n e^* c_\tT (\delta_{\mu a} A^\imath_\ft  - \delta_{\mu \ft} A^\imath_a).
\end{align}
Here the label $^\imath$ denotes that the temporal components are rescaled by a factor of $\ti$ to get Euclidean products in imaginary time, e.g. $V^\imath = - \ti V$ for the scalar Coulomb potential. Furthermore $n$ is the density of (charged) particles and $e^*$ is the charge of the constituent particles (e.g., $e^*= 2e$ for Cooper pairs of electrons). 

In the EM coupling term, the usual electromagnetic gauge invariance $A^{\imath}_{\mu} \to A^{\imath}_{\mu}+ \partial_{\mu} \lambda$ is equivalent to conservation of particle number $\partial_{\mu} j^{\imath}_{\mu} = 0$, which we already identified as the glide constraint, i.e. the conservation of charge for a charged elastic medium. Therefore EM gauge invariance is guaranteed in the path integral for the dual gauge fields via the glide constraint, even in the dislocation-condensed phases.

For a charged elastic medium, the EM interaction term Eq.~\eqref{eq:EM coupling} simply has to be added
to the original quantum-elasticity action Eq.~\eqref{eq:solid action}.
After this, the EM field $A_\mu$ will just be `carried along' in the strain-stress duality transformation explained in Sec.~\ref{sec:Dual elasticity in three dimensions}.
In this way, the EM field $A_\mu$ will just be `carried along' in the duality. This results in an expression describing a simple linear coupling between the stress fields and the EM gauge fields~\cite{QLC2D}. The dual Lagrangian is
\begin{align}
\mathcal{L}^\mathrm{EM}_\mathrm{dual} &= \mathcal{L}^\mathrm{neutral}_\mathrm{dual} + \mathcal{L}_\mathrm{Meissner} + \mathcal{L}_\mathrm{int} + \mathcal{L}_\mathrm{Maxw},\label{eq:general dual EM Lagrangian}\\
 \mathcal{L}_{\rm Meissner}&=\tfrac{1}{2}  \mathcal{A}^{a \dagger}_m C^{-1}_{mn ab} \mathcal{A}^b_n  + \frac{1}{2} \frac{1}{\mu}\mathcal{A}^{a \dagger}_\ft \mathcal{A}^a_\ft, \nonumber\\
 &= \tfrac{1}{2} \varepsilon_0\omega_\mathrm{p}^2 \Big(  (A^\imath_a)^2 + \tfrac{\mu}{\kappa} (A^\imath_\ft)^2 \Big).\label{eq:Meissner in hiding}\\
 \mathcal{L}_\mathrm{int} &=
 - \sigma^a_m  C^{-1}_{mn a b} \mathcal{A}_n^b - \frac{1}{\mu} \sigma^a_\ft \mathcal{A}^a_\ft \nonumber\\
 &=  -\frac{n e^* c_\tT}{D \kappa} \sigma^a_a A^\imath_\ft + \frac{ne^* c_\tT}{\mu} \sigma^a_\ft A_a. \label{eq:EM-stress interaction}
 \end{align}
 Here $\mathcal{L}^\mathrm{neutral}_\mathrm{dual}$ is just the stress Lagrangian of the neutral elastic medium, for instance Eq.~\ref{eq:dual solid Lagrangian}, and $\mathcal{L}_\mathrm{Maxw} = \frac{1}{4\mu_0} (\partial_\mu A_\nu - \partial_\nu A_\mu)^2$, where $\mu_0$ is the magnetic constant. The term $\mathcal{L}_\mathrm{Meissner}$, which has the form of a Meissner term giving a mass to the photon field, arises automatically in the duality construction; in the solid it is exactly canceled by elastic terms while in the liquid crystals it can remain to cause the real Meissner effect indicative of superconductivity. In this sense, the Meissner effect is already ``lying in wait'' in the crystal to become  manifest when shear rigidity is destroyed~\cite{ZaanenNussinovMukhin04,QLC2D}. 
 Eq.~\eqref{eq:EM-stress interaction} is valid in all dimensions. However, when we wish to express the stress tensors in terms of dual stress gauge fields, in 3+1D one encounters the two-form fields: 
 $\sigma^a_\mu = \epsilon_{\mu\nu\kappa\lambda}\partial_\nu \tfrac{1}{2} b^a_{\mu\nu}$. Upon substitution one finds,
\begin{align}
\mathcal{L}_\mathrm{int} &= \tfrac{1}{4} b^{a\dagger}_{\mu\nu} g^a_{\mu\nu,\lambda} A^\imath_\lambda +  \tfrac{1}{4} A^{\imath\dagger}_\lambda g^{a\dagger}_{\lambda,\mu\nu} b^a_{\mu\nu}\nonumber\\
&= \tfrac{1}{4}\tilde{b}^{a\dagger}_{\mu\nu} \tilde{g}^a_{\mu\nu,\lambda} A^\imath_\lambda +  \tfrac{1}{4} A^{\imath\dagger}_\lambda \tilde{g}^{a\dagger}_{\lambda,\mu\nu} \tilde{b}^a_{\mu\nu}
\end{align}

where $g^{a\dagger}_{\lambda,\mu\nu} = (g^a_{\mu\nu,\lambda})^*$, and as explained in Eq.~\eqref{eq:tilde fields definition} the variables with a tilde are rescaled with the dislocation velocity $c_\td$. The $\tilde{g}^{a}_{\lambda,\mu\nu}$ are momentum-dependent coefficients tabulated in matrix form as
\begin{widetext}
\begin{align}
 \tilde{b}^{a\dagger}_{\mu\nu} \tilde{g}^a_{\mu\nu,\lambda} A^\imath_\lambda &= 
 \begin{pmatrix} \tilde{b}^{ \dagger}_{1-} &  \tilde{b}^{\tL\dagger}_{\tR \tS} &
 \tilde{b}^{\tR \dagger}_{\tR \tS } & \tilde{b}^{\tS\dagger}_{\tR\tS}
 \end{pmatrix}
ne^* \begin{pmatrix}
  -\frac{\sqrt{2}}{D\kappa} c_\td \tilde{p}  & 0 & 0 & 0 \\
  -\ti \frac{1}{D\kappa} \omega_n & \frac{1}{\mu} c_\tT q & 0 & 0 \\
  0 & 0 & \frac{1 }{\mu} c_\tT q & 0 \\
  0& 0 & 0 & \frac{1}{\mu} c_\tT q 
 \end{pmatrix}
\begin{pmatrix}
 A_\ft \\ A_\tL \\ A_\tR \\ A_\tS
\end{pmatrix}.\label{eq:b-A coupling matrix tilde}
\end{align}
\end{widetext}
This follows the pattern set in the 2+1D case: collecting the two-form indices $\mu,\nu$ in a vector, the coupling to the EM field is captured by just a matrix, and is linear. The other components of $b^a_{\mu\nu}$ do not couple to $A_\mu$. From this form, we can immediately infer that the $b^\tR_{1\tR}$--$b^\tS_{1\tS}$ sector ($\tT 3$) does not couple to electromagnetic fields. The `longitudinal'
rotational Goldstone mode of the quantum nematic with its vanishing propagator in the solid resides in this sector and is therefore completely invisible to electromagnetic means. The matrices $g^a_{\mu\nu,\lambda}$ are all that is necessary to calculate the EM response given the dual stress propagators of the neutral elastic medium, as we shall see now. 

To this end, observe that the general form of the stress part of the Lagrangian is 
$\mathcal{L} = \tfrac{1}{8}  \tilde{b}^{a\dagger}_{\mu\nu} ( G^{-1})^{ab}_{\mu\nu,\kappa\lambda}  \tilde{b}^b_{\kappa\lambda}$. Define the matrix $G$ via $(G^{-1})^{ab}_{\mu\nu,\kappa\lambda} G^{bc}_{\nu\rho,\lambda\sigma} = \delta_{ac}\delta_{\mu\rho}\delta_{\kappa\sigma}$. The stress gauge fields can now be integrated out, and we learn how these dress the electromagnetic fields: 
  \begin{align}\label{eq:stress contribution to EM fields}
  \mathcal{L}_{\rm para}  &=  \tfrac{1}{8}  \tilde{b}^{a\dagger}_{\mu\nu} (G^{-1})^{ab}_{\mu\nu,\kappa\lambda}  \tilde{b}^b_{\kappa\lambda} + \tfrac{1}{4}  \tilde{b}^{a\dagger}_{\mu\nu}  \tilde{g}^a_{\mu\nu,\lambda} A^{\imath}_\lambda + \tfrac{1}{4} A^{\imath\dagger}_\lambda  \tilde{g}^{\dagger a}_{\lambda,\mu\nu}  \tilde{b}^a_{\mu\nu} \nonumber\\
  &= - \tfrac{1}{8} A^{\imath \dagger}_\rho  \tilde{g}^{a\dagger}_{\rho,\mu\nu} G^{ab}_{\mu\nu,\kappa\lambda}  \tilde{g}^b_{\kappa \lambda,\sigma} A^\imath_\sigma.
 \end{align}
This must be added to the `diamagnetic' or `Meissner' contribution already present in the dual Lagrangian Eq.~\eqref{eq:Meissner in hiding}. As usual, the shear velocity $c_\tT^2 = \mu / \rho$ while the plasmon frequency $\omega_\mathrm{p}$ is defined as
\begin{equation}
 \omega_\mathrm{p}^2  = \frac{(n e^*)^2}{\rho \varepsilon_0}.
\end{equation}
Finally, $\varepsilon_0$ is the dielectric constant in units of $[\varepsilon_0] = \frac{\mathrm{C}^2}{\mathrm{J}\;\mathrm{m}}$. 

Using Eq.~\eqref{eq:stress contribution to EM fields} the only extra input that is needed are the stress propagators $G^{ab}_{\mu\nu,\kappa\lambda}$ for the {\em neutral} elastic medium and we can directly compute the
electromagnetic propagator $\langle A^\dagger_\mu A_\nu \rangle$ enumerating the electromagnetic response, where the elastic medium just translates into the usual photon self-energy. 
The effective EM action due to the medium is of  the form $\mathcal{L}^{\mathrm{EM}}_{\mathrm{medium}} = \mathcal{L}_\mathrm{para} + \mathcal{L}_\mathrm{Meissner}$, to be added to the vacuum Maxwell action which in our units reads:
\begin{equation}
 \mathcal{L}_\mathrm{Maxw} = \tfrac{1}{2}\varepsilon_0 c_\tT^2 q^2 \lvert A^\imath_\ft \rvert^2 + \tfrac{1}{2}\varepsilon_0 (\omega_n^2 + c_l^2 q^2) \big( \lvert A^\imath_\tR \rvert^2 + \lvert A^\imath_\tS \rvert^2 \big).
 \label{eq:Maxwell action}
\end{equation}
Here we have taken the Coulomb gauge fix for the EM field, removing $A_\tL$. The shear velocity $c_\tT$ shows up because of our definition of the temporal components, and $c_l$ is the speed of light. The full effective EM action $\mathcal{L}^\mathrm{EM}_{\mathrm{eff,medium}} = \mathcal{L}_\mathrm{Maxw} +  \mathcal{L}^{\mathrm{EM}}_{\mathrm{medium}}$ and it is straightforward to compute the full portfolio of EM response from this action, as we will now show. 

With the knowledge that the electromagnetic response can be straightforwardly derived from the neutral elastic propagator, we can translate these results into observable quantities. We will work in real time and real frequencies $\omega$ for the remainder of this section. Given the photon propagator that follows from $\mathcal{L}^\mathrm{EM}_{\mathrm{eff,medium}}$, we can define the photon self-energy $\Pi_{mn}(\omega,q)$ via:
 \begin{align}
 \langle A^{\dagger}_m(\omega, q) A_n(-\omega,-q) \rangle = \frac{1}{\varepsilon_0}\frac{1}{\omega^2 - c_l^2 q^2 -\Pi_{mn} (\omega, q)}.
\label{photonpropdef}
\end{align}
For the diagonal components of the self-energy we write $\Pi_m = \Pi_{mm}$ (no sum).
The conductivity tensor $\hat{\sigma}_{mn} (\omega,q)$ and the dielectric function $\hat{\varepsilon}_{mn}(\omega,q)$ are defined as:
\begin{align}
 \hat{\sigma}_{mn} (\omega,q) &= -\ti \omega \left(\hat{\varepsilon}_{mn}(\omega,q) -\varepsilon_0\right),\nonumber\\
 &= \varepsilon_0\frac{\ti}{\omega} \Pi_{mn} (\omega,q).
\label{eq:optical conductivity}
\end{align}
Furthermore, in QLC2D we showed it is useful to define an energy- and momentum-dependent penetration depth $\lambda_m(\omega,q)$ which characterizes the screening (exponential decay) of the photon component $A_m$ in the medium, which is either due to the skin effect or to the Meissner effect, as follows:
\begin{equation} \label{eq:penetration depth from self-energy}
  \lambda_m(\omega,q) = \frac{c_l}{\mathrm{Im} \sqrt{\omega^2 - \Pi_{mm}}}.
  \end{equation}
When the penetration depth stays finite in the limit $\omega \to 0$ we will interpret this as a genuine Meissner effect indicating the presence of superconductivity. 

For the longitudinal EM response, we calculate the photon propagator in the Coulomb gauge to find from Eq.~\eqref{eq:stress contribution to EM fields} and \eqref{eq:stress gauge field components} that $ \langle A^{\dagger}_{\ft}(\omega,q) A_{\ft}(-\omega,-q) \rangle_\mathrm{para} = \varepsilon_0 \omega_\mathrm{p}^2 \frac{1}{(D \kappa)^2}\langle \sigma^a_a \; \sigma^b_b \rangle$. In fact the total propagator including the Meissner term is simply proportional to the longitudinal propagator. This is immediately derived from Eq.~\eqref{eq:EM coupling}, and we find~\cite{QLC2D}
\begin{equation}
 \langle A^{\dagger}_{\ft}(\omega,q) A_{\ft}(-\omega,-q) \rangle = \varepsilon_0 \omega_\mathrm{p}^2 \mu G_\tL.
\end{equation}
This implies that the longitudinal EM response just coincides with the longitudinal elastic propagator of the medium. For this reason we shall not address this response much further below. We only wish to emphasize that the special features of quantum liquid crystals, such as the appearance in the longitudinal dielectric function of a second, gapped pole due to the dislocation condensate next to the ordinary plasmon, is only noticeable at finite momentum. All these poles have vanishing spectral weight for momentum going to zero. For this reason, finite-momentum spectroscopy such as {\em electron energy-loss spectroscopy} (EELS) is the only way to observe these features. In fact, new machines which have proper resolution at finite momentum seem to be coming online at present~\cite{VigEtAl15}.

For the transverse EM response, all the quantities in Eqs.~\eqref{eq:optical conductivity},\eqref{eq:penetration depth from self-energy} are simply related. We shall give results in the form of poles of the photon propagator itself, the transverse conductivity or the penetration depth, whichever provides the clearest picture.

\subsection{The electrodynamics of the Wigner crystal}

As we already alluded to above, with our initial assumptions, the theory of quantum elasticity of charged matter describes a generic isotropic ``Wigner'' crystal formed from charged bosonic constituents. Nevertheless, in the ordered crystalline phase, the wisdoms in this section are entirely general: they apply as well to a `conventional' Wigner crystal of electrons as to any other medium of this kind. Surely, the dualization can be carried out only for a crystal formed out of bosons. In the condensed-matter context one could envisage a crystal formed from ``preformed'' Cooper pairs that are subsequently subjected to quantum melting. This may be of relevance to e.g. the `stripy' charge order found in cuprate superconductors~\cite{FradkinKivelsonTranquada15,BergFradkinKivelson09,HamidianEtAl16}.

In QLC2D we derived the electrodynamics of the isotropic Wigner crystal using the duality transformation detailed in Sec.~\ref{subsec:Charged dual elasticity}.  It is a rather classic subject but to our surprise we found several 
novelties, it seems all related to the remarkably efficiency of the stress formalism.  The longitudinal channel is according to expectations: the longitudinal phonon acquires a plasmon gap --- it is not quite the standard plasmon
because it is still propagating with the longitudinal phonon velocity which is larger than the sound velocity of a liquid ($c_\tL$ vs. $c_\kappa$). The longitudinal optical conductivity is characterized by an infinitely sharp Drude peak at zero frequency, indicating
that the Wigner crystal is a perfect conductor (but not a superconductor, see below). This has to be the case since our elastic medium is supposed to live in a perfect Galilean continuum where all matter at finite density is perfectly conducting since momentum is conserved. In impurity language: the crystal as a whole is `unpinned' and can be `set in motion' by an infinitesimal external potential.

The surprises are in the transverse response. The finite-frequency EM photon exerts a shear force on the crystal which responds by a reactive restoring force encapsulated by the transverse phonons/stress photons
expelling the photon: the Wigner crystal is characterized by a skin
depth $\lambda(\omega,\mathbf{q})$. What is the difference with a superconductor where the same photon ``acquires a mass''? The stress formalism exhibits an elegant view on this affair. As we already announced in the previous section, a literal Meissner term is present in the effective action: Eq.~\eqref{eq:Meissner in hiding}. However, in order to determine the response of the medium there is a second term due to the stress photons in Eq.~\eqref{eq:stress contribution to EM fields}. One finds that at precisely zero 
frequency $\omega \to 0$, this produces a term that exactly cancels the bare Meissner term. In the DC limit the penetration (skin) depth becomes infinite and static magnetic fields can penetrate the whole solid. In the quantum liquid crystals on the other hand, the transverse stress photons acquire a mass because of the dual dislocation condensate. The effect is that the cancellation is no longer complete and static magnetic fields are expelled, proving that these are indeed conventional superconductors as well exhibiting a regular Meissner effect. 

Last but not least, what is happening to the transverse phonons of the neutral crystal when we couple in electromagnetism? The transverse motions are surely exempted from the plasmon mechanism, but these do correspond to fluctuating electrical dipoles. Surprisingly, the dispersion of the `dressed' transverse phonons turns out to become {\em quadratic} in the charged crystal.  Keeping in mind that we consider $D+1$-dimensional electromagnetism coupled to a $D+1$-dimensional medium one expects that these behaviors should not depend on dimensionality. This is indeed the case, as we now will demonstrate.  Consider the 3+1D elasticity Lagrangian 
Eq.~\eqref{eq:solid stress gauge field Lagrangian} and the $G^{ab}_{\mu\nu,\kappa\lambda}$ follow immediately by inverting the Gaussian kernel.  By substituting these in Eq.~\eqref{eq:stress contribution to EM fields}we obtain
$\mathcal{L}_\mathrm{para}$. Adding  $\mathcal{L}_\mathrm{Meissner}$ from Eq.~\eqref{eq:Meissner in hiding}  the contribution of the elastic medium to the EM action is obtained which reads in the Coulomb gauge  ($A_\tL = 0$),
\begin{align}
 \mathcal{L}^\mathrm{EM}_\mathrm{solid}
 &= \tfrac{1}{2}\varepsilon_0 \omega_\mathrm{p}^2 \Big(  \frac{c_\tT^2 q^2}{\omega_n^2 + c_\tL^2 q^2} \lvert A^\imath_\ft \rvert^2 \nonumber\\
 &\phantom{mmmmmm}+ \frac{\omega_n^2 }{\omega_n^2 + c_\tT^2 q^2} \big( \lvert A^\imath_\tR\rvert^2 +\lvert A^\imath_\tS\rvert^2\big) \Big).
\end{align}
Taking into account the Maxwell action Eq.~\eqref{eq:Maxwell action} the full photon propagator is straightforwardly obtained. This yields two degenerate transverse propagators $\langle A^{\imath\dagger}_\tR\; A^\imath_\tR \rangle$ and $\langle A^{\imath\dagger}_\tS\; A^\imath_\tS \rangle$. These are like in 2+1D characterized by two poles with dispersion relations at small momentum:
\begin{align}
\omega_{1} & = \sqrt{ (\omega^2_\mathrm{p} + (c_l^2+c^2_\tT)q^2} \simeq \sqrt{\omega_\mathrm{p}^2+c_l^2 q^2} + \mathcal{O}(q^2),\nonumber \\
\omega_{2} & = \frac{c_\tT c_l q^2}{\sqrt{ (\omega^2_\mathrm{p} + (c_l^2+c^2_\tT)q^2)}} \simeq \frac{c_l c_\tT}{\omega_\mathrm{p}} q^2 + \mathcal{O}(q^4).
\label{Xtallongwave}
\end{align}
assuming $c_l \gg c_\tT$. The first mode is the familiar plasma-polariton characterized the plasmon energy $\omega_\mathrm{p}$.  The second mode is massless and characterized by a quadratic dispersion, the surprise 
we already encountered in 2+1D. The take home message is that except for the fact that there are two photon polarizations 3+1D (instead a single one in 2+1D) everything else is independent of dimensionality.

\subsection{The superconducting nematic}\label{subsec:The superconducting nematic}

This dimensionality-independence of the electromagnetic effects extends also to the quantum liquid crystals. We refer to QCL2D for a detailed discussion of the electromagnetism in the quantum nematics. As we 
already discussed above, a highlight is the way that the Meissner effect shows up, unambiguously proving that the charged dual stress superconductor is at the same time a regular superconductor. This important matter is easy to verify to hold in 3+1D as well. The penetration depth can be computed from Eq.~\eqref{eq:penetration depth from self-energy}. The full expression is lengthy but 
the main interest is in the $\omega =0$ limit that determines whether we are dealing with a Meissner phase: 
\begin{equation}
 \lambda_{\tR,\tS}^\mathrm{nem} (0,q) = \frac{c_l}{\omega_\mathrm{p}} \sqrt{1 + 2 c_\tT^2 q^2/\Omega^2} \equiv \lambda_\tL \sqrt{1 + \lambda_\mathrm{s}^2 q^2}.
\end{equation}
Here we have defined the London penetration depth as $\lambda_\tL = c_l / \omega_\mathrm{p}$ and the `shear penetration depth'  as $\lambda_\mathrm{s} = \sqrt{2} c_\tT / \Omega$. At $q= 0$, the penetration depth is 
exactly $\lambda_\tL$,  demonstrating that the quantum nematic is a superconductor. These results are identical to those obtained in 2+1D\cite{QLC2D}. Since the quantum nematic is isotropic, the result holds for both $\tR,\tS$-polarizations of the photon field.

There is a lot  more going on, such as the way that the massive shear photons acquire electromagnetic weight and thereby becoming measurable in principle by electromagnetic experiments. However, this is by and large rather independent
of dimensionality as we just discussed for the Wigner crystal itself. 
These propagating massive modes are of course special to our ``maximally-correlated limit''. This is different for the massless modes which behave universally, independent of the degree of the microscopic correlations. We already saw this at work in the Wigner crystal: the quadratic dispersion of the dressed transverse phonon has to apply universally.  Besides the sound mode, the neutral quantum nematic is characterized by the rotational Goldstone bosons/torque modes
as we demonstrated in Sec.~\ref{subsec:Collective modes in the quantum nematic}. As we repeatedly emphasized the sound mode acts as the phase mode of the superconductor ``eating the photon'', and this works in the same way as in 2+1D. However, more is going on in the rotational  Goldstone sector in 3+1D as compared to the 2+1D case. Instead of the single transverse torque mode of the latter, in three dimensions symmetry just imposes that there are three modes. As we discussed, these are polarized like phonons: there are two degenerate transverse modes but in addition there is now also a `longitudinal rotational phonon'.  The qualitative novelty is that this longitudinal mode does {\em not} couple to electromagnetic fields, as we mentioned 
below Eq.~\eqref{eq:b-A coupling matrix tilde}. Imagining that our quantum liquid crystals would be formed from electrons, this has the interesting consequence that this mode cannot be observed directly since the only way to apply external forces is through its electrical charge. The two 'transverse rotational phonons' are coupled to the two transverse photon polarizations, and show up as massless poles in the photon propagator, identical to the 2+1D case. For instance, in the optical conductivity one finds a zero-frequency Drude peak, the massless rotational phonon and a gapped shear mode~\cite{QLC2D}.

\begin{figure*}[t]
 \includegraphics[width=7cm]{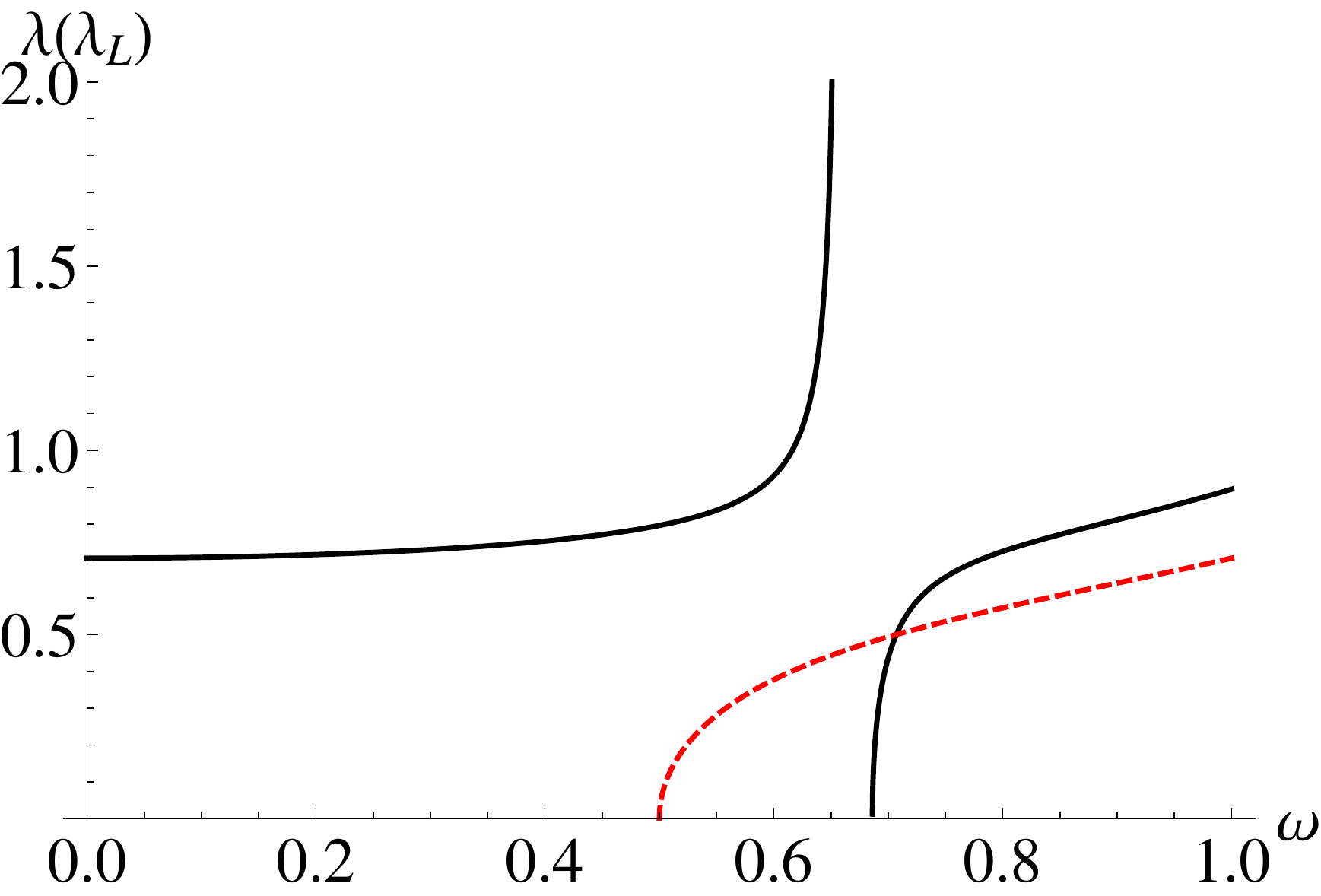}
 \hspace{1cm}
 \includegraphics[width=7cm]{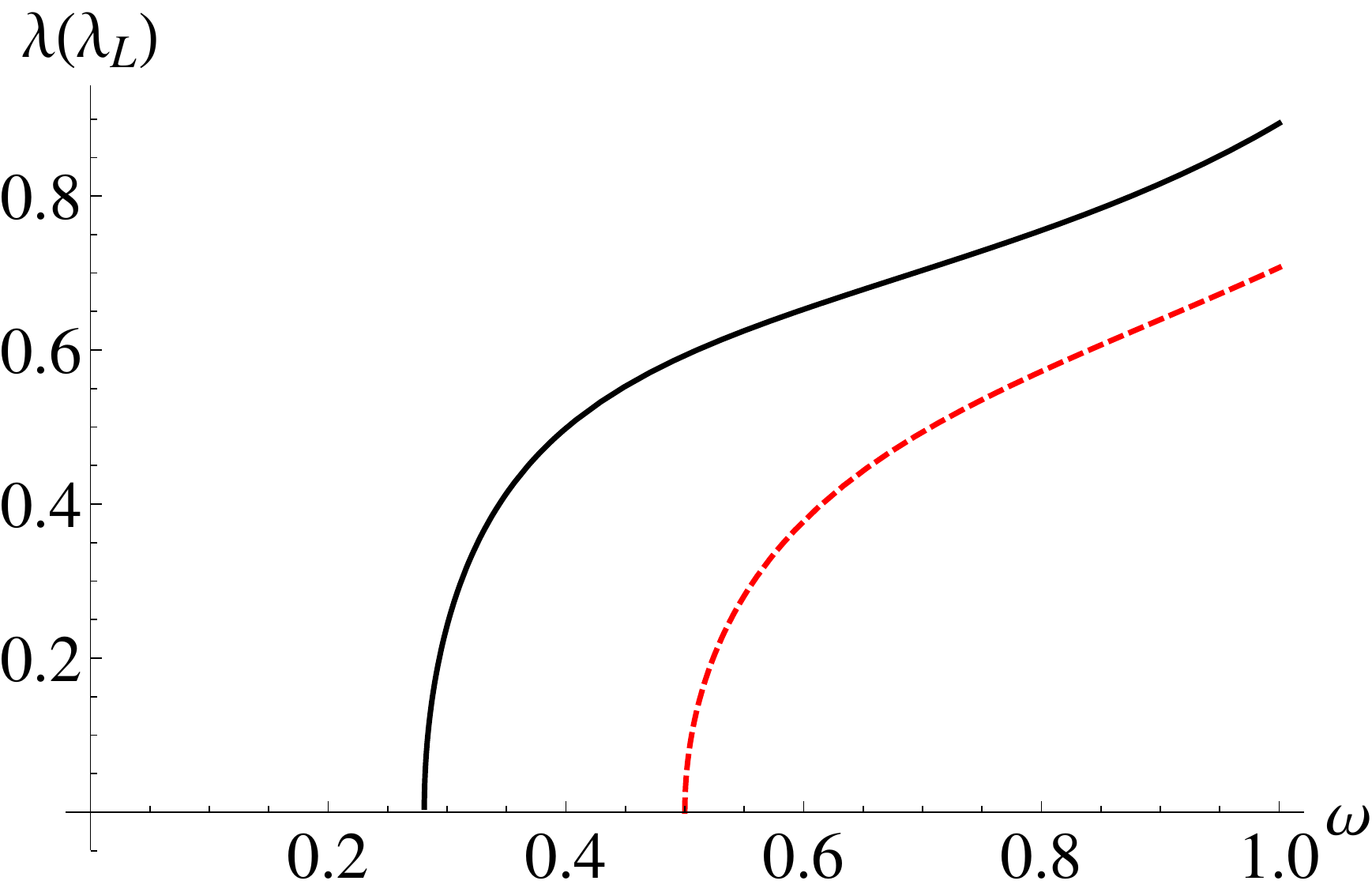}
 \caption{Penetration depth in the smectic (solid black lines) at $q=0.5$ as a function of energy $\omega$ for $\eta = 0$, left: $\tR$-polarization; right: $\tS$-polarization. For comparison we have plotted the same quantity for the Wigner crystal (dashed red line).
 For $\eta = 0$, the $\tR$-polarization is within the liquid plane, and we find a finite penetration depth at $\omega \to 0$, indicative of the Meissner effect. Conversely, the $\tS$-polarization is perpendicular to the liquid plane, and the penetration depth appears only at finite energy, like the screening of AC magnetic fields in an ordinary conductor (skin effect). Nevertheless, the screening in this direction is stronger resulting in a shorter penetration depth than that of the Wigner crystal. The smectic shows very anisotropic superconductivity. As an aside, the divergence in the left plot is the resonance of the rotational Goldstone mode --- in general all poles of the photon propagator also show up in this `dynamic' penetration depth.
 }\label{fig:xy smectic Meissner}
\end{figure*}

\subsection{2D superconductivity in the smectics}\label{subsec:2D superconductivity in the smectics}

The story of the above repeats itself in the smectic and the columnar phases. In QCL2D we analyzed the electromagnetic responses of the 2+1D quantum smectics in detail, finding out that the already intricate mixture of solid-like and 
liquid-like behaviors gets further enriched when electromagnetism is added. Like for the solid and the nematic, the 3+1D case is by and large a further variation on the same theme, where the novelties are related to the fact that we now are dealing with either a `2+1D solid' with an `extra quantum-liquid direction' (the columnar phase), or a `2+1D quantum liquid' with an `extra solid direction': the smectic phase. There is a lot to explore here but we have not found anything that is qualitatively new. We leave it therefore for follow-up work in the future.

However, there is one aspect that deserves closer consideration given that it is highly relevant for the gross physics of these systems. As we discussed in Sec.~\ref{sec:smectic}, the liquid planes of the 3+1D smectics behave in many regards like 2+1D quantum nematics. The important implication is that in these directions the charged quantum smectics should be like two-dimensional superconductors while a Wigner crystal, i.e. perfect metal, behavior should occur in the `solid direction'. Let us demonstrate here that is indeed the case.   

Let us first focus on the transverse conductivity, given by Eq.~\eqref{eq:optical conductivity}. We use the same simplifications as in Sec.~\ref{sec:smectic} and we first consider momentum within the liquid plane, i.e. $\eta = 0$. The transverse conductivities are independent of the angle $\zeta$ within the plane due to the axial symmetry and read:
\begin{align}
 \hat{\sigma}_{\tR} &= \ti \omega_\mathrm{p}^2 \frac{1}{\omega} \frac{ \omega^2 (\omega^2 - c_\td^2 q^2) - \Omega^2 (\omega^2 - \frac{1}{2} c_\td^2 q^2)}{(\omega^2 - c_\tT^2 q^2 )(\omega^2 - c_\td^2 q^2) - \Omega^2 (\omega^2 - \frac{1}{2} c_\td^2 q^2)},\\
  \hat{\sigma}_{\tS} &= \ti \omega_\mathrm{p}^2  \frac{ \omega (\omega^2 - c_\td^2 q^2  - \Omega^2)}{(\omega^2 - c_\tT^2 q^2 )(\omega^2 - c_\td^2 q^2) - \frac{1}{2}\Omega^2 \omega^2 }.
\end{align}
This gives away the mixed character of the smectic. The first line addresses the response within the liquid plane and it is identical to that of the nematic showing three poles: the $\omega =0$ perfect conductor Drude peak, the gapped shear mode and the massless rotational Goldstone mode. The second line is the response orthogonal to the liquid plane. The transverse Drude peak is lacking, instead showing  the presence of the gapped condensate mode  and in addition a quadratically dispersing undulation mode.

What happens to the superconductivity? We computed the penetration depth for photons with polarization $\tR$ (in the liquid plane) and $\tS$ (orthogonal to the liquid plane), respectively, and the result is plotted in Fig.~\ref{fig:xy smectic Meissner}. For the $\tR$-polarization we find a finite penetration depth at $\omega =0$, and therefore the system is in a Meissner phase. On the other hand, the penetration depth for the $\tS$-polarization only sets in at finite energy as is usual for the ordinary, metallic skin effect, such that static magnetic fields can penetrate. This gives away that this smectic behaves like a strictly 2D superconductor where photon polarizations are screened in the liquid direction but not along 
the solid direction. Nevertheless, the skin effect for the $\tS$-polarization is enhanced as compared to the Wigner crystal with no dislocation condensate at all. 

The anisotropic superconductivity can also be studied as function of the interrogation angle $\eta$, showing that again there is only a Meissner response precisely  in the liquid plane. Upon tilting away from the planar direction the penetration depth diverges at zero frequency, while the skin depth decreases rapidly at finite frequencies. All of this confirms our earlier statements, that superfluidity and superconductivity exist whenever shear rigidity is lost.  A similar effect takes place in the columnar phases, where there are one-dimensional superconducting lines.

%% file: sec_conclusions.tex
Having arrived at the end of this long story, the question arises: ``what is it good for?'' It is a long story mainly because we had to explore an extended landscape riddled with a surprisingly rich set of phenomena, resting on a mathematical machinery that is remarkably efficient in charting out every corner of it.
The secret is that in the limit of `maximally correlated' quantum fluids all the physics is captured in terms of propagating collective modes. The seemingly intractable problem of a uniform superfluid (or superconductor) characterized by extremely strong crystalline correlations that continue to break the isotropy of space, turns out to be captured in a language of free fields subjected to linear mode couplings in addition to a run-of-the-mill Higgs mechanism. 

This gross picture was already established in 2+1D. The main novelty of this paper is that we addressed the question of principle whether such a theory of maximal solid-like correlations can also be formulated in 3+1D. This is in turn is rooted in a foundational affair: do Abelian weak--strong dualities actually make sense in 3+1D? Although well established in 1+1/2D and 2+1/3D, it is even controversial whether the very notion of a dual disorder field theory makes any sense in 3+1 (and higher) dimensions.  By just assuming the 
two-form minimal coupling form for the Higgs term due to the dislocation condensate, we found out that the theory is internally consistent while being a natural generalization of the 2+1D case. The way this works is far from trivial; the case in point is the way that the rotational Goldstone bosons of the quantum nematic as well as the massless modes of the smectic and columnar phases roll out of the formalism. A fine-tuned clockwork is here at work, producing answers that precisely satisfy a set of intricate symmetry requirements.      
These outcomes are critically dependent on the way that the `stringy' dislocation condensate is incorporated, and we assess this as evidence for the correctness of the two-form minimal coupling Higgs construction after Rey and others~\cite{Rey89,MotrunichSenthil05,Franz07,YeGu15}. We emphasize that here in elasticity, there is a way to probe the degrees of freedom of the stringy (dislocation) condensate itself through torque stress (see Sec.~\ref{subsec:Torque stress in the quantum nematic}), while this is not possible in for instance the ordinary Abelian-Higgs model employing only the original gauge fields.

Focusing in on the physics, as compared to the 2+1D case the third spatial dimension opens up space for more phenomena. In the first place, the much richer nature  of the 
three dimensional rotational (point-group) symmetry breaking as compared to the simple 2D case comes into play. An universal outcome that does not depend on specific assumptions
of the duality construction is the prediction that a superconducting nematic characterized by a truly 3D point group will carry two transverse and one longitudinal rotational Goldstone modes. Perhaps the most striking ramifications of the third dimension are found in the smectic and columnar phases, with their very rich spectrum of `intertwined' solid-like and liquid-like collective excitations.   
 
There is surely an issue with the relevance and timeliness of our findings in the present epoch of physics. At first sight it seems to deal with an old subject ---
the quantum melting of solids into superfluids/-conductors while the idea of quantum liquid crystals is rooted in a straightforward 
extension of classic soft-matter wisdoms to the zero-temperature realms.  Is it of experimental pertinence? This is problematic as well
because presently there is nothing available in the laboratories that appears to be obviously related to the forms of matter that 
we dealt with in this paper. Our substances are formed from bosons living in the spatial continuum. Waiting further progress in cold atom 
experimentation, the only physical system that is obviously of this kind is $^4$He. The microscopy of helium --- in  essence a quantized Van der Waals liquid ---
is just too simple to have room for vestigial quantum phases: it forms either a solid or an isotropic superfluid, separated by a hard first-order transition. 

The superconductors in strongly-interacting electron systems such as the cuprates and iron pnictides may be a different story. 
There is good evidence for electronic liquid-crystalline order in these systems~\cite{AndoEtAl02, Vojta09, OganesyanKivelsonFradkin01, BorziEtAl07,HinkovEtAl08, FradkinEtAl10,FradkinKivelson10,Fradkin12,ChuangEtAl10,ChuEtAl12,FernandesChubukovSchmalian14}. However, the trouble is now that electron liquids are formed in the 
background of an ionic lattice that breaks the translational and orientational symmetries of space explicitly.  The nematic order parameters get thereby ``pinned to the lattice'' with the effect that the massless modes as discussed in this paper will acquire anisotropy gaps. These may well be so large that the 
continuum limit results of this paper are of no relevance to the electronic nematics realized in the laboratory.  

Considering the ``massive stress photons'' that we predict, the situation is even more precarious. These occur only at finite energies, and their literal existence is critically dependent on the central assumption underlying this whole development: the `solid correlation length' $l_\mathrm{sol}$  in the liquid is assumed to be infinitely large compared to the lattice constant $a$. This limit is unphysical: it is just impossible to reach it departing from the microscopic building blocks of condensed matter physics 
(electrons, atoms). The problem we have not addressed yet is: how to perturb away from this `maximal correlation' limit? The obvious small parameter is $a/l_\mathrm{sol}$ and 
is also clear that in a way the limit is singular. For finite $l_\mathrm{sol}$ one has to accommodate the interstitials (loose constituent bosons).  Departing from the maximally-correlated limit it is in principle clear how to proceed, namely by activating perturbation theory: loop corrections associated with (anti)dislocations have to be inserted, realizing that interstitials are associated with tightly bound dislocation--antidislocation pairs. These perturbative corrections have to be relevant: their effect has to be that the propagating massive stress photons will get overdamped at sufficiently small momenta. 

The significance of this theory of ``dual stress superconductivity'' is therefore not in its literal applicability, but in the sense that in physics it is without exception very useful to know the limits. The textbooks revolve around the opposite limit: the quantum `kinetic gas' theory departing from weakly-interacting, nearly free constituents. The case in point is the Bogoliubov treatment of the Bose condensate. In order to get into the regime where $l_\mathrm{sol}$ becomes a couple of times the lattice constant, perturbative treatment already becomes hazardous. The case in point is once again $^4$He, specifically the roton minimum in its density excitation spectrum. According to Bogoliubov theory this is associated 
with the cross over from the linear collective (second) sound to the free (quadratic) propagation of the gaseous bosons. However, resting on the single-mode approximation Feynman already put forward that it is associated with solid-like correlations in the liquid. Viewing it from the stress-superconductivity limit,  one observes the presence of the massive shear stress photon next to the second sound in the longitudinal channel (Fig.~\ref{fig:nematic spectral functions}).  As we argued in the previous paragraph, this will get corrupted by the interstitials in the regime $l_\mathrm{sol} \sim 2$--$3 a$ characterizing superfluid helium, but could it be that the roton can be alternatively viewed as the remnant of this shear stress mode? Now would like to interrogate how the roton reacts to dynamical shear stress, an experiment that has to the best of our knowledge never been carried out. The situation in the candidate electron system 
is actually much worse. As we discussed in detail in Sec.~\ref{sec:Charged quantum liquid crystals}, when there would exist such strong solid-like correlations in the superconductors, they are very well hidden from the eye of the experimentalist for any conventional form of experimentation. Presently optical spectroscopy is the only source of information on the electromagnetic response and since 
this is restricted to very small momenta it is completely insensitive to the solid-like correlations. High-resolution electron loss spectroscopy is in this regard very promising and the experiments
that are just coming on line may shed light on these matters in the near future~\cite{VigEtAl15}.

We have laid the groundwork for the description of strongly-correlated  bosonic quantum liquid crystals, but it also suggests a wealth of further problems to study:
\begin{itemize}
\item
One would like to get beyond the `isotropic' limit we have assumed all along. We do not surmise that any problem of principle will arise: lowering the  point group symmetry will 
just introduce more complicated details without affecting the gross workings of the formalism. In a first step, one should depart from the elasticity theory associated with a particular 
space group: these are available in tabulated form. One then should take care that the dislocation condensate itself is appropriately modified by the point group anisotropy, introducing 
further parameters. The outcomes for the effective rotational elasticity theory should then be matched with the known results for this theory pending the particular point groups.
\item
Perhaps most pressing, one would like to formulate the perturbation theory associated with the finite $l_\mathrm{sol}/a$ regime. As we just argued, the limit where $l_\mathrm{sol}/a \rightarrow \infty$
might well be singular with regard to the propagating nature of the massive stress photons. More generally, in order to apply wisdoms obtained in the limits to the intermediate-coupling regime where nature resides, one needs to know also what happens in the vicinity of the limits. It is clear on should start by formulating the diagrammatics, but it may well turn into quite a challenge given the already complex structure of just the `tree level' theory of the stress superconductor.
\item
It is an interesting challenge to find out how to dualize `backwards', i.e. describing the solid as a condensate of the topological excitations of the nematic superconductor. It is well 
understood how to accomplish this in the case of the boson--vortex duality, but it is a bit of conceptual challenge in the context of the present elasticity duality. Surely it cannot be a proliferation of disclinations, which would lead to a more-disordered state, namely the isotropic superfluid, see also the next point. Instead, since quantum liquid crystals are also genuine superfluids/superconductors, it is interesting to study the $U(1)$-vortices in the superfluid/superconducting order by themselves, which has not been done in the quantum liquid-crystals context. The main question is how to wire in the information required to recover the crystal? 
\item
Even more challenging: how to formulate the weak--strong duality associated with the melting of the quantum nematic into an isotropic superconductor? We have not yet attempted to formulate 
such a duality in the 2+1D case. In principle this does not appear to be particular challenging~\cite{Kleinert83,Kleinert89b}. The rotational parts of 2+1D point groups are Abelian with the consequence that the  disclinations behave as Abelian defects as well: these will form a simple dual Bose condensate that will restore the full rotational $O(2)$ symmetry. However, in 3+1D hell breaks loose. The generic 3D point groups are non-Abelian.
If the disclinations were point particles (and they are not) one would face complications such as a non-Abelian braiding. But on top of that, in 3+1D disclinations are non-Abelian {\em strings} and very little is known regarding their basic topological properties let alone that it is clear how to construct `stringy' condensate field theories. 
\item 
Even in the case of the Abelian dislocation condensates in 3+1D, we had to rely on guesswork related to the nature of the effective field theory describing the properties of this condensate. 
However these can be studied from first principles. Lacking any form of controlled analytical mathematics one can just mobilize the computer. The 2+1D superfluid--superconductor duality 
has been studied with much success in this way, addressing subtle quantitative issues even in the difficult critical regime~\cite{NguyenSudbo99,HoveSudbo00,HoveMoSudbo00,SmisethSmorgravSudbo04,SmisethEtAl05,SmorgravEtAl05}. The basic numerical methodology used in 2+1D can be straightforwardly 
generalized to 3+1D, to just find out what the computer has to tell about the nature of the universal long-wavelength theory.     
 \end{itemize}
 
 Perhaps the most interesting challenge is to find out a formulation of this `elastic' weak--strong duality that is so simple that it can be incorporated in the elementary text books of condensed 
  matter physics. We are of the strong opinion that this is most desired, as an antidote against the misleading aspects associated with the weak-coupling `gaseous'  physics that monopolizes 
 present day wisdom.

%% file: sec_fouriercoordinates.tex
We often use coordinate systems in Fourier--Matsubara space, in which the components are parallel or orthogonal to 1) the momentum $\mathbf{q}$ or 2) the Euclidean spacetime momentum $p_\mu = (\omega_n , q_m)$. These coordinate systems are only valid at finite momentum. 

The first system has the time coordinate $\tau$, a {\em longitudinal} component $\tL$ parallel to the momentum $\mathbf{q}$, and two mutually orthogonal {\em transverse} components $\tR,\tS$ which are perpendicular to the momentum. The spatial projectors on the longitudinal (1-dimensional) resp. transverse (2-dimensional) subspaces are
\begin{align}
 P^\tL_{mn} &= \frac{q_m q_n}{q^2}, &
 P^\tT_{mn} &= \delta_{mn} - \frac{q_m q_n}{q^2}.
\end{align}

The second system has the same two transverse component $\tR,\tS$, while the temporal and longitudinal components are combined into a component $0$ parallel to the spacetime momentum $p_\mu$ and a component $1$ orthogonal to it. This nomenclature stems from the helical (0,+1,-1) coordinate system in three dimensions~\cite{Kleinert89b,ZaanenNussinovMukhin04}.

The transformation between a vector $A_\mu$ in the original Cartesian $(\tau,x,y,z)$-coordinates and the same vector $A_\alpha$ in the new coordinates can be defined through a vierbein-field $e^\alpha_\mu$:
\begin{equation}
 A_\mu = e^\alpha_\mu(p) A_\alpha(p).
 \end{equation}
Although these fields are real valued, due the properties of the Fourier transformation we demand that the `Hermitian conjugate' field $A^\dagger_\alpha$ satisfy $A^\dagger_\alpha(p) = A(p)$~\cite{ZaanenNussinovMukhin04,Cvetkovic06,QLC2D}. This implies that we need to insert factors of $\ti$ in several places. To be specific, we have \cite{ZaanenNussinovMukhin04,Cvetkovic06,QLC2D}:
\begin{align}
 A_\mu 
 &= e^\tau_\mu A_\tau + \ti e^\tL_\mu A_\tL + e^\tR_\mu A_\tR + \ti e^\tS_\mu A_\tS  \\
 &= e^0_\mu A_0 + e^1_\mu A_1 + e^\tR_\mu A_\tR + \ti e^\tS_\mu A_\tS.
\end{align}
Below, these factors of $\ti$ are already taken into account. They have some perhaps unexpected consequences. For instance, the compression stress reads
\begin{equation}
 \sigma^x_x + \sigma^y_y + \sigma^z_z = - \sigma^\tL_\tL + \sigma^\tR_\tR - \sigma^\tS_\tS.\end{equation}

Here, we make one arbitrary choice of the coordinate system, which is valid for all momenta except when $(q_x,q_y,q_z) = (q,0,0)$, in which case one should perform a spatial rotation first. We define $p = \sqrt{\omega_n^2 + q^2}$, $q = \sqrt{q_x^2 + q_y^2 + q_z^2}$ and $\wp = \sqrt{q_y^2 + q_z^2}$. Note that $\omega_n, q_x, q_y , q_z$ change sign under $p_\mu \to - p_\mu$ while $p ,q , \wp$ do not. 

The explicit coordinate transformations are:

{\bf(i)} $A_{\ft,x,y,z} \leftrightarrow A_{\ft,L,\tR,\tS}$ 
\begin{align}
 \begin{pmatrix} 
  A_\ft \\ A_x \\ A_y \\ A_z
 \end{pmatrix}
&=
\begin{pmatrix}
 1 & 0 & 0 & 0\\
 0 & \ti \frac{q_x}{q} & -\frac{\wp }{ q }& 0 \\
 0 & \ti\frac{ q_y }{ q }&\frac{ q_x q_y }{ \wp q} & \ti \frac{q_z }{ \wp} \\
 0 & \ti\frac{ q_z }{ q} & \frac{ q_x q_z}{ \wp q} & - \ti \frac{q_y }{ \wp}
\end{pmatrix}
\begin{pmatrix}
 A_\ft \\ A_\tL \\ A_\tR \\ A_\tS
\end{pmatrix},\\
\begin{pmatrix}
 A_\ft \\ A_\tL \\ A_\tR \\ A_\tS
\end{pmatrix}
&=
\begin{pmatrix}
 1 & 0 & 0 & 0 \\
 0 & -\ti \frac{q_x }{ q} & -\ti \frac{q_y }{ q} & -\ti \frac{q_z }{ q} \\
 0 & - \frac{\wp }{ q} &  \frac{q_x q_y }{ \wp q }&  \frac{q_x q_z }{ \wp q} \\
 0 & 0&  -\ti\frac{ q_z }{ \wp }& \ti \frac{q_y }{ \wp} \\
\end{pmatrix}
\begin{pmatrix} 
  A_\ft \\ A_x \\ A_y \\ A_z
 \end{pmatrix}.
\end{align}

{\bf(ii)} $A_{\ft,x,y,z} \leftrightarrow A_{0,1,\tR,\tS}$ 
\begin{align}
 \begin{pmatrix} 
  A_\ft \\ A_x \\ A_y  \\ A_z
 \end{pmatrix}
&=
\begin{pmatrix}
 \ti \frac{\omega_n}{cp} & -\frac{q}{p} & 0 & 0\\
 \ti \frac{q_x}{p} & \frac{\omega_n q_x}{cqp} & -\frac{\wp}{q} & 0 \\
 \ti \frac{q_y}{p} & \frac{\omega_n q_y}{c qp} & \frac{q_x q_y}{\wp q} & \ti \frac{q_z}{ \wp} \\
 \ti \frac{q_z}{p} & \frac{\omega_n q_z}{c q p}  &  \frac{q_x q_z}{ \wp q} & - \ti \frac{q_y }{\wp}
\end{pmatrix}
\begin{pmatrix}
 A_0 \\ A_{1} \\ A_\tR \\ A_\tS
\end{pmatrix},\\
\begin{pmatrix}
 A_0 \\ A_{1} \\ A_\tR \\ A_\tS
\end{pmatrix}
&=
\begin{pmatrix}
 -\ti \frac{ \omega_n }{ cp }& -\ti \frac{ q_x }{ p }& -\ti \frac{ q_y }{ p} & -\ti\frac{  q_z }{ p} \\
 - \frac{q}{p} & \frac{  \omega_n q_x }{ cq p} &  \frac{ \omega_n q_y  }{ cq p } & \frac{ \omega_n q_z }{c q p }\\
 0 & - \frac{ \wp }{ q }&  \frac{ q_x q_y }{ \wp q} & \frac{  q_x q_z }{ \wp q} \\
 0 & 0&  -\ti \frac{ q_z }{ \wp} & \ti\frac{  q_y }{ \wp} \\
\end{pmatrix}
\begin{pmatrix} 
  A_\ft \\ A_x \\ A_y \\ A_z
 \end{pmatrix}.
\end{align}

{\bf(iii)} $A_{\ft,\tL,\tR,\tS} \leftrightarrow A_{0,1,\tR,\tS}$ 
\begin{align}
 \begin{pmatrix} 
  A_\ft \\ A_\tL \\ A_\tR \\ A_\tS
 \end{pmatrix}
&=
\begin{pmatrix}
 \ti  \frac{ \omega_n }{ cp} & - \frac{  q }{p} & 0 & 0\\
 \frac{ q }{ p } & -\ti \frac{ \omega_n}{cp} & 0 &0 \\
 0 & 0 &  1 & 0 \\
 0 & 0 & 0 & 1
\end{pmatrix}
\begin{pmatrix}
 A_0 \\ A_1 \\ A_\tR \\ A_\tS
\end{pmatrix},  \label{eq:transf tLT 0+1-1}\\
\begin{pmatrix}
 A_0 \\ A_1 \\ A_\tR \\ A_\tS
\end{pmatrix}
&=
\begin{pmatrix}
 -\ti  \frac{ \omega_n }{ cp}  &   \frac{  q }{ p} & 0 &0\\
 -  \frac{ q}{ p} & \ti \frac{  \omega_n }{cp}  & 0 & 0 \\
 0 & 0 & 1 & 0 \\
  0 & 0 & 0 & 1
\end{pmatrix}
\begin{pmatrix} 
  A_\ft \\ A_\tL \\ A_\tR \\ A_\tS
 \end{pmatrix}.\label{eq:tLRS to 01RS transformation}
\end{align}

Instead of using $q_x, q_y,q_z$ explicitly, we can also express this using the total momentum $q$ and two angles $\eta,\zeta$. Note that these transformations are well defined even if $\mathbf{q} = (q_x, 0,0)$ although we still have to make an arbitrary choice for $\zeta$.
\begin{equation}
 \begin{pmatrix} q_x \\ q_y \\ q_z \end{pmatrix}
 =
 q \begin{pmatrix}\label{eq:eta zeta angles}
    \cos \eta \\ \sin \eta \cos \zeta \\ \sin \eta \sin \zeta
   \end{pmatrix}.
\end{equation}
Then the transformations are

{\bf(i)} $A_{\ft,x,y,z} \leftrightarrow A_{\ft,L,\tR,\tS}$ 
\begin{align}
 \begin{pmatrix} 
  A_\ft \\ A_x \\ A_y \\ A_z
 \end{pmatrix}
&=
\begin{pmatrix}
 1 & 0 & 0 & 0\\
 0 & \ti \scriptstyle \cos \eta & -\scriptstyle \sin \eta & 0 \\
 0 & \ti\scriptstyle  \sin \eta \cos \zeta &\scriptstyle  \cos \eta \cos \zeta  & \ti \scriptstyle \sin \zeta \\
 0 & \ti \scriptstyle \sin \eta \sin \zeta &  \scriptstyle \cos \eta \sin \zeta  & - \ti \scriptstyle \cos \zeta
\end{pmatrix}
\begin{pmatrix}
 A_\ft \\ A_\tL \\ A_\tR \\ A_\tS
\end{pmatrix},\\
\begin{pmatrix}
 A_\ft \\ A_\tL \\ A_\tR \\ A_\tS
\end{pmatrix}
&=
\begin{pmatrix}
 1 & 0 & 0 & 0 \\
 0 & -\ti\scriptstyle  \cos \eta & -\ti\scriptstyle  \sin \eta \cos \zeta  & -\ti \scriptstyle \sin \eta \sin \zeta \\
 0 & -\scriptstyle  \sin \eta & \scriptstyle \cos \eta \cos \zeta  & \scriptstyle  \cos \eta \sin \zeta \\
 0 & 0&  -\ti \scriptstyle \sin \zeta & \ti \scriptstyle \cos \zeta  \\
\end{pmatrix}
\begin{pmatrix} 
  A_\ft \\ A_x \\ A_y \\ A_z
 \end{pmatrix}.
\end{align}

{\bf(ii)} $A_{\ft,x,y,z} \leftrightarrow A_{0,1,\tR,\tS}$ 
\begin{align}
 \begin{pmatrix} 
  A_\ft \\ A_x \\ A_y  \\ A_z
 \end{pmatrix}
&=
\begin{pmatrix}
 \ti \frac{\omega_n}{cp} & -\frac{q}{p} & 0 & 0\\
 \ti \frac{q \cos \eta}{p }& \frac{\omega_n\cos \eta}{cp} & - \scriptstyle \sin \eta  & 0 \\
 \ti \frac{q \sin \eta \cos \zeta}{p} & \frac{\omega_n  \sin \eta \cos \zeta}{cp }&\scriptstyle  \cos \eta \cos \zeta & \ti \scriptstyle \sin \zeta  \\
 \ti \frac{q \sin \eta \sin \zeta }{p} & \frac{\omega_n   \sin \eta \sin \zeta}{cp } & \scriptstyle \cos\eta \sin \zeta  & - \ti \scriptstyle \cos \zeta
\end{pmatrix}
\begin{pmatrix}
 A_0 \\ A_{1} \\ A_\tR \\ A_\tS
\end{pmatrix},\\
\begin{pmatrix}
 A_0 \\ A_{1} \\ A_\tR \\ A_\tS
\end{pmatrix}
&=
\begin{pmatrix}
 -\ti \frac{\omega_n }{ cp} & -\ti \frac{q \cos \eta }{ p} & -\ti \frac{q \sin \eta \cos \zeta}{ p} & -\ti \frac{q \sin \eta \sin \zeta }{ p }\\
 - \frac{q}{p} &  \frac{\omega_n \cos \eta }{ c p} & \frac{ \omega_n \sin \eta \cos \zeta  }{ c p } & \frac{\omega_n q \sin \eta \sin \zeta }{c  p} \\
 0 & -  \scriptstyle\sin \eta  &  \scriptstyle \cos \eta \cos \zeta  &  \scriptstyle\cos \eta \sin \zeta  \\
 0 & 0&  -\ti  \scriptstyle\sin \zeta  & \ti \scriptstyle \cos \zeta  \\
\end{pmatrix}
\begin{pmatrix} 
  A_\ft \\ A_x \\ A_y \\ A_z
 \end{pmatrix}.
\end{align}
The transformations $A_{\ft,\tL,\tR,\tS} \leftrightarrow A_{0,1,\tR,\tS}$ do not depend on $\eta,\zeta$.